\documentclass[lettersize,journal]{IEEEtran}

\usepackage{amsmath,amsfonts}
\usepackage{amsthm}
\usepackage{array}
\usepackage{textcomp}
\usepackage{stfloats}
\usepackage{url}
\usepackage{makecell}
\usepackage{ulem}
\usepackage{verbatim}
\usepackage{graphicx}
\usepackage{cite}
\usepackage{booktabs}
\usepackage{multirow}
\usepackage{multicol}% http://ctan.org/pkg/multicols
\usepackage{diagbox}
\usepackage{mathtools}
\usepackage{tabularx}
\usepackage{colortbl}
\usepackage{hhline}
\usepackage{float}
\usepackage{xcolor} % Ensure color is available
\usepackage{times}
\usepackage{listings}
\usepackage{setspace}
\usepackage{subcaption}
\usepackage{enumitem}
\usepackage{algorithm}
\usepackage{algpseudocode}
\usepackage[hidelinks]{hyperref}
\usepackage{placeins}

\newcommand{\vecx}{\mathbf{x}}
\newcommand{\vecm}{\mathbf{m}}

\newcommand{\vecX}{\mathbf{X}}
\newcommand{\vecG}{\mathbf{G}}
\newcommand{\vecU}{\mathbf{U}}

\newcommand{\vecT}{\mathbf{T}}

\newcommand{\vecN}{\mathbf{N}}
\newcommand{\vecC}{\mathbf{C}}

\newcommand{\vecz}{\mathbf{z}}

\newcommand{\vecu}{\mathbf{u}}

\newcommand{\vecg}{\mathbf{g}}
\newcommand{\vecv}{\mathbf{v}}
\newcommand{\bbS}{\mathbb{S}}

\newcommand{\argmax}{\arg\max}
\usepackage{tikz}
\usetikzlibrary{arrows.meta,
                chains,
                positioning,
                shapes.geometric,matrix, backgrounds}
\begin{document}

\title{Depth-Based Local Center Clustering: A Framework for Handling Different Clustering Scenarios}

\author{Siyi~Wang, Alexandre~Leblanc, Paul~D.~McNicholas
        % <-this % stops a space
%\thanks{This paper was produced by the IEEE Publication Technology Group. They are in Piscataway, NJ.}% <-this % stops a space
\IEEEcompsocitemizethanks{\IEEEcompsocthanksitem 
Code is available at \href{https://github.com/lytgysrn/dlcc}{https://github.com/lytgysrn/dlcc}}}

\markboth{}%
{Fu \MakeLowercase{\textit{et al.}}: Depth-based Local Center Clustering}


\maketitle

\begin{abstract}
Cluster analysis, or clustering, plays a crucial role across numerous scientific and engineering domains. Despite the wealth of clustering methods proposed over the past decades, each method is typically designed for specific scenarios and presents certain limitations in practical applications. In this paper, we propose depth-based local center clustering (DLCC). This novel method makes use of data depth, which is known to produce a center-outward ordering of sample points in a multivariate space. However, data depth typically fails to capture the multimodal characteristics of {data}, something of the utmost importance in the context of clustering. To overcome this, DLCC considers data depth based on depth-defined subsets of {data}. From this, local centers can be identified as well as clusters of varying shapes. Overall, DLCC is a flexible clustering approach that seems to overcome some limitations of traditional clustering methods, thereby enhancing data analysis capabilities across a wide range of application scenarios.
\end{abstract}

\begin{IEEEkeywords}
Clustering, statistical depth, similarity matrix, spatial depth
\end{IEEEkeywords}

\newtheorem{thm}{\bf Definition}

%%%%%%%%%%%%%%%%%%%%%%%%%%%%%%%%%%%%%%%%%%%%%%%%%%%%
\section{Introduction}

Clustering, or cluster analysis, is one of the most important topics in data science. A cluster is usually seen as a set of similar objects that have been gathered or grouped together \cite{jain1988algorithms}. Generally speaking, clustering is the task of partitioning a collection of objects, into separate clusters, in such a way that objects in the same cluster are as similar as possible and objects in separate clusters as distinct as possible \cite{xu2015comprehensive}. Research on clustering methods has a long history in several disciplines, including biology, psychology, geography, and marketing \cite{jain1999data}. In recent years, the rapid development of sensing and storage technology has resulted in the collection of vast quantities of high-dimensional, large-scale, data. In turn, this has increased the need for flexible and computationally efficient clustering methods \cite{jain2010data} in many fields, such as pattern recognition \cite{hu2006system}, computer vision \cite{yang2008unsupervised, tron2007benchmark} and image processing \cite{sharma2016review}.

The complexity of real-world data distributions has given rise to numerous clustering algorithms with different assumptions and objectives.
Across those methodologies, many clustering approaches rely on identifying representative points (exemplars), such as centers, density peaks, or anchors, and forming clusters by grouping these exemplars and assigning remaining points accordingly. In the following, we review representative clustering methods under this unified viewpoint.

%%%%%%%%%%%%%%%%%%%%%%%%%%%%%%%%%%%%%%%%%%%%%%%%%%%%
\subsection{Clustering Methods}\label{sec:CM}

\subsubsection{Center-based Clustering} 
Center-based clustering methods form clusters around a set of representative points defined as cluster centers, typically obtained by minimizing within-cluster dispersion under a prescribed distance measure. The classical k-means algorithm \cite{macqueen1967some} exemplifies this paradigm, where exemplars are defined as arithmetic means of cluster members. While computationally efficient and conceptually simple,  k-means is sensitive to outliers, biased toward spherical cluster shapes, and highly dependent on initialization, often leading to suboptimal local minima.

Many variants of k-means have been proposed to address those limitations and refine the definition of exemplars. Partitioning around medoids (PAM; \cite{kaufman1990partitioning}) replaces means with medoids, improving robustness by restricting exemplars to sample points. Weighted k-means (W-k-means; \cite{huang2005automated}) and lasso weighted k-means (LW-k-means; \cite{chakraborty2020detecting}) provide a more flexible k-means approach by incorporating a weight vector for each variable, where LW-k-means enables feature selection in high-dimensional data by adding a lasso penalty, partially alleviating the spherical symmetry assumption. 

Beyond hard assignments, fuzzy extensions of k-means reinterpret exemplars as prototypes that attract data points with varying membership strengths. Classical fuzzy C-means \cite{bezdek1984fcm} and its variants define cluster centers as membership-weighted averages, thereby softening hard boundaries while retaining a center-based exemplar structure. Equilibrium k-means (EKM; \cite{he2025equilibrium}), a recently proposed method, addresses the limitations of fuzzy clustering under class imbalance by modifying the interaction between data points and exemplars through an attraction–repulsion mechanism, resulting in more stable and representative centroid estimation.

Despite these advances, center-based methods fundamentally rely on distance-driven exemplar definitions and a one-exemplar-per-cluster assumption, which makes them sensitive to metric choice and limits their ability to model complex and nonconvex data geometries.

%%%%%%%%%%%%%%%%%%%%%%%%%%%%%%%%%%%%%%%%%%%%%%%%%%%%
\subsubsection{Density-based Clustering}
Density-based clustering assumes that clusters are regions of high density separated by low-density areas, making it well-suited for detecting arbitrarily shaped clusters. In this paradigm, exemplars are implicitly or explicitly defined as density maxima or points located within dense regions. A classical example is density-based spatial clustering of applications with noise (DBSCAN; \cite{ester1996density}), which characterizes clusters as density-connected components, where each component is represented by a ``core point''. While effective for arbitrary shapes, its reliance on a fixed threshold limits its ability to handle clusters with varying densities.

Mean shift clustering \cite{comaniciu2002mean} further adopts a density-driven exemplar definition by iteratively shifting points toward local modes of an estimated density function. The resulting modes act as exemplars that summarize local structure without requiring a predefined number of clusters. However, the effectiveness of mean shift depends critically on hyperparameters for estimating density (e.g., the kernel bandwidth).

Density peak clustering (DPC; \cite{rodriguez2014clustering}) explicitly defines exemplars as density peaks that are well separated from higher-density points. A key limitation of it is that spurious (satellite) peaks may be promoted to exemplars, which degrades robustness. Recent work has therefore incorporated hierarchical ideas to improve exemplar grouping. In particular, fast main density peak clustering within relevant regions (R-MDPC; \cite{guan2024fast}) introduces the notion of relevant regions by first forming single-peak clusters led by density peaks and then merging density-connected single-peak clusters into larger regions, within which main peaks are identified. This enhancement better handles complex cluster shapes and reducing satellite peak interference. More broadly, such hierarchical schemes provides a generic mechanism for grouping representatives via linkage criteria. See also \cite{ran2023comprehensive, xu2016denpehc, xia2025gbct} for related clustering methods that adopt hierarchical schemes to merge over-fragmented components.

Density-based approaches still depend on density estimation and/or distance-based relations among representatives, which can be unstable under strong overlap between clusters or in high-dimensional spaces. 
\subsubsection{Graph-based Clustering}
Graph-based clustering forms clusters by partitioning a similarity graph, where vertices represent samples and edge weights encode pairwise affinities. Spectral clustering (SC; \cite{ng2001spectral, von2007tutorial}) solves a relaxed graph-cut problem via Laplacian eigen-decomposition and typically applies a discretization step (e.g., k-means) to obtain hard labels.

To avoid potential information loss from post-processing discretization and to couple graph construction with label learning, constrained Laplacian rank (CLR; \cite{nie2016constrained}) approaches directly produce discrete labels from an optimally structured graph. However, both SC and CLR-style formulations can be computationally prohibitive on large-scale data due to full-graph construction.

Exemplars in graph-based clustering are used not only to summarize data structure but also to reduce the computational burden of graph construction and partitioning. This has motivated anchor-based graph clustering, which selects $m\!\ll\! n$ representative anchors and builds a sparse bipartite graph to approximate the original affinity graph \cite{zhang2022fast,shu2022self,ma2025large}. Although anchors improve scalability, the resulting graph can remain sensitive to the choice of distance metric, graph-construction hyperparameters, and the anchor selection strategy.

%%%%%%%%%%%%%%%%%%%%%%%%%%%%%%%%%%%%%%%%%%%%%%%%%%%%
\subsection{Motivation}
Viewed through an exemplar-based lens, many clustering methods succeed or fail primarily due to how \emph{representativeness} is defined. Most existing approaches tie exemplars to distance- or density-driven criteria (and their associated hyperparameters). Such criteria can work well when a single notion of ``closeness'' or ``density'' is appropriate, but may yield misleading exemplars when the chosen metric or density scale is mismatched to the data (e.g., Euclidean distance under highly skewed distributions, or DBSCAN-style density reachability in high-dimensional spaces).

On the one hand, center-based methods typically define a single exemplar per cluster through distance minimization, an assumption that can be overly restrictive for irregular or nonconvex structures. On the other hand, density-based methods often induce multiple exemplars (e.g., core points or density peaks), but the resulting exemplar quality depends on reliable density definition.

These limitations motivate an exemplar definition that remains consistent across dimensionalities and geometries, while allowing the number of exemplars to adapt to local structure. In response, we propose the depth-based local center clustering (DLCC) algorithm. DLCC addresses this by leveraging statistical data depth, a nonparametric tool for multivariate analysis. Tukey \cite{Tukey1975} pioneered the idea of center-outward ordering in multivariate data, with the introduction of the first depth definition, half-space depth. Data depth enables the center-outward ordering of data, making it a potent tool for locating a center for any dataset. When considering subsets of the data, data depth with respect to (w.r.t.) each subset aids in defining multiple exemplars. Naturally, as the subset size grows, the number of exemplars diminishes, and vice versa.

To inherit the simplicity of center-based clustering and the shape adaptivity of density-based approaches, DLCC combines depth-defined local exemplars with two grouping strategies for exemplar aggregation. Here, the term ``depth-based'' encompasses two aspects: (1) data depth is used to construct a similarity matrix instead of relying on distance, adaptively capturing the geometric characteristics of each point's location; (2) data depth measures the centrality of points within each point's neighborhood, helping to identify exemplars.

%%%%%%%%%%%%%%%%%%%%%%%%%%%%%%%%%%%%%%%%%%%%%%%%%%%%%%%%%%%%%%%%%%%%%%%%
\subsection{Preliminaries about depth and related work in clustering}
A depth function quantifies how ``central'' a point is within a multidimensional data distribution, providing a meaningful center-outward ordering of points. The term ``depth'' reflects the analogy to physical depth: points closer to the ``central area'' of a distribution are considered ``deeper'', while points farther away are ``shallower''. It is worth to note that a point with high depth is central but does not necessarily coincide with a region of high density. Consequently, clustering approaches based on depth are conceptually distinct from those based on density estimates. This centrality-based ordering is particularly useful in multivariate data analysis, where traditional univariate notions of order are not well-defined. 

Serfling and Zuo \cite{zuo2000general} formulated several desirable properties of depth functions: affine invariance, maximality at center, monotonicity on rays, and vanishing at infinity. A deepest point $\vecm=\text{argmax}_{\vecz \in \mathbb{R}^d} D(\vecx \mid F)$ is usually referred to as a \emph{depth median}. In applications, where the inherent distribution $F$ is usually unknown, depth is calculated w.r.t. the empirical distribution $F_n$ of $\mathbf{X}$, that is $D(\vecz \mid F_n)$ is the depth of $\vecz$ w.r.t. the sample $\mathbf{X}$. Hereinafter, we will denote the latter as $D(\vecz\mid \mathbf{X})$ to emphasize the role of the sample on calculating the depth of the point $\vecz$. \textcolor{black}{For a broader overview of various depth notions, see Mosler and Mozharovskyi \cite{mosler2022choosing}.}

Previous studies have explored the application of data depth in clustering. Jornsten \cite{jornsten2004clustering} proposed DDclust, an algorithm that refines the output of other clustering methods like PAM by recalculating observation depths within each cluster and reallocating points for better results. Jeong et al. \cite{jeong2016data} introduced data depth-based clustering analysis (DBCA), an alternative to DBSCAN. DBCA replaces the sample Mahalanobis depth (MD) function
\begin{equation}
D_{\mathrm{MD}}(\vecz \mid \mathbf{X})=\left[1+(\vecz-\bar{\mathbf{X}})^{\prime} \hat{\mathbf\Sigma}^{-1}(\vecz-\bar{\mathbf{X}})\right]^{-1}
\label{eq:maha1}
\end{equation}
with fixed depth median versions
\begin{equation}
D_{\mathrm{MD}}^i(\vecz)=\left[1+(\vecz-\vecx_i)^{\prime} \hat{\mathbf\Sigma}^{-1}(\vecz-\vecx_i)\right]^{-1},
\label{eq:maha2}
\end{equation}
where $\vecz$ denotes a point in $\mathbb{R}^d$, $\bar{\mathbf{X}}$ is the mean vector, $\vecx_i$ is an observation in $\mathbf{X}$, while $\hat{\mathbf\Sigma}$ is the estimated covariance matrix of dataset $\mathbf{X}$. Unlike DBSCAN, which defines neighborhoods via a fixed Euclidean radius, DBCA uses \eqref{eq:maha2} to induce elliptical, affine-invariant neighborhoods. Subsequent extensions further developed this depth-based framework by refining robustness of covariance matrix estimation and rules of core points \cite{huang2017crad, mckenney2025statistical}. 

Another recent development in depth-based clustering was proposed by Francisci et al.~\cite{francisci2023analytical}, who build upon the concept of local depth introduced by Agostinelli and Romanazzi~\cite{agostinelli2011local}. Their work generalizes local depth to lens depth and $\beta$-skeleton depth, and exploits depth as a surrogate for density estimation to develop a mean-shift like clustering algorithm. Beyond density-based paradigms, data depth has also been explored for identifying representative centroids in center-based clustering, particularly for k-means initialization across diverse datasets \cite{torrente2021initializing, albert2022band}.

So far, the notion of depth serves an auxiliary role to the more traditional clustering methods and is not employed directly in examining the multimodality of data. In contrast, DLCC places data depth at the core of both exemplar identification and similarity construction, thereby using depth directly to examine local and global cluster structure. 

The DLCC framework itself is not restricted to a specific depth function, and in principle can be instantiated with a broad class of depth notions. In this work, we adopt spatial depth (SD; \cite{serfling2002depth}) as a representative choice, as it offers a favorable trade-off between theoretical interpretability and computational simplicity. Although SD is invariant only under similarity transformations, its straightforward computation makes it well suited for high-dimensional applications. The sample version of SD is given by
\begin{equation}
D_{\mathrm{SD}}(\vecz\mid \mathbf{X})=
%1-\left\|\mathrm{E}_{\mathbf{X}}\left(\frac{\veczboldsymbol{z}-\mathbf{X}}{\|\vecz \boldsymbol{z}-\mathbf{X} \|_1}\right)\right\|}| =
1-\left\| \sum_{i=1}^n \frac{\vecz-\vecx_i}{n\| \vecz-\vecx_i \|} \right\|,
\label{eq:sd}
\end{equation}
where $n$ is the total number of observations.

DLCC requires a similarity matrix to determine the neighbors of each observation. \textcolor{black}{Traditional distance measures, such as Euclidean distance, are typically based on assumptions about the underlying cluster shapes (e.g., spherical or linear structures), which may not hold in more complex scenarios. In contrast, data depth offers a nonparametric, robust, and adaptive alternative.} As the depth \eqref{eq:sd} provide a center-outward ordering and their value ranges from $0$ to $1$, the depth values of other points can be viewed as a measure of similarity to the depth median. Thus, similarity matrices can be constructed based on data depth. For this, however, one requires an approach capable of defining neighbors of any point $\vecx_i$ in $\mathbf{X}$, i.e., the observations in $\mathbf{X}$ that are most similar to $\vecx_i$. It should be clear that the usual notion of depth cannot be applied directly here, as that can only provide similarity measurements related to the global median. 

Now, consider a point $\vecx_i$ in $\mathbf{X}$ whose neighbors need to be identified and introduce the set 
$$\mathbf{X}_{Ri} = \mathbf{X} \cup \{ 2\vecx_i - \vecx_j \, | \, j \neq i, j = 1, 2, \ldots, n\}.$$
We note that this new set is comprised of all the points belonging to the original sample $\mathbf{X}$ as well as of the reflection of all the points of $\mathbf{X}$ through $\vecx_i$. Crucially, $\vecx_i$ is a depth median of $\mathbf{X}_{Ri}$ as it is a center of symmetry for this set. It is also the case that $D(\vecx_i|\mathbf{X}_{Ri})=1$ and that $D(\vecx_j|\mathbf{X}_{Ri})$ can be used to quantify the similarity of any point $\vecx_j$ ($j \neq i$) to the point of interest $\vecx_i$. 
Subsequently, by repeatedly computing the depth values for points in $\mathbf{X}$ w.r.t. $\mathbf{X}_{R1}, \mathbf{X}_{R2},\ldots,\mathbf{X}_{Rn}$,  a $n\times n$ non-mutual depth-based similarity matrix can be defined. We denote it as $\mathbb{S}$, where $\bbS_{i,j}=D(\vecx_j|\mathbf{X}_{Ri})$ represents the entry in the $i$th row and $j$th column. As will be seen shortly, this can cause some computational challenges.

This approach to localizing the notion of depth was firstly introduced by Paindaveine and Van Bever~\cite{paindaveine2013}, who proposed a form of local depth that differs from the one of Agostinelli and Romanazzi~\cite{agostinelli2011local}.  In particular, their construction was shown to yield non-spherical depth neighborhoods in some contexts. 
The same authors \cite{paindaveine2015nonparametrically} used this concept and introduced an affine-invariant k-nearest neighbor classification algorithm that adapts to the local geometry of the sample and was shown to be consistent. 

%%%%%%%%%%%%%%%%%%%%%%%%%%%%%%%%%%%%%%%%%%%%%%%%%%%%
\subsection{Main Contributions and Organization of the Paper}

This paper makes the following main contributions:
\begin{itemize}
    \item \textbf{Efficient Depth-Based Similarity Construction:}
    We improve the computational efficiency of constructing depth-based similarity matrices under spatial depth by developing a matrix-based implementation.

    \item \textbf{DLCC Clustering Framework:}
     We propose DLCC, a depth-based clustering framework that uses data depth to measure similarities, define robust neighborhoods, and extract exemplars (``local centers''). DLCC then forms clusters by grouping these local centers and propagating labels to all observations.

    \item \textbf{A Complete Procedure and Empirical Evaluation:}
    We provide a practical end-to-end procedure for DLCC, including filtering,  grouping local centers and cluster formation mechanisms, together with guidance on strategy/parameter selection. We also provide extensive experiments on synthetic and real-world data sets.
\end{itemize}

The remainder of the paper is organized as follows: Section \ref{sec:DSM} introduces the depth-based similarity matrix. Section \ref{sec:DLCC} provides a detailed description of the DLCC framework. In Section \ref{sec:EPS}, we present details about parameter selections. Section \ref{sec:ee} showcases experimental results of DLCC using both synthetic and real-world datasets. Lastly, Section \ref{sec:conclu} concludes the paper.

Figure \ref{fig:flowchart} provides the flowchart of the DLCC method, and Table \ref{tab:notations} summarizes the key notations used in this paper to help readers navigate its contents more easily. Additional graphs providing further insights into our methods are included in the supplementary file.
\begin{figure*}[!ht]
\centering
\resizebox{0.8\textwidth}{!}{\begin{tikzpicture}[scale=0.4,
    node distance = 5mm and 7mm,
    start chain = going below,
    arr/.style = {-Stealth},
    box/.style = {rectangle, draw, align=center, on chain},
    auto
]
  \node (n0) [box]{Dataset};
 \node (n1) [box, right=of n0, xshift=0.5cm] {Depth-based\\similarity matrix\\\textit{\textcolor{black}{Section \ref{sec:DSM}}}};
  \node (n2) [box, right=of n1, xshift=1cm] {Neighborhoods \&\\Local centers\\\textit{\textcolor{black}{Section \ref{sec:DLCC}
}}}; 
  \node (n3) [box, right=of n2, xshift=2cm] {Filtered centers\\\textit{\textcolor{black}{Section \ref{sec:lcs}}}}; 
  \node (n4) [box, below=of n3, yshift=-1cm] {Groups of \\filtered centers\\\textit{\textcolor{black}{Section \ref{sec:lcs} $\&$ Section \ref{sec:EPSMAX}}}}; 
  \node (n5) [box, left=of n4, xshift=-2cm] {Temporary\\clusters\\\textit{\textcolor{black}{Section \ref{sec:clinDLCC}}}}; 
  \node (n6) [box, below=of n5,yshift=-0.5cm] {Unlabelled points\\\textit{\textcolor{black}{Section \ref{sec:clinDLCC}}}};
  \node (n7) [box, left=of n5, xshift=-1cm] {Final clusters}; 

\draw[arr] (n0) -- node[above, midway] {} (n1);
\draw[arr] (n1) -- node[above, midway] {} (n2);
\draw[arr] (n2) -- node[above, midway] {} (n3);
\draw[arr] (n3) -- node[above, midway, sloped] {} (n4);
\draw[arr] (n4) -- node[above, midway] {} (n5);
\draw[arr] (n5) -- node[above, midway] {} (n7);
\draw[arr] (n4) |- node[below, midway] {} (n6);
\draw[arr] (n6) -| node[midway,below] {classification} (n7);

\end{tikzpicture}
}
\caption{Flowchart for the DLCC algorithm, detailing the sections that include the key contents represented in the nodes.}
\label{fig:flowchart}
\end{figure*}

\begin{table}[ht]
\caption{Summary of the main notations}
\centering
\resizebox{0.5\textwidth}{!}{ 
\begin{tabular}{p{3cm} p{10cm}} 
\toprule
\textbf{Notation} & \textbf{Description} \\ 
\midrule
$n$ and $d$ & Number of observations and dimensions, respectively; \\
$D_{\mathrm{SD}}(\vecz|\mathbf{X})$ & Spatial depth of $\vecz$ w.r.t. $\mathbf{X}$; \\
$\mathbf{X}$ & A given dataset; \\
$\mathbf{X}_{Ri}$ & Dataset reflected around $\vecx_i$, i.e., $\mathbf{X}_{Ri} = \mathbf{X} \cup \{ 2\vecx_i - \vecx_j \, | \, j \neq i\}$; \\
$\mathbf{E}$, $\mathbf{\Tilde{E}}$ & Matrices of unit vector sums from $\mathbf{X}$ and reflected points to $\mathbf{X}$; \\
$\mathbb{S}$ & Depth-based similarity matrix, with $\mathbb{S}_{i,j}$ its $ij$th entry; \\
$s$ & Neighborhood size; \\
$\vecN_i$ & Depth-based neighborhood of $\vecx_i$; \\
$r_i(\vecx_j)$ & Depth rank of $\vecx_j$ within $\vecN_i$; \\
$c$, $c$, $\vecC$ & A local center, a certain set of local centers, and the set of all filtered centers, respectively; \\
$\mathbb{M}$ & Local center neighborhoods' similarity matrix, with $\mathbb{M}_{i,j}$ its $ij$th entry; \\
$\delta$ & Similarity threshold for local center grouping ($\mathbb{M}$-based); \\
$\vecg$ & A group of local centers; \\
$\mathbf{U}_\mathbf{g}$ & Unique neighbors of group $\vecg$; \\
$K$ and $k$ & Number of clusters and cluster index; \\
$\mathcal{X}_k$ & Observations in cluster $k$; \\
$f$ & Frequency of a local center; \\
$\mathcal{P}$ & Cumulative proportion of neighbors of local centers to all points; \\
$\mathcal{S}_k$, $\mathcal{\hat{S}}_k$ & Score pools for accepting and leaving points unlabelled in cluster $k$; \\

\bottomrule
\end{tabular}
}
\label{tab:notations}
\end{table}

%%%%%%%%%%%%%%%%%%%%%%%%%%%%%%%%%%%%%%%%%%%%%%%%%%%%
\section{Spatial Depth based Similarity Matrix} \label{sec:DSM}

A spatial-depth-based similarity matrix can be constructed from calculating $D_{\mathrm{SD}}(\vecx_j|\mathbf{X}_{Ri})$ for all $i,j = 1,2, \ldots,n$. One advantage of using this depth measure is that the shape of the resulting neighborhoods is determined by the local geometry of the sample and does not rely on any parameters. This is a potentially very interesting advantage, but it does entail the computationally intensive process of repeatedly calculating depth values for the $n$ sample points w.r.t. $n$ distinct datasets $\mathbf{X}_{R1}, \mathbf{X}_{R2}, \ldots, \mathbf{X}_{Rn}$. Hence, we delve into certain properties of spatial depth in the current context (working with reflected samples) and introduce an efficient approach to calculate the similarity matrix. 
 
In order to compute the spatial depth of a point $\vecx$ from \eqref{eq:sd}, we note that it is necessary to find the vectors from all sample points to $\vecx$ and their corresponding norms. In any of the combined datasets $\mathbf{X}_{Ri}$, the $n$ points from the original dataset $\mathbf{X}$ remain unchanged. Consequently, we can retain a $n \times d$ matrix, denoted as $\mathbf{E}$, to prevent redundant computations. Here, the $q$th row is the sum of the unit vectors from other observations in $\mathbf{X}$ to $\vecx_q$, expressed as:
\begin{equation*}
\mathbf{E}_q=\sum_{\vecx\in\mathbf{X}\setminus\{\vecx_q\}}\frac{\vecx_q-\vecx}{\|\vecx_q-\vecx\|}.
\end{equation*}
Taking the combined dataset $\mathbf{X}_{Ri}$ as an example, the subsequent step is to derive a similar $n \times d$ matrix $\mathbf{\Tilde{E}}_i$. In this matrix, the $q$th row is represented by 
\begin{equation*}
\mathbf{\Tilde{E}}_{i,q}=\sum_{\vecx\in\mathbf{X}_{Ri}\setminus\mathbf{X}}\frac{\vecx_q-\vecx}{\|\vecx_q-\vecx\|},
\end{equation*}
and corresponds to the the sum of the unit vectors from all the reflected points in $\mathbf{X}_{Ri}$ to $\vecx_q$.

\newtheorem{theorem}{Theorem}
\newtheorem{lemma}{Lemma}
\newtheorem{prop}{Proposition}

\begin{prop}[Symmetry]\label{the:sym}
Consider two distinct points $\vecu$ and $\vecv$, and a point $\vecx_i \in \mathbf{X}$. Also, let $\vecu^i$ and $\vecv^i$ denote the points obtained by reflecting $\vecu$ and $\vecv$, respectively, about $\vecx_i$, that is $\vecu^i = 2\vecx_i-\vecu$ and $\vecv^i = 2\vecx_i-\vecv$.
Then, we have that
\begin{equation*}
\vecu - \vecv^i=\vecv - \vecu^i,
\end{equation*}
and that 
\begin{equation*}
\lVert \vecv^i - \vecu \rVert^2 =
 2\lVert \vecv - \vecx_i \rVert^2 + 2\lVert \vecu - \vecx_i \rVert^2 - \lVert \vecu - \vecv \rVert^2.
\end{equation*}
\end{prop}

The above result implies that we can express the squared norm of a vector from any point after reflection to an arbitrary point in $\mathbf{X}$, as a linear combination of the squared norms of vectors linking points in $\mathbf{X}$. Therefore, if the norm information is concurrently stored while generating the matrix $\mathbf{E}$, the norm of any vector from the reflected point to a point in $\mathbf{X}$ can be directly computed. Let us denote the $n\times n$ square norm matrix as $\mathbf{L}$, where the entry in position $(i,j)$ stores the norm of the vector between point $\vecx_i$ and $\vecx_j$.   This computation no longer involves calculating the root of the sum of squares of the vector coordinates, thus making the norm computation independent of the dimension $d$. 

From Proposition~\ref{the:sym}, it can also be observed that the vector from any reflected point $\vecv^i$ to a vector $\vecu$ can be expressed as $\vecu+\vecv-2\vecx_i$. Then, the sum of unit vectors from all reflection points in $\mathbf{X}_{Ri}$ to $\vecx_q$ can be reformulated as follows:
%\begin{eqnarray}\label{eq: sunit}
%\sum_{j\neq i} \frac{\vecu+\vecx_j-2\vecx_i}{C_j}=\vecu\sum _{j\neq i}\frac{1}{C_j}+\sum_{j\neq i} \frac{\vecx_j-2\vecx_i}{C_j},
%\end{eqnarray}
\begin{equation*}\begin{split}
\sum_{\vecx \in \mathbf{X}_{Ri} \setminus \mathbf{X}} &\frac{\vecx_q-\vecx}{C_{\vecx}}  =
\sum_{\vecx \in \mathbf{X} \setminus \{\vecx_i\}} \frac{\vecx_q+\vecx-2\vecx_i}{C_{2\vecx_i-\vecx}} \nonumber \\
&\quad= \vecx_q \sum_{\vecx \, \in \, \mathbf{X} \setminus \{\vecx_i\}} \frac{1}{C_{2\vecx_i-x}} \ \ +\sum_{\vecx \, \in \, \mathbf{X} \setminus \{x_i\}} \frac{\vecx-2\vecx_i}{C_{2\vecx_i-\vecx}}, 
%\label{eq: sunit}
\end{split}\end{equation*}
%where $C_j$ is a constant representing the norm of the vector from $\vecx^i_j$ to $\vecu$. By separating $\vecu$ from the summation, we can save some vector addition computations. Even though $-2\vecx_i$ could also be separated from the summation, we maintain $\vecx_j-2\vecx_i$ in this form for interpretability, as it represents $-\vecx^i_j$.
where $C_{\vecx}= \|\vecx_q-\vecx\|$ and, by separating $\vecx_q$ from the summation, we save some vector addition computations. 

%Even though $-2\vecx_i$ could also be separated from the summation, we maintain $\vecx_j-2\vecx_i$ in this form for interpretability, as it represents $-\vecx^i_j$.

The matrix $\mathbf{\Tilde{E}}_i$ can now be constructed using matrix computations. The necessary notations to understand this process are summarized below:
\begin{itemize}
\item The $(n-1) \times n$ matrix $\mathbf{\Tilde{L}}_i$ stores the norms of vectors from $\vecx^i_j$ to $\boldsymbol{x_q}$ at each entry $\Tilde{L}_{i,jq}$.
\item The $(n-1) \times n$ matrix $\mathbf{R}_i$ is obtained by taking the element-wise reciprocal of $\mathbf{\Tilde{L}}_i$, i.e., $R_{i,jq}=1/\Tilde{L}_{i,jq}$.
\item The $(n-1) \times d$ matrix $\mathbf{\Tilde{X}}_i$ contains each reflected point (excluding $\vecx_i$ itself) as a row.
\item The $n \times d$ matrix $\mathbf{B}_i$ consists of points in $\mathbf{X}$, each multiplied by the corresponding column sum of $\mathbf{R}_i$.
\end{itemize}
With these,  $\mathbf{\Tilde{E}}_i$ is expressed as:
%\begin{eqnarray}\label{eq:refE}
$\mathbf{\Tilde{E}}_i = \mathbf{B}_i - \mathbf{R}^\prime_i \mathbf{\Tilde{X}}_i.$
%\end{eqnarray}
Through the above procedure, $\mathbf{\tilde E}_i$ can be computed efficiently. For example, on a $2000 \times 600$ dataset, computing $\mathbf{E}$ takes approximately $2$ seconds, whereas each $\mathbf{\tilde E}_i$ requires about $0.1$ seconds per loop. All experiments in this paper were conducted on a PC equipped with an Intel(R) Core(TM) i7-11700K CPU (3.6~GHz) and sufficient physical memory, using \textsc{Matlab} R2023a. Having obtained $\mathbf{E}$ and $\mathbf{\Tilde{E}}$,  the reflection spatial depth values of any $\vecx_q$ in $\mathbf{X}$ based on $\mathbf{X}_{Ri}$ can be computed via
\begin{eqnarray} \label{eq:rsd}
    D_{\mathrm{SD}}(\vecx_q|\mathbf{X}_{Ri})=1-\left\|\frac{\mathbf{E}_q+\mathbf{\Tilde{E}}_{i,q}}{2n-1}\right\|.
\end{eqnarray}

From a theoretical perspective, computing depth-based similarities requires $O(n^3)$ operations, since evaluating spatial depth for $n$ points incurs $O(n^2)$ cost and is repeated over $n$ reflected datasets. While the proposed implementation does not reduce this complexity, it substantially accelerates computation in practice through efficient reuse of the distance relationship and matrix-level computations.

\begin{figure}
    \centering
    \includegraphics[width=\linewidth]{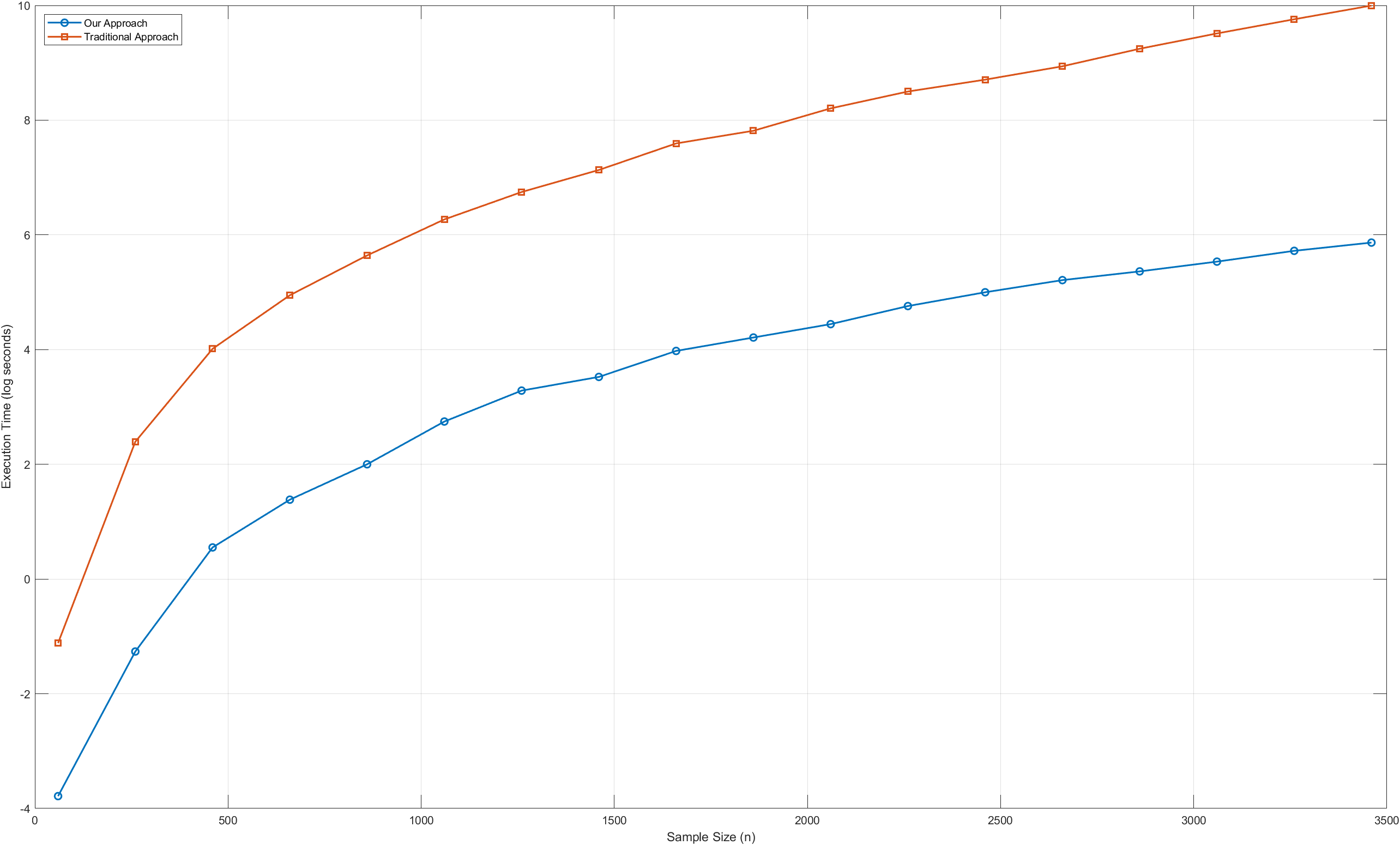}
    \caption{Running time comparison (log scale) between the proposed method and the traditional construction.}
    \label{fig:runtime}
\end{figure}

Figure~\ref{fig:runtime} illustrates the empirical runtime comparison on synthetic datasets with fixed dimension $d=500$ and sample size ranging from $60$ to $3600$, reported on a logarithmic scale. For a fair comparison, we implemented the baseline in \textsc{Matlab} following the same algorithmic logic as the local depth in the \texttt{DepthProc} \textsf{R} package~\cite{depthproc}. As shown, the proposed approach consistently achieves significant runtime reductions.

%%%%%%%%%%%%%%%%%%%%%%%%%%%%%%%%%%%%%%%%%%%%%%%%%%%%
\section{Depth-based Local Center Clustering} \label{sec:DLCC}

In this section, we first introduce some concepts that are key to DLCC and then 
provide a detailed explanation of the clustering procedure.

As explained in Section~\ref{sec:DSM}, it is necessary to use an appropriately localized version of data depth for clustering applications. In the framework of DLCC, we establish two foundational concepts: \emph{neighborhoods} and \emph{local ranks}. These concepts are formalized in the following definition:

\begin{thm}[DLCC Neighborhoods and Local Ranks]\label{def:dnlr}
%For any point $\vecx_i$ in a dataset $\vecX$:
\textbf{Neighbors and Neighborhoods:}
Given an integer parameter~$s$, the neighbors of $\vecx_i \in \vecX$ are defined as the set of $s-1$ points $\vecx_q \in \vecX \setminus \{\vecx_i\}$ with the highest depth values w.r.t. the dataset $\mathbf{X}_{Ri}$, i.e., the points with the largest $D\left(\vecx_{q} \mid \vecX_{Ri} \right)$. The neighborhood of $\vecx_i$, denoted as $\vecN_i$, is thus formed by including $\vecx_i$ and its $s-1$ neighbors, constituting a subset of $\vecX$ with size $s$.

\textbf{Local Ranks:}
The local rank of an observation $\vecx_j$ within the neighborhood $\vecN_i$ is defined as:
\begin{eqnarray}\label{eq:rank}
r_{i}(\vecx_j)=1+ \sum _{\vecx_q\in \vecN_i} \mathbb{I}\left[D\left(\vecx_{j} \mid \vecN_i\right) < D\left(\vecx_{q}\mid \vecN_i\right)\right],
%\#\left\{\vecx_{q}: D\left(\vecx_{j} \mid \vecN_i\right) < D\left(\vecx_{q}\mid \vecN_i\right), \vecx_{q}\in \vecN_i \right\}+1
\end{eqnarray}
where $\mathbb{I}$ denotes the indicator function.
\end{thm}
\normalsize
If $r_{i}(\vecx_j)=1$, then $\vecx_j$ is characterized as the center of the subset $\vecN_i$. We define the central point of $\vecN_i$ as a local center. These local centers can be understood as representative points for the entire dataset $\mathbf{X}$. 

\begin{figure}[ht]
\centering
    \includegraphics[scale=0.33]{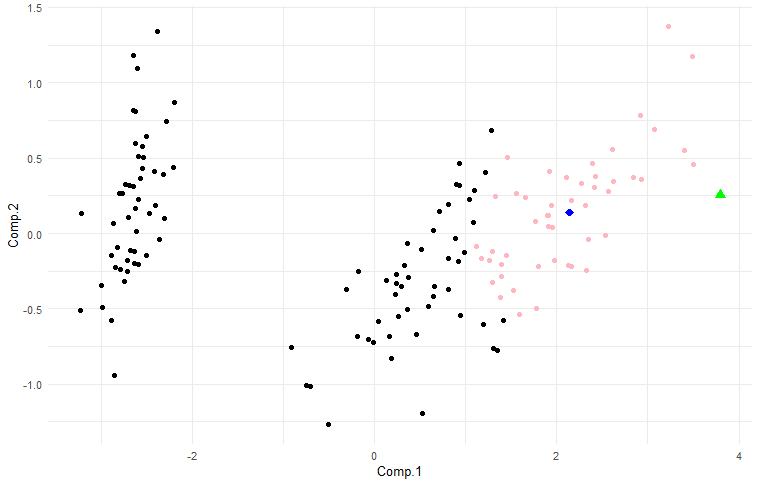}
    \caption{Local center example ($s=50$) in the two–PC visualization of the Iris dataset 
(see Section~\ref{sec:ee} for dataset description). A focus point (green triangle), 
its neighbors (pink points), and the resulting local center (blue diamond) are highlighted.}
    \label{fig:irisLC}
\end{figure}

\noindent\textbf{Notes:}
\begin{enumerate}
    \item While $\vecx_i$ is deepest in $\vecX_{Ri}$ by construction, it is not generally deepest in $\vecN_i$, as illustrated in Figure \ref{fig:irisLC}.
    \item There should be less than $n$ unique local centers due to the occurrence of shared local centers among the sets $\vecN_i$. The \emph{frequency} of a local center, denoted by $f$, is defined as the number of neighborhoods that share this local center.
\end{enumerate}

The DLCC method uncovers clusters through analyzing and grouping local centers. However, as $n$ subsets might yield a large number of local centers, many of them might not be useful for clustering. For example, an outlier residing between two clusters can be the deepest point in a subset containing observations from both clusters. Hence, it is crucial to filter out unnecessary local centers before grouping them. The local centers that remain after going through this filtering process are denoted as \emph{filtered centers}. The filtering of local centers will be discussed in details in Section~\ref{sec:lcs}.

DLCC introduces two strategies for grouping filtered local centers, designed to address different types of clustering challenges. The conditions for grouping the filtered centers under these two strategies are summarized as follows.

\begin{thm}\label{Def:minmax}
Let ${c_1,\ldots,c_A}$ denote a finite set of filtered centers and ${\vecN_{c_1},\ldots,\vecN_{c_A}}$ their corresponding neighborhoods. Define the similarity between any two neighborhoods as the proportion of overlapping observations among them.
For any two neighborhoods $\vecN_{c}$ and $\vecN_{c'}$, define their similarity as
\begin{eqnarray}\label{eq:sim_lc}
\mathrm{sim}(\vecN_{c}, \vecN_{c'}) 
= \frac{|\vecN_{c} \cap \vecN_{c'}|}{s},
\end{eqnarray}
where $|\cdot|$ denotes set cardinality and $s$ is the neighborhood size. A grouping is a partition $\{\vecg_1,\ldots,\vecg_G\}$ of the centers $\{c_1,\ldots,c_A\}$. Two grouping strategies are defined as follows:
\begin{itemize}
    \item $\textbf{Min strategy}$: for any $c \in \vecg_i$:
\begin{eqnarray}\label{eq:min1}
   \min_{c^* \in \vecg_i} \mathrm{sim}(\vecN_{c}, \vecN_{c^*}) 
    &\!\!\!\!>\!\!\!\!& \max_{j \neq i} \min_{c^\dagger \in \vecg_j} \mathrm{sim}(\vecN_{c}, \vecN_{c^\dagger})\\\label{eq:min2}
    \max_{c^* \in \vecg_i \setminus \{c\}} \mathrm{sim}(\vecN_{c}, \vecN_{c^*}) 
    &\!\!\!\!>\!\!\!\!& \max_{c^\dagger \notin \vecg_i} \mathrm{sim}(\vecN_{c}, \vecN_{c^\dagger}).
\end{eqnarray}
\item  $\textbf{Max strategy}$: Given a parameter $\delta$, termed as the \emph{similarity threshold}, $\forall$ $c$ in group $\mathbf{g}$, $\exists c^*$ in the same group s.t.  $\mathrm{sim}( \,\vecN_{c},\vecN_{c^*})\,>Q$, and for $\forall c^\dagger \notin \mathbf{g}$,\\$\mathrm{sim}( \,\vecN_{c},\vecN_{c^\dagger})\,\leq \delta.$
\end{itemize}
\end{thm}

\textcolor{black}{Note that, ``group'' refers specifically to a set of filtered centers and should not be confused with ``cluster''.} Some insights into the strategy chosen for grouping local centers are described in Section~\ref{sec:strachos}.

\begin{thm}[Unique neighbor]\label{Def:un}
Define $\bigcup_{c_i \in \mathbf{g}} \vecN_{c_i}$ and $\bigcup_{c_j \notin \mathbf{g}} \vecN_{c_j}$ as the unions of neighborhoods for filtered centers within and outside of group $\mathbf{g}$, respectively. The set of unique neighbors for $\mathbf{g}$, $\mathbf{U}_\mathbf{g}$, is the largest subset of $\bigcup_{c_i \in \mathbf{g}} \vecN_{c_i}$ satisfying:
\[
\mathbf{U}_\mathbf{g} \cap \bigcup_{c_j \notin \mathbf{g}} \vecN_{c_j} = \emptyset.
\]
\end{thm}
The number of clusters, denoted by $K$, aligns with the grouping of the filtered centers. Temporary clusters are derived from the unique neighbors of each group. The details of constructing these temporary clusters based on the unique neighbors and filtered centers will be further elaborated later in this section.

With these temporary clusters, it is expected that most observations have been allocated to a particular cluster. Nevertheless, there may still be some points that remain unclustered. In such cases, the clustering problem can be reframed as a classification problem. Intuitively, classification methods can be used to assign the remaining points to a cluster. It should be noted that the choice of classification method is very flexible here. Indeed, we could potentially consider any classification method at this stage, depending on the specific requirements of the task at hand. DLCC incorporates two built-in classification methods:  $k$-nearest neighbors ($k$NN) based on the similarity matrix $\bbS$, and random forests (RF).

\subsection{\textcolor{black}{Local Center Selection and Grouping}}\label{sec:lcs}
Locating and filtering local centers is at the core of DLCC. For neighborhoods $\vecN_1, \vecN_2, \ldots, \vecN_n$ of size $s$, local centers are identified. To filter local centers, two benchmarks are suggested: 
\begin{itemize}
    \item a local center should be central to its neighborhood; and
    \item a local center with higher frequency $f$ is more reliable.
\end{itemize}
The first step of filtering local centers is to pick up local centers satisfying $r_{i}(\vecx_i)\leq2$ (Definition \ref{def:dnlr}), which aims to meet the first benchmark. The requirement of a central position is relaxed from $r_{i}(\vecx_i)=1$ to $r_{i}(\vecx_i)\leq2$\ for adding flexibility. Local centers with 
$f=1$ are discarded next, as they are less reliable. The remaining steps in filtering local centers differ between the min and max strategies.  
\subsubsection{Min strategy}
After the first filtering step, assume there are $T$ local centers remaining. Index these local centers in the decreasing order of their frequencies, i.e., $\{c_{1}, \ldots, c_{T}\},$ which corresponds to $f_{1}\geq f_{2}\geq\cdots\geq f_T$. With this order, the cumulative proportion of neighbors for the local centers down the list can be calculated as
\begin{eqnarray}\label{eq:cp}
\mathcal{P}_t=\frac{1}{n}\left|\bigcup_{p=1}^{t} \vecN_{c_{p}}\right|,
\end{eqnarray}
where $|\cdot|$ represents the cardinality of a set.

The filtering procedure under the min strategy consists of two stages: 
(1) selecting a frequency cutoff and forming an grouping, and 
(2) refining the grouping to satisfy the min strategy in Definition~\ref{Def:minmax}. 
For a fixed set of local centers, the max strategy yields a unique grouping, whereas the min strategy may admit multiple feasible groupings. Thus, we use a heuristic procedure that performs filtering and grouping simultaneously. 
Stage (1) is implemented by Algorithms~\ref{alg:local_center_filtering} with grouping detailed in Algorithm~\ref{alg:and_grouping}, and 
stage (2) by the group trimming process in Algorithm~\ref{alg:GTP}. Implementation details are given next and may be skipped on a first reading without loss of the main ideas of the paper.

Our method leverages a stability criterion for evaluating a local center $c_i$, termed average neighbor-induced depth (AND). Recall $\bbS_{j,i}=D_{\mathrm{SD}}(\vecx_i\mid \mathbf{X}_{Rj})$ denote the depth-based similarity from $\vecx_j$ to $\vecx_i$, we define AND as 
\begin{equation}\label{eq:o2c}
\mathrm{AND}(\vecx_i) \;=\; \frac{1}{s}\sum_{j\in \vecN_i} \bbS_{j,i}.
\end{equation}
Intuitively, $\mathrm{AND}(i)$ aggregates the depth of $\vecx_i$ computed w.r.t. reflected datasets centered at its neighbors. It tends to be large only when $\vecx_i$ is consistently central relative to the collection of reflected data sets induced by its neighborhood. 

One might ask why local depth is not adopted here. As noted in the literature~\cite{paindaveine2013, wang2025beta}, local depth may assign relatively high values to points located between two distributions, which can be central in a geometric sense but not necessarily lie in regions of high data concentration. In clustering tasks, such behavior may lead to the selection of misleading representative centers.

We next identify stable centers from the frequency-sorted local-center candidates.
For each candidate $c_t$, we consider its top $\lfloor s/2\rfloor$ depth-based neighbors (i.e., half points in $\vecN_{c}$), and two candidates are regarded as mutual neighbors if they appear in each other’s top $\lfloor s/2\rfloor$ neighbors. The set of mutual neighbors of local center $c$ is denoted as $\mathrm{MN}(c)$

A candidate $c_t$ is declared a stable center if its AND is not smaller than that of any of its mutual neighbors. Let $\mathrm{SC}$ be the set of stable centers. We take the cutoff index as the last stable center in the frequency order and retain the local centers up to this index. To avoid premature truncation, we further check the cumulative proportion $\mathcal{P}_t$. If $\mathcal{P}_t<0.75$, we either (1) simply extend the frequency cutoff until the coverage first exceeds $0.75$ when the gap is large, or (2) when the coverage is close to $0.75$, greedily add a small number of additional candidates that maximize the product of marginal coverage gain and AND, thereby efficiently increasing coverage without introducing many low-quality centers.

Let $\vecC$ denote the retained local-center set, and let $\mathbb{M}$ collect the pairwise neighborhood similarities among centers in $\vecC$, with entries $\mathbb{M}_{i,j}=\mathrm{sim}(\vecN_{c_i},\vecN_{c_j})$ computed by \eqref{eq:sim_lc}. 
For each stable center $c_t$, we form a group by aggregating centers whose similarities to $c_t$ exceed an adaptive threshold $\delta_t$. Throughout the algorithm, all such similarity thresholds are constrained to lie in $[\delta_{\min},\delta_{\max}]$ and in all experiments, we set $(\delta_{\min},\delta_{\max})=(0.4,0.7)$ to avoid overly loose or overly stringent grouping.
\begin{equation}
\label{eq:delta_t}
\begin{aligned}
\delta_t
=
\max\!\Bigl\{
&\min_{c_u\in \mathrm{MN}(c_t)\cap\vecC}
\mathbb{M}_{t,u}
\;\text{s.t.}\;
\mathbb{M}_{t,u}
>
\max_{c_v\in\vecC\setminus\mathrm{MN}(c_t)}
\mathbb{M}_{t,v},
\\
&\delta_{\min}
\Bigr\}.
\end{aligned}
\end{equation}
This definition ensures that $\delta_t$ retains only mutual neighbors of $c_t$ while excluding non-mutual candidates, subject to a conservative lower bound that prevents overly loose thresholds.

The initial grouping may not cover all local centers in $\vecC$. 
We therefore perform a second pass on the unassigned candidates, applying the same rules for defining extra stable centers but restricting mutual-neighbor relations to the unassigned set.  When a candidate center $c_t$ has no remaining mutual neighbors, we assess its stability by whether its neighborhood $\vecN_{c_t}$ contributes sufficient new coverage beyond the neighborhoods of all other local centers in $\vecC$. Specifically, we define $c_t$ as a stable center only if
\begin{equation}
\label{eq:checkc}
\left|\vecN_{c_t}\setminus \bigcup_{c_u\in\vecC\setminus\{c_t\}}\vecN_{c_u}\right|>\frac{s}{3}.
\end{equation}
We then use the same rule for creating groups of newly defined stable centers.

Preliminary groups formed around stable centers may overlap. We resolve such overlaps and decide possible merges using a lightweight two-stage heuristic. 
Consider two groups $\vecg_p$ and $\vecg_q$ with nonempty intersection (the extension to overlaps involving more than two groups is similar). We define their cores by removing the overlapped centers:
\[
\mathrm{core}(\vecg_p)=\vecg_p\setminus(\vecg_p\cap\vecg_q),\qquad 
\mathrm{core}(\vecg_q)=\vecg_q\setminus(\vecg_p\cap\vecg_q).
\]
Each overlapped center $c\in \vecg_p\cap\vecg_q$ is hard-assigned to the group with the larger minimum neighborhood similarity between $c$ and the corresponding core; if both are below $\delta_{\min}$, $c$ is discarded for ambiguity.

After this, we test whether two previous overlapped groups should be merged. We merge $\vecg_p$ and $\vecg_q$ if
\begin{equation}
\max_{c_u\in \vecg_p,\,c_v\in \vecg_q}\mathbb{M}_{u,v}
\;\ge\;
\min\!\{
\min_{c_u,c_v\in \vecg_p}\mathbb{M}_{c_u,c_v},\;
\min_{c_u,c_v\in \vecg_q}\mathbb{M}_{c_u,c_v}
\},
\label{eq:merge_rule}
\end{equation}
and
\begin{eqnarray} \label{eq:test2}
    \min_{c_u,c_v\in \vecg_p\cup \vecg_q}\mathbb{M}_{u,v}\ge \delta_{\min}.
\end{eqnarray}
In implementation, the within-group minima in \eqref{eq:merge_rule} are clipped to $[\delta_{\min},\delta_{\max}]$.

Lastly, if some local centers in $\vecC$ remain unassigned, we assign each such center $c_t$ to
\begin{equation}
\begin{aligned}
g^*_t
=\arg\max_{g\in\mathcal{G}(c_t)}\;\min_{c_u\in \vecg_g}\mathbb{M}_{t,u},
\end{aligned}
\label{eq:final_assign}
\end{equation}
where
\[
\mathcal{G}(c_t)=\left\{g:\min_{c_u\in \vecg_g}\mathbb{M}_{t,u}\ge 
\min\!\left(\min_{c_u,c_v\in \vecg_g}\mathbb{M}_{u,v},\,\delta_{\max}\right)\right\},
\]
otherwise $c$ is dropped from $\vecC$.

\algrenewcommand\algorithmicrequire{\textbf{Input:}}
\algrenewcommand\algorithmicensure{\textbf{Output:}}

\begin{algorithm}[t]
\small
\caption{AND-based stable-center selection and grouping}
\label{alg:and_grouping}
\begin{algorithmic}[1]
\Require Candidate local centers, neighborhoods $\{\vecN_{c}\}$ (size $s$), pointwise similarities $\bbS$, AND $\mathrm{AND}(\cdot)$ in \eqref{eq:o2c}, threshold bounds $[\delta_{\min},\delta_{\max}]$, and sample size $n$.
\Ensure Retained centers $\vecC$ and groups $\vecG=\{\vecg_1,\ldots,\vecg_{\hat K}\}$.
\State \textbf{Mutual-neighbor graph:} for each candidate $c$, keep its top $\lfloor s/2\rfloor$ candidate neighbors under $\bbS$; define $\mathrm{MN}(c)$ as mutual neighbors.
\State \textbf{Stable centers:} $\mathrm{SC}\gets\{c_t:\mathrm{AND}(c_t)\ge \mathrm{AND}(c_u),\ \forall c_u\in\mathrm{MN}(c_t)\}$.
\State \textbf{Frequency cutoff:} $t_0\gets \max\{t: c_t\in\mathrm{SC}\}$; initialize $\vecC\gets\{c_1,\ldots,c_{t_0}\}$.
\State Compute $\mathcal{P}(\vecC)=\frac{1}{n}\big|\cup_{c\in\vecC}\vecN_c\big|$ and if $\mathcal{P}(\vecC)<0.75$ update $\vecC$ using the two-case rule in the text until $\mathcal{P}(\vecC)\ge 0.75$.
\State Build neighborhood similarity matrix $\mathbb{M}$ over $\vecC$.
\State $\vecG\gets\emptyset$.
\For{each $c_t\in \mathrm{SC}$}
    \State Compute $\delta_t$ by \eqref{eq:delta_t} and form $\vecg_{t}\gets\{c_u\in\vecC:\mathbb{M}_{t,u}\ge \delta_t\}$.
    \State $\mathcal{G}\gets \mathcal{G}\cup\{\vecg_t\}$.
\EndFor
\State \textbf{Second pass:} $\mathcal{L}\gets \vecC\setminus \bigcup_{\vecg\in\vecG}\vecg$.
\If{$\mathcal{L}\neq\emptyset$}
\State Obtain $\mathrm{SC}_2$ by repeating Line~2 on $\mathcal{L}$ (mutual neighbors are defined within $\mathcal{L}$ only); if a candidate $c\in\mathcal{L}$ has no mutual neighbor in $\mathcal{L}$, add it to $\mathrm{SC}_2$ only if \eqref{eq:checkc} holds.
\State For each $c\in\mathrm{SC}_2$, repeat Lines~7--10 to form groups and append them to $\vecG$.
\EndIf
\State \textbf{Check overlap:} apply the two-stage heuristic in the text (core-based hard assignment, then merge test \eqref{eq:merge_rule} and \eqref{eq:test2}).
\State If any centers in $\vecC$ remain unassigned, assign them by \eqref{eq:final_assign}.
\State $\hat K\gets|\vecG|$ and return $\vecC,\vecG$.
\end{algorithmic}
\end{algorithm}
In the above procedure, we target a candidate set whose cumulative neighborhood coverage is at least $0.75$. When $\{c_1,\ldots,c_T\}$ yields $\mathcal{P}_T<0.75$, the retained local centers may be overly concentrated in overlapping regions, resulting in insufficient coverage. 
To avoid discarding informative but low-frequency candidates, we expand the candidate set by iteratively adding dropped centers from the initial filtering step (i.e., those with $f = 1$ or $r_i(\vecx_i) > 2$) until the target coverage is reached. Details are summarized in Algorithm~\ref{alg:local_center_filtering}.
\begin{algorithm}[t]
\small
\caption{Local-center filtering for the min strategy}
\label{alg:local_center_filtering}
\begin{algorithmic}[1]
\Require Frequency-sorted candidates $\{c_1,\ldots,c_T\}$, full local-center set $\mathcal{C}_{\mathrm{all}}$, neighborhoods $\{\vecN_c\}$ (size $s$), AND scores $\mathrm{AND}(\cdot)$, and sample size $n$.
\Ensure Grouping results $\vecG$ and retained centers $\vecC$.

\State Compute $\mathcal{P}_{\mathrm{all}}=\mathcal{P}(\mathcal{C}_{\mathrm{all}})$ and set $\tau\gets \min\{0.75,\;0.9\,\mathcal{P}_{\mathrm{all}}\}$.
\If{$\mathcal{P}(\vecC)<\tau$}
    \State Initialize $\vecC\gets\{c_1,\ldots,c_T\}$.
 and $\mathcal{N}\gets \bigcup_{c\in\vecC}\vecN_c$.
    \While{$\mathcal{P}(\vecC)<\tau$}
        \State For each $u\in\mathcal{C}_{\mathrm{all}}\setminus\vecC$, compute $\Delta(u)=|\vecN_u\setminus \mathcal{N}|$.
        \State Select $u^*=\arg\max_{u}\ \Delta(u)\cdot \mathrm{AND}(u)$.
        \State Update $\vecC\gets\vecC\cup\{u^*\}$ and $\mathcal{N}\gets \mathcal{N}\cup \vecN_{u^*}$.
    \EndWhile
    \State Apply Algorithm~\ref{alg:and_grouping} with the candidate set fixed to $\vecC$ (i.e., skip frequency cutoff stage).
\Else
    \State Apply Algorithm~\ref{alg:and_grouping} on $\{c_1,\ldots,c_T\}$.
\EndIf
\end{algorithmic}
\end{algorithm}

After Algorithm~\ref{alg:local_center_filtering}, we obtain groups of filtered centers, denoted by $\vecG=\{\vecg_1,\ldots,\vecg_{\hat{K}}\}$. 
If any centers in $\vecG$ violate the min-strategy conditions \eqref{eq:min1} or \eqref{eq:min2}, we apply the group trimming process in Algorithm~\ref{alg:GTP}. 
The number of clusters is not required; when a user specifies $K$, Algorithm~\ref{alg:GTP} first adjusts the number of groups to $K$ by either dropping less reliable groups (when $\hat{K}>K$) or splitting a weakly cohesive group (when $\hat{K}<K$).  For notational simplicity, we denote the left- and right-hand sides of \eqref{eq:min1} by $\mathrm{mw}(c)$ and $\mathrm{mb}(c)$, respectively.
\begin{algorithm}[t]
\small
\caption{Group Trimming Process}
\label{alg:GTP}
\begin{algorithmic}[1]
\Require Groups of filtered centers $\vecG=\{\vecg_1,\ldots,\vecg_{\hat{K}}\}$, neighborhood-similarity matrix $\mathbb{M}$, optional $K$, and optional stability scores $\mathrm{AND}(\cdot)$ for centers.
\Ensure Trimmed groups $\vecG$ satisfying the min strategy.
\If{$K$ is provided}
    \If{$\hat{K}>K$}
        \State For each group $\vecg$, compute $t(\vecg)=\min\{t: c_t\in\vecg\}$.
        \State Drop the $(G-K)$ groups from $\vecG$ with the largest $t(\vecg)$.
    \ElsIf{$\hat{K}<K$}
        \While{$\hat{K}<K$}
      \State Select the weakest group $\vecg_g$ where
$$
g=\arg\min_{\ell}\{\min_{c_u,c_v\in \vecg_\ell}\mathbb{M}_{u,v}\}.
$$
            \State Choose the first prototype center $p\in\vecg_{g}$ as the most stable one (i.e., it has the highest $\mathrm{AND}$ value within $\vecg_{g}$).
            \State Choose the second prototype $q\in\vecg_{g}\setminus\{p\}$ that maximizes
            $$
           \left\{
            \min_{c_u,c_v\in \vecg^{(1)}(q')}\mathbb{M}_{u,v},\ 
            \min_{c_u,c_v\in \vecg^{(2)}(q')}\mathbb{M}_{u,v}
            \right\},
            $$
            where $\vecg^{(1)}(q')$ and $\vecg^{(2)}(q')$ are obtained by assigning each $c\in\vecg_{g}$ to $\{p,q'\}$ with larger neighborhood similarity.
            \State Replace $\vecg_{g}$ by the two split groups induced by $\{p,q\}$.
            \State Update $\vecG$, set $\hat{K}\gets \hat{K}+1$.
        \EndWhile
    \EndIf
\EndIf

\While{violations of \eqref{eq:min1} or \eqref{eq:min2} exist}
    \While{violations of \eqref{eq:min1} exist}
        \For{each group $\vecg_l\in\vecG$ with violations of \eqref{eq:min1}}
            \State Drop the center $c\in\vecg_l$ that maximizes
            \[
            \frac{1}{|\vecg_l\setminus\{c\}|}\sum_{c'\in\vecg_l\setminus\{c\}}\bigl(\mathrm{mw}(c')-\mathrm{mb}(c')\bigr).
            \]
        \EndFor
        \State Remove any empty groups and update $\vecG$.
    \EndWhile
    \If{violations of \eqref{eq:min2} exist}
        \State Drop all centers violating \eqref{eq:min2} and update $\vecG$.
    \EndIf
\EndWhile
\end{algorithmic}
\end{algorithm}

\subsubsection{Max Strategy}
\begin{thm}[Isolated Local Center]
A local center $c_i$ is considered isolated if, for every other local center $c_j \neq c_i$,
the neighborhood similarity satisfies
\[
\mathbb{M}_{i,j}=0.
\]
Here, $\mathbb{M}$ denotes the neighborhood similarity matrix over local-center candidates, where $\mathbb{M}_{i,j}$ records the similarity between the neighborhoods associated with $c_i$ and $c_j$.
\end{thm}

In other words, an isolated local center is one whose neighborhood does not overlap with any other neighborhoods. As the max strategy assumes each cluster consists of 
various convex shapes, isolated local centers are discarded to avoid defining 
extremely small clusters that may in fact correspond to outliers.
\begin{thm}[Everywhere Non-Convex Cluster]
    An everywhere non-convex cluster is characterized by all of its points having non-convex neighborhoods. 
\end{thm}
Examples of everywhere non-convex cluster include one-dimensional manifolds, such as spirals, and self-crossing curves. A region containing such a cluster is termed a locally non-convex region. The max strategy fails to identify everywhere non-convex clusters because as the depth functions used herein only suit convex shapes, preventing recognition of local centers satisfying $r_i(\vecx_i)\leq 2$ in a non-convex neighborhood. To assess the presence of everywhere non-convex clusters and mitigate their impact, we categorize local centers into three sets based on the outcome of prior filtering:
\begin{itemize}
    \item Set $\mathcal{C}_1$ includes all local centers that remain after initial filtering, characterized by non-isolated local centers with $r_i(\vecx_i)\leq 2$ and $f\neq 1$.
    \item Set $\mathcal{C}_2$ contains non-isolated local centers $c_v$ for which $\forall c_i \in \mathcal{C}_1$, \textcolor{black}{$c_v$ satisfies} $\mathbb{M}_{v,i} = 0$. 
    \item Set $\mathcal{C}_3$ consists of local centers $c_v \notin \mathcal{C}_1$ satisfying $\exists c_i \in \mathcal{C}_1$ such that $\mathbb{M}_{v,i} > 0$.
\end{itemize}
The presence of everywhere non-convex clusters is inferred if $\mathcal{C}_2$ is not empty. In such cases, Algorithm \ref{alg:re_filtering} is used to segregate locally non-convex regions from other clusters.

\begin{algorithm}[bht]
\small
\caption{Segregating the locally non-convex region}
\label{alg:re_filtering}
\begin{algorithmic}[1]
\Require Sets of local centers $\mathcal{C}_1$, $\mathcal{C}_2$, and $\mathcal{C}_3$, neighborhood similarity matrix $\mathbb{M}$
\Ensure Set of filtered centers $\vecC$
\If{$\mathcal{C}_2 \neq \emptyset$}
    \For{$v$ in the indices of local centers in $\mathcal{C}_3$}
     \If{ $\max_{j:c_j \in \mathcal{C}_2} \mathbb{M}_{v,j}>\max_{i:c_i \in \mathcal{C}_1} \mathbb{M}_{vi}$   
        }
            \State $\mathcal{C}_2 = \mathcal{C}_2 \cup \{c_v\}$
%            \State $c_3 = c_3 \setminus \{c_v\}$
    \For{$i$ in the indices of local centers in $\mathcal{C}_1$}
        \If{$\mathbb{M}_{v,i} > 0$}
 %           \State $c_3 = c_3 \cup \{c_i\}$
            \State $\mathcal{C}_1 = \mathcal{C}_1 \setminus \{c_i\}$
        \EndIf
    \EndFor
   \EndIf
   \EndFor
   
\State $\vecC = \mathcal{C}_1 \cup \mathcal{C}_2$
\If{any isolated local centers exist within the updated set $\vecC$}
    \State Remove isolated local centers from $\vecC$
\EndIf
\Else
\State $\vecC = \mathcal{C}_1$
\EndIf
\end{algorithmic}
\end{algorithm}

This Algorithm ensures that local centers in $\mathcal{C}_2$, which capture the essence of locally non-convex regions, remain disjoint with the local centers in other clusters, i.e., local centers in $\mathcal{C}_1$.
\subsection{Clustering in DLCC} \label{sec:clinDLCC}
 The grouping results $\vecG=\{\vecg_1,\ldots,\vecg_K\}$ for the filtered centers are at hand and $\vecC$ represent the set of filtered centers. Consequently, the number of clusters is inherently defined to be equal to $K$, as each group’s unique neighbors directly initiate a corresponding temporary cluster. The subsequent step involves defining a score for each observation w.r.t. each cluster in order to update these temporary clusters.

The score accounts for both clustering and assessing if each observation's clustering result is acceptable, and it is computed using the similarity matrix
 $\mathbb{S}$. Since $\mathbb{S}$ is asymmetric, it is updated by replacing $\mathbb{S}_{j,i}$ and $\mathbb{S}_{i,j}$ with their average value. For clarity, we continue to denote the symmetrized matrix as $\mathbb{S}$, ensuring that $\mathbb{S}_{i,j} = \mathbb{S}_{j,i}$. We define the score function for an observation $\vecx_i$ w.r.t. a temporary cluster $k$ under as
\begin{eqnarray}\label{eq:scoremin}
\operatorname{score}_{i|k}=\frac{\max_{j \in \mathcal{I}_k} \bbS_{j,i} - \max_{j \in \bigcup_{z\neq k} \mathcal{I_z}} \bbS _{j,i}}{\max\{ \max_{j \in \mathcal{I}_k} \bbS_{j,i} , \max_{j \in \bigcup_{z\neq k} \mathcal{I_z}} \bbS _{j,i}\} \,},
\end{eqnarray}
where $\mathcal{I}_k$ is the set of indices for filtered centers within the $k$th group, i.e., $\mathcal{I}_k=\{j: \vecx_j \in \vecg_k\}$. The score \eqref{eq:scoremin} takes a value between $-1$ and $1$ and a positive value indicates that, compared with the nearest filtered center in other groups, the similarity between $\vecx_i$  and the nearest filtered center in group $k$ is larger.

For each cluster, two pairs of sets are generated: $\vecT_k$, containing the unique neighbors within the cluster ($\vecU_{\vecg_k}$), and $\mathcal{S}_k$, the corresponding scores for points in $\vecT_k$ w.r.t. the $k$th cluster; as well as $\hat{\vecT}_k$, holding unassigned observations with positive scores toward the $k$th cluster, and $\hat{\mathcal{S}}_k$, their associated scores. During the update process, observations in $\hat{\vecT}_k$ with sufficiently high scores in $\hat{\mathcal{S}}_k$ may be moved to $\vecT_k$ and $\mathcal{S}_k$, while those in $\vecT_k$ with low scores in $\mathcal{S}_k$ may be moved to $\hat{\vecT}_k$ and $\hat{\mathcal{S}}_k$.  After refinement, $\vecT_k$ contains the observations confidently assigned to the $k$th cluster, while $\hat{\vecT}_k$ collects those remaining unassigned. We refer to $\{\vecT_k\}_{k=1}^K$ as the set of temporary clusters.

The detailed steps are outlined in Algorithm~\ref{alg:utc}, with only a minor difference between the min and max strategies. Under the max strategy, we use a single mean-based cutoff for each cluster, applied uniformly to points in $\vecT_k$ and $\hat{\vecT}_k$ and take $\gamma=0.5$, yielding a looser threshold for admitting points into temporary clusters. Under the min strategy, the additional quantile-based cap in Line~9 enforces at least $s/2$ observations per temporary cluster, preventing overly small clusters.

\begin{figure*}[th!]
     \centering
		\includegraphics[width=0.9\textwidth]{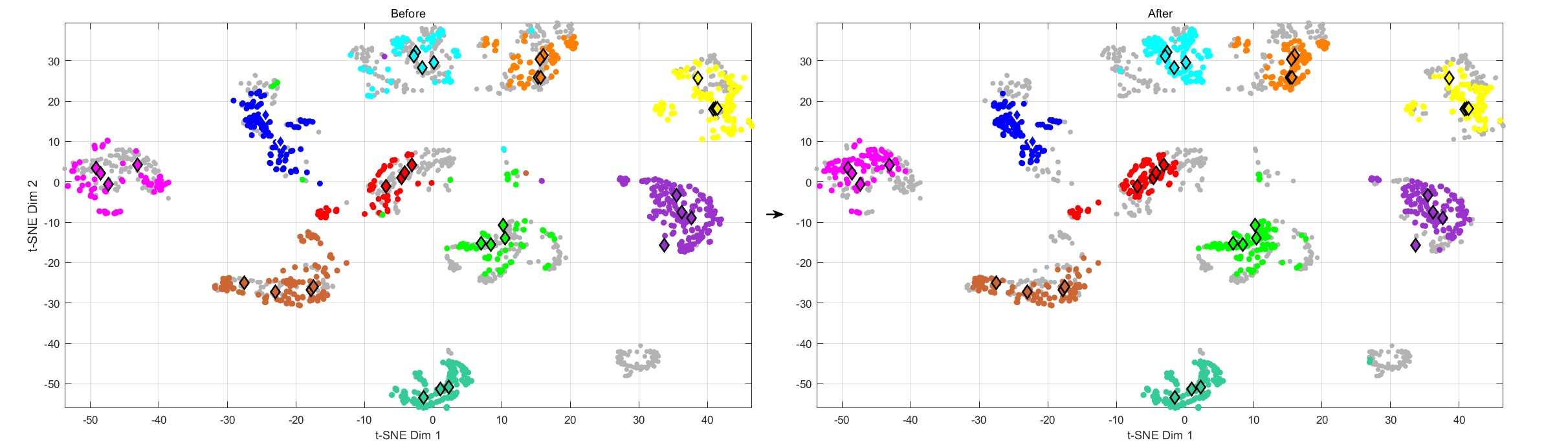}    
     \caption{t-SNE visualization \cite{van2008visualizing} of the Yale~B dataset in two-dimensional space (see Section~\ref{sec:ee} for dataset description). 
Colors indicate the temporary labels of the points, i.e., membership in 
$\{\vecT_k\}_{k=1}^K$. 
From left to right, the panels illustrate the progression from the initial state 
(unique neighbors) to the temporary clusters after Algorithm~\ref{alg:utc}. 
Points enclosed by diamond-shaped boxes denote filtered centers.}
     \label{Fig:tsnescore}
\end{figure*}

Figure~\ref{Fig:tsnescore} presents an example based on the Yale~B face recognition dataset \cite{georghiades2001few}. Algorithm~\ref{alg:utc} updates the temporary clusters by discarding unique neighbors whose scores relative to their currently assigned clusters are insufficient, while simultaneously assigning labels to non–unique neighbors that are sufficiently close to the filtered centers of their respective groups. As shown in the figure, this procedure removes visually misclassified points while correctly labeling most points near local centers.

\begin{algorithm}
\small
\caption{Updating Temporary Clusters}
\label{alg:utc}
\begin{algorithmic}[1] 
\Require For $k=1,\ldots,K$, temporary clusters $\vecT_k$, scores $\mathcal{S}_k$ for points in $\vecT_k$; Remaining observations for each cluster $\hat{\vecT}_k$ and their scores $\hat{\mathcal{S}}_k$; neighborhood size $s$;  $\texttt{strategy}\in\{\texttt{min},\texttt{max}\}$.
\Ensure Updated $\vecT_k$ and $\mathcal{S}_k$ for $k=1,\ldots,K$.
\State Set $\gamma\gets 1$ if $\texttt{strategy}=\texttt{min}$, otherwise $\gamma\gets 0.5$.
\For{$k = 1$ to $K$}
    \State $\alpha_{1}\gets \gamma\cdot \mathrm{mean}(\hat{\mathcal{S}}_k)$ \Comment{if $\hat{\mathcal{S}}_k=\emptyset$, use $\mathrm{mean}(\hat{\mathcal{S}}_{\mathrm{all}})$, where $\hat{\mathcal{S}}_{\mathrm{all}}=\bigcup_{\ell:\hat{\mathcal{S}}_\ell\neq\emptyset}\hat{\mathcal{S}}_\ell$}
    \For{each $i \in \vecT_k$ with score $< \alpha_{1}$}
        \State $\vecT_k \gets \vecT_k \setminus \{i\}$,\quad $\hat{\vecT}_k \gets \hat{\vecT}_k \cup \{i\}$; update $\mathcal{S}_k$ and $\hat{\mathcal{S}}_k$.
    \EndFor
      \If{\texttt{strategy = min}}
    \State $SS \gets$ sorted scores in $\mathcal{S}_k$ (descending)
    \State $idx \gets \argmax_{j = \mathrm{floor}(|\mathcal{S}_k|/2) \text{ to } |\mathcal{S}_k|-1} (SS[j] - SS[j+1])$
    \State $\alpha_2 \gets \min(SS[idx],  1 - \frac{s/2 - |\mathcal{S}_k|}{|\hat{\mathcal{S}}_k|}$th quantile of $\hat{\mathcal{S}}_k)$
       \Else \Comment{\texttt{strategy = max}}
        \State $\alpha_2 \gets \alpha_1$.
    \EndIf
     \For{each $u \in \hat{\vecT}_k$ with score $> \hat{\alpha}_2$}
        \State $\hat{\vecT}_k \gets \hat{\vecT}_k \setminus \{u\}$,\quad $\vecT_k \gets \vecT_k \cup \{u\}$.
      \EndFor
\EndFor
\end{algorithmic}
\end{algorithm}

\subsubsection{Classification }
After obtaining temporary clusters, we further assign the remaining unlabeled observations via a classification step. 
The motivation is that, for exemplar-based clustering methods, assigning ambiguous points solely according to their similarity to an exemplar can be unreliable, as such points are often weakly associated with any single exemplar. Instead, incorporating information from confidently assigned samples provides a more robust basis for assigning these ambiguous points. This idea has also been explored in other clustering researches. For example, Shi et al.~\cite{shi2022global} noted that, beyond center-based allocation, assignments of determined samples can be used to infer labels of undetermined ones.

Two built-in options for classifying the remaining observations in DLCC are depth-based $k$NN and RF. In both cases, the temporary-cluster labels serve as the training labels, and we intentionally avoid additional hyperparameter tuning to keep the procedure lightweight and reproducible.

The $k$NN option is a simple nonparametric classifier that can accommodate non-convex decision boundaries. Since DLCC already provides the depth-based similarity matrix $\bbS$, neighbors are obtained by sorting $\bbS_{i,\cdot}$, avoiding additional distance evaluations. Moreover, the choice of $k$ can be made data-adaptive via the notion of natural neighbors~\cite{zhu2016natural}. 
Concretely, we take the smallest $k$ such that at least $95\%$ of observations have at least one \emph{mutual} neighbor within their top-$k$ neighbor lists, where the $95\%$ threshold serves as a safeguard against potential outliers.

As for RF, it is a widely used nonparametric classifier that makes no explicit shape assumptions and often remains effective in high-dimensional spaces~\cite{breiman2001random}. Again, we do not perform hyperparameter tuning and fix the number of trees at $100$.

\subsection{Complexity Analyses}
The overall computational cost of DLCC is dominated by constructing the depth-based similarity matrix $\bbS$ and the associated local-center estimation. 
As discussed in Section~\ref{sec:DSM}, computing $\bbS$ in its exact form requires $O(n^3)$ operations. Estimating local centers over $n$ neighborhoods of size $s$ incurs an additional cost of $O(n s^2)$.  Subsequent filtering and grouping steps operate only on the $T$ local-center candidates, where $T\ll n$, and thus contribute negligibly to the overall complexity.

The temporary clustering step and the depth-based $k$NN classification are performed on top of the precomputed similarity matrix $\bbS$ and incur at most linear cost in $n$.  For the RF option, training a forest with $100$ trees has a typical cost roughly $O(100\,n\log n)$.

Consequently, the overall complexity is dominated by the $O(n^3)$ cost of constructing $\bbS$. When the data dimension $d$ is large, this complexity is expressed as $O(n^3 d)$, since the computation of spatial depth involves the evaluation of $L^2$ norms of averaged unit vectors.

\section{Exploring Parameter Selection}\label{sec:EPS}
\subsection{Strategy Chosen}\label{sec:strachos}
Under the min strategy, DLCC attempts to find centroids of clusters from filtered centers. The variation in the number of filtered centers across different clusters allows for better flexibility than center-based clustering methods. The min strategy is more suitable when the sizes of all clusters are comparable. Moreover, because it groups centers more conservatively, it is often more reliable for overlapping clusters.

The max strategy views each cluster as a union of multiple finite convex regions, where each region is approximated by the convex hull of neighborhood points associated with a filtered center. Filtered centers are grouped via connectivity. Specifically, two filtered centers $c_y$ and $c_z$ are assigned to the same group if there exists a chain $c_{p_1},c_{p_2},\ldots,c_{p_h}$ with $c_{p_1}=c_y$, $c_{p_h}=c_z$, such that the similarity between neighborhoods of any adjacent pair in the chain exceeds $\delta$. Consequently, the max strategy is typically more effective for non-convex structures or highly unbalanced cluster sizes.

When no prior information about the dataset is available, we recommend selecting the strategy via visualization. For high-dimensional data, a 2D embedding (e.g., t-SNE~\cite{van2008visualizing}) is often informative: if clusters appear well separated and roughly balanced, the min strategy is recommended; otherwise, the max strategy is preferred.
\subsection{Max Strategy}\label{sec:EPSMAX}
The choice of the parameters $s$ (neighborhood size) and $\delta$ (neighborhood similarity threshold) affect the resulting clustering differently. The neighborhood size $s$ directly affects the defined neighborhoods for each observation and hence impacts the construction of the set $\vecC$ of filtered centers. Under the max strategy, for any given~$s$, the value of $\delta$ further influences how the filtered centers are grouped. Basically, a smaller value of $\delta$ results in fewer, and larger, groups of filtered centers.
\subsubsection{Hierarchical Structure of Grouping}
Given a certain $s$, all possible values of the number of groups can be derived. By starting at $\delta=\delta_0$ and progressively reducing $\delta$,  all $\delta$'s at the time that the number of groups changes can be recorded. 

The initial choice for $\delta$ can be based on the following value:
\begin{eqnarray}\label{eq:ub}
 \delta_U := \min_{c} \max_{c^\dagger}\{\mathrm{sim}( \,\vecN_{c},\vecN_{c^\dagger}): c \neq c^\dagger \in \vecC\}.
\end{eqnarray}
For any $\delta \geq \delta_U$, there will be at least one group with a single filtered center. It is thus desirable to start with a smaller value, say $\delta_0 = 0.99 \delta_U$, so that multiple smaller subsets are connected to form clusters.

Suppose $\delta_0$ is chosen, partitioning $\vecC$ into groups $\vecg_1, \vecg_2, \ldots$ \textcolor{black}{Now, if we reduce the $\delta$ value from $\delta_0$ to a value below the maximum similarity among neighborhoods of centers between any two groups $\vecg_i$ and $\vecg_j$, as described by:
}
\begin{eqnarray} \label{eq:stm}
\delta<\max \{\mathrm{sim}( \,\vecN_{c},\vecN_{c^\dagger}): c \in \vecg_i, c^\dagger \in \vecg_j\},
\end{eqnarray}
then $\vecg_i$ and $\vecg_j$ will merge. The value in \eqref{eq:stm} is referred to as the \emph{grouping critical point} for $\vecg_i$ and $\vecg_j$. Observing how grouping results evolve as $\delta$ decreases from $\delta_0$ reveals the hierarchical clustering structure.

\begin{figure*}[h!]
\centering
\begin{tikzpicture}
\small
\tikzset{box/.style={draw, rectangle, thick, text centered, minimum height=3em, minimum width=3em, fill=orange!30},
          line/.style={-Stealth, thick}}

% Nodes
\node[box] (m1) {
    $\begin{pmatrix}
    1 & 0.6 & 0 & 0 & 0 & 0 \\ 
    0.6 & 1 & 0 & 0 & 0 & 0 \\ 
    0 & 0 & 1 & 0.5 & 0.3 & 0.2 \\ 
    0 & 0 & 0.5 & 1 & 0.4 & 0.1 \\ 
    0 & 0 & 0.3 & 0.4 & 1 & 0.6 \\ 
    0 & 0 & 0.2 & 0.1 & 0.6 & 1 
    \end{pmatrix}$
};
\node[box, right=of m1, xshift=0.7cm] (m2) {
    $\begin{pmatrix}
    1 & 0 & 0 & 0 \\
    0 & 1 & 0.5 & 0.3 \\
    0 & 0.5 & 1 & 0.4 \\
    0 & 0.3 & 0.4 & 1
    \end{pmatrix}$
};
\node[box, right=of m2, xshift=0.7cm] (m3) {
    $\begin{pmatrix}
    1 & 0 & 0 \\
    0 & 1 & 0.4 \\
    0 & 0.4 & 1
    \end{pmatrix}$
};
\node[box, right=of m3, xshift=0.7cm] (m4) {
    $\begin{pmatrix}
    1 & 0 \\
    0 & 1
    \end{pmatrix}$
};
  % Additional Info
\node[box, align=center, above=0.5cm of m4,fill=blue!10] (info) {%
    $\boldsymbol{G} = \{6, 4, 3, 2\}$ \\
    $\boldsymbol{\delta} = \{0.6, 0.5, 0.4, 0\}$
  };
% Lines
\draw[line] (m1) -- (m2);
\draw[line] (m2) -- (m3);
\draw[line] (m3) -- (m4);

\end{tikzpicture}
\caption{A toy example illustrating the hierarchical structure in grouping filtered centers, alongside the relationship between the threshold $\delta$ and the number of groups $G$. This process generates two lists, $\boldsymbol{G}$ and $\boldsymbol{\delta}$, which store the number of groups and the corresponding threshold values at each stage, respectively. The elements of these lists are displayed in the top right.}
\label{fig:tegrouping}
\end{figure*}
We use a toy example to illustrate the process, shown in Figure \ref{fig:tegrouping}. Each square matrix is referred as a ``grouping matrix''. The order of the matrix denotes the number of groups, and its entries identify the grouping critical points among these groups for a given $\delta$ value. In each step, the grouping matrix is obtained from the previous one by merging \textcolor{black}{columns and rows} that contain the largest non-trivial (not $0$ or $1$) value. The merging process retains the larger of the values in the corresponding entries of the \textcolor{black}{columns and rows.} Eventually, we are left with completely disjoint groups, represented by the identity matrix as the grouping matrix. 

The process of grouping generates two key lists: one for the number of groups and another for the corresponding threshold values. More precisely, in the toy example, \textcolor{black}{the potential numbers of groups are $6$, $4$, $3$, and $2$, with the $\delta$ values ranging between the intervals $[0.6, \delta_U)$, $[0.5, 0.6)$, $[0.4, 0.5)$, and $[0, 0.4)$ at each stage.}
% \textcolor{red}{\begin{eqnarray}\notag
% \delta_U > \delta \geq 0.6 \longrightarrow 0.6 > \delta \geq 0.5 \longrightarrow 0.5 > \delta \geq 0.4 \longrightarrow\\
% \notag
% 0.4 > \delta \geq 0.
% \end{eqnarray}}%
The hierarchical structure \textcolor{black}{ensures that if the number of clusters is known or can be estimated using any methods, a suitable $\delta$ can be directly determined from the relationship between the lists $\boldsymbol{G}$ and $\boldsymbol{\delta}$.}
\subsubsection{Parameter selection}
For the max strategy, clusters are formed by connecting multiple local convex regions. Thus, the neighborhood size $s$ should be relatively small. In our experiments, we recommend $s \in [20,60]$. For each $s$, we generate a corresponding candidate list of thresholds $\boldsymbol{\delta}$ and evaluate the resulting solutions accordingly. Note that the construction of the similarity matrix is independent of the parameters. Therefore, it can be computed once in advance before the parameter selection stage.

Without access to ground truth, clustering is inherently exploratory. 
As noted by Hennig~\cite{hennig2015clustering}, there is in general no unique ``true'' clustering, and validation often rely on a combination of internal validation indices (IVI) and subjective assessment, where visualization plays an important supporting role. Our parameter selection rule follow this philosophy.

In terms of IVIs, classical criteria such as the silhouette width \cite{rousseeuw1987silhouettes} and the Calinski-Harabasz index \cite{calinski1974dendrite} are primarily designed for compact and approximately convex clusters, and are not suitable for clusters with arbitrary shapes. To better reflect arbitrary geometries, we adopt the cluster validity index based on density-involved distance (CVDD) \cite{hu2019internal}.\footnote{\url{https://github.com/hulianyu/CVDD}}
CVDD replaces the usual notion of separation by a density-aware connectivity distance defined on a $k$NN graph. Two points are regarded as close only if they can be connected by a path that stays in relatively high-density regions. Consequently, CVDD penalizes clustering results that effectively require ``crossing'' low-density gaps to maintain within-cluster connectivity, while remaining insensitive to non-convex shapes as long as they are density-connected. The resulting score balances density-aware between-cluster separation against within-cluster connectivity; further details are referred to Hu and Zhong \cite{hu2019internal}.

However, IVIs inevitably encode particular assumptions about ``separation'' and ``cohesion'', and tuning based on a single IVI alone is therefore often insufficient in practice. Indeed, Lewis et al.~\cite{lewis2012human} reported that human assessments can be more consistent than existing IVIs when judging clusterings. We therefore combine CVDD with visualization for model selection. Specifically, we run DLCC over the grid of $(s,\boldsymbol{\delta})$, which yields solutions with varying estimated numbers of clusters $\hat{K}$. For each distinct $\hat{K}$, we retain the solution that attains the largest CVDD value as its representative. We then visualize the representatives of the top three $\hat{K}$ values ranked by CVDD and choose the final model by visual assessments.

In experiments shown in Figure~\ref{fig:dlcc_synth_vis}, although CVDD identifies suitable parameter pairs for most datasets, it may not always favor the most appropriate $\hat{K}$. For instance, on the Bainba dataset (\ref{subfig:bainba}), CVDD actually prefers merging the two upper groups due to their proximity and similar local densities; visually, however, they form two distinct clusters and should remain separated.
\subsection{Min strategy}
\begin{figure*}[t]
    \centering
    \setlength{\tabcolsep}{3pt}
    \renewcommand{\arraystretch}{1.0}

    \begin{tabular}{cccc}
         \subfloat[Iris\label{subfig:digiiris}]{
            \includegraphics[width=0.24\textwidth]{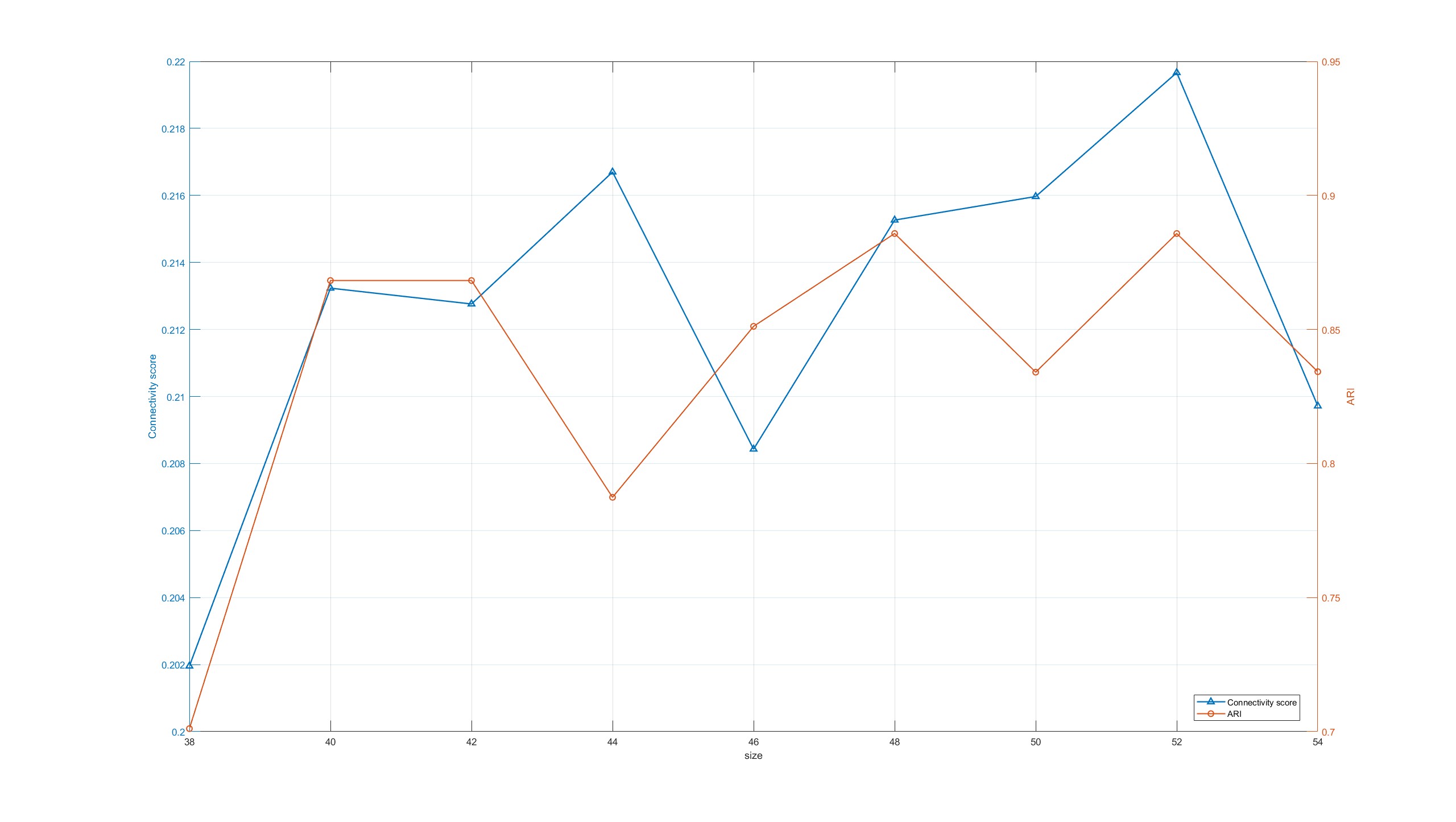}
        } 
        &
        \subfloat[Optidigits\label{subfig:digi}]{
            \includegraphics[width=0.24\textwidth]{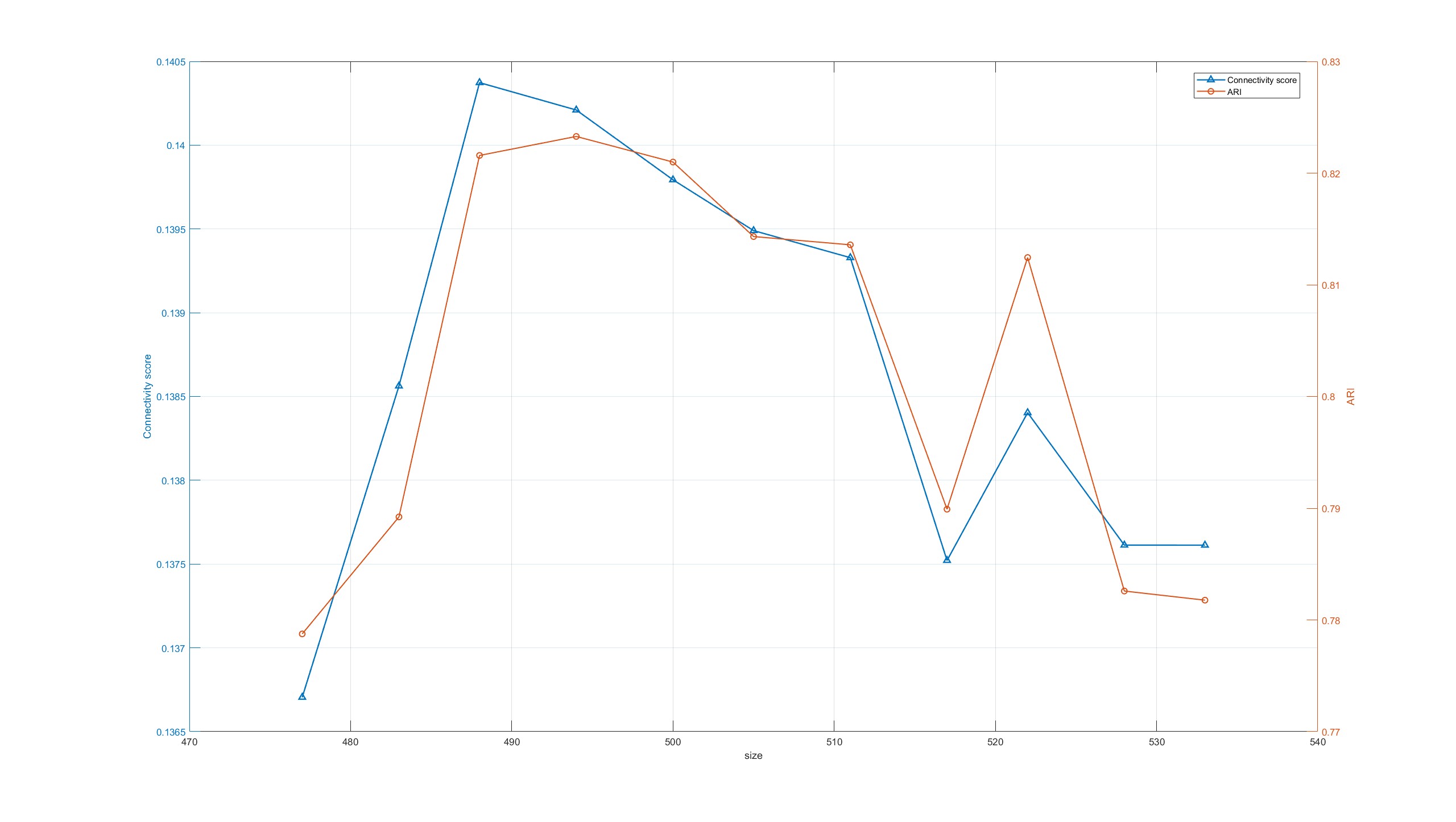}
        } 
        &
        \subfloat[Segmentation\label{subfig:dseg}]{
            \includegraphics[width=0.24\textwidth]{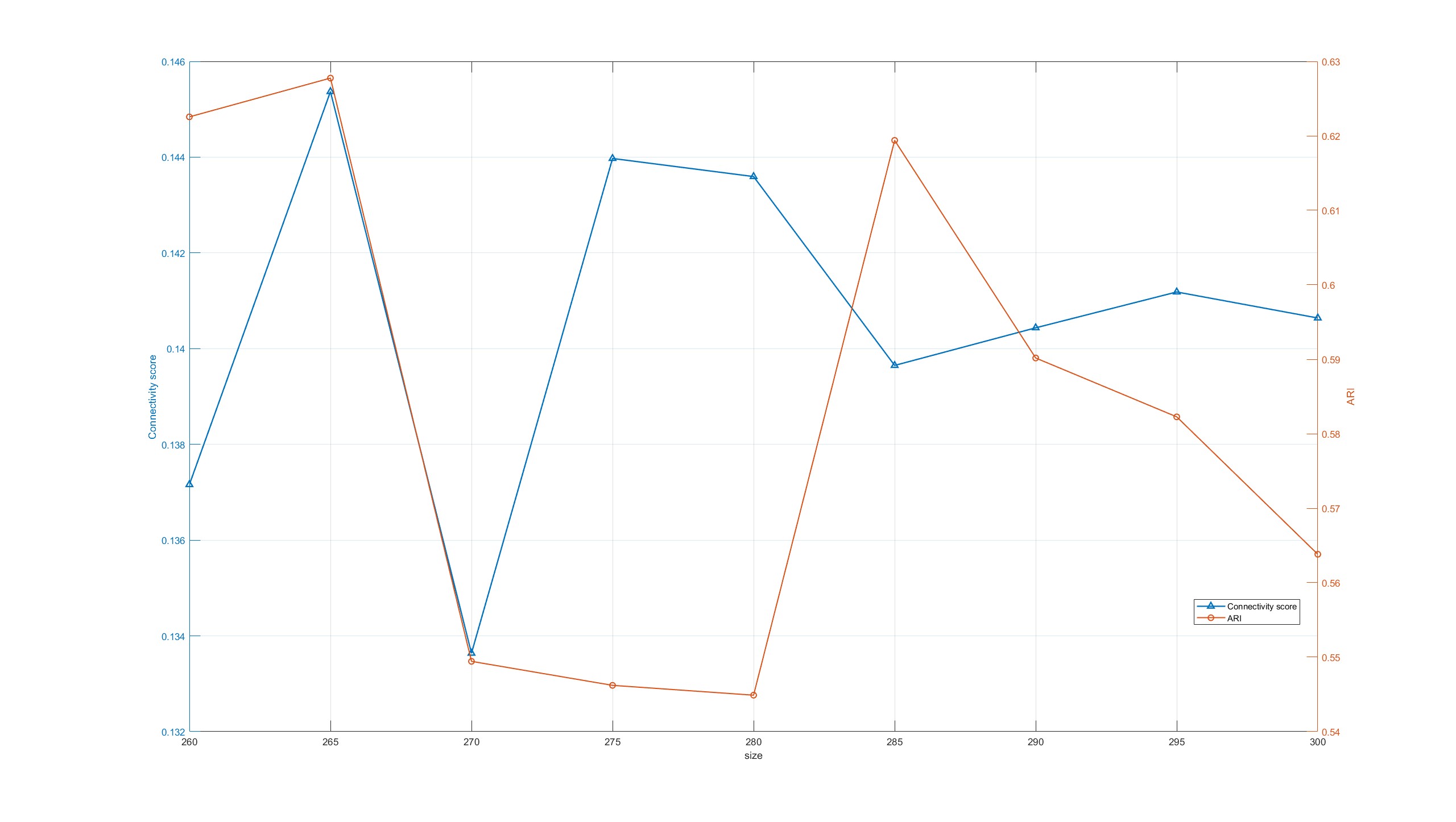}
        } &
       \subfloat[Yale B\label{subfig:con1}]{
            \includegraphics[width=0.24\textwidth]{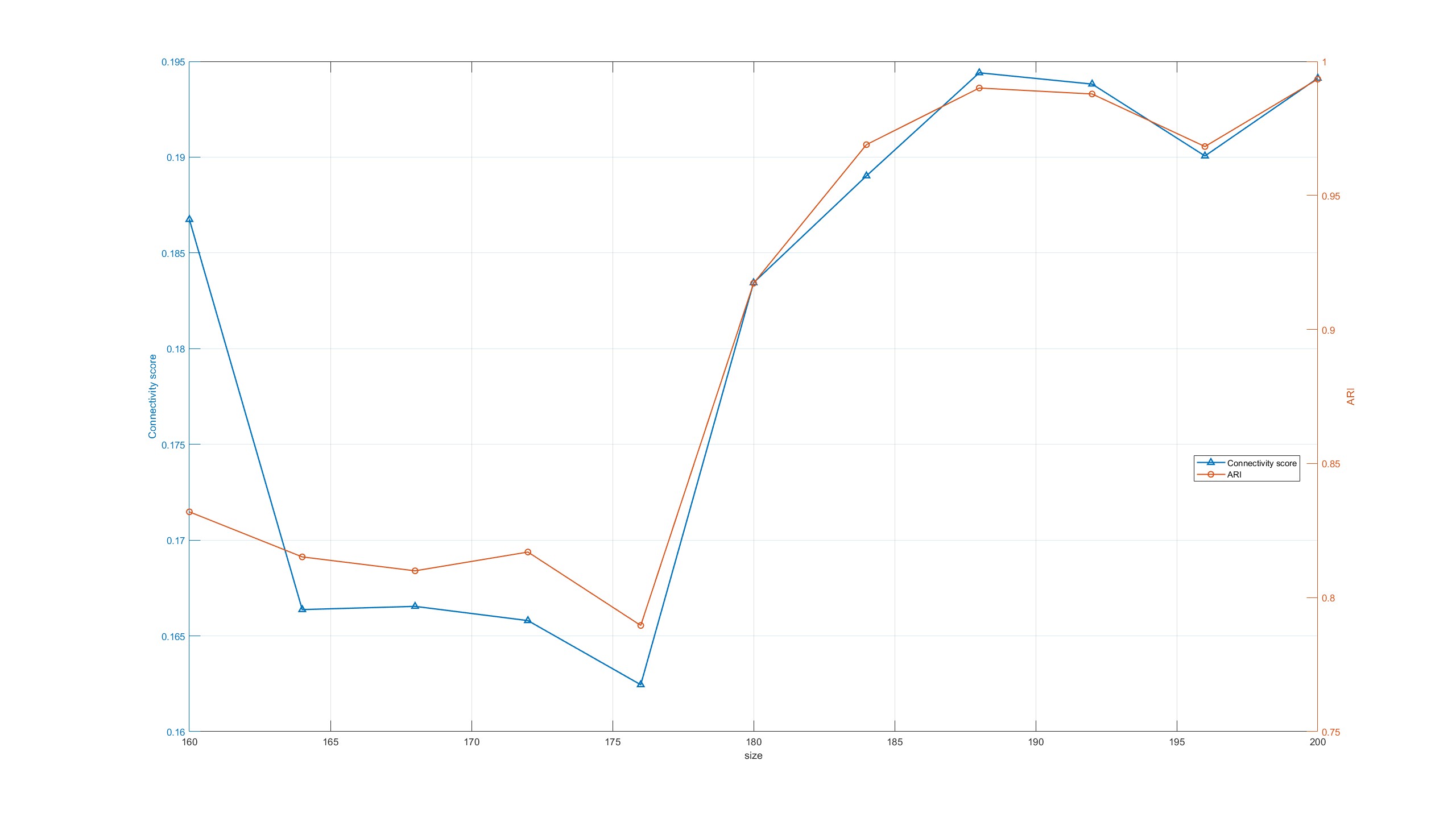}
        } \\
    \end{tabular}

    \caption{Empirical association between the proposed depth-connectivity score (blue; left axis) and the ARI (orange; right axis) as the neighborhood size $s$ varies under the min strategy with a fixed $K$.}
    \label{fig:connect_ari}
\end{figure*}
Under the min strategy, particularly for real-world high-dimensional data (e.g., $d>50$), selecting parameters solely using a single IVI is often challenging. Nevertheless, IVIs (e.g., CVDD) remain useful for narrowing down reasonable ranges of $s$ and estimate $\hat{K}$. With a known or an estimated $K$, we recommend choose $s \in [0.6\,n/K,\, n/K]$. To further evaluate candidate clusterings, we then exploit the symmetrized depth-based similarity matrix $\bbS$. Before proceeding, we introduce the reachable similarity, a max-min (bottleneck-path) connectivity measure in graph theory \cite{gabow1988algorithms,chebotarev2011graph}.
\begin{thm}[Reachable similarity]\label{def:rs}
Let $S$ be a symmetric similarity matrix with entries in $[0,1]$. The \emph{reachable similarity} between nodes $i$ and $j$ is
\[
\mathrm{rs}(i,j)=\max_{\boldsymbol{p}: i\to j}\ \min_{(u,v)\in \boldsymbol{p}} S_{u,v},
\]
i.e., the strongest bottleneck similarity over all paths connecting $i$ and $j$.
\end{thm}
Given a candidate partition $\{\pi_g\}_{g=1}^{K}$, we quantify connectivity using two pointwise quantities: the intra-cluster reachable similarity (IR) and the between-cluster reachable similarity (BR). For each point with label $\widehat{y}_i$, its IR is defined as the average reachable similarity from $\vecx_i$ to other points within the same cluster,
where $\mathrm{rs}(\cdot,\cdot)$ is computed on the subgraph induced by nodes in $\pi_{\widehat{y}_i}$.

To define BR, we first identify the nearest competing cluster for $\vecx_i$ by comparing $\vecx_i$ to filtered centers from other clusters
\[
\tilde{g}_i=\arg\max_{g\neq \widehat{y}_i}\ \max_{\vecx_u\in\vecg_g}\ \bbS_{i,u},
\]
where $\vecg_g$ is the $g$th group of filtered centers. We then set BR$_i$ to be the average reachable similarity from $\vecx_i$ to points in the competing cluster $\pi_{\tilde{g}_i}$, with $\mathrm{rs}(\cdot,\cdot)$ computed on the subgraph induced by nodes in $\pi_{\tilde{g}_i}$.

Finally, we summarize a clustering result by the mean gap
\begin{equation}\label{eq:connect}
\texttt{Connectivity}=\frac{1}{n}\sum_{i=1}^{n}\big(\texttt{IR}_i-\texttt{BR}_i\big),
\end{equation}
where larger values indicate stronger within-cluster connectivity relative to the most competitive cross-cluster connectivity. We observe that when $K$ is fixed and ground truth labels are available, this depth-connectivity score exhibits a strong positive association with external agreement measures such as the adjusted Rand index (ARI; \cite{hubert1985comparing}) across a range of datasets; representative examples are shown in Fig.~\ref{fig:connect_ari}. This observation motivates using \eqref{eq:connect} as a practical tool for ranking candidate neighborhood sizes $s$ under the min strategy when ground truth labels are unavailable.

\section{Experiments and Evaluations}\label{sec:ee}
In this section, we compare DLCC under both strategies with several competing clustering methods on synthetic and real-world data sets. In DLCC, the final-stage classifier is selected via CVDD; when RF is chosen, we report the median ARI, NMI, and CE over $100$ independent runs to account for its stochasticity. Detailed records of the selected classifiers are provided in the supplementary material.

\subsection{External Metric}
To compare the performance of different clustering algorithms on the same datasets, we use the clustering error (CE; \cite[p.~2774]{elhamifar2013sparse}), ARI and normalized mutual information (NMI; \cite{strehl2002cluster}). The CE simply computes the proportion of mismatched points. This makes CE intuitive, with a value range of $0$ (perfect match) to $1$. The ARI, on the other hand, measures pairwise similarity between clustering results and ground truth. The expected value fo the ARI under random class assignment is $0$ and it takes a value $1$ under perfect agreement. NMI measures the shared information between the predicted and true partitions and is normalized to $[0,1]$, with $1$ indicating perfect agreement and larger values implying better alignment.

\subsection{Methods for Comparison }
We compare DLCC with representative clustering methods that rely on exemplars or explicitly group local sub-regions/exemplars which show resemblance of DLCC, including LW-k-means \cite{chakraborty2020detecting}, EKM \cite{he2025equilibrium}, R-MDPC \cite{guan2024fast}, granular-ball clustering (GBCT; \cite{xia2025gbct}), and ACLR \cite{ma2025large}. We also include a recent depth-based alternative to mean-shift \cite{francisci2023analytical}, denoted as DMS. For all those methods except DMS, we use the authors' released implementations; DMS is implemented by us using lens depth \cite{liu2011lens} since no existing implementation can be found.

All benchmark datasets considered in this paper are equipped with ground-truth labels. Since a single solution is generally not expected to simultaneously optimize all three external metrics (ARI, NMI, and CE), we adopt ARI as the primary selection metric throughout. Whenever a method involves hyperparameter tuning, we select and report the solution with the best ARI, and report the corresponding NMI and CE for the same solution.

For fair comparisons, we assume the true number of clusters $K$ is known, and set $K$ accordingly for methods that require it. For LW-k-means, we report the median performance over $100$ random initializations. For EKM, we set the number of replications to $1000$ and report the solution with the smallest objective value (as in the original implementation). For R-MDPC, we follow the authors' recommendation by fixing $cv=1$ and scanning neighborhood sizes in $[0.5\sqrt{n},\,2\sqrt{n}]$. For ACLR, we scan the hierarchy of binary trees $h\in\{7,\ldots,11\}$ and the anchor-graph neighborhood size in $[10,100]$. For DMS, we scan the quantile level used to determine the localization parameter $\tau$ over $[0.01,0.5]$. The direction parameter is tuned as follows: we first evaluate two initial values ($20$ and $500$) and choose the better one as the starting point. We then perform a one-sided search by increasing from $20$ (or decreasing from $500$) and stop when the ARI drops by more than $0.05$ for two consecutive steps. 

\subsection{Synthetic Datasets}
\begin{figure*}[t]
    \centering
    \setlength{\tabcolsep}{2pt} % tighter spacing between columns
    \renewcommand{\arraystretch}{1.0}

 \begin{tabular}{cccc}
    \subfloat[3 blobs]{\includegraphics[width=0.24\textwidth]{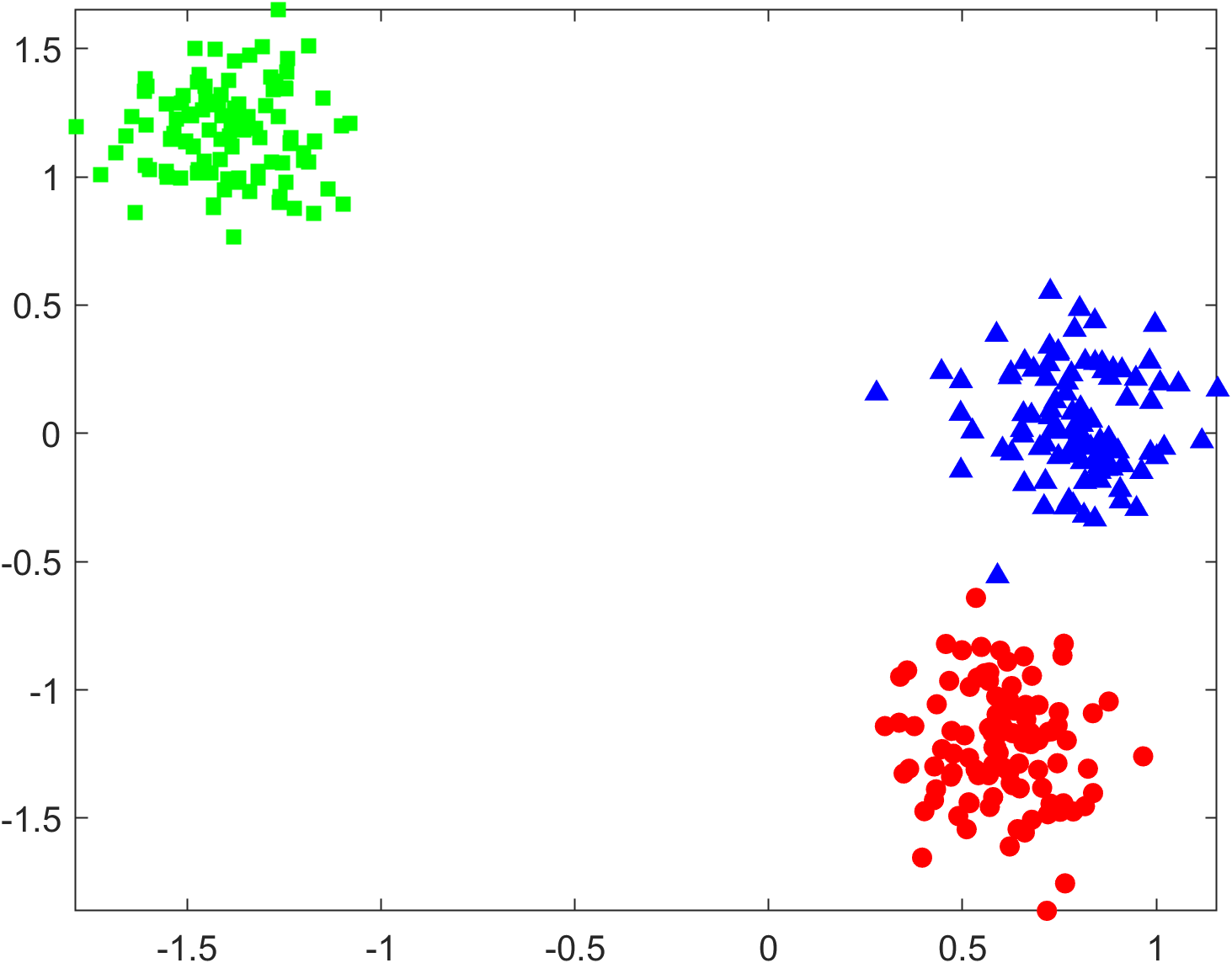}} &
    \subfloat[Aggregation]{\includegraphics[width=0.24\textwidth]{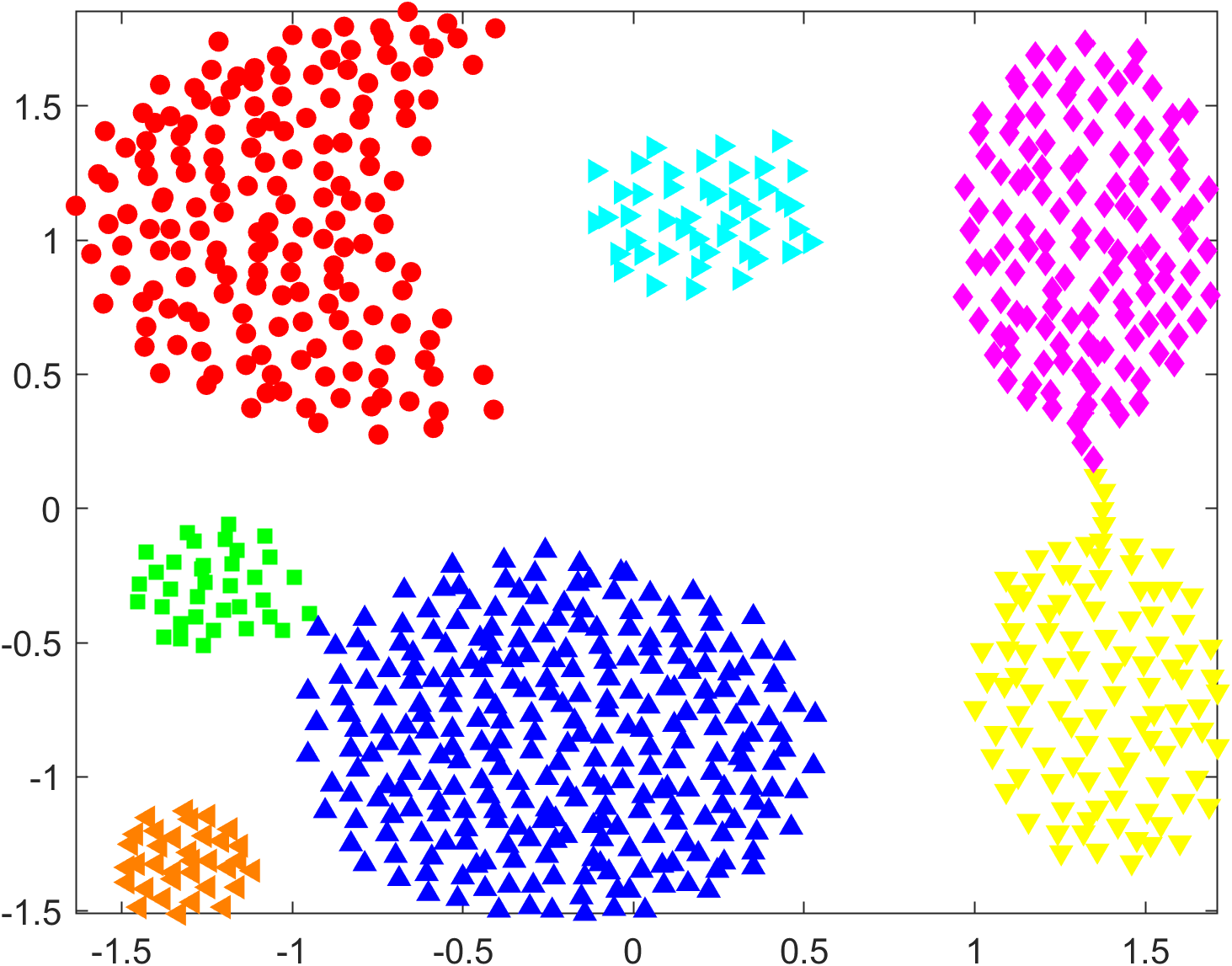}} &
    \subfloat[Atom]{\includegraphics[width=0.24\textwidth]{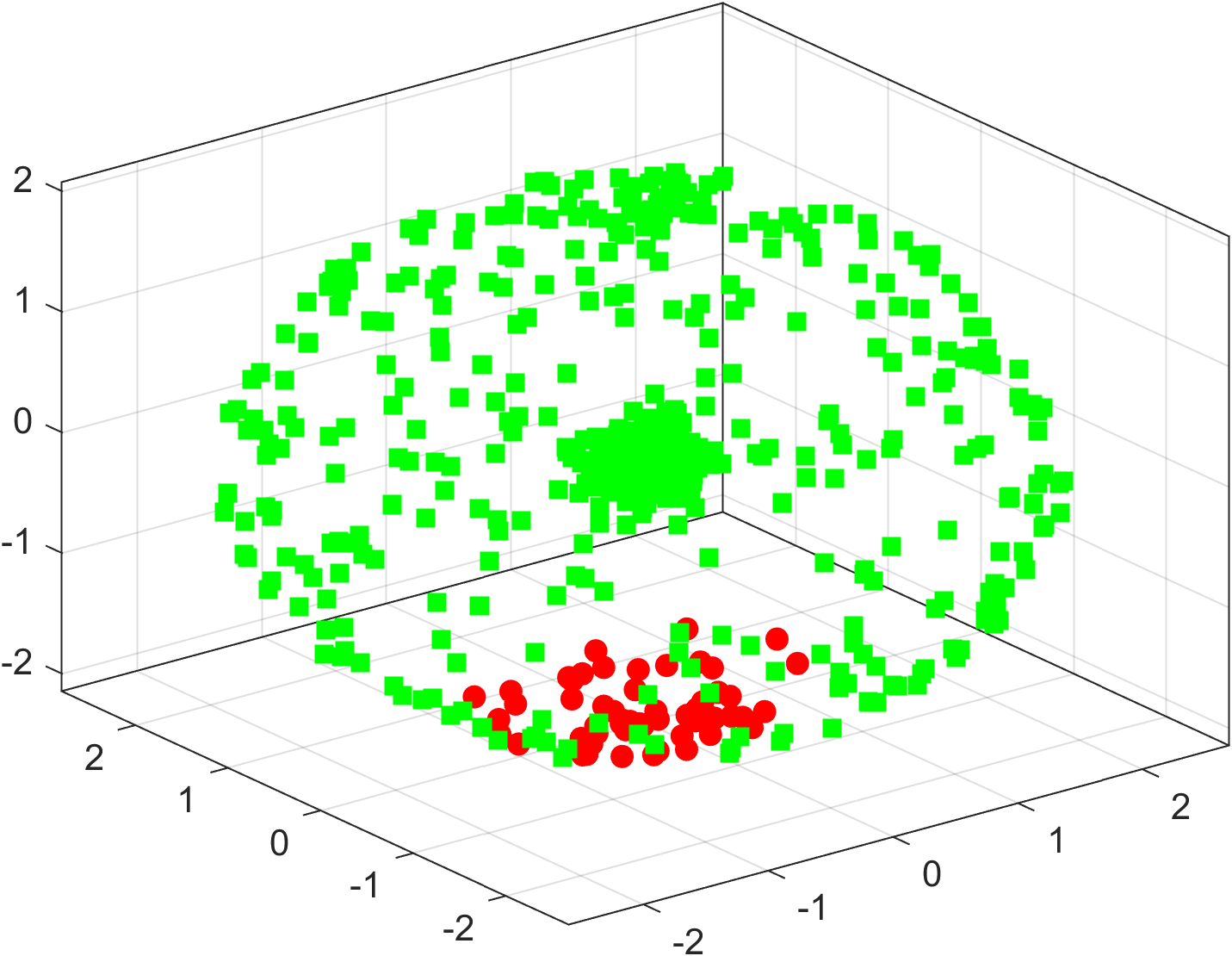}} &
    \subfloat[Bainba\label{subfig:bainba}]{\includegraphics[width=0.24\textwidth]{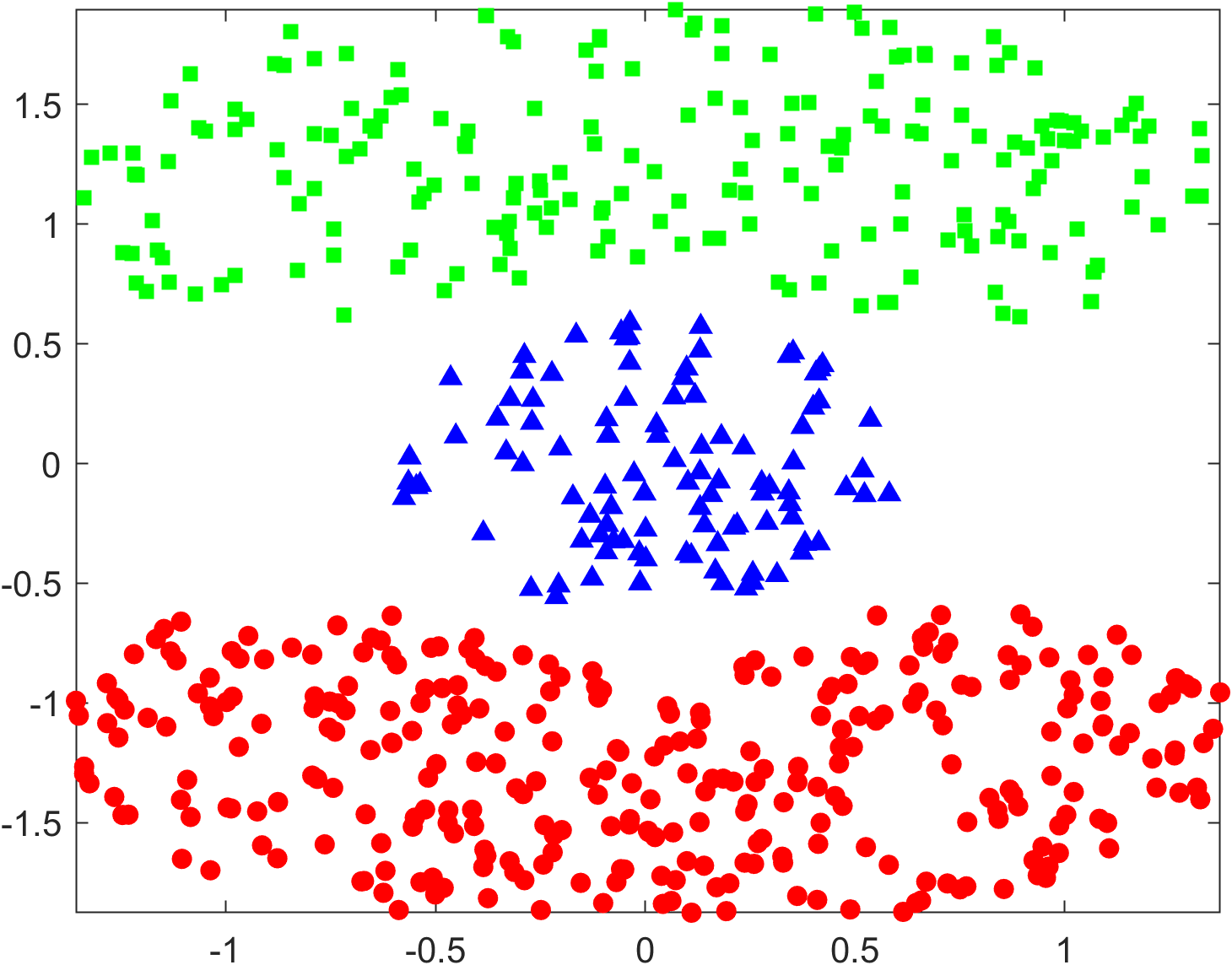}} \\

    \subfloat[Blend]{\includegraphics[width=0.24\textwidth]{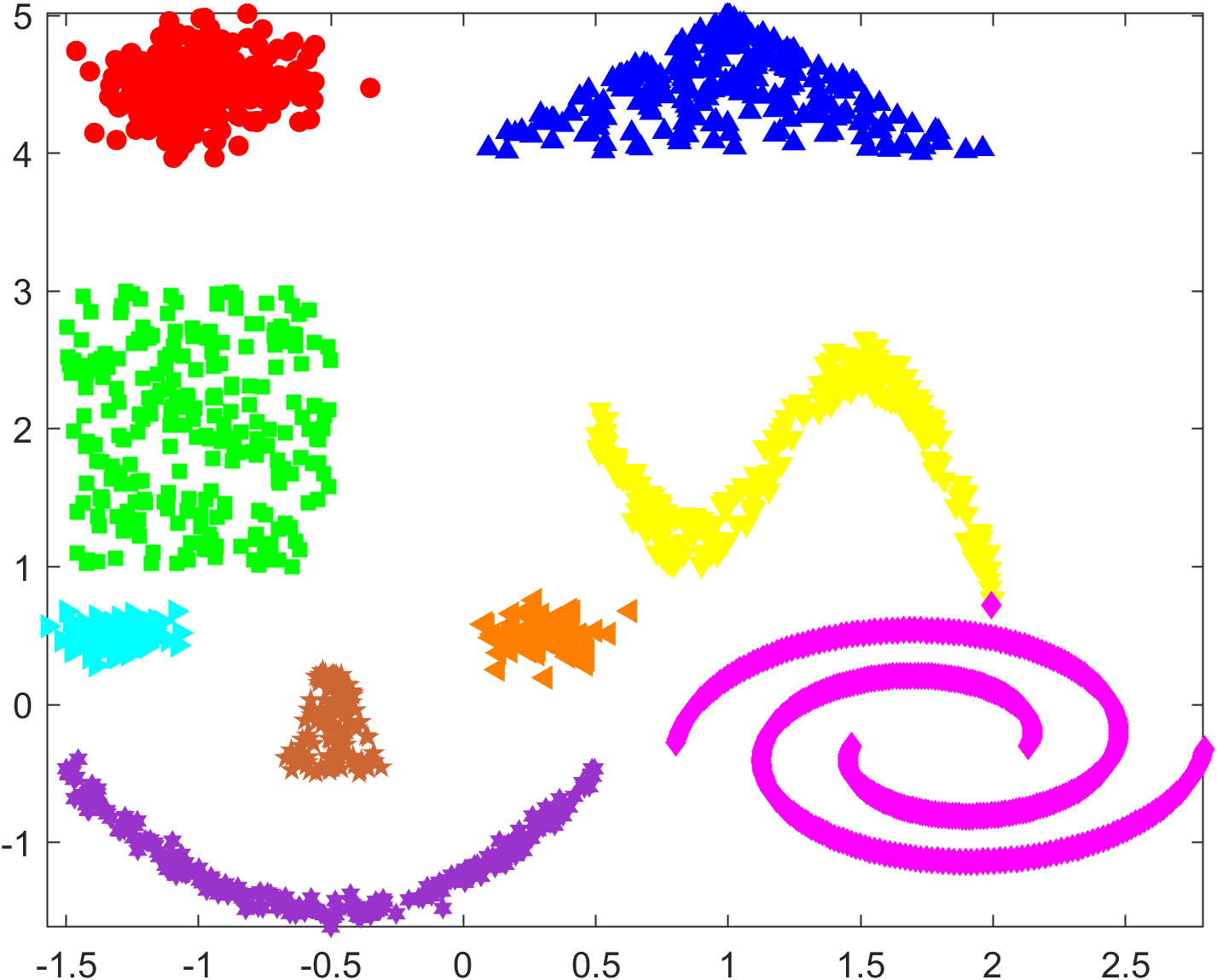}} &
    \subfloat[Chainlink]{\includegraphics[width=0.24\textwidth]{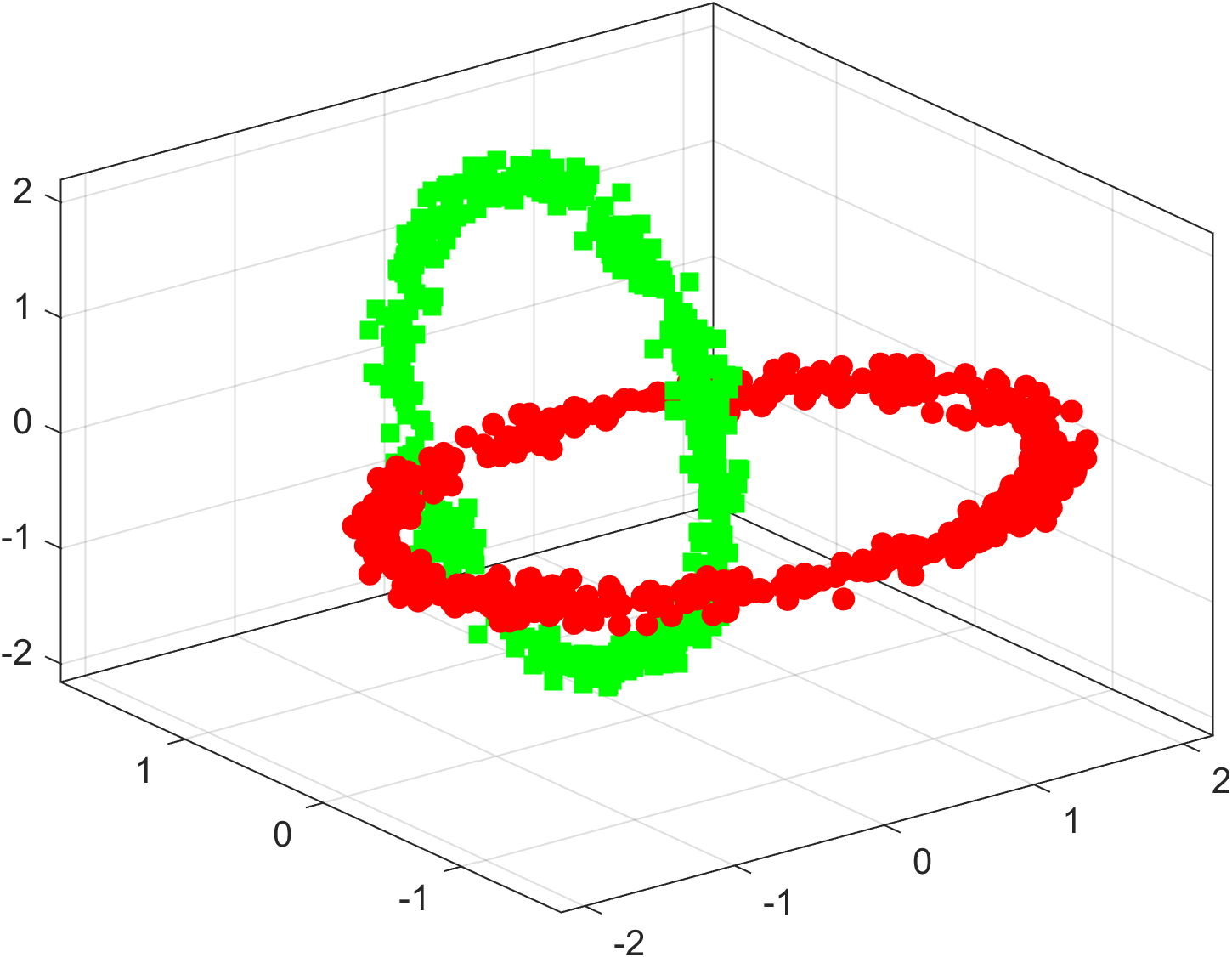}} &
    \subfloat[Cuboids]{\includegraphics[width=0.24\textwidth]{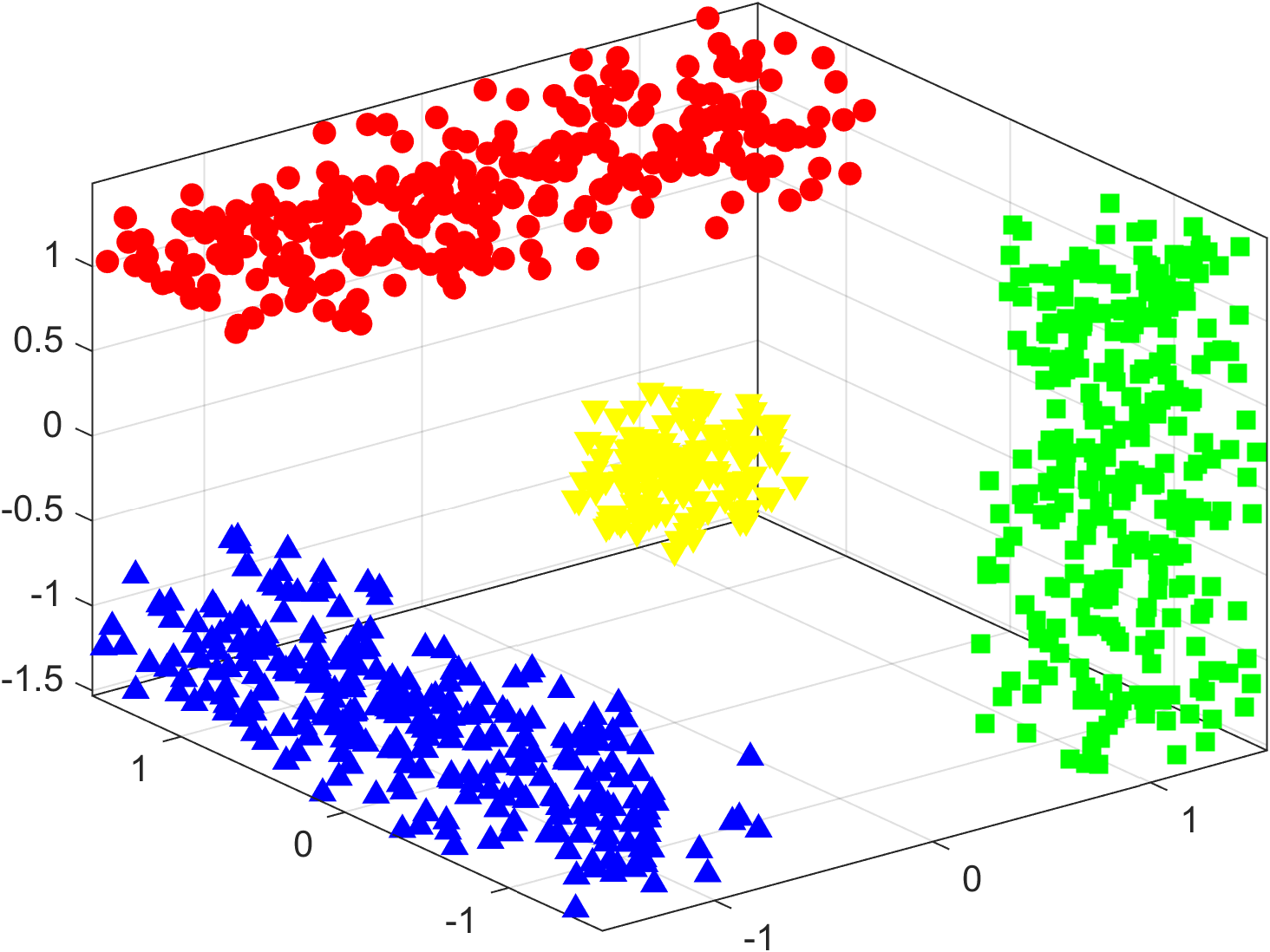}} &
    \subfloat[Diamond]{\includegraphics[width=0.24\textwidth]{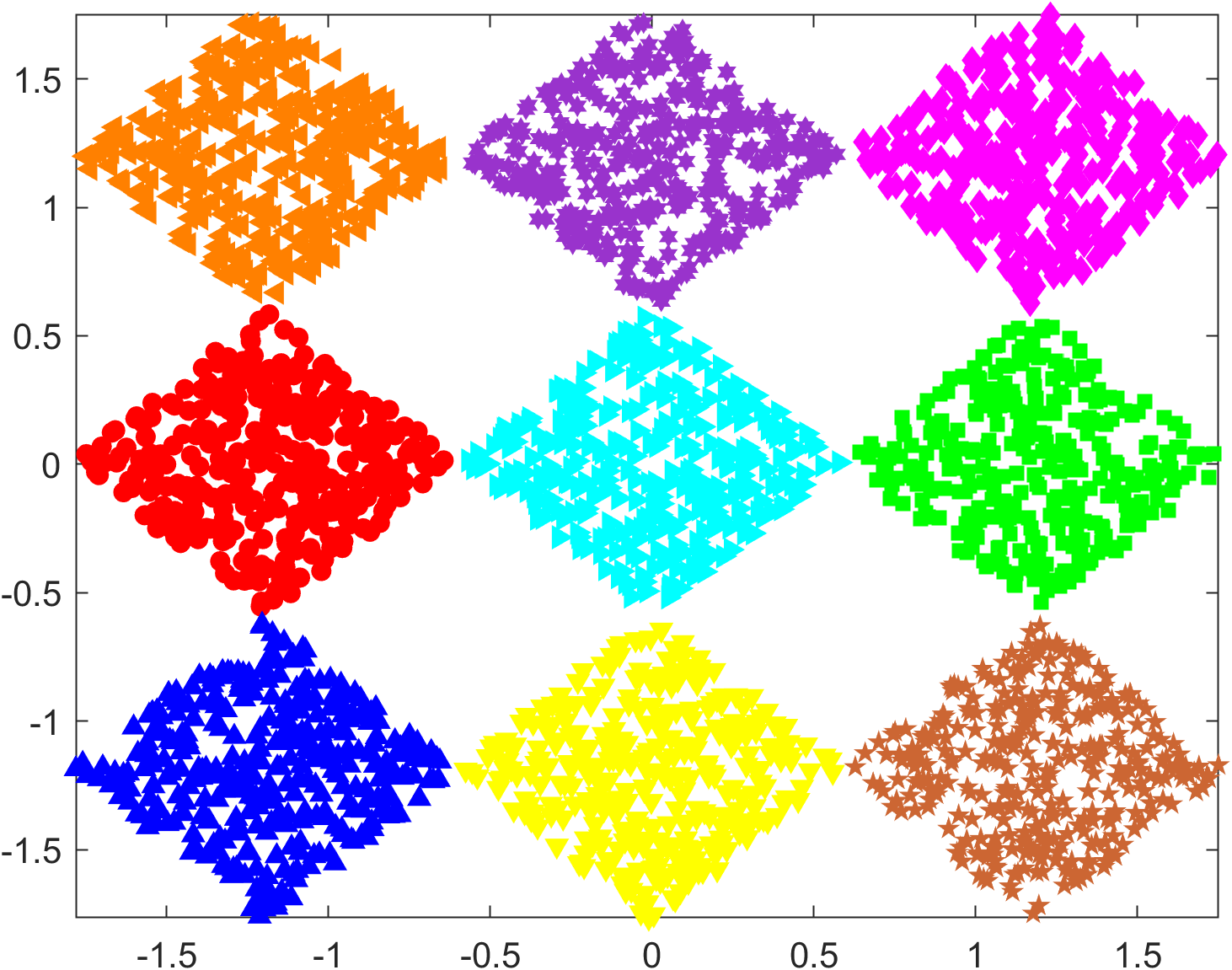}} \\

    \subfloat[Flame]{\includegraphics[width=0.24\textwidth]{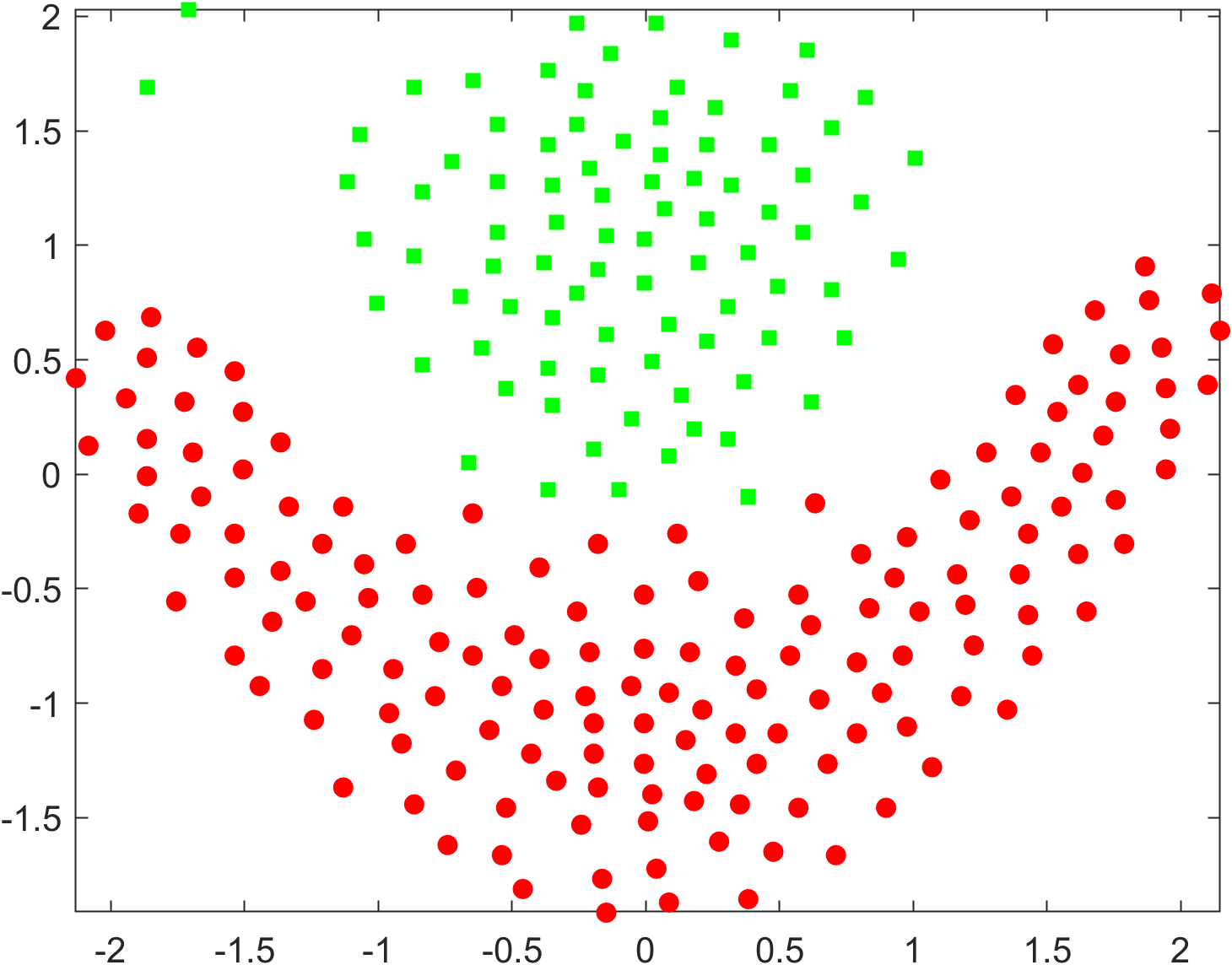}} &
    \subfloat[Jain]{\includegraphics[width=0.24\textwidth]{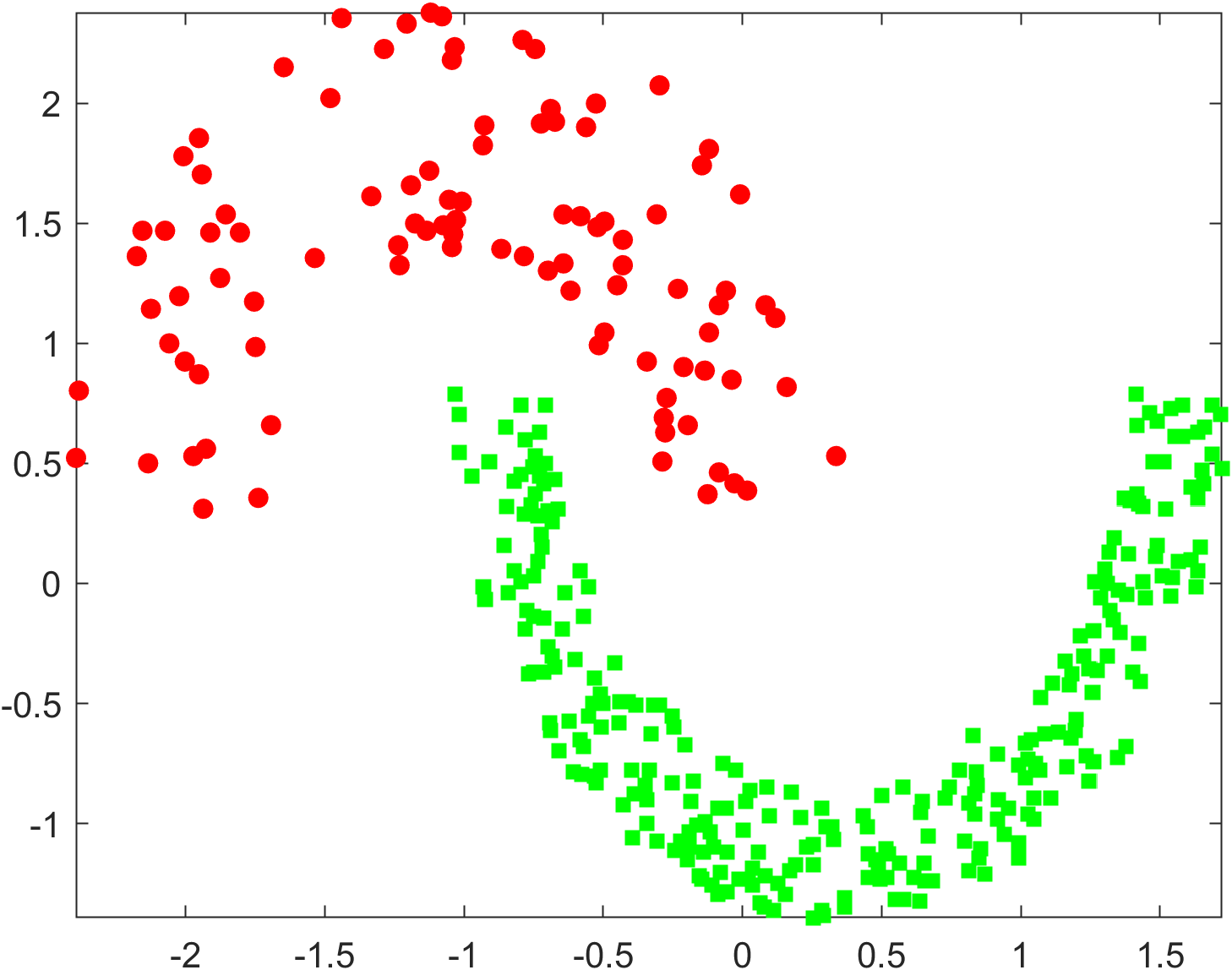}} &
    \subfloat[Ray]{\includegraphics[width=0.24\textwidth]{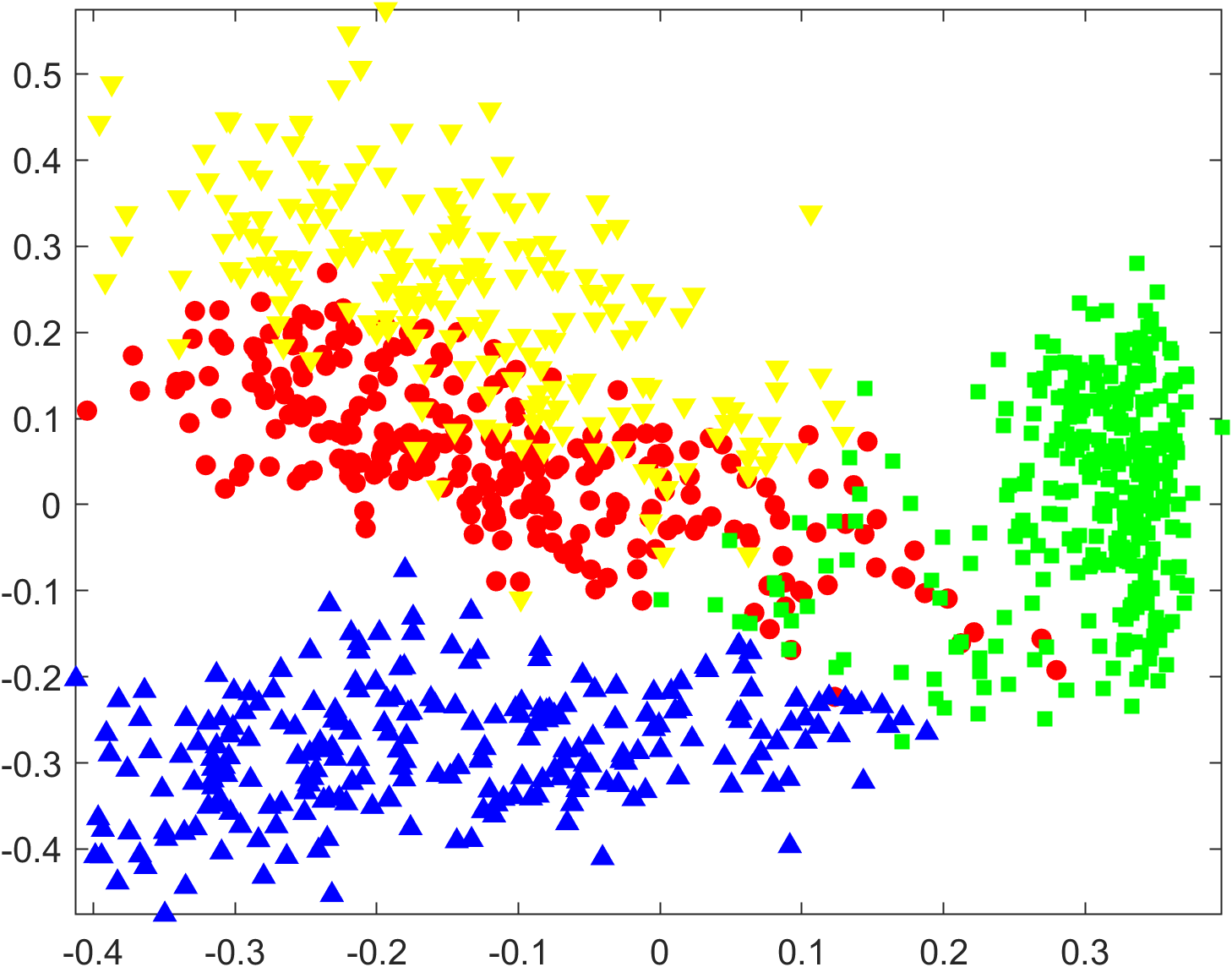}} &
    \subfloat[Starbeam]{\includegraphics[width=0.24\textwidth]{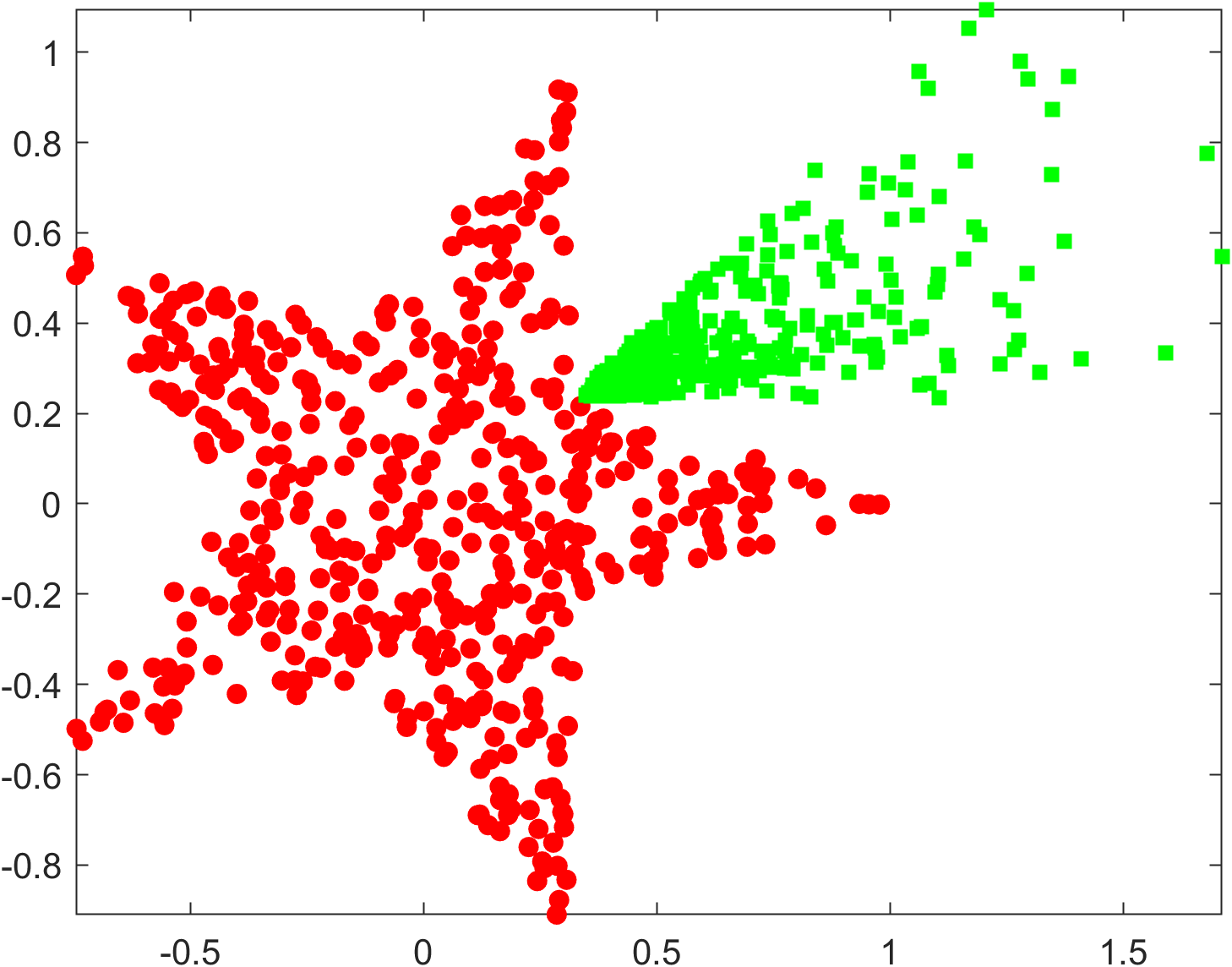}} \\
\end{tabular}
    \caption{DLCC clustering visualizations on 12 synthetic datasets. For the Ray dataset ($d=5$), the plot is produced by projecting the data onto the first two principal components.}
    \label{fig:dlcc_synth_vis}
\end{figure*}

We evaluate DLCC on $12$ synthetic datasets designed to cover a broad range of clustering scenarios and to give insights on both the strengths and limitations of the proposed DLCC algorithm. Figure~\ref{fig:dlcc_synth_vis} visualizes the DLCC partitions. The datasets 3 blobs, Aggregation, Atom, Chainlink, Cuboids, Diamond, Flame, and Jain are benchmark examples from~\cite{fernandes2021towards}. Bainba and Blend are generated using the {\tt mlbench} package~\cite{Rmlbench2021} in {\sf R}~\cite{R2021}. Ray is adapted from a spectral-clustering study~\cite{passino2022spectral}, and Starbeam is constructed by us.

Following the order in Figure~\ref{fig:dlcc_synth_vis}, the synthetic datasets represent the following clustering challenges:
\begin{enumerate}[label=(\alph*), leftmargin=*, itemsep=0pt, topsep=2pt]
    \item well-separated, convex shapes.
    \item bridge-connected shapes with varying sizes.
    \item  a non-convex component enclosing a convex core.
    \item proximity-driven ambiguity under strong size imbalance.
    \item arbitrary shapes with substantial heterogeneity.
    \item non-convex interlocked rings.
    \item well-separated shapes with different orientations.
    \item close convex shapes.
    \item close non-convex shapes with similar density.
    \item two curved non-convex shapes separated by a narrow gap, with varying densities.
    \item  overlapping/entangled shapes.
    \item  non-convex connected shapes with varying densities.
\end{enumerate}
In these experiments, we adopt the min strategy for 3 blobs, Diamond, and Ray, and the max strategy for the remaining datasets (refer to Section~\ref{sec:strachos} for reasons).

\newlength{\cellWidth}
\setlength{\cellWidth}{\textwidth/10-2\tabcolsep} 
\newlength{\cellWidthFirst}
\setlength{\cellWidthFirst}{3.5\cellWidth} 
\newlength{\cellWidthv}
\setlength{\cellWidthv}{\textwidth/12-2\tabcolsep} 

\begin{table*}[ht]
\caption{Comparative analysis of selected clustering algorithms on synthetic datasets (For methods without a fixed $K$, when the estimated $\hat{K}$ differs from the true $K$, the CE is not meaningful and we report $\hat{K}$ instead).}
\centering
{\setlength{\tabcolsep}{3pt}\renewcommand{\arraystretch}{0.92}%
\scalebox{1}[0.95]{
\begin{tabularx}{\textwidth}{
>{\raggedright\arraybackslash}m{\cellWidthFirst}
>{\raggedright\arraybackslash}m{\cellWidth}
>{\raggedright\arraybackslash}m{\cellWidth}
>{\raggedright\arraybackslash}m{\cellWidth}
>{\raggedright\arraybackslash}m{\cellWidth}
>{\raggedright\arraybackslash}m{\cellWidth}
>{\raggedright\arraybackslash}m{\cellWidth}
>{\raggedright\arraybackslash}m{\cellWidth}
>{\raggedright\arraybackslash}m{\cellWidth}}
\toprule
\textbf{Datasets} & \textbf{Metric} & \textbf{DLCC} & \textbf{LW-k-means} & \textbf{EKM} & \textbf{R-MDPC} & \textbf{GBCT} & \textbf{ACLR} & \textbf{DMS} \\
\midrule
\shortstack[l]{3 blobs\\ $n=300, d=2, K=3$} &
\shortstack[l]{ARI\\NMI\\CE} &
\shortstack[c]{$\boldsymbol{1}$\\$\boldsymbol{1}$\\$\boldsymbol{0.00\%}$} &
\shortstack[c]{$0.9900$\\$0.9830$\\$0.33\%$} &
\shortstack[c]{$\boldsymbol{1}$\\$\boldsymbol{1}$\\$\boldsymbol{0.00\%}$} &
\shortstack[c]{$0.9900$\\$0.9830$\\$0.33\%$} &
\shortstack[c]{$0.9900$\\$0.9830$\\$0.33\%$} &
\shortstack[c]{$\boldsymbol{1}$\\$\boldsymbol{1}$\\$\boldsymbol{0.00\%}$} &
\shortstack[c]{$\boldsymbol{1}$\\$\boldsymbol{1}$\\$\boldsymbol{0.00\%}$} \\
\midrule
\shortstack[l]{Aggregation\\ $n=788, d=2, K=7$} &
\shortstack[l]{ARI\\NMI\\CE} &
\shortstack[c]{$0.9927$\\$0.9883$\\$0.38\%$} &
\shortstack[c]{$0.7434$\\$0.8409$\\$13.07\%$} &
\shortstack[c]{$0.6325$\\$0.7646$\\$29.31\%$} &
\shortstack[c]{$0.8415$\\$0.9091$\\$14.09\%$} &
\shortstack[c]{$0.7359$\\$0.8378$\\$21.57\%$} &
\shortstack[c]{$\boldsymbol{0.9956}$\\$\boldsymbol{0.9923}$\\$\boldsymbol{0.25\%}$} &
\shortstack[c]{$0.9109$\\$0.9503$\\$\hat{K}=6$} \\
\midrule
\shortstack[l]{Atom\\ $n=800, d=3, K=2$} &
\shortstack[l]{ARI\\NMI\\CE} &
\shortstack[c]{$0.0193$\\$0.1220$\\$43.00\%$} &
\shortstack[c]{$0.1929$\\$0.3044$\\$28.00\%$} &
\shortstack[c]{$0.1066$\\$0.2332$\\$33.63\%$} &
\shortstack[c]{$\boldsymbol{1}$\\$\boldsymbol{1}$\\$\boldsymbol{0.00\%}$} &
\shortstack[c]{$\boldsymbol{1}$\\$\boldsymbol{1}$\\$\boldsymbol{0.00\%}$} &
\shortstack[c]{$\boldsymbol{1}$\\$\boldsymbol{1}$\\$\boldsymbol{0.00\%}$} &
\shortstack[c]{$0.7587$\\$0.8201$\\$20.25\%$} \\
\midrule
\shortstack[l]{Bainba\\ $n=600, d=2, K=3$} &
\shortstack[l]{ARI\\NMI\\CE} &
\shortstack[c]{$\boldsymbol{1}$\\$\boldsymbol{1}$\\$\boldsymbol{0.00\%}$} &
\shortstack[c]{$0.9814$\\$0.9686$\\$0.67\%$} &
\shortstack[c]{$0.4600$\\$0.5016$\\$38.50\%$} &
\shortstack[c]{$\boldsymbol{1}$\\$\boldsymbol{1}$\\$\boldsymbol{0.00\%}$} &
\shortstack[c]{$0.7578$\\$0.7764$\\$19.67\%$} &
\shortstack[c]{$\boldsymbol{1}$\\$\boldsymbol{1}$\\$\boldsymbol{0.00\%}$} &
\shortstack[c]{$0.9600$\\$0.9362$\\$\hat{K}=4$} \\
\midrule
\shortstack[l]{Blend\\ $n=2000, d=2, K=10$} &
\shortstack[l]{ARI\\NMI\\CE} &
\shortstack[c]{$0.8579$\\$0.9588$\\$\hat{K}=9$} &
\shortstack[c]{$0.7310$\\$0.8432$\\$19.58\%$} &
\shortstack[c]{$0.5793$\\$0.7363$\\$32.50\%$} &
\shortstack[c]{$0.8337$\\$0.9448$\\$16.35\%$} &
\shortstack[c]{$0.9656$\\$0.9795$\\$1.45\%$} &
\shortstack[c]{$\boldsymbol{1}$\\$\boldsymbol{1}$\\$\boldsymbol{0.00\%}$} &
\shortstack[c]{$0.8218$\\$0.8985$\\$\hat{K}=14$} \\
\midrule
\shortstack[l]{Chainlink\\ $n=1000, d=3, K=2$} &
\shortstack[l]{ARI\\NMI\\CE} &
\shortstack[c]{$\boldsymbol{1}$\\$\boldsymbol{1}$\\$\boldsymbol{0.00\%}$} &
\shortstack[c]{$-0.0010$\\$0.0000$\\$49.80\%$} &
\shortstack[c]{$-0.0010$\\$0.0000$\\$49.70\%$} &
\shortstack[c]{$\boldsymbol{1}$\\$\boldsymbol{1}$\\$\boldsymbol{0.00\%}$} &
\shortstack[c]{$\boldsymbol{1}$\\$\boldsymbol{1}$\\$\boldsymbol{0.00\%}$} &
\shortstack[c]{$\boldsymbol{1}$\\$\boldsymbol{1}$\\$\boldsymbol{0.00\%}$} &
\shortstack[c]{$\boldsymbol{1}$\\$\boldsymbol{1}$\\$\boldsymbol{0.00\%}$} \\
\midrule
\shortstack[l]{Cuboids\\ $n=1002, d=3, K=4$} &
\shortstack[l]{ARI\\NMI\\CE} &
\shortstack[c]{$\boldsymbol{1}$\\$\boldsymbol{1}$\\$\boldsymbol{0.00\%}$} &
\shortstack[c]{$\boldsymbol{1}$\\$\boldsymbol{1}$\\$\boldsymbol{0.00\%}a$} &
\shortstack[c]{$0.6713$\\$0.7390$\\$28.14\%$} &
\shortstack[c]{$\boldsymbol{1}$\\$\boldsymbol{1}$\\$\boldsymbol{0.00\%}$} &
\shortstack[c]{$\boldsymbol{1}$\\$\boldsymbol{1}$\\$\boldsymbol{0.00\%}$} &
\shortstack[c]{$\boldsymbol{1}$\\$\boldsymbol{1}$\\$\boldsymbol{0.00\%}$} &
\shortstack[c]{$\boldsymbol{1}$\\$\boldsymbol{1}$\\$\boldsymbol{0.00\%}$} \\
\midrule
\shortstack[l]{Diamond\\ $n=3000, d=2, K=9$} &
\shortstack[l]{ARI\\NMI\\CE} &
\shortstack[c]{$\boldsymbol{1}$\\$\boldsymbol{1}$\\$\boldsymbol{0.00\%}$} &
\shortstack[c]{$\boldsymbol{1}$\\$\boldsymbol{1}$\\$\boldsymbol{0.00\%}$} &
\shortstack[c]{$0.7855$\\$0.8600$\\$16.33\%$} &
\shortstack[c]{$0.9992$\\$0.9990$\\$0.03\%$} &
\shortstack[c]{$0.8758$\\$0.9473$\\$10.27\%$} &
\shortstack[c]{$\boldsymbol{1}$\\$\boldsymbol{1}$\\$\boldsymbol{0.00\%}$} &
\shortstack[c]{$\boldsymbol{1}$\\$\boldsymbol{1}$\\$\boldsymbol{0.00\%}$} \\
\midrule
\shortstack[l]{Flame\\ $n=240, d=2, K=2$} &
\shortstack[l]{ARI\\NMI\\CE} &
\shortstack[c]{$0.9666$\\$0.9269$\\$0.83\%$} &
\shortstack[c]{$0.4880$\\$0.4422$\\$15.00\%$} &
\shortstack[c]{$0.3988$\\$0.3544$\\$18.33\%$} &
\shortstack[c]{$\boldsymbol{1}$\\$\boldsymbol{1}$\\$\boldsymbol{0.00\%}$} &
\shortstack[c]{$0.9501$\\$0.8991$\\$1.25\%$} &
\shortstack[c]{$0.9666$\\$0.9269$\\$0.83\%$} &
\shortstack[c]{$0.9666$\\$0.9269$\\$0.83\%$} \\
\midrule
\shortstack[l]{Jain\\ $n=373, d=2, K=2$} &
\shortstack[l]{ARI\\NMI\\CE} &
\shortstack[c]{$\boldsymbol{1}$\\$\boldsymbol{1}$\\$\boldsymbol{0.00\%}$} &
\shortstack[c]{$0.6352$\\$0.5908$\\$9.92\%$} &
\shortstack[c]{$0.5681$\\$0.5088$\\$12.06\%$} &
\shortstack[c]{$\boldsymbol{1}$\\$\boldsymbol{1}$\\$\boldsymbol{0.00\%}$} &
\shortstack[c]{$0.2563$\\$0.2463$\\$19.03\%$} &
\shortstack[c]{$\boldsymbol{1}$\\$\boldsymbol{1}$\\$\boldsymbol{0.00\%}$} &
\shortstack[c]{$0.7146$\\$0.6315$\\$\hat{K}=3$} \\
\midrule
\shortstack[l]{Ray\\ $n=1000, d=5, K=4$} &
\shortstack[l]{ARI\\NMI\\CE} &
\shortstack[c]{$\boldsymbol{0.7577}$\\$\boldsymbol{0.7541}$\\$\boldsymbol{9.65\%}$} &
\shortstack[c]{$0.7090$\\$0.7439$\\$11.90\%$} &
\shortstack[c]{$0.4189$\\$0.4847$\\$31.40\%$} &
\shortstack[c]{$0.5364$\\$0.6175$\\$37.70\%$} &
\shortstack[c]{$0.6117$\\$0.6595$\\$21.10\%$} &
\shortstack[c]{$0.6794$\\$0.7013$\\$13.10\%$} &
\shortstack[c]{$0.2814$\\$0.4245$\\$\hat{K}=14$} \\
\midrule
\shortstack[l]{Starbeam\\ $n=780, d=2, K=2$} &
\shortstack[l]{ARI\\NMI\\CE} &
\shortstack[c]{$\boldsymbol{1}$\\$\boldsymbol{1}$\\$\boldsymbol{0.00\%}$} &
\shortstack[c]{$0.4822$\\$0.5072$\\$15.26\%$} &
\shortstack[c]{$0.4680$\\$0.4982$\\$15.77\%$} &
\shortstack[c]{$0.5957$\\$0.4671$\\$11.28\%$} &
\shortstack[c]{$-0.0248$\\$0.0379$\\$39.23\%$} &
\shortstack[c]{$0.9948$\\$0.9859$\\$0.13\%$} &
\shortstack[c]{$0.2823$\\$0.4351$\\$\hat{K}=4$} \\
\bottomrule
\end{tabularx}
}}
\label{table:comparison_synthetic}
\end{table*}

Table~\ref{table:comparison_synthetic} summarizes the clustering performance on the $12$ synthetic benchmarks. Overall, DLCC delivers consistently strong results and remains competitive across a broad range of challenges. In particular, DLCC attains perfect recovery on $7$ datasets, next to ACLR (an advanced graph-based method), and achieves favorable ARI/NMI with low CE on most of the remaining examples, illustrating its ability to handle both convex and non-convex structures as well as varying densities and close proximity.

Nevertheless, DLCC faces difficulties on Atom and on the spiral structures in Blend. As discussed in Section~\ref{sec:lcs}, the local neighborhoods used to define local centers are expected to be approximately convex. This assumption may fail for manifold-like structures, where the neighborhood geometry around most points is non-convex. A representative failure case is Atom, in which the depth-defined local centers fail to adequately capture this geometry. A similar issue arises for the spiral components in Blend. This limitation of depth-defined local centers explains the observed performance degradation and highlights an inherent weakness of the current DLCC model.

\subsection{Real-world Datasets}
As for the real-world benchmarks, we evaluate DLCC on $13$ datasets. Specifically, Iris, Seed, Wine, Parkinsons (Pa), Breast Cancer (BC), Segmentation (Seg), Optidigits, and Pendigits are obtained from the UCI Machine Learning Repository~\cite{Dua:2019}. A highly imbalanced dataset, Anuran Calls (Anuran) is also included, whose cluster sizes ranging from as few as $68$ to as many as $3478$~\cite{diaz2012compressive}. We further include high-dimensional datasets ($d \geq 100$) with $n > d$, namely Yale~B~\cite{georghiades2001few} and HAR~\cite{Anguita2013APD}, representing face images and human-activity signals, respectively. Among these, Yale~B, Optidigits, and Pendigits are accessed via the {\sf R} package {\tt PPCI}~\cite{hofmeyr2019ppci}. In addition, we consider two microarray datasets, Leukemia~\cite{golub1999molecular} and SuCancer~\cite{su2001molecular}, which are characterized by extremely high dimensionality and small sample sizes. Overall, this benchmark suite is representative, spanning tabular, image, and time-series datasets, as well as a wide range of dimensionalities (low-dimensional to microarray-scale) and sample sizes.

For these real-world datasets, we adopt the max strategy for Pa, Anuran, Leukemia, and SuCancer due to severe class imbalance, and use the min strategy for all remaining datasets. Table~\ref{table:comparison_real} shows that DLCC is consistently competitive across the $13$ real-world benchmarks. It achieves the best performance on $6$ datasets and ranks within the top three on all datasets. DLCC also proves highly adaptable for high-dimensional data applications, attaining the strongest results on Yale~B and HAR; on SuCancer it is still among the best, although the absolute performance is modest for all methods.

The table further indicates that competing methods are often dataset-dependent: some methods can be outstanding on particular datasets (e.g., EKM on Wine, R-MDPC on Optidigits, and ACLR on several benchmarks), yet no single method dominates uniformly. This is largely because different algorithms have different notions of what constitutes a ``cluster'' (e.g., centroid-based compactness, graph connectivity, or density/shape regularity), which may align well with some datasets but not with others. This variability underscores the practical value of DLCC as a robust alternative with strong performance across different real-world scenarios.

\begin{table*}[ht]
\caption{Comparative analysis of selected clustering algorithms on real datasets.}
\centering
{\setlength{\tabcolsep}{3pt}\renewcommand{\arraystretch}{0.92}%
\scalebox{1}[0.95]{
\begin{tabularx}{\textwidth}{
>{\raggedright\arraybackslash}m{\cellWidthFirst}
>{\raggedright\arraybackslash}m{\cellWidth}
>{\raggedright\arraybackslash}m{\cellWidth}
>{\raggedright\arraybackslash}m{\cellWidth}
>{\raggedright\arraybackslash}m{\cellWidth}
>{\raggedright\arraybackslash}m{\cellWidth}
>{\raggedright\arraybackslash}m{\cellWidth}
>{\raggedright\arraybackslash}m{\cellWidth}
>{\raggedright\arraybackslash}m{\cellWidth}}
\toprule
\textbf{Datasets} & \textbf{Metric} & \textbf{DLCC} & \textbf{LW-k-means} & \textbf{EKM} & \textbf{R-MDPC} & \textbf{GBCT} & \textbf{ACLR} & \textbf{DMS} \\
\midrule

\shortstack[l]{Iris\\ $n=150, d=4, K=3$} &
\shortstack[l]{ARI\\NMI\\CE} &
\shortstack[c]{$\boldsymbol{0.8857}$\\$\boldsymbol{0.8642}$\\$\boldsymbol{4.00\%}$} &
\shortstack[c]{$\boldsymbol{0.8857}$\\$\boldsymbol{0.8642}$\\$\boldsymbol{4.00\%}$} &
\shortstack[c]{$0.6101$\\$0.6526$\\$17.33\%$} &
\shortstack[c]{$\boldsymbol{0.8857}$\\$\boldsymbol{0.8642}$\\$\boldsymbol{4.00\%}$} &
\shortstack[c]{$0.5638$\\$0.7038$\\$26.00\%$} &
\shortstack[c]{$0.7592$\\$0.7960$\\$9.33\%$} &
\shortstack[c]{$0.6956$\\$0.7711$\\$12.67\%$} \\
\midrule

\shortstack[l]{Wine\\ $n=178, d=13, K=3$} &
\shortstack[l]{ARI\\NMI\\CE} &
\shortstack[c]{$0.8975$\\$0.8759$\\$3.37\%$} &
\shortstack[c]{$0.8516$\\$0.8267$\\$5.06\%$} &
\shortstack[c]{$\boldsymbol{0.9134}$\\$\boldsymbol{0.8920}$\\$\boldsymbol{2.81\%}$} &
\shortstack[c]{$0.7269$\\$0.7435$\\$9.55\%$} &
\shortstack[c]{$0.7033$\\$0.7397$\\$10.67\%$} &
\shortstack[c]{$0.4041$\\$0.4127$\\$26.97\%$} &
\shortstack[c]{$0.2748$\\$0.3933$\\$36.52\%$} \\
\midrule

\shortstack[l]{Seed\\ $n=210, d=7, K=3$} &
\shortstack[l]{ARI\\NMI\\CE} &
\shortstack[c]{$0.7868$\\$0.7409$\\$\boldsymbol{7.62}\%$} &
\shortstack[c]{$0.6687$\\$0.6728$\\$12.86\%$} &
\shortstack[c]{$0.7715$\\$0.7315$\\$8.10\%$} &
\shortstack[c]{$0.7289$\\$0.6946$\\$10.00\%$} &
\shortstack[c]{$0.1614$\\$0.3061$\\$49.52\%$} &
\shortstack[c]{$0.7132$\\$0.6841$\\$10.48\%$} &
\shortstack[c]{$\boldsymbol{0.7870}$\\$\boldsymbol{0.7423}$\\$\boldsymbol{7.62}\%$} \\
\midrule

\shortstack[l]{Pa\\ $n=195, d=22, K=2$} &
\shortstack[l]{ARI\\NMI\\CE} &
\shortstack[c]{$\boldsymbol{0.4226}$\\$\boldsymbol{0.3265}$\\$\boldsymbol {14.36}\%$} &
\shortstack[c]{$-0.0937$\\$0.1189$\\$44.10\%$} &
\shortstack[c]{$-0.0710$\\$0.0558$\\$31.79\%$} &
\shortstack[c]{$0.1256$\\$0.2776$\\$31.79\%$} &
\shortstack[c]{$-0.0515$\\$0.0360$\\$29.23\%$} &
\shortstack[c]{$0.2890$\\$0.1965$\\$17.44\%$} &
\shortstack[c]{$0.0499$\\$0.1767$\\$\hat{K}=3$} \\
\midrule

\shortstack[l]{BC\\ $n=569, d=30, K=2$} &
\shortstack[l]{ARI\\NMI\\CE} &
\shortstack[c]{$0.7361$\\$0.6352$\\$7.03\%$} &
\shortstack[c]{$0.7177$\\$0.6215$\\$7.56\%$} &
\shortstack[c]{$0.6466$\\$0.5532$\\$9.67\%$} &
\shortstack[c]{$0.5175$\\$0.5009$\\$13.71\%$} &
\shortstack[c]{$-0.0302$\\$0.0553$\\$42.18\%$} &
\shortstack[c]{$\boldsymbol{0.8241}$\\$\boldsymbol{0.7324}$\\$\boldsymbol{4.57\%}$} &
\shortstack[c]{$0.0024$\\$0.0188$\\$37.08\%$} \\
\midrule

\shortstack[l]{Seg\\ $n=2086, d=18, K=7$} &
\shortstack[l]{ARI\\NMI\\CE} &
\shortstack[c]{$\boldsymbol{0.6278}$\\$\boldsymbol{0.6971}$\\$\boldsymbol{21.86}\%$} &
\shortstack[c]{$0.4782$\\$0.6141$\\$41.04\%$} &
\shortstack[c]{$0.4862$\\$0.6839$\\$43.29\%$} &
\shortstack[c]{$0.5832$\\$0.7265$\\$34.42\%$} &
\shortstack[c]{$0.2407$\\$0.5731$\\$53.40\%$} &
\shortstack[c]{$0.3829$\\$0.5133$\\$48.95\%$} &
\shortstack[c]{$0.4981$\\$0.6178$\\$\hat{K}=8$} \\
\midrule

\shortstack[l]{Yale B\\ $n=2000, d=600, K=10$} &
\shortstack[l]{ARI\\NMI\\CE} &
\shortstack[c]{$\boldsymbol{0.9846}$\\$\boldsymbol{0.9851}$\\$\boldsymbol{0.70\%}$} &
\shortstack[c]{$0.8461$\\$0.9306$\\$13.25\%$} &
\shortstack[c]{$0.1549$\\$0.4090$\\$60.20\%$} &
\shortstack[c]{$0.7427$\\$0.8765$\\$17.35\%$} &
\shortstack[c]{$0.1725$\\$0.5722$\\$61.10\%$} &
\shortstack[c]{$0.8421$\\$0.9097$\\$14.65\%$} &
\shortstack[c]{$0.5882$\\$0.7744$\\$33.80\%$} \\
\midrule

\shortstack[l]{Optidigits\\ $n=5620, d=64, K=10$} &
\shortstack[l]{ARI\\NMI\\CE} &
\shortstack[c]{$0.8233$\\$0.8394$\\$8.83\%$} &
\shortstack[c]{$0.4923$\\$0.6486$\\$35.96\%$} &
\shortstack[c]{$0.0144$\\$0.1290$\\$82.99\%$} &
\shortstack[c]{$\boldsymbol{0.8543}$\\$\boldsymbol{0.8978}$\\$\boldsymbol{8.75\%}$} &
\shortstack[c]{$0.0003$\\$0.0426$\\$88.19\%$} &
\shortstack[c]{$0.7326$\\$0.7900$\\$22.14\%$} &
\shortstack[c]{$0.5671$\\$0.6549$\\$\hat{K}=11$} \\
\midrule

\shortstack[l]{Pendigits\\ $n=10992, d=16, K=10$} &
\shortstack[l]{ARI\\NMI\\CE} &
\shortstack[c]{$0.6649$\\$0.7433$\\$18.31\%$} &
\shortstack[c]{$0.5749$\\$0.6912$\\$27.85\%$} &
\shortstack[c]{$0.5174$\\$0.6484$\\$30.04\%$} &
\shortstack[c]{$0.6739$\\$0.7840$\\$19.17\%$} &
\shortstack[c]{$0.0013$\\$0.0754$\\$86.14\%$} &
\shortstack[c]{$\boldsymbol{0.7695}$\\$\boldsymbol{0.8368}$\\$\boldsymbol{12.50\%}$} &
\shortstack[c]{$0.6696$\\$0.7497$\\$\hat{K}=17$} \\
\midrule

\shortstack[l]{HAR\\ $n=10299, d=561, K=6$} &
\shortstack[l]{ARI\\NMI\\CE} &
\shortstack[c]{$\boldsymbol{0.6693}$\\$\boldsymbol{0.7248}$\\$\boldsymbol{17.74}\%$} &
\shortstack[c]{$0.4494$\\$0.5873$\\$42.81\%$} &
\shortstack[c]{$0.2932$\\$0.4633$\\$55.20\%$} &
\shortstack[c]{$0.4995$\\$0.6468$\\$35.51\%$} &
\shortstack[c]{$0.3279$\\$0.5455$\\$64.21\%$} &
\shortstack[c]{$0.5994$\\$0.6355$\\$35.74\%$} &
\shortstack[c]{$0.4075$\\$0.5633$\\$\hat{K}=9$} \\
\midrule

\shortstack[l]{Anuran\\ $n=7195, d=22, K=10$} &
\shortstack[l]{ARI\\NMI\\CE} &
\shortstack[c]{$0.8592$\\$0.7146$\\$\hat{K}=11$} &
\shortstack[c]{$0.5859$\\$0.6551$\\$34.37\%$} &
\shortstack[c]{$0.5145$\\$0.6229$\\$58.21\%$} &
\shortstack[c]{$0.7895$\\$0.6489$\\$25.70\%$} &
\shortstack[c]{$0.0289$\\$0.0727$\\$50.02\%$} &
\shortstack[c]{$\boldsymbol{0.8737}$\\$\boldsymbol{0.7559}$\\$\boldsymbol{24.04\%}$} &
\shortstack[c]{$0.8296$\\$0.6857$\\$\hat{K}=9$} \\
\midrule

\shortstack[l]{Leukemia\\ $n=72, d=3571, K=2$} &
\shortstack[l]{ARI\\NMI\\CE} &
\shortstack[c]{$0.8897$\\$0.8056$\\$2.78\%$} &
\shortstack[c]{$0.8897$\\$0.8056$\\$2.78\%$} &
\shortstack[c]{$0.2127$\\$0.2178$\\$26.39\%$} &
\shortstack[c]{$0.7851$\\$0.6790$\\$5.56\%$} &
\shortstack[c]{$0.0095$\\$0.0036$\\$34.72\%$} &
\shortstack[c]{$\boldsymbol{0.9440}$\\$\boldsymbol{0.8954}$\\$\boldsymbol{1.39\%}$} &
\shortstack[c]{$0.5905$\\$0.5536$\\$11.11\%$} \\
\midrule

\shortstack[l]{SuCancer\\ $n=174, d=7909, K=2$} &
\shortstack[l]{ARI\\NMI\\CE} &
\shortstack[c]{$\boldsymbol{0.1154}$\\$\boldsymbol{0.2317}$\\$\boldsymbol{32.76}\%$} &
\shortstack[c]{$-0.0030$\\$0.0041$\\$47.70\%$} &
\shortstack[c]{$-0.0045$\\$0.0006$\\$48.28\%$} &
\shortstack[c]{$\boldsymbol{0.1154}$\\$\boldsymbol{0.2317}$\\$\boldsymbol{32.76}\%$} &
\shortstack[c]{$-0.0022$\\$0.0290$\\$49.43\%$} &
\shortstack[c]{$\boldsymbol{0.1154}$\\$0.1808$\\$\boldsymbol{32.76\%}$} &
\shortstack[c]{$0.0169$\\$0.0214$\\$42.53\%$} \\
\bottomrule
\end{tabularx}
}}
\label{table:comparison_real}
\end{table*}

\section{Conclusions and Discussions}\label{sec:conclu}
In this paper, we have proposed an innovative clustering framework, DLCC, which first defines a depth-based similarity matrix to provide orderings of neighbors for each observation. DLCC then uses depth to define representative points, or ``local centers'', with the number of these local centers controlled by the parameter indicating the neighborhood size, $s$. Our results show that DLCC delivers superior clustering performances under a wide range of scenarios. This includes handling both convex and non-convex shapes, managing clusters of balanced and unbalanced sizes, working with well-separated or overlapping clusters, and analyzing clustering problems in both low and high-dimensional spaces. 

Despite these advantages, DLCC has several limitations that deserve further investigation. First, although DLCC can handle datasets with on the order of $10^4$ observations, its computational cost is dominated by constructing the depth-based similarity matrix, which can limit scalability to substantially larger datasets. One potential direction is to incorporate fast approximate neighborhood search to estimate neighborhoods of points for defining local centers (e.g.,~\cite{fu2016efanna}), at the cost of weakening some desirable properties of depth-induced neighborhoods (e.g., invariance under similarity transformations for SD). Another direction is to adopt a sampling or landmark strategy i.e., estimate local centers from a representative subset. Then compute depth-based similarities only between each observation and these local centers, thereby avoiding the full $n\times n$ similarity matrix construction.

Second, DLCC currently requires user intervention in selecting parametes, the neighborhood size $s$ (and, under the max strategy, the threshold $\delta$), typically based on evaluations of clustering outputs. This reduces ease of use in practice. A promising future direction is an adaptive DLCC that automatically determines observation-specific neighborhood sizes, performs grouping in a data-driven manner, and reduces or eliminates manual strategy selection.

Finally, while this paper focuses on spatial depth due to its computational simplicity in high-dimensional spaces, the DLCC framework itself is not tied to a single depth notion. It would be of interest to explore alternative depth definitions, such as integrated rank-weighted depth~\cite{ramsay2019integrated}, and to study how different depth choices affect the performance of DLCC across diverse data sets.

\section*{Acknowledgments}
This work is supported by the Canada Research Chairs program (PM), a Dorothy Killiam Fellowship (PM), and respective NSERC Discovery Grants (AL, PM).

%\section{References}

%\bibliography{references}
%\bibliographystyle{IEEEtran}


\section*{Supplementary materials}

\subsection*{Classification method selection}
The last stage of our proposed DLCC algorithm is classification based on pseudo-labels (temporary clusters). For the synthetic data sets under the max strategy, we simply apply k-nearest neighborhood (kNN) on the similarity matrix $\bbS$, since kNN is well suited to non-convex and arbitrarily shaped clusters by relying only on local neighborhood information. For the min-strategy synthetic data sets and all real-world data sets, we consider both kNN and random forests (RF), and select the classification method using the CVDD criterion. Specifically, we compute the CVDD score under each candidate classifier and choose the one with the larger value. In the event of a tie, we select kNN, as it is deterministic (no randomness) and computationally more efficient than RF.

Table~\ref{tab:cvdd_select} reports the CVDD values for kNN and RF across all data sets, along with the selected classifier.
\begin{table}[ht]
\centering
\caption{CVDD-based selection of the classifier in the final DLCC stage. For each data set, we report the CVDD scores under kNN and RF, and select the classifier with the larger CVDD. Ties are resolved in favor of kNN.}
\label{tab:cvdd_select}
\begin{tabular}{lccc}
\hline
Dataset & CVDD (kNN) & CVDD (RF) & Selected \\
\hline
3\_blobs              & $3459.50$  & $3243.30$  & kNN \\
diamond9              & $12451.00$ & $12531.00$ & RF  \\
simubyPassino         & $529.22$   & $564.42$   & RF  \\
Iris                  & $142.55$   & $154.87$   & RF  \\
seed                  & $83.14$   & $65.80$   & kNN \\
seg                   & $1.69$   & $1.48$   & kNN \\
wine                  & $30.75$   & $27.66$   & kNN \\
bc                    & $9.10$   & $7.22$   & kNN \\
leukemia              & $3.25$   & $3.25$   & kNN \\
SuCancer              & $1.69$   & $1.69$   & kNN \\
pa                    & $6.86$   & $7.96$   & RF  \\
yale                  & $9.48$   & $10.44$  & RF  \\
digit                 & $7.22$   & $6.82$   & kNN \\
pendigi               & $124.07$ & $73.86$  & kNN \\
har                   & $375.40$ & $562.92$ & RF  \\
frog                  & $3.15$   & $1.68$   & kNN \\
\hline
\end{tabular}
\end{table}
\section*{Graphs}
Here, we provide visualizations for all $25$ data sets described in the paper, including the ground truth labels, grouped local centers, the resulting temporary clusters, and the final DLCC clustering outputs. These graphs are intended to give readers additional insight into the mechanics of DLCC, as well as to highlight the differing behaviors of the min and max strategies. Specifically, the min strategy tends to act more like center-based approaches, whereas the max strategy more closely resembles density-based methods.

For data sets with dimension greater than three, we use tSNE to embed the observations into two dimensions prior to visualization. We caution that inter-point distances in tSNE plots should not be interpreted quantitatively, as the embedding may distort global geometry and can be visually misleading.

% -------- 3_blobs --------
\begin{figure*}[ht]
\centering
\begin{subfigure}[H!]{0.48\textwidth}
\centering
\includegraphics[width=\textwidth]{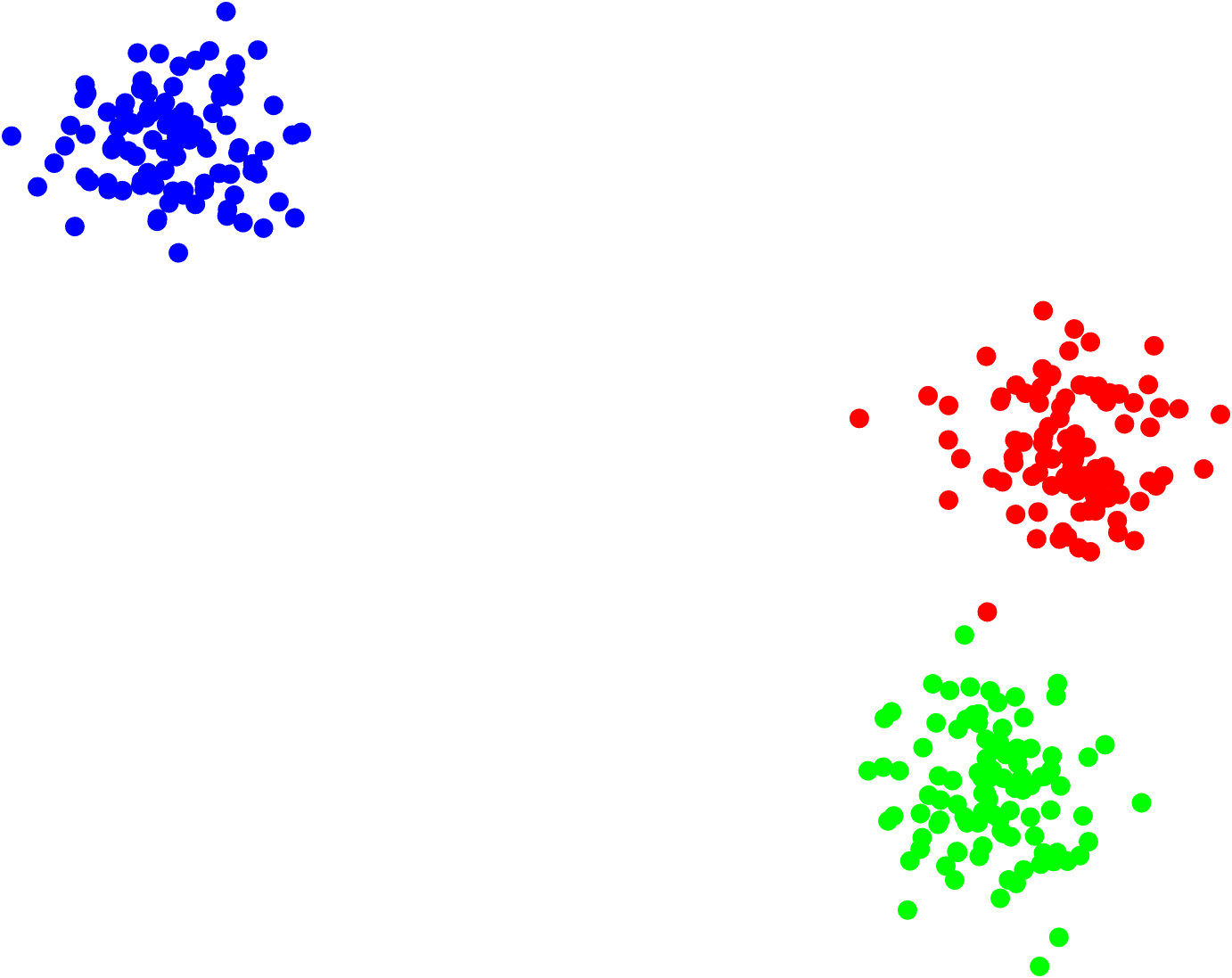}
\caption{}
\end{subfigure}
\hfill
\begin{subfigure}[H!]{0.48\textwidth}
\centering
\includegraphics[width=\textwidth]{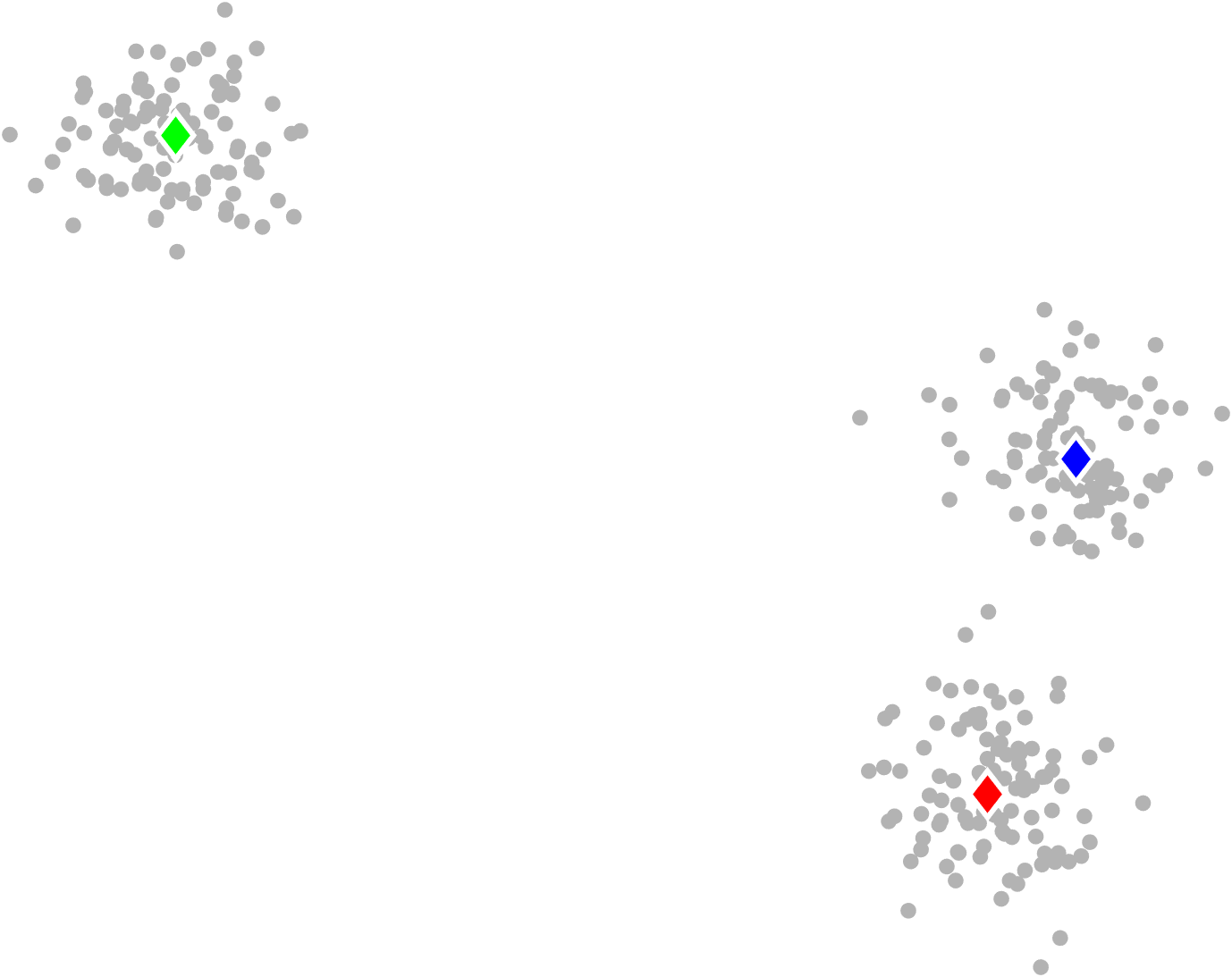}
\caption{}
\end{subfigure}

\begin{subfigure}[H!]{0.48\textwidth}
\centering
\includegraphics[width=\textwidth]{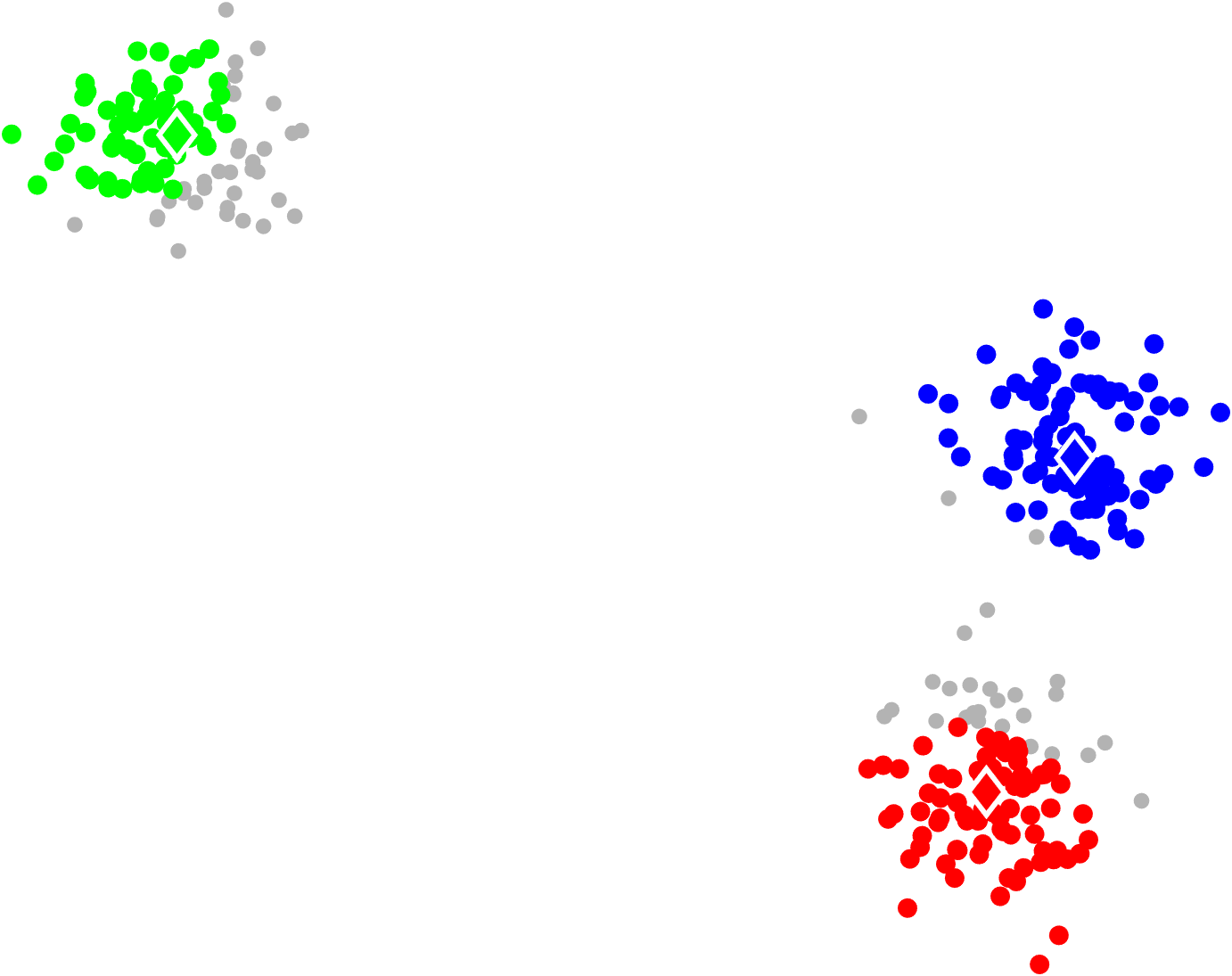}
\caption{}
\end{subfigure}
\hfill
\begin{subfigure}[H!]{0.48\textwidth}
\centering
\includegraphics[width=\textwidth]{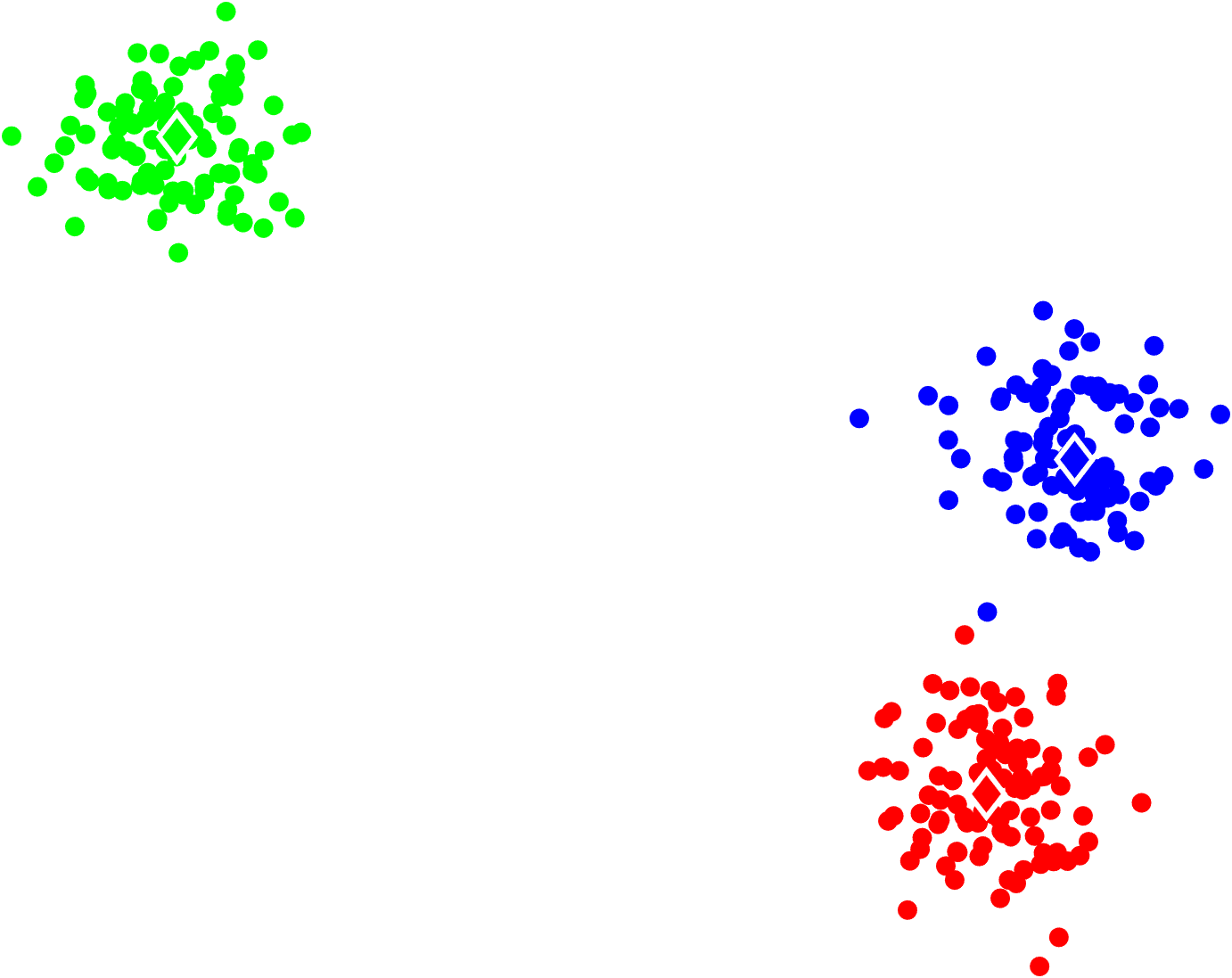}
\caption{}
\end{subfigure}
\caption{DLCC visualization on the \texttt{3 Blobs} dataset. (a) Ground truth labels; (b) Grouped local centers; (c) Temporary clusters; (d) Final DLCC clustering result.}
\label{fig:dlcc-3_blobs}
\end{figure*}

% -------- bc --------
\begin{figure*}[ht]
\centering
\begin{subfigure}[H!]{0.48\textwidth}
\centering
\includegraphics[width=\textwidth]{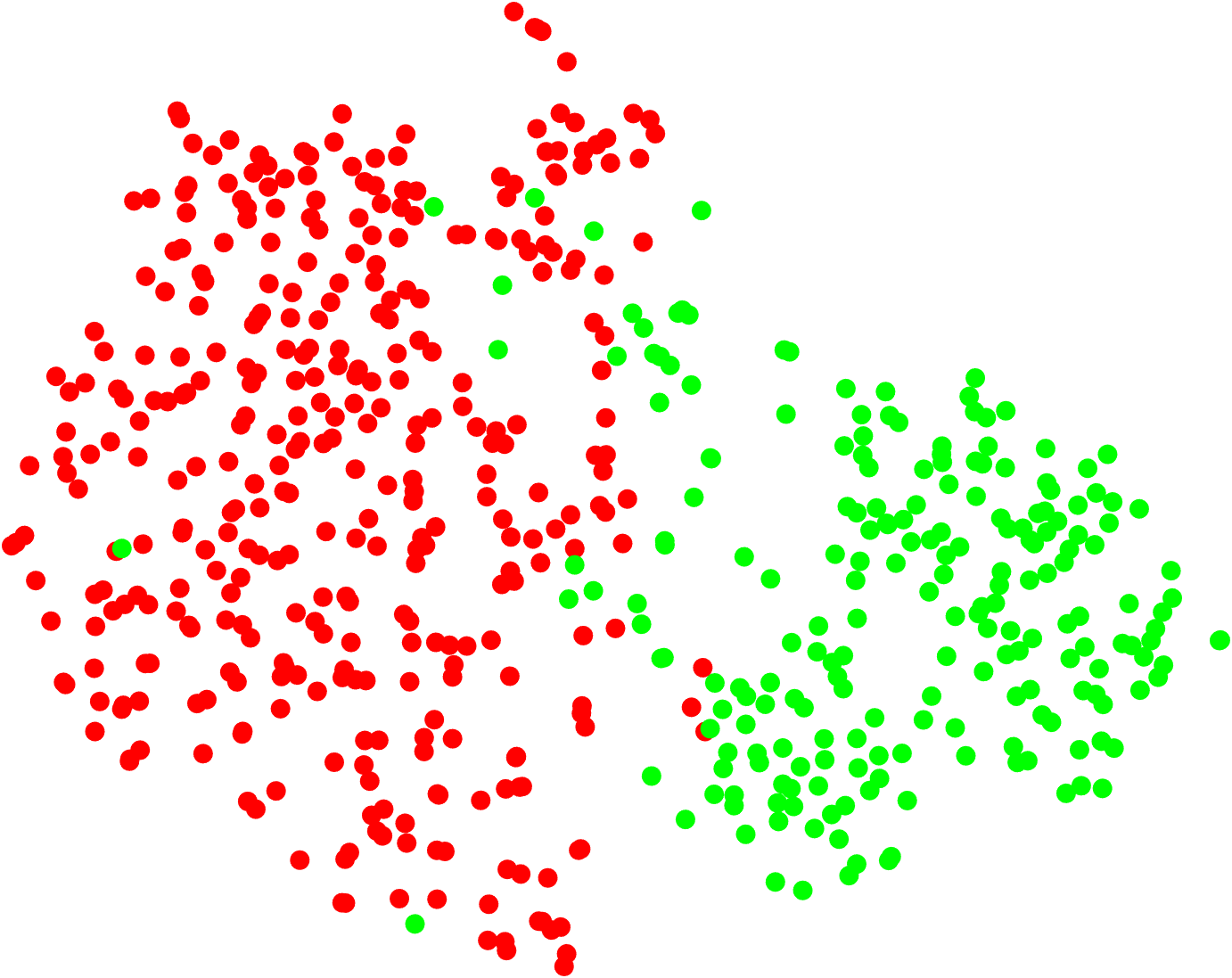}
\caption{}
\end{subfigure}
\hfill
\begin{subfigure}[H!]{0.48\textwidth}
\centering
\includegraphics[width=\textwidth]{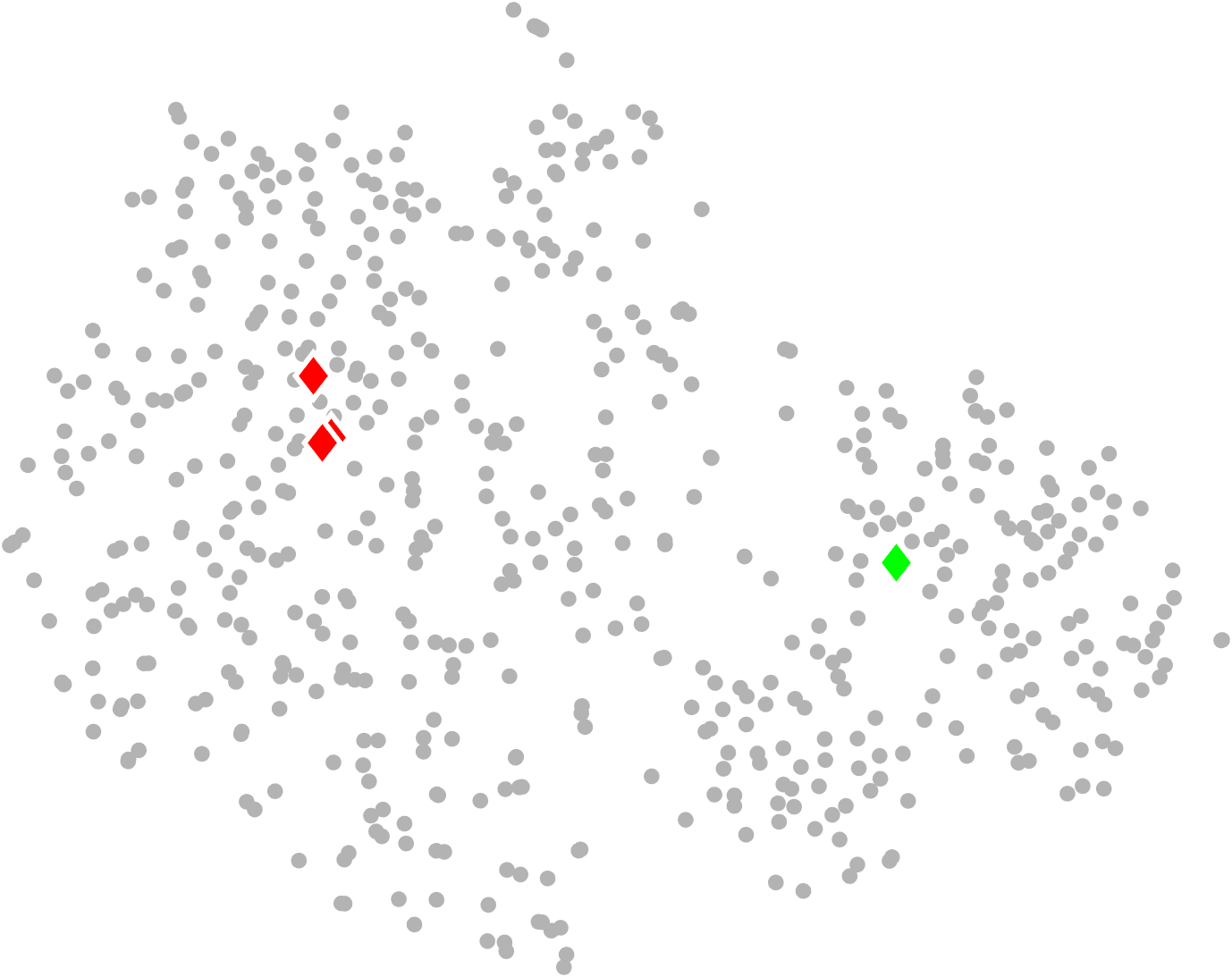}
\caption{}
\end{subfigure}

\begin{subfigure}[H!]{0.48\textwidth}
\centering
\includegraphics[width=\textwidth]{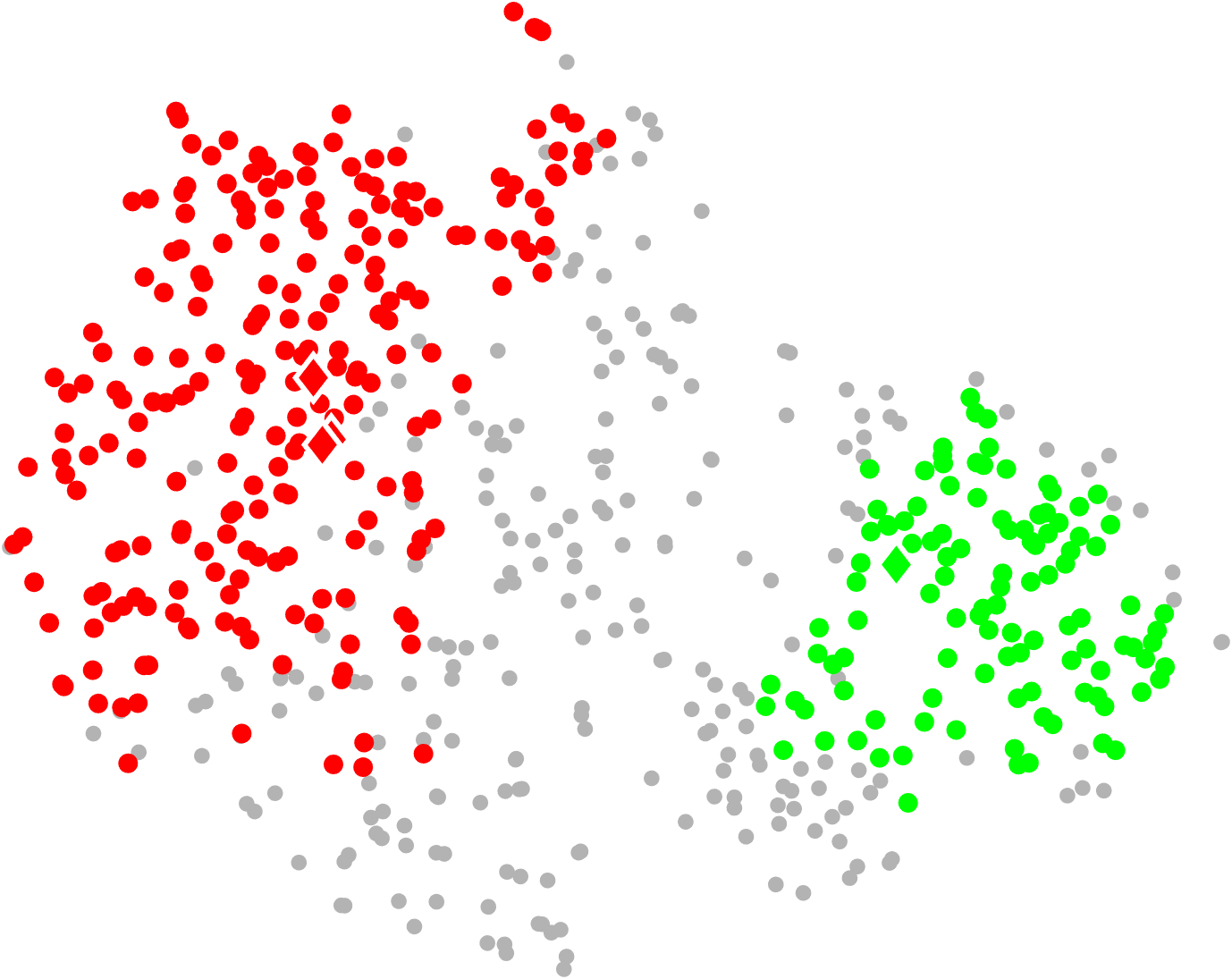}
\caption{}
\end{subfigure}
\hfill
\begin{subfigure}[H!]{0.48\textwidth}
\centering
\includegraphics[width=\textwidth]{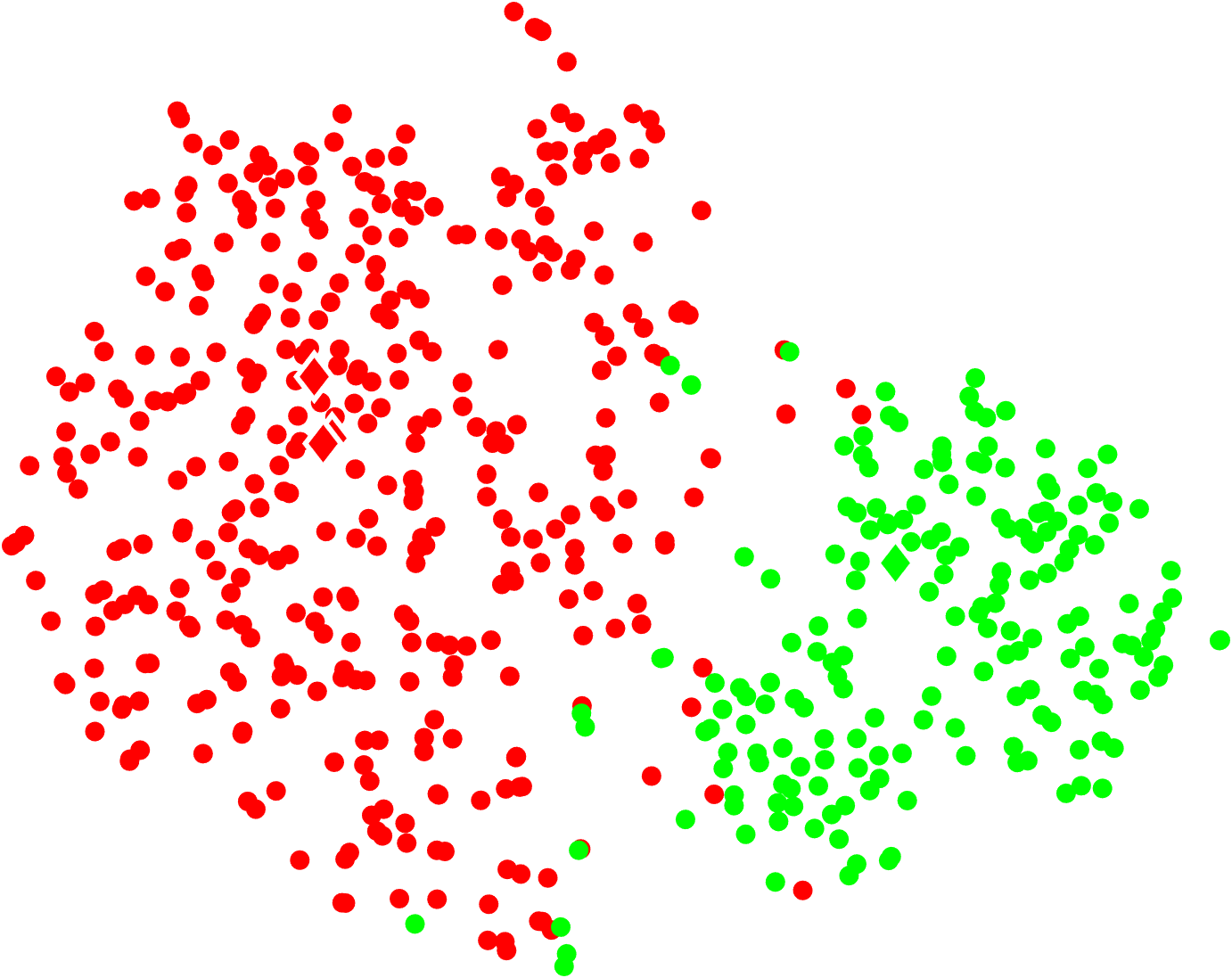}
\caption{}
\end{subfigure}
\caption{DLCC visualization on the \texttt{Breast Cancer} dataset. (a) Ground truth labels; (b) Grouped local centers; (c) Temporary clusters; (d) Final DLCC clustering result.}
\label{fig:dlcc-bc}
\end{figure*}

% -------- diamond9 --------
\begin{figure*}[ht]
\centering
\begin{subfigure}[H!]{0.48\textwidth}
\centering
\includegraphics[width=\textwidth]{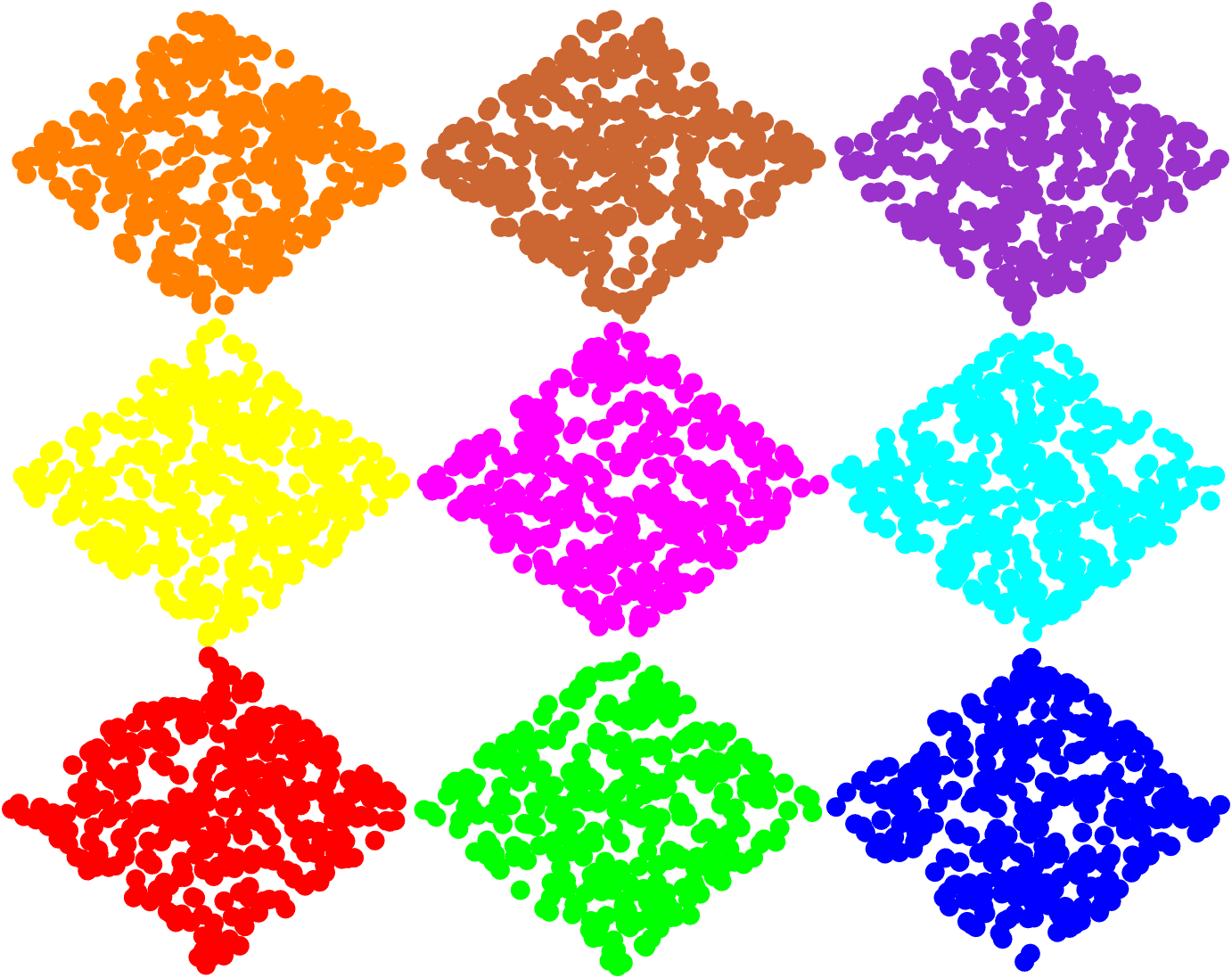}
\caption{}
\end{subfigure}
\hfill
\begin{subfigure}[H!]{0.48\textwidth}
\centering
\includegraphics[width=\textwidth]{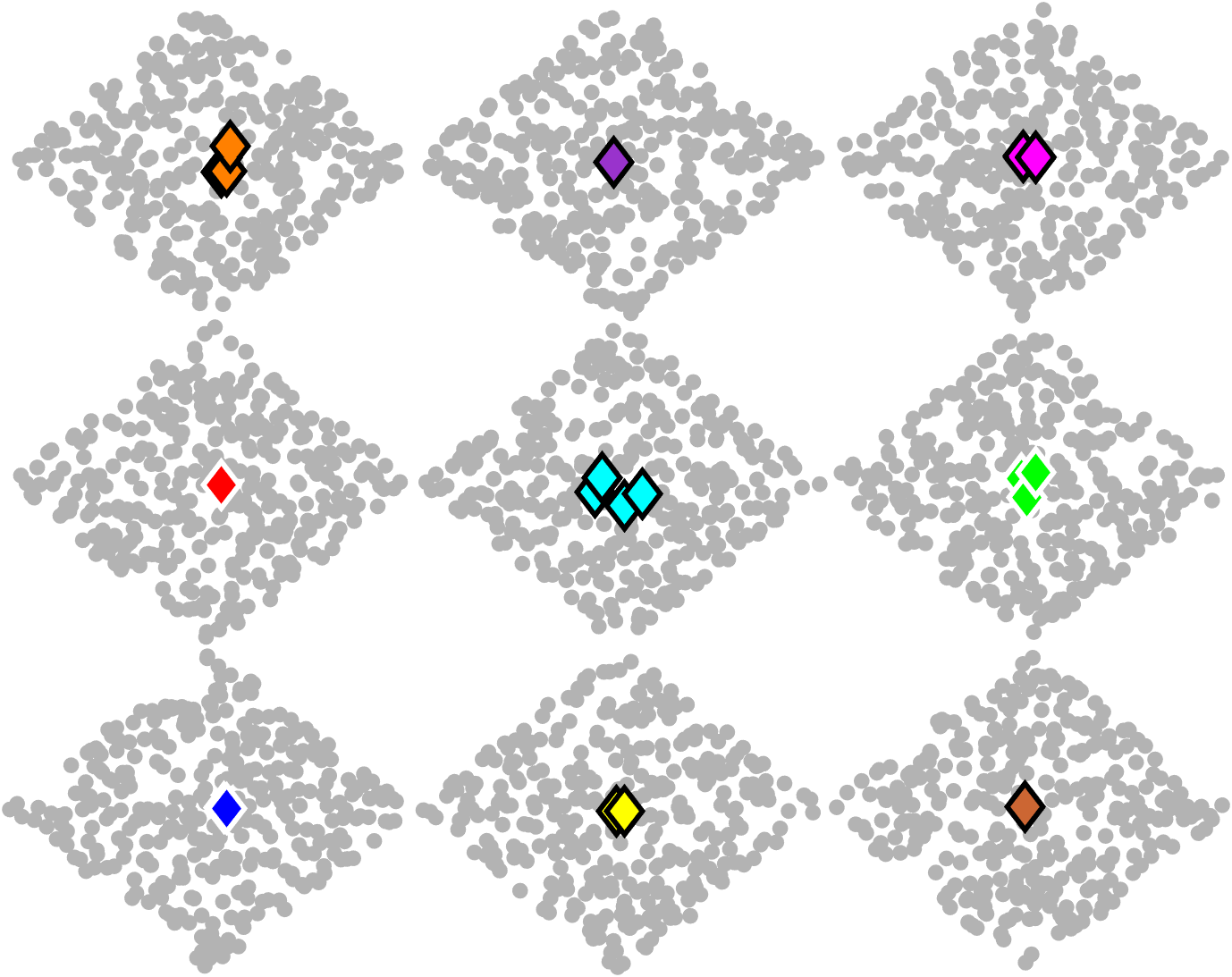}
\caption{}
\end{subfigure}

\begin{subfigure}[H!]{0.48\textwidth}
\centering
\includegraphics[width=\textwidth]{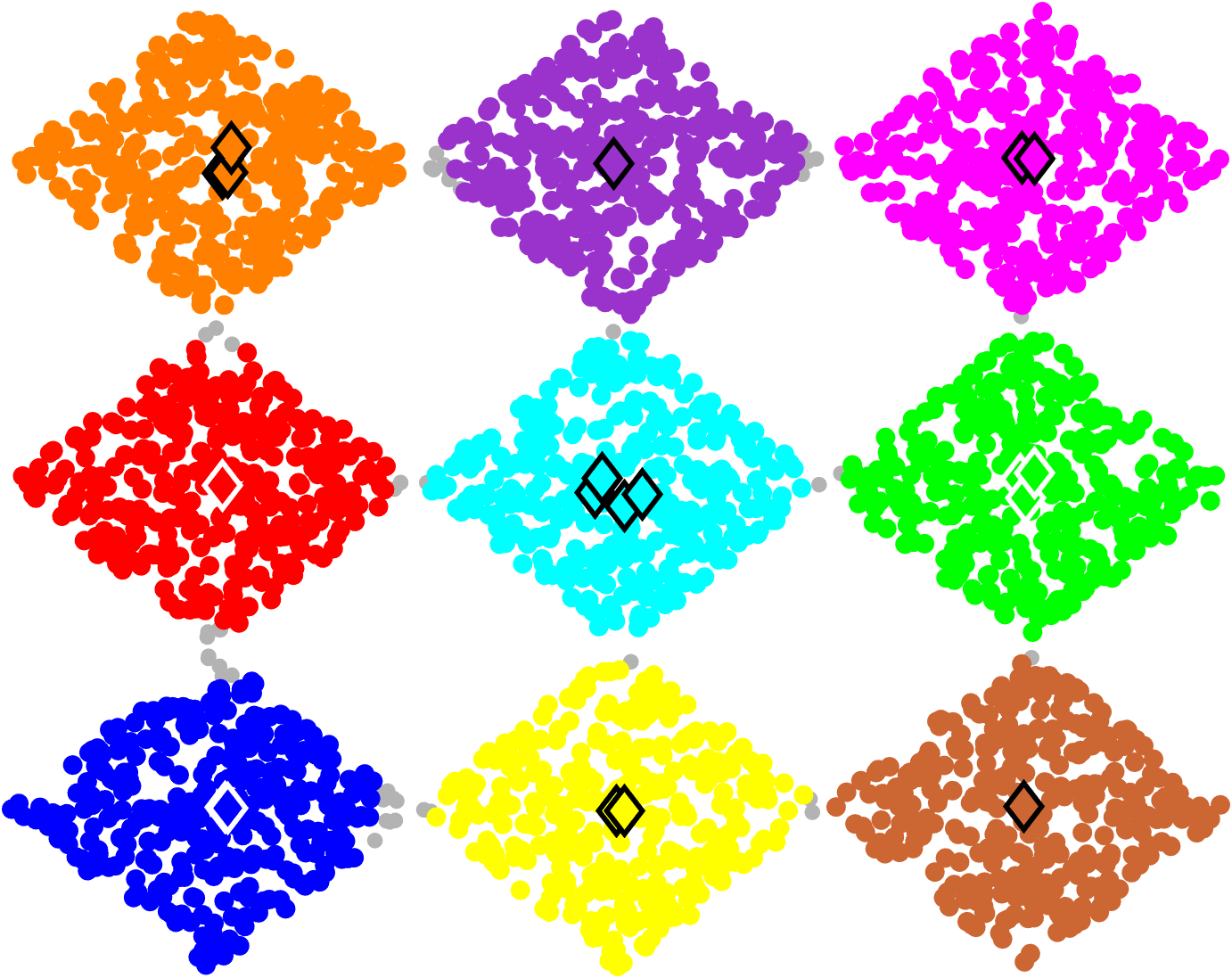}
\caption{}
\end{subfigure}
\hfill
\begin{subfigure}[H!]{0.48\textwidth}
\centering
\includegraphics[width=\textwidth]{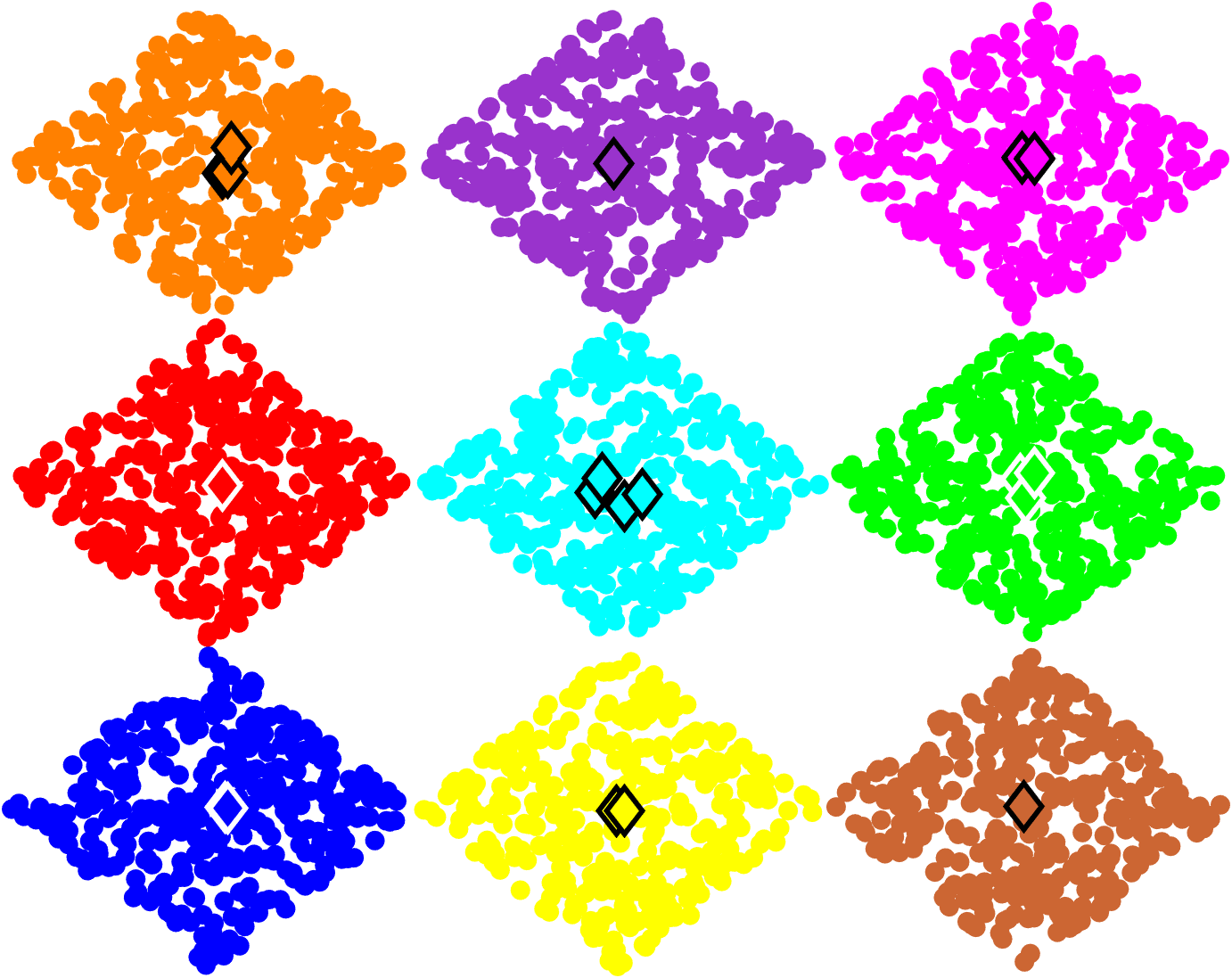}
\caption{}
\end{subfigure}
\caption{DLCC visualization on the \texttt{Diamond} dataset. (a) Ground truth labels; (b) Grouped local centers; (c) Temporary clusters; (d) Final DLCC clustering result.}
\label{fig:dlcc-diamond9}
\end{figure*}

% -------- HAR --------
\begin{figure*}[ht]
\centering
\begin{subfigure}[H!]{0.48\textwidth}
\centering
\includegraphics[width=\textwidth]{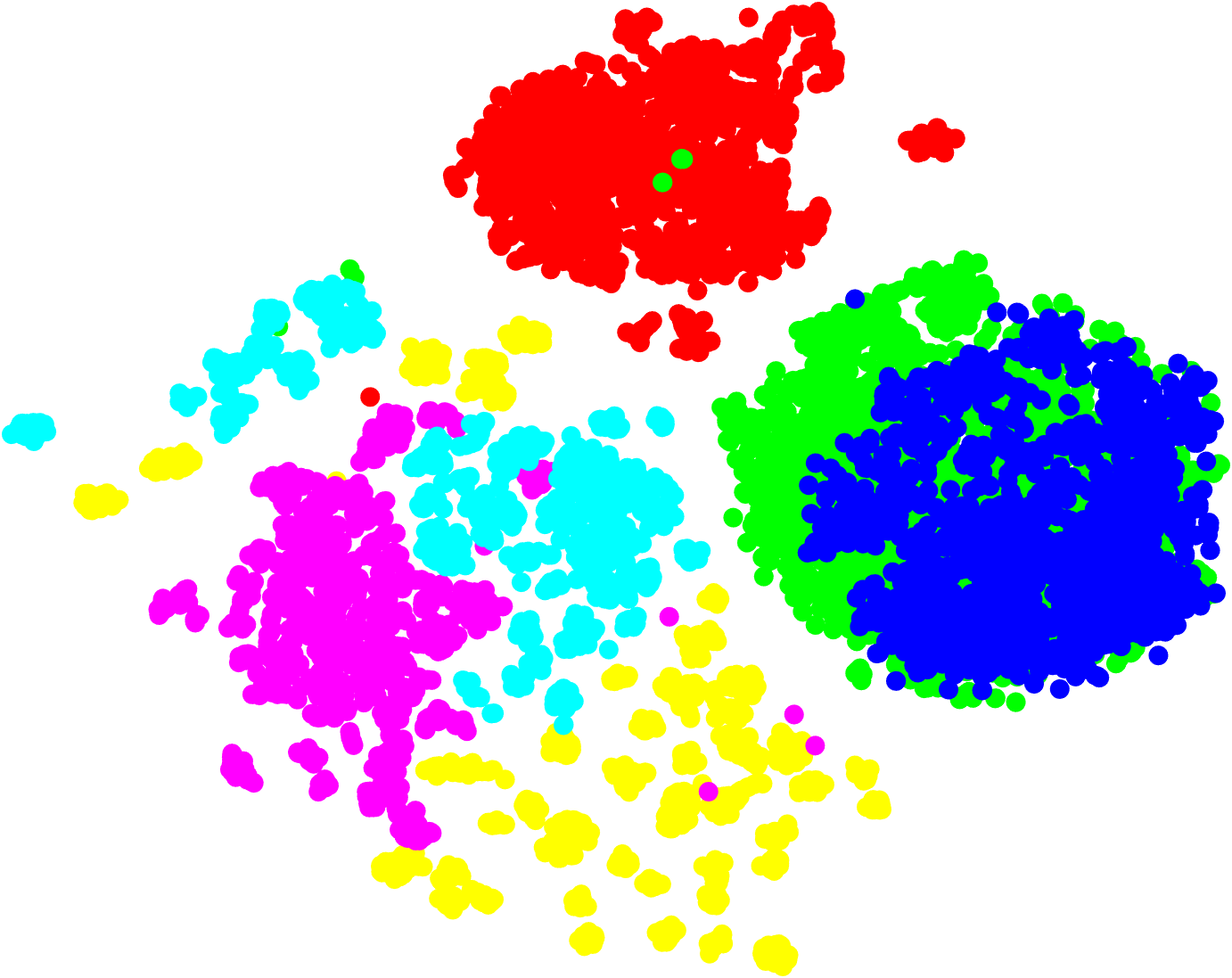}
\caption{}
\end{subfigure}
\hfill
\begin{subfigure}[H!]{0.48\textwidth}
\centering
\includegraphics[width=\textwidth]{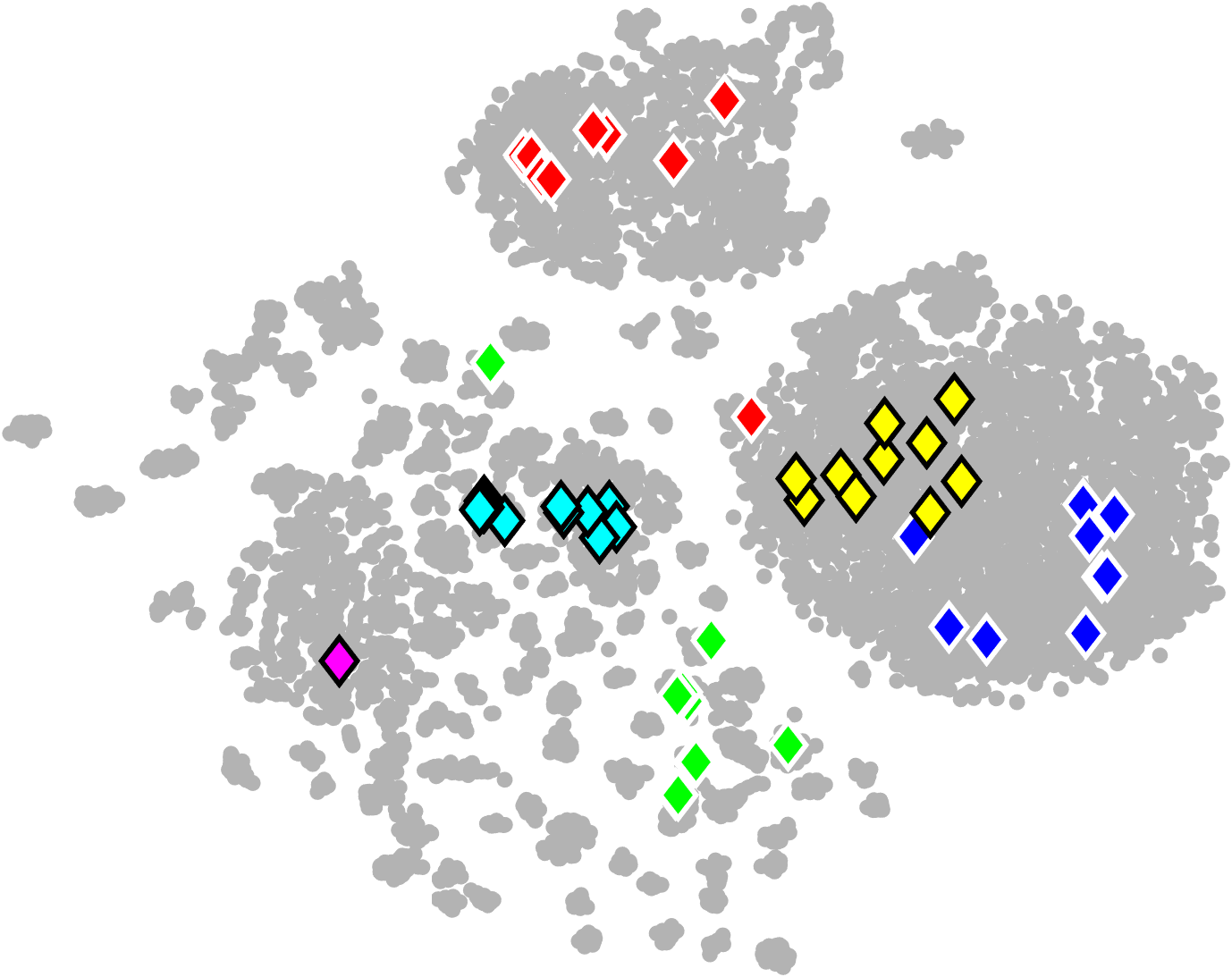}
\caption{}
\end{subfigure}

\begin{subfigure}[H!]{0.48\textwidth}
\centering
\includegraphics[width=\textwidth]{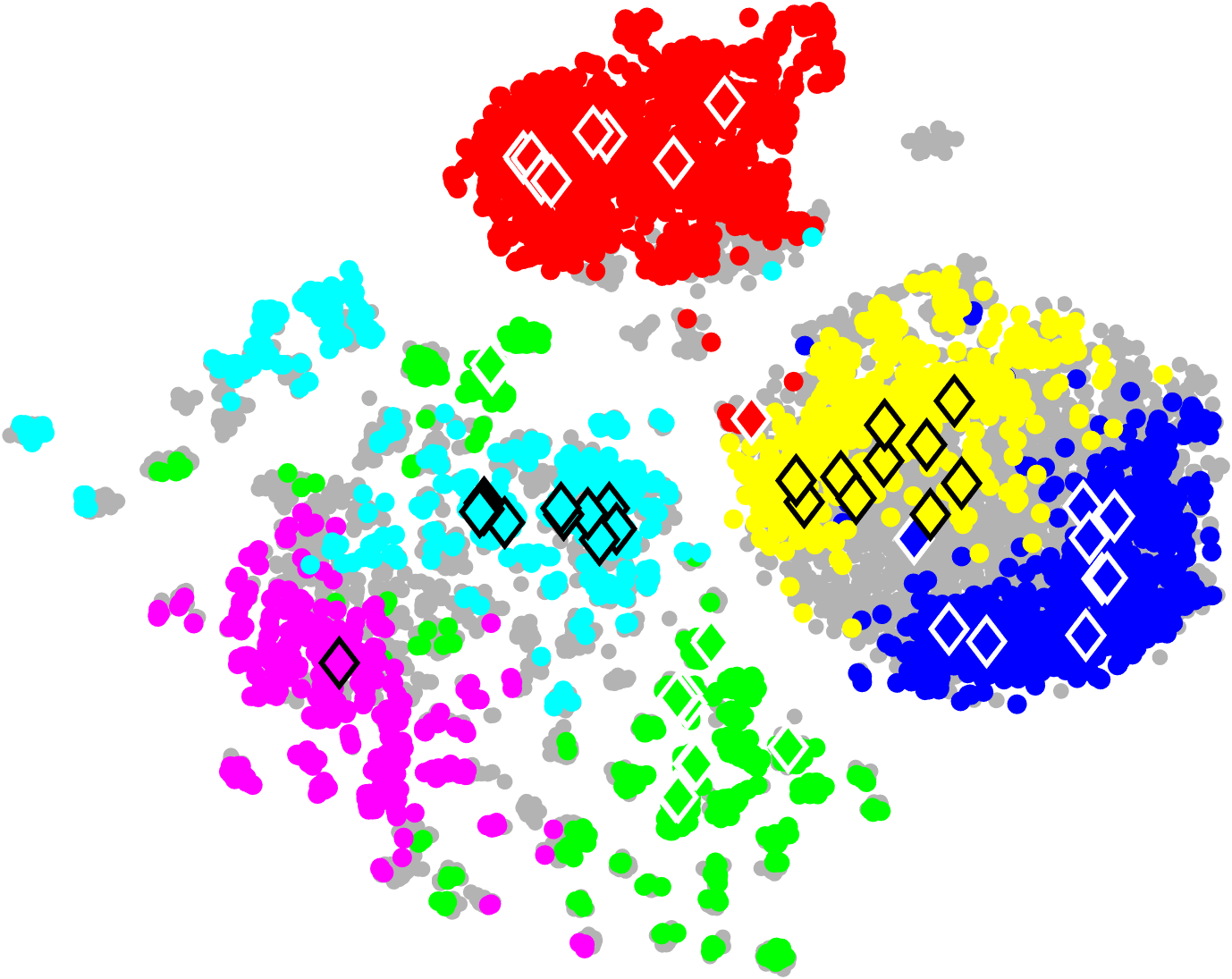}
\caption{}
\end{subfigure}
\hfill
\begin{subfigure}[H!]{0.48\textwidth}
\centering
\includegraphics[width=\textwidth]{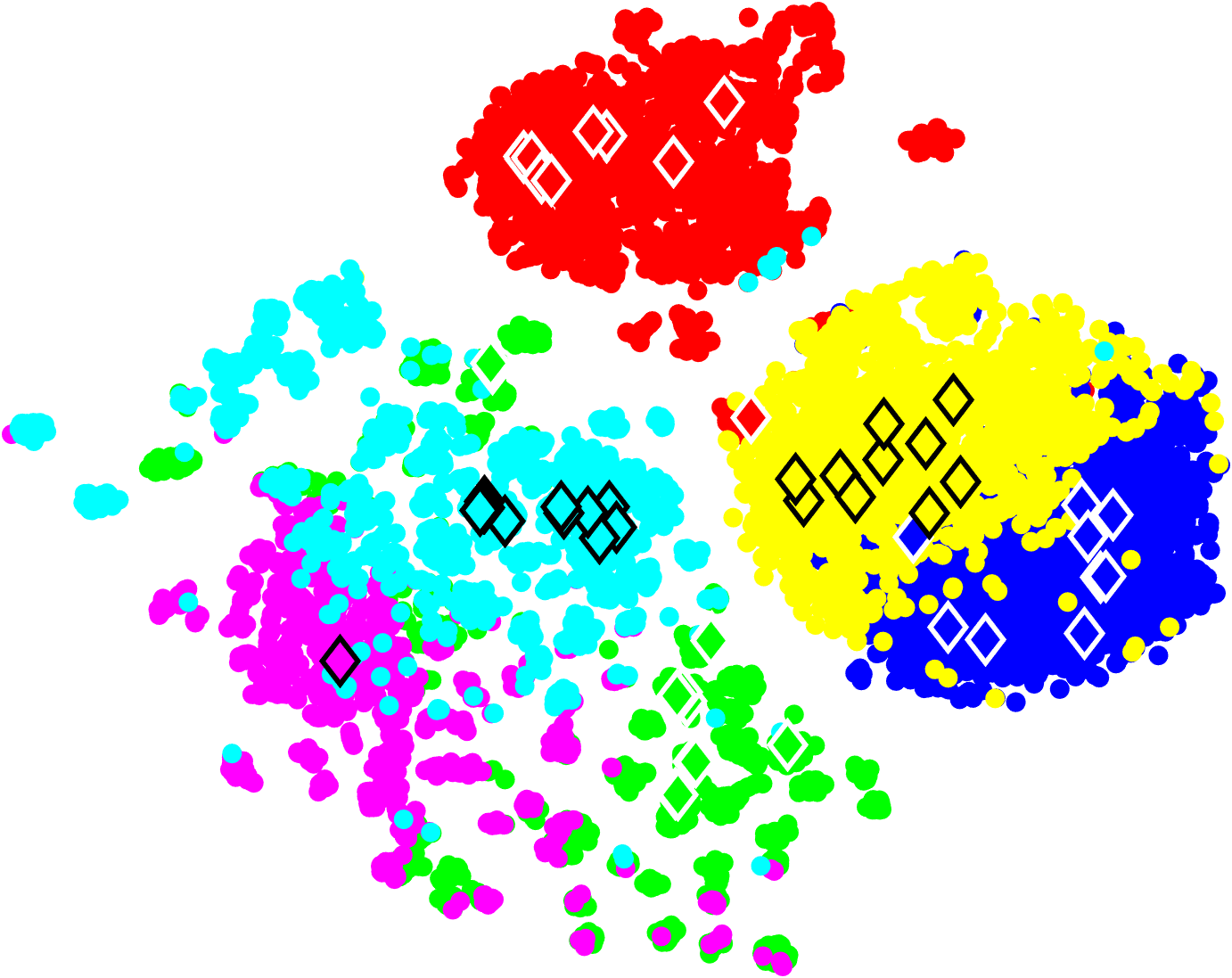}
\caption{}
\end{subfigure}
\caption{DLCC visualization on the \texttt{HAR} dataset. (a) Ground truth labels; (b) Grouped local centers; (c) Temporary clusters; (d) Final DLCC clustering result.}
\label{fig:dlcc-HAR}
\end{figure*}

%----iris-----
\begin{figure*}[ht]
\centering
\begin{subfigure}[H!]{0.48\textwidth}
\centering
\includegraphics[width=\textwidth]{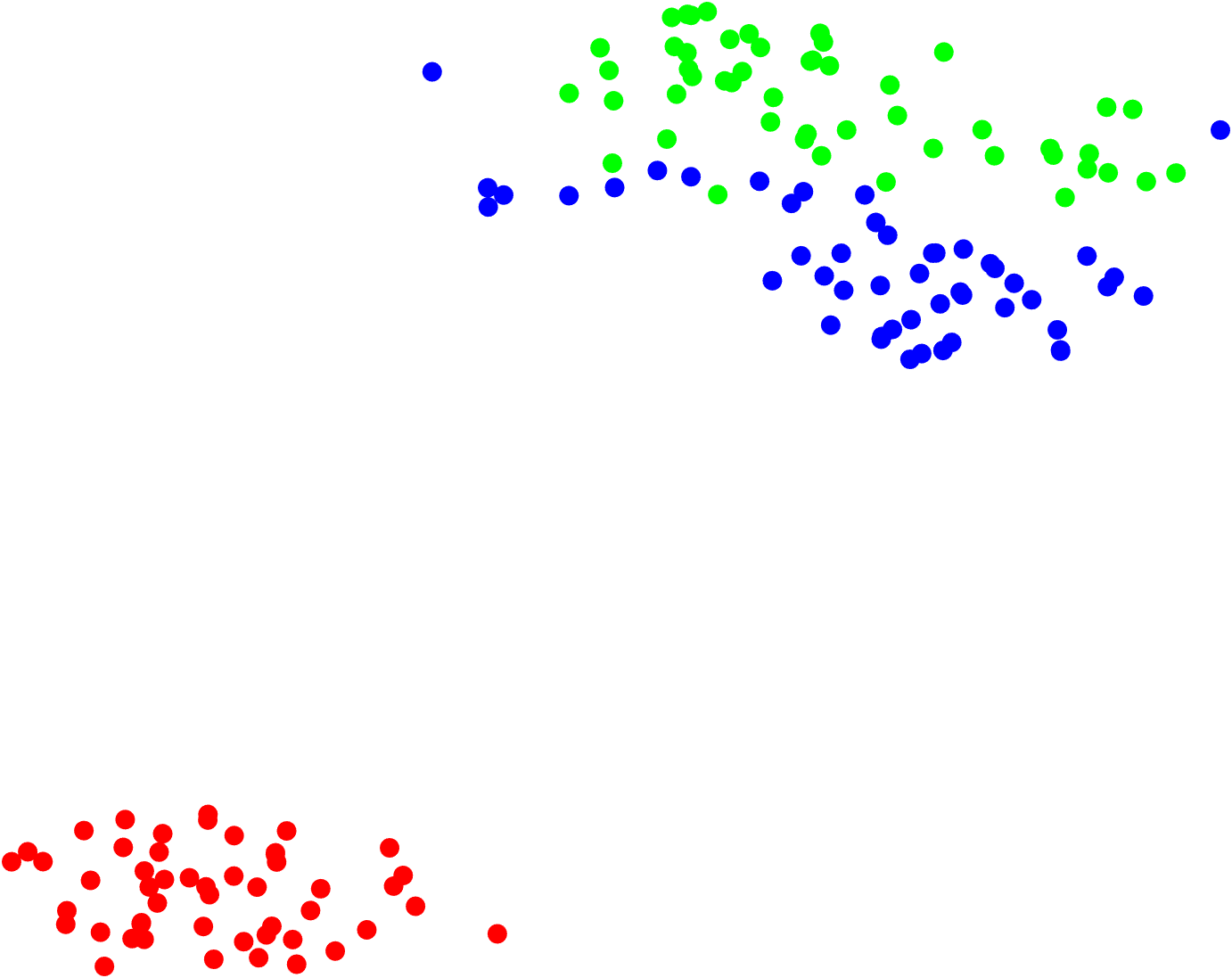}
\caption{}
\end{subfigure}
\hfill
\begin{subfigure}[H!]{0.48\textwidth}
\centering
\includegraphics[width=\textwidth]{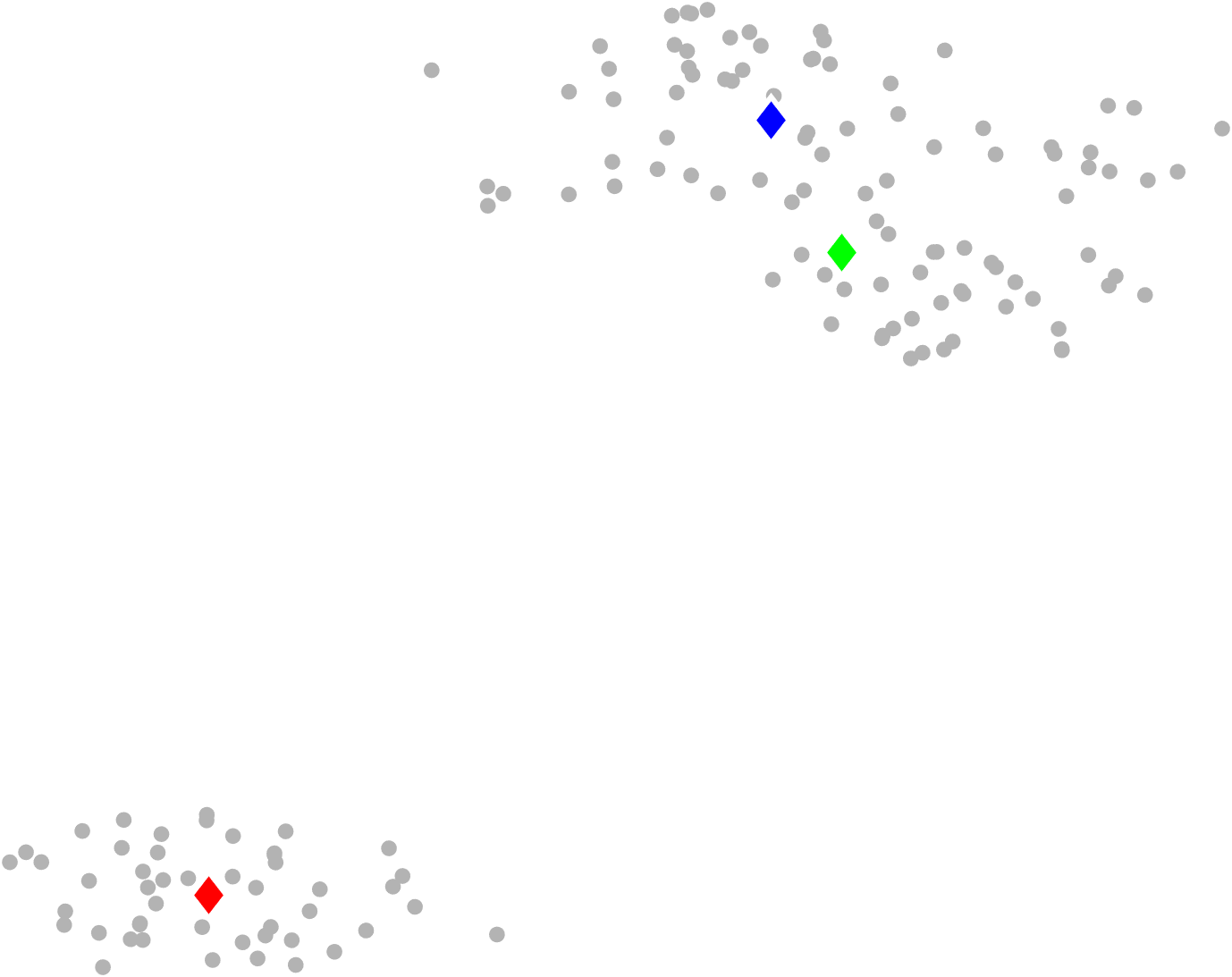}
\caption{}
\end{subfigure}

\begin{subfigure}[H!]{0.48\textwidth}
\centering
\includegraphics[width=\textwidth]{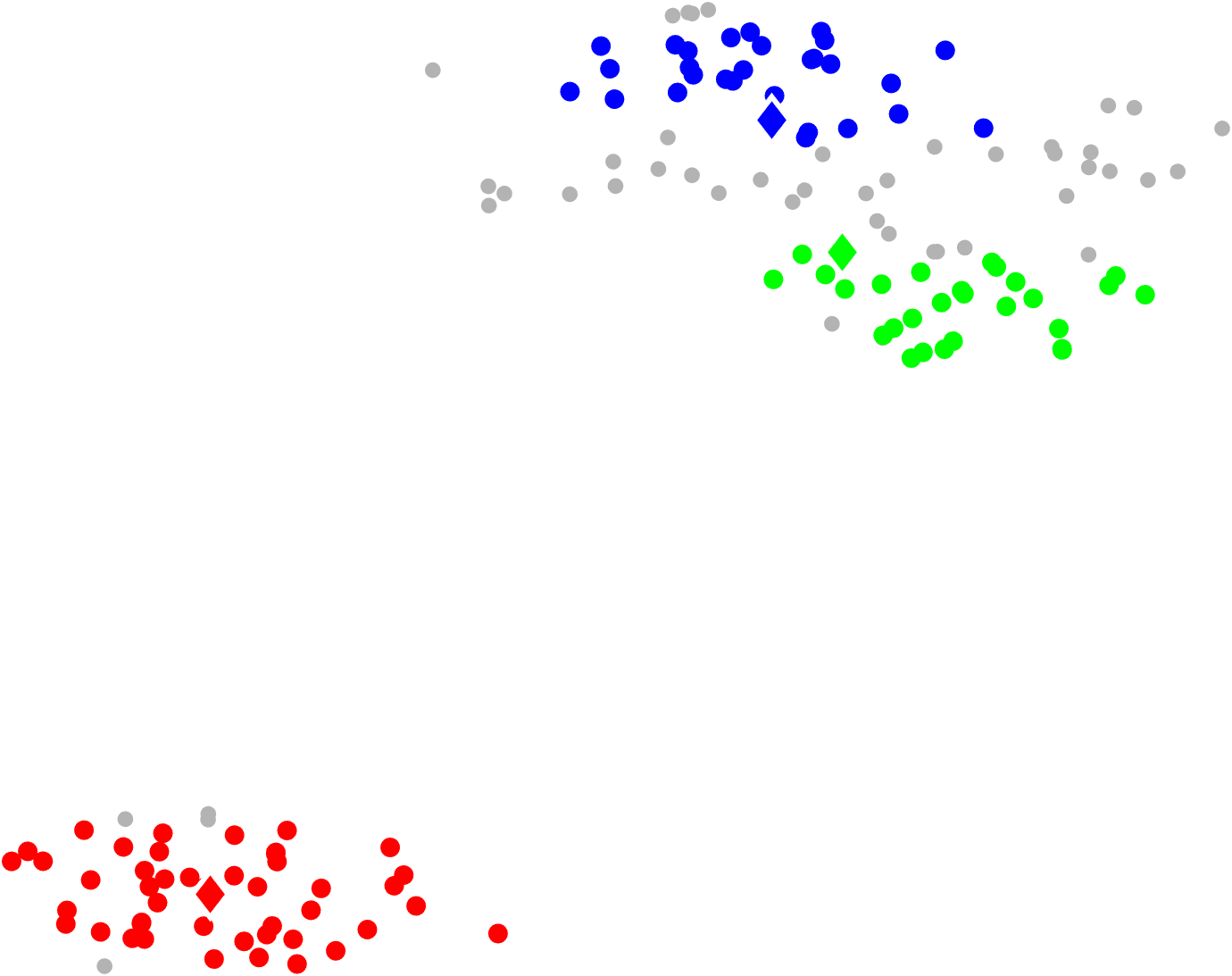}
\caption{}
\end{subfigure}
\hfill
\begin{subfigure}[H!]{0.48\textwidth}
\centering
\includegraphics[width=\textwidth]{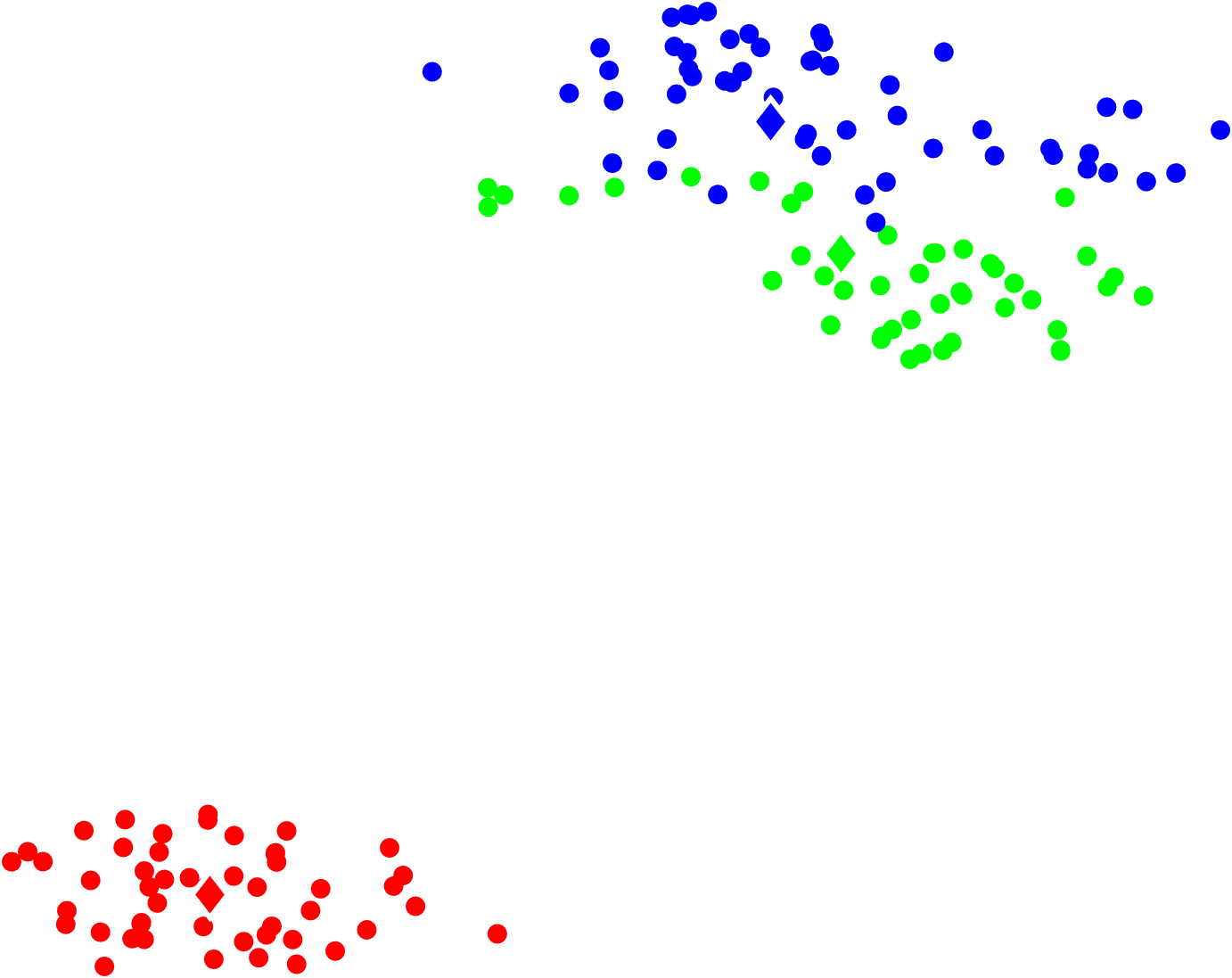}
\caption{}
\end{subfigure}
\caption{DLCC visualization on the \texttt{Iris} dataset. (a) Ground truth labels; (b) Grouped local centers; (c) Temporary clusters; (d) Final DLCC clustering result.}
\label{fig:dlcc-meas}
\end{figure*}

\begin{figure*}[ht]
\centering
\begin{subfigure}[H!]{0.48\textwidth}
\centering
\includegraphics[width=\textwidth]{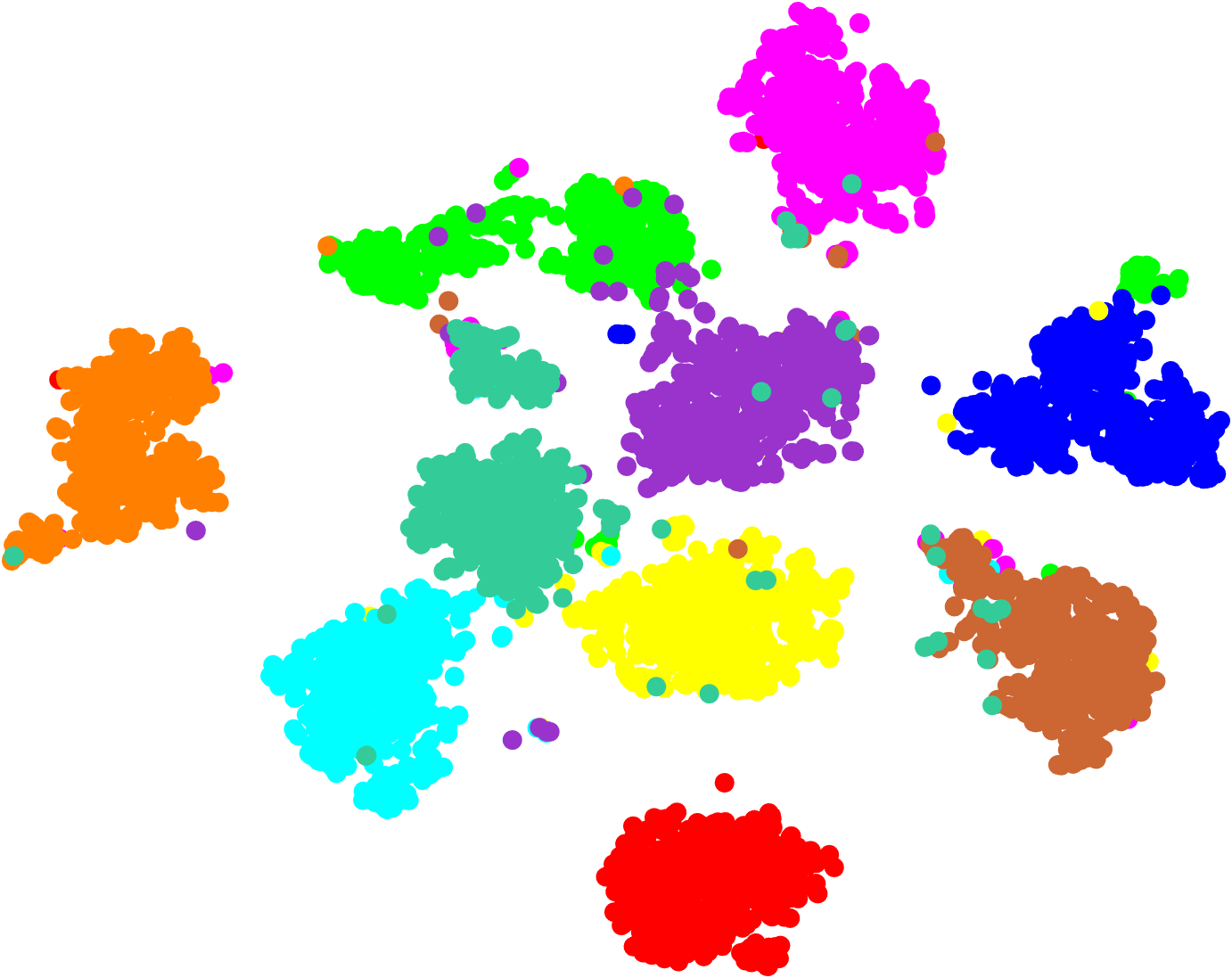}
\caption{}
\end{subfigure}
\hfill
\begin{subfigure}[H!]{0.48\textwidth}
\centering
\includegraphics[width=\textwidth]{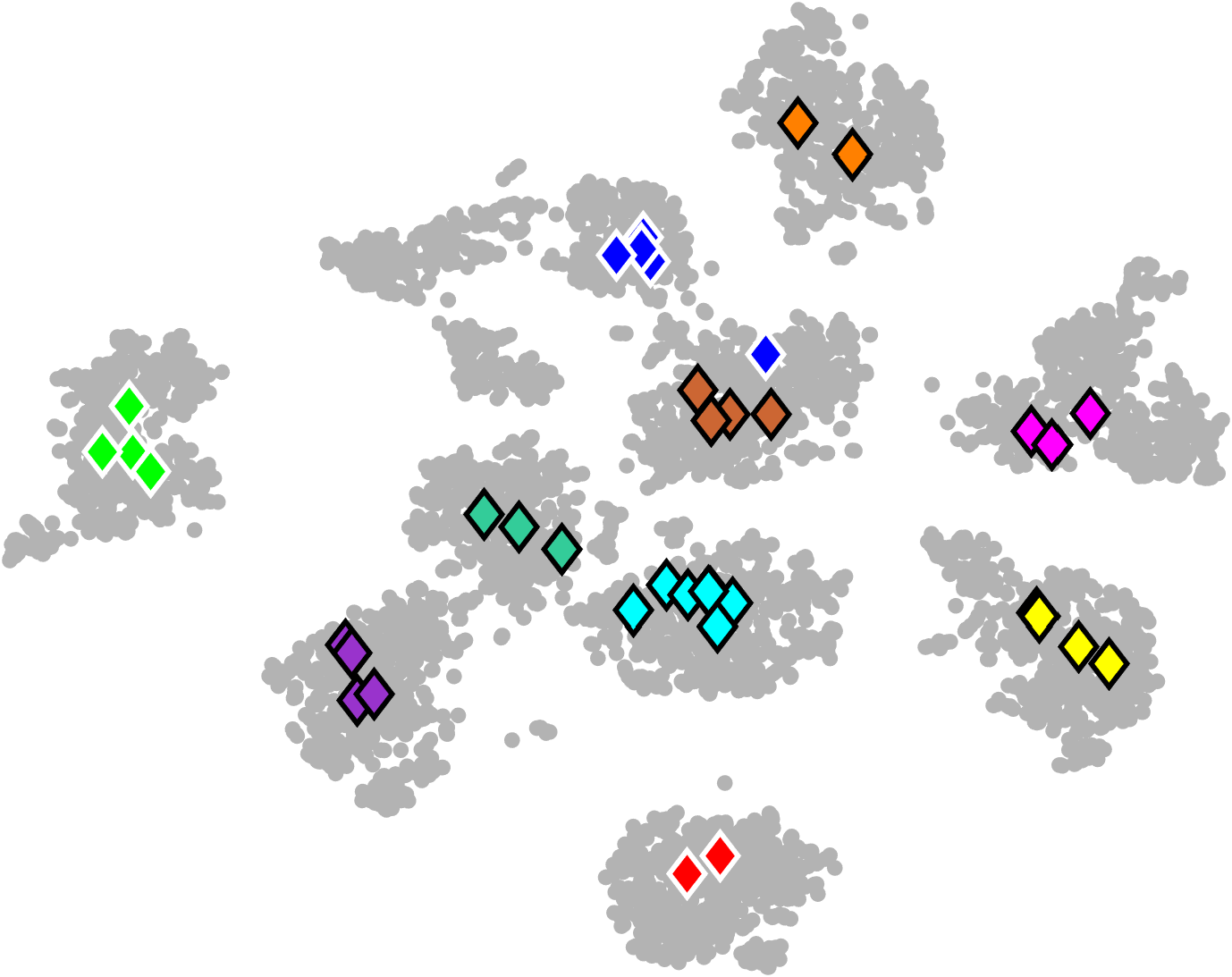}
\caption{}
\end{subfigure}

\begin{subfigure}[H!]{0.48\textwidth}
\centering
\includegraphics[width=\textwidth]{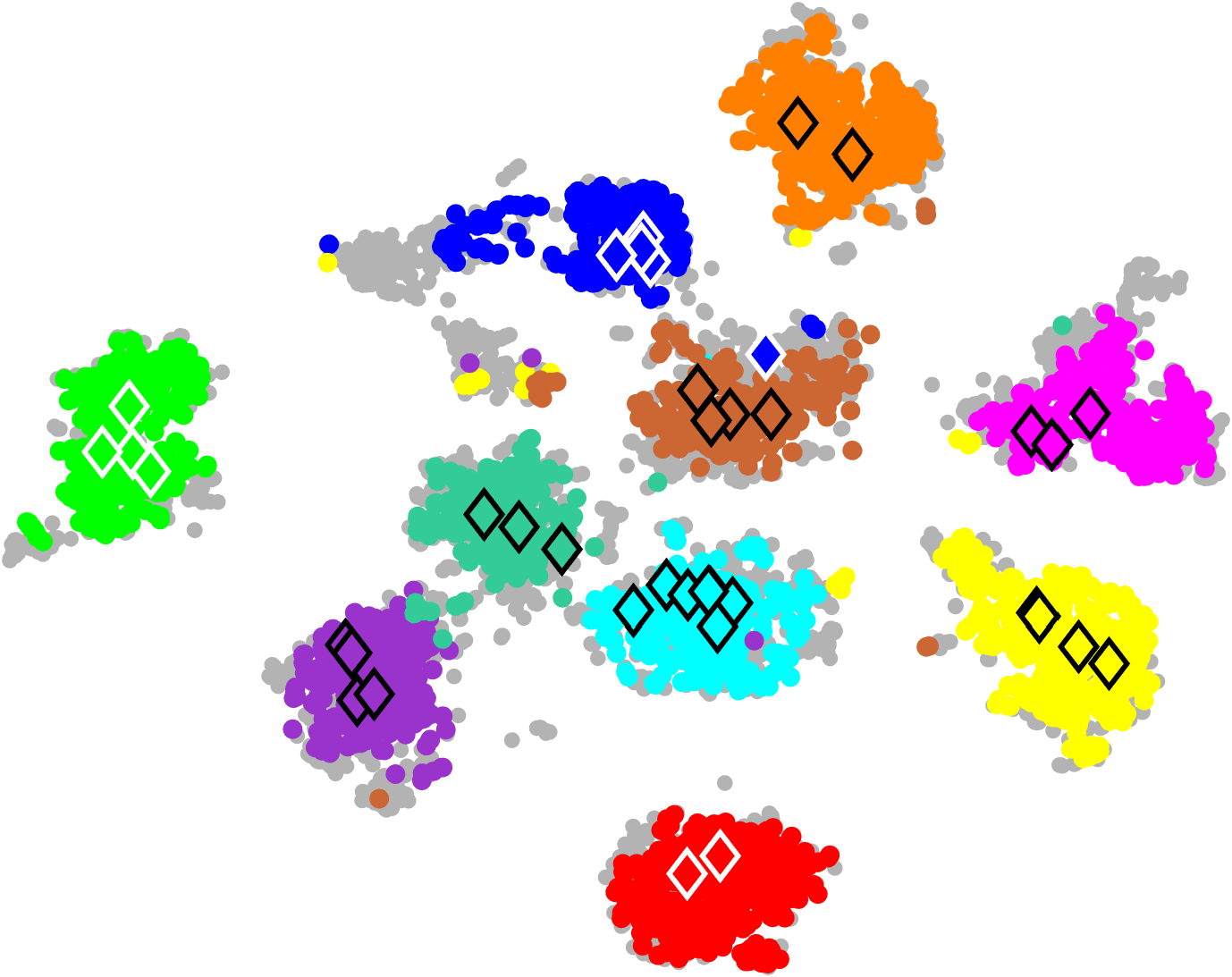}
\caption{}
\end{subfigure}
\hfill
\begin{subfigure}[H!]{0.48\textwidth}
\centering
\includegraphics[width=\textwidth]{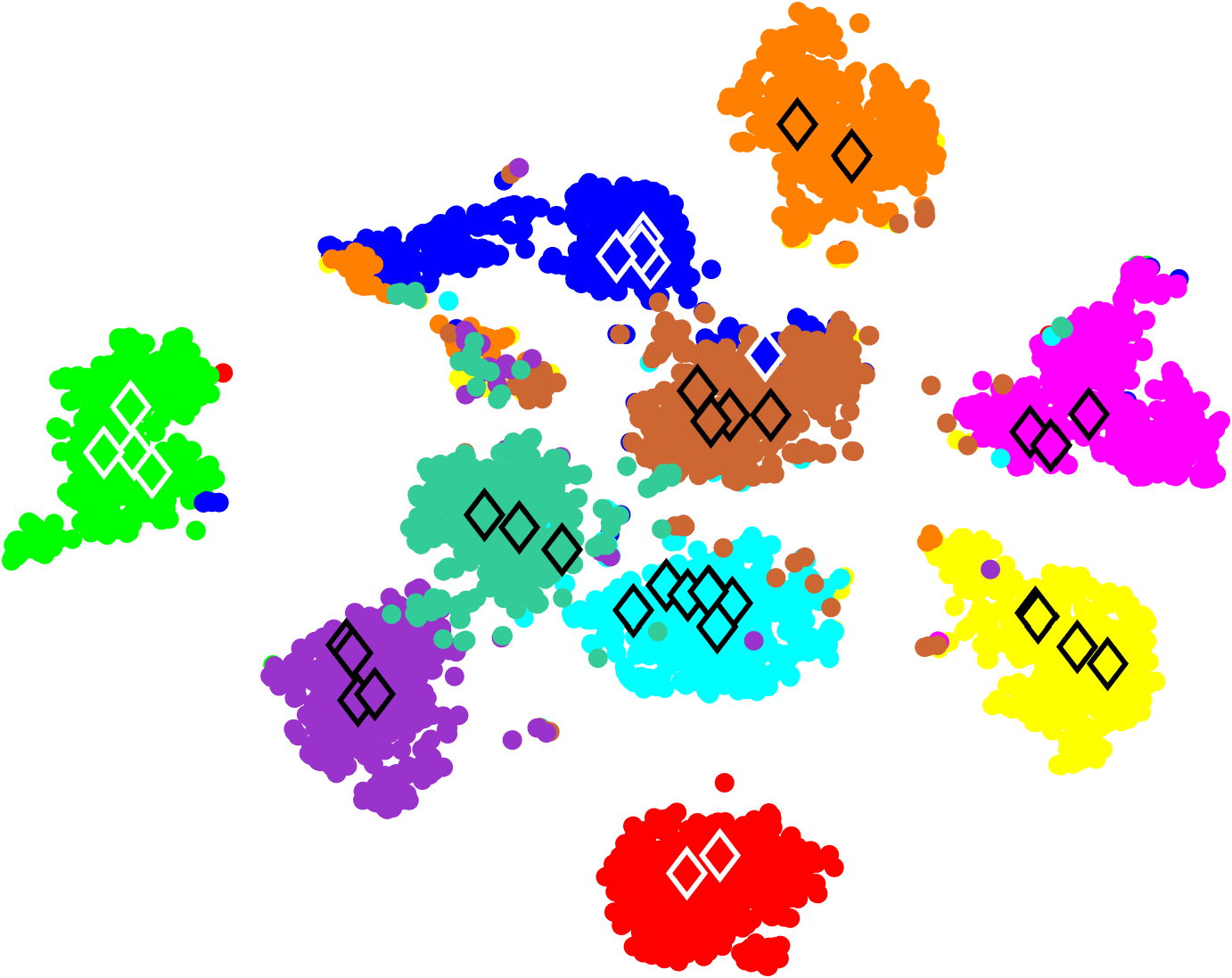}
\caption{}
\end{subfigure}
\caption{DLCC visualization on the \texttt{Optidigits} dataset. (a) Ground truth labels; (b) Grouped local centers; (c) Temporary clusters; (d) Final DLCC clustering result.}
\label{fig:dlcc-optidigits}
\end{figure*}

% -------- pendigit --------
\begin{figure*}[ht]
\centering
\begin{subfigure}[H!]{0.48\textwidth}
\centering
\includegraphics[width=\textwidth]{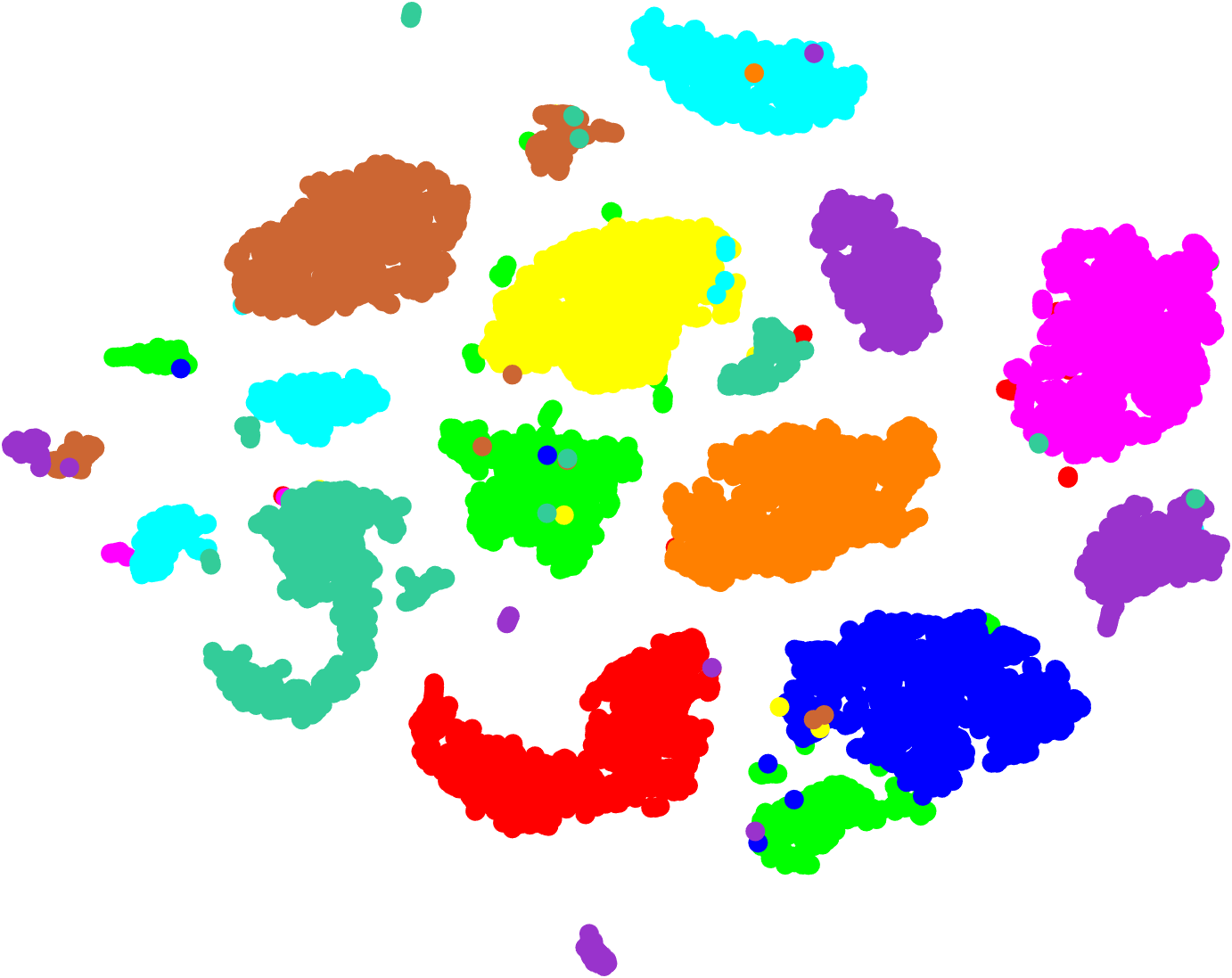}
\caption{}
\end{subfigure}
\hfill
\begin{subfigure}[H!]{0.48\textwidth}
\centering
\includegraphics[width=\textwidth]{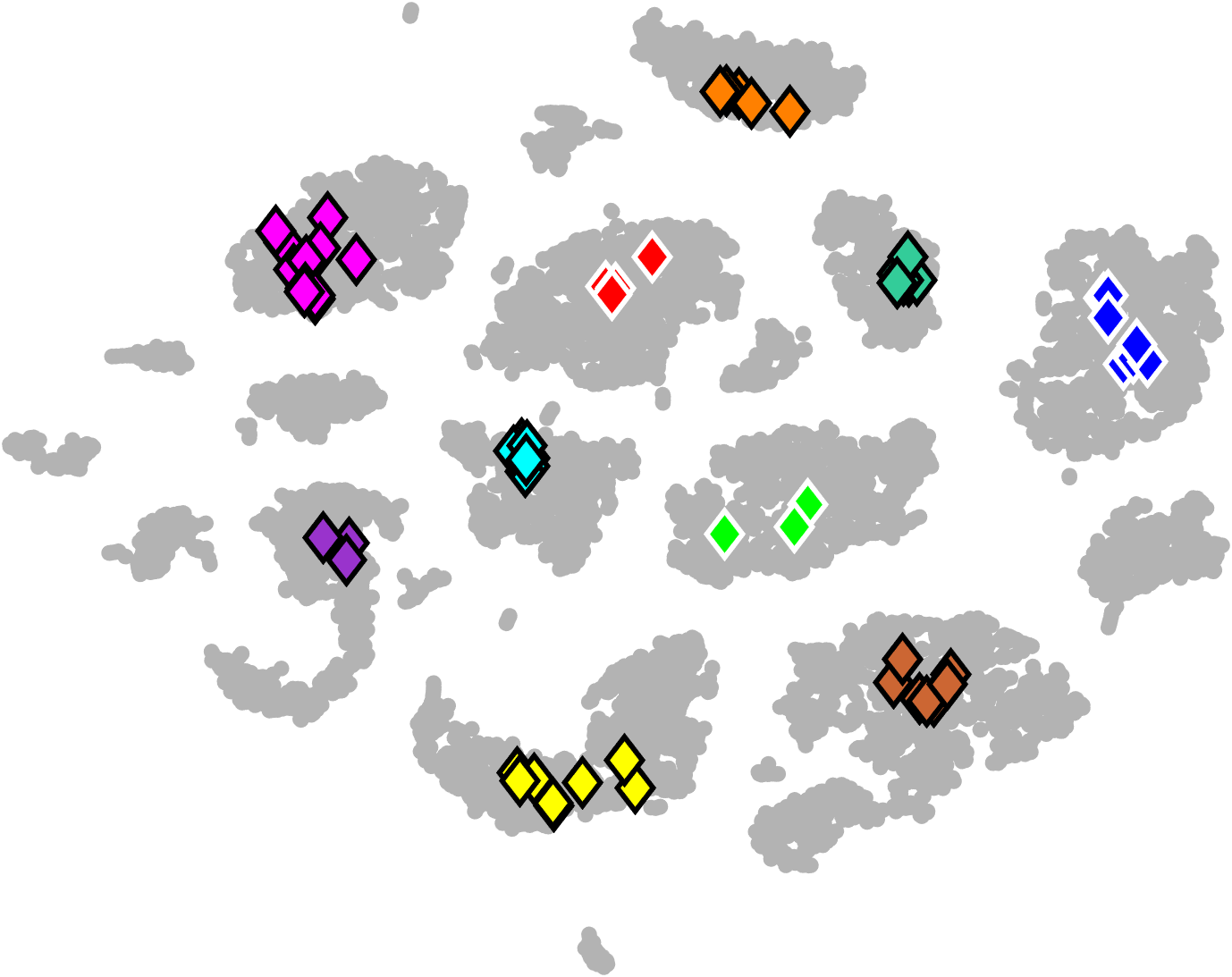}
\caption{}
\end{subfigure}

\begin{subfigure}[H!]{0.48\textwidth}
\centering
\includegraphics[width=\textwidth]{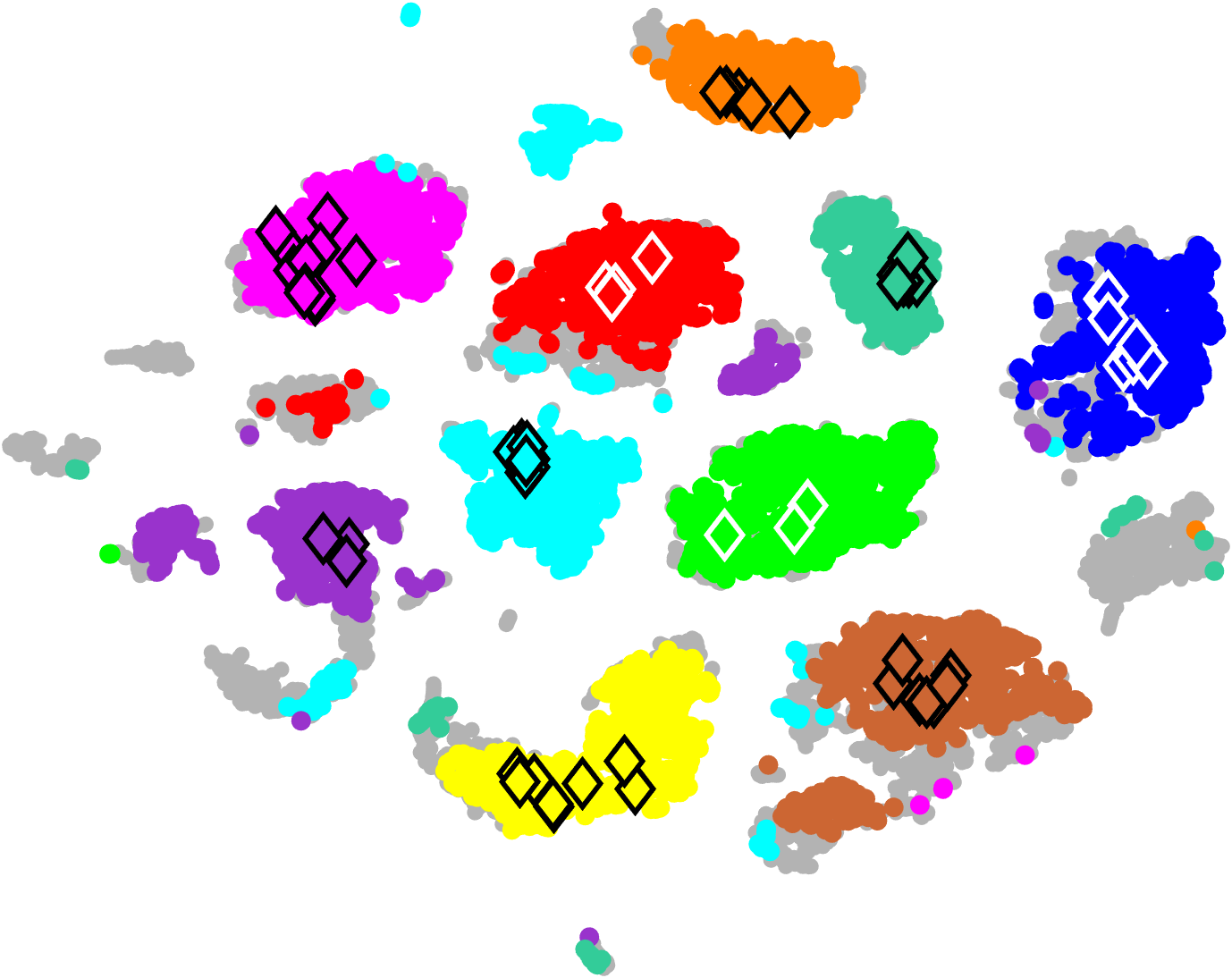}
\caption{}
\end{subfigure}
\hfill
\begin{subfigure}[H!]{0.48\textwidth}
\centering
\includegraphics[width=\textwidth]{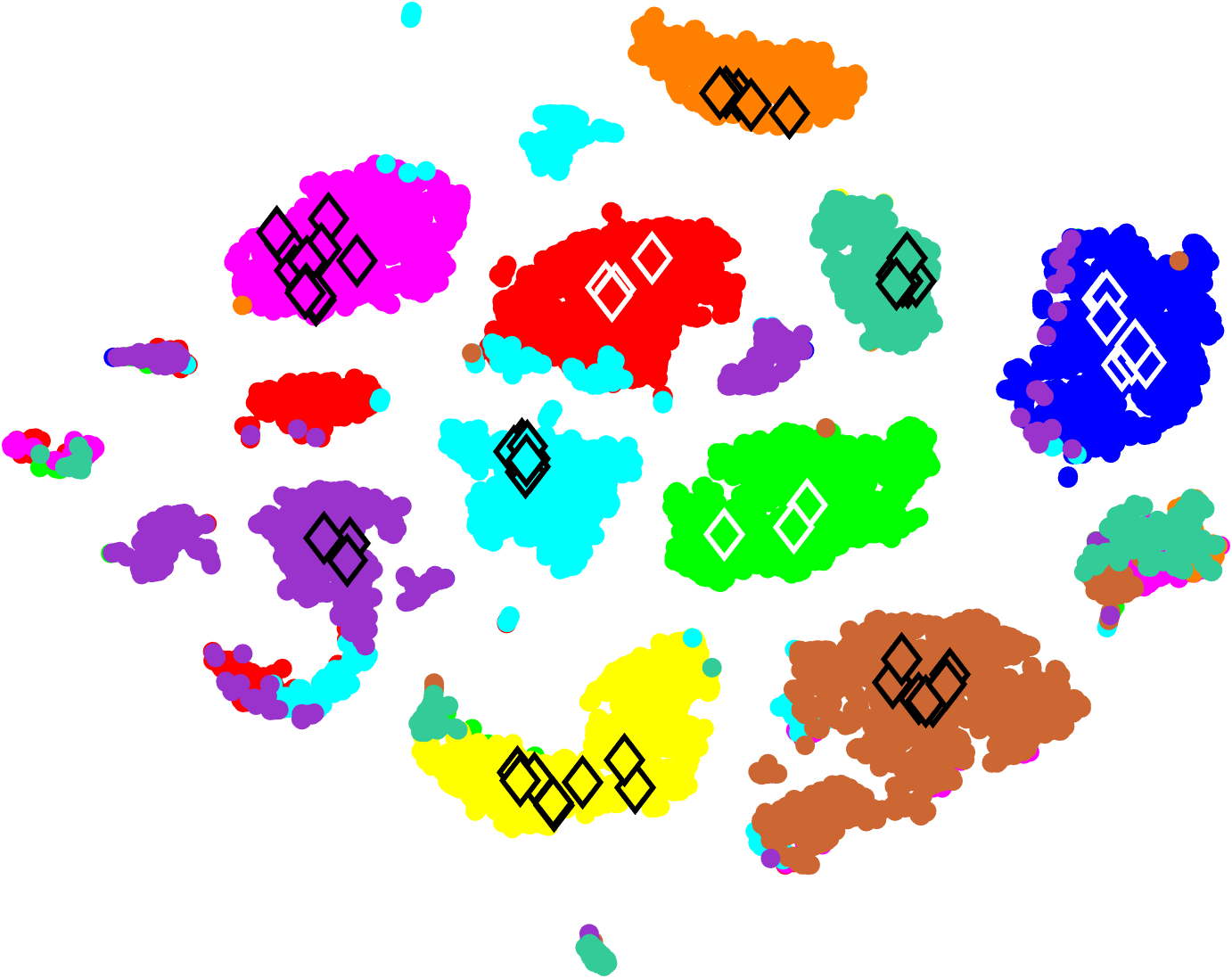}
\caption{}
\end{subfigure}
\caption{DLCC visualization on the \texttt{Pendigits} dataset. (a) Ground truth labels; (b) Grouped local centers; (c) Temporary clusters; (d) Final DLCC clustering result.}
\label{fig:dlcc-pendigit}
\end{figure*}

% ------ ray --------
\begin{figure*}[ht]
\centering
\begin{subfigure}[H!]{0.48\textwidth}
\centering
\includegraphics[width=\textwidth]{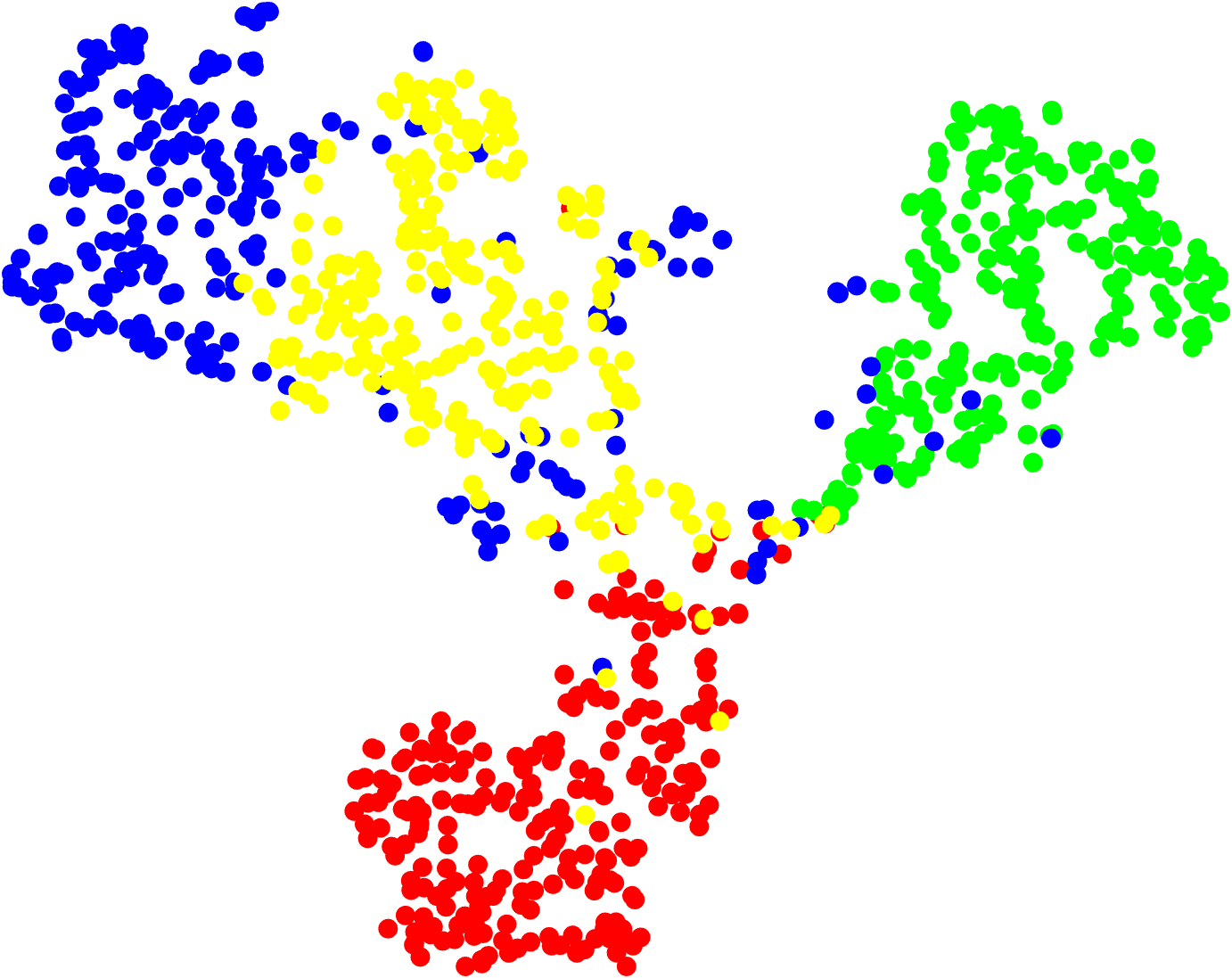}
\caption{}
\end{subfigure}
\hfill
\begin{subfigure}[H!]{0.48\textwidth}
\centering
\includegraphics[width=\textwidth]{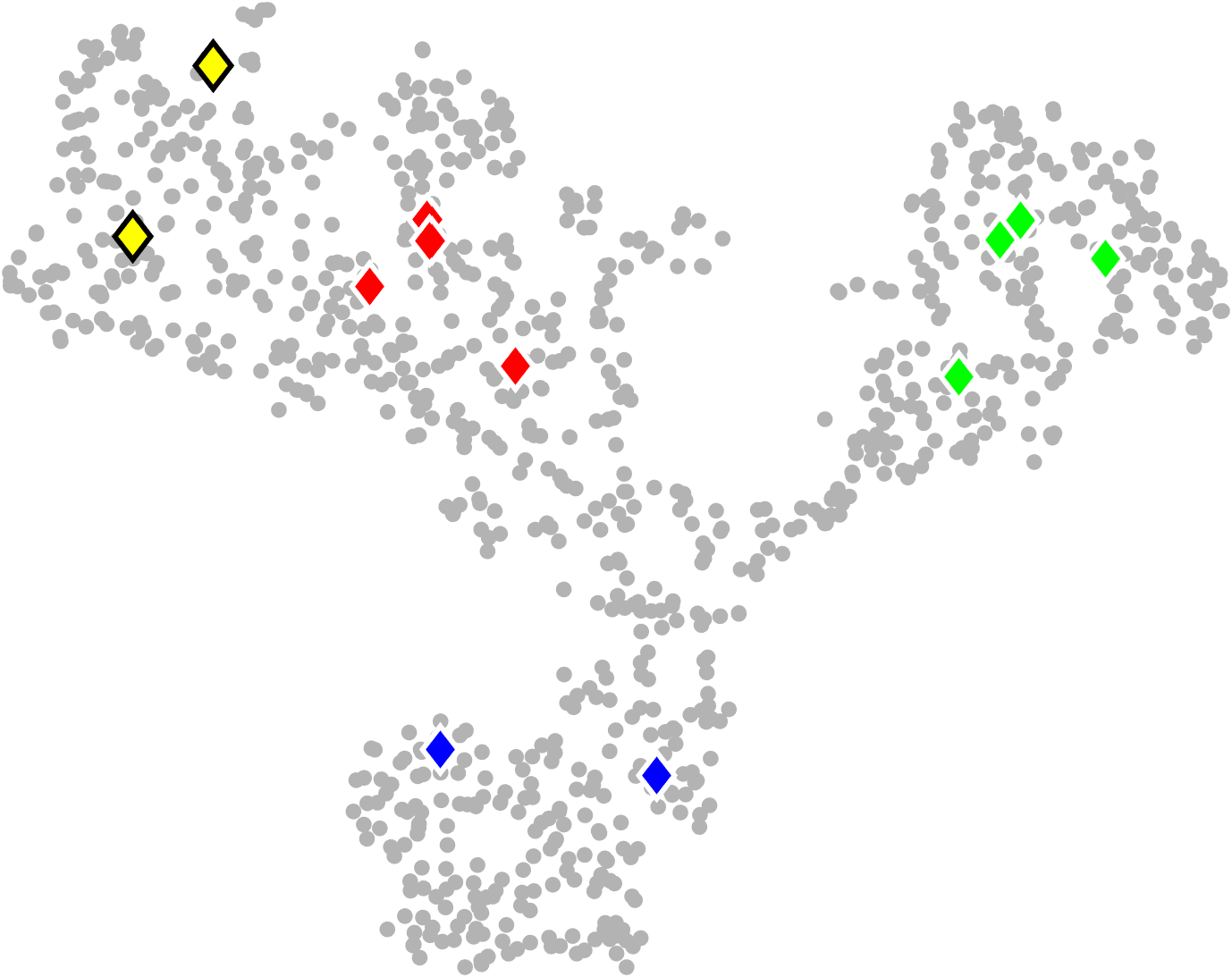}
\caption{}
\end{subfigure}

\begin{subfigure}[H!]{0.48\textwidth}
\centering
\includegraphics[width=\textwidth]{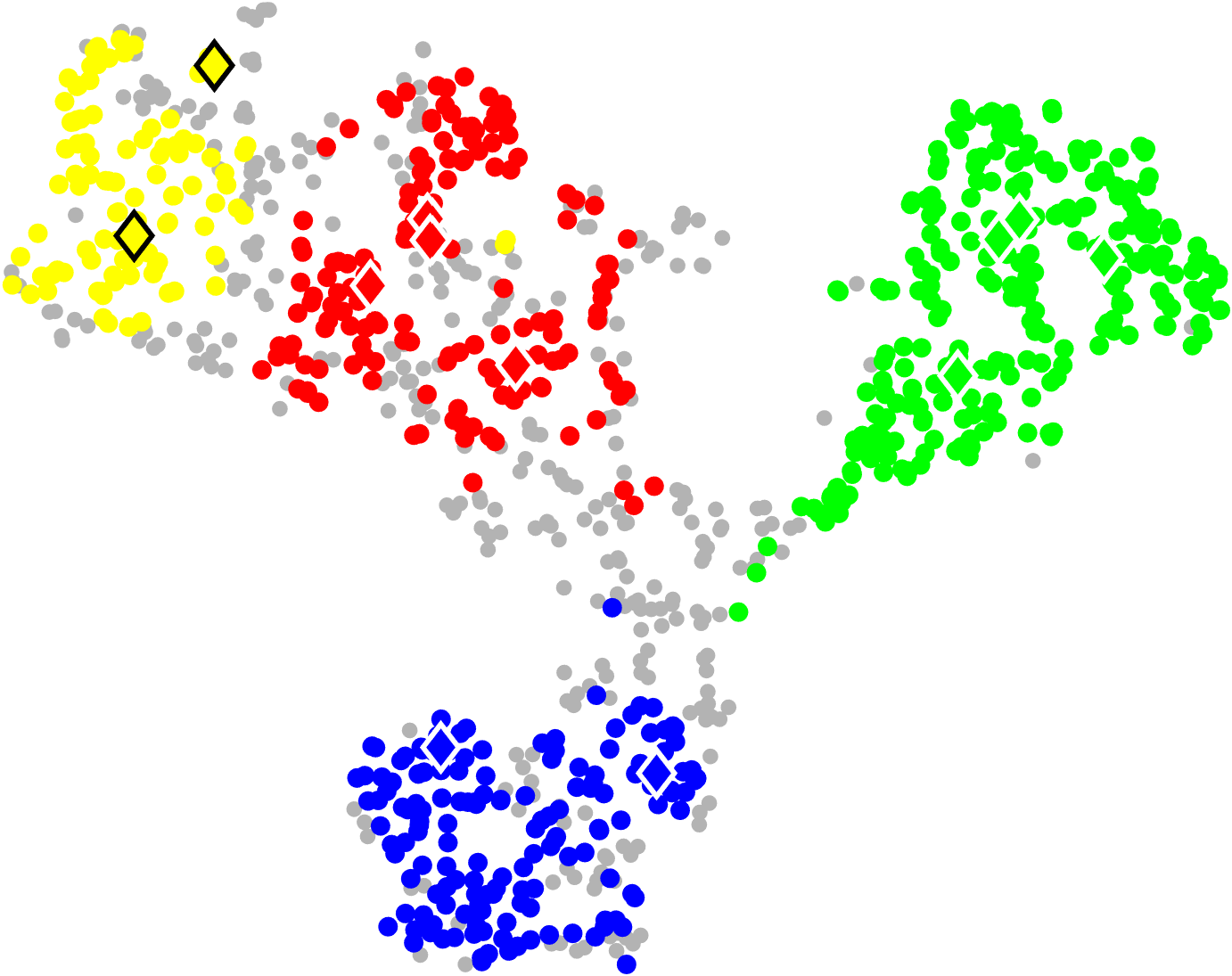}
\caption{}
\end{subfigure}
\hfill
\begin{subfigure}[H!]{0.48\textwidth}
\centering
\includegraphics[width=\textwidth]{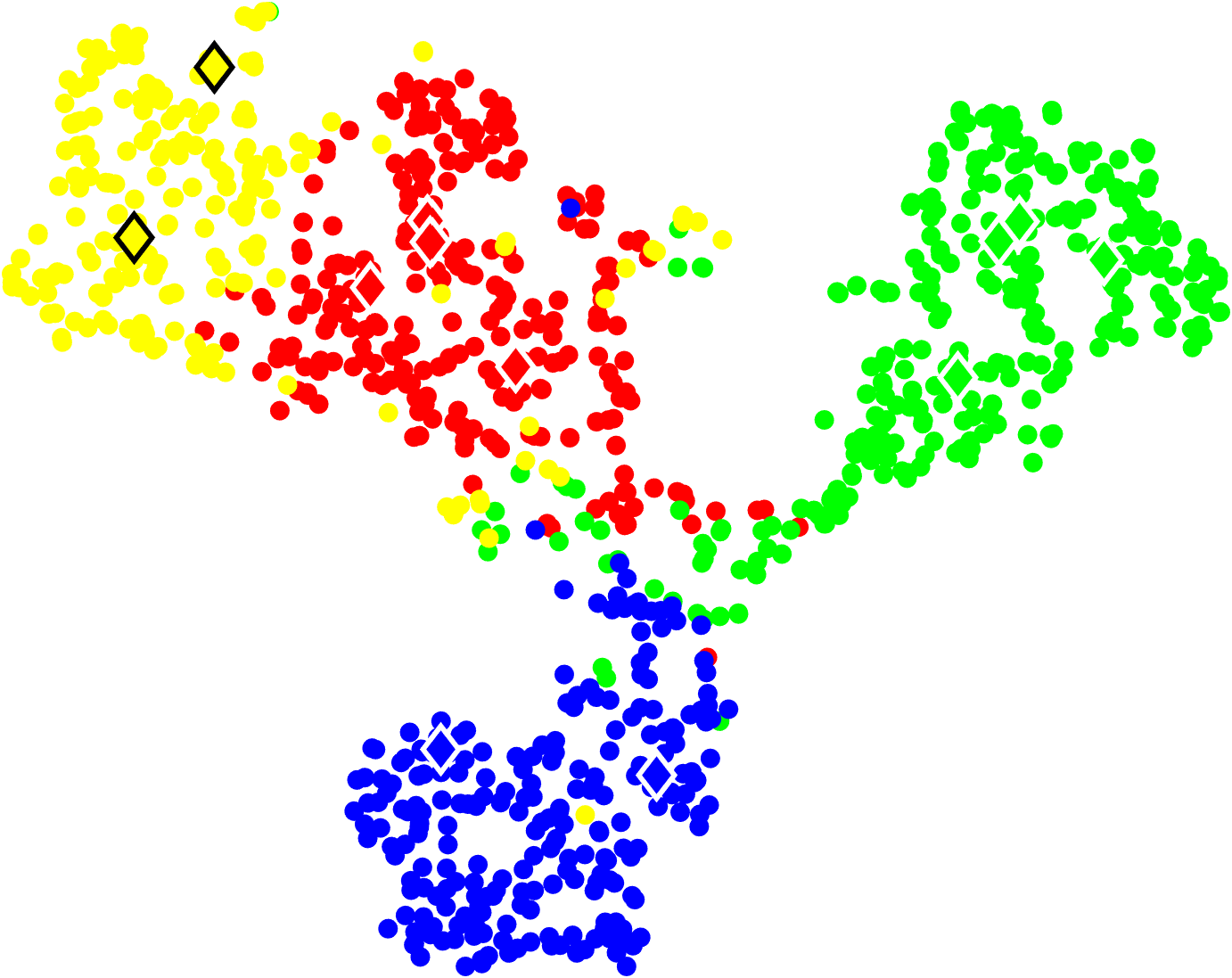}
\caption{}
\end{subfigure}
\caption{DLCC visualization on the \texttt{Ray} dataset. (a) Ground truth labels; (b) Grouped local centers; (c) Temporary clusters; (d) Final DLCC clustering result.}
\label{fig:dlcc-simubyPassino}
\end{figure*}

% -------- seed --------
\begin{figure*}[ht]
\centering
\begin{subfigure}[H!]{0.48\textwidth}
\centering
\includegraphics[width=\textwidth]{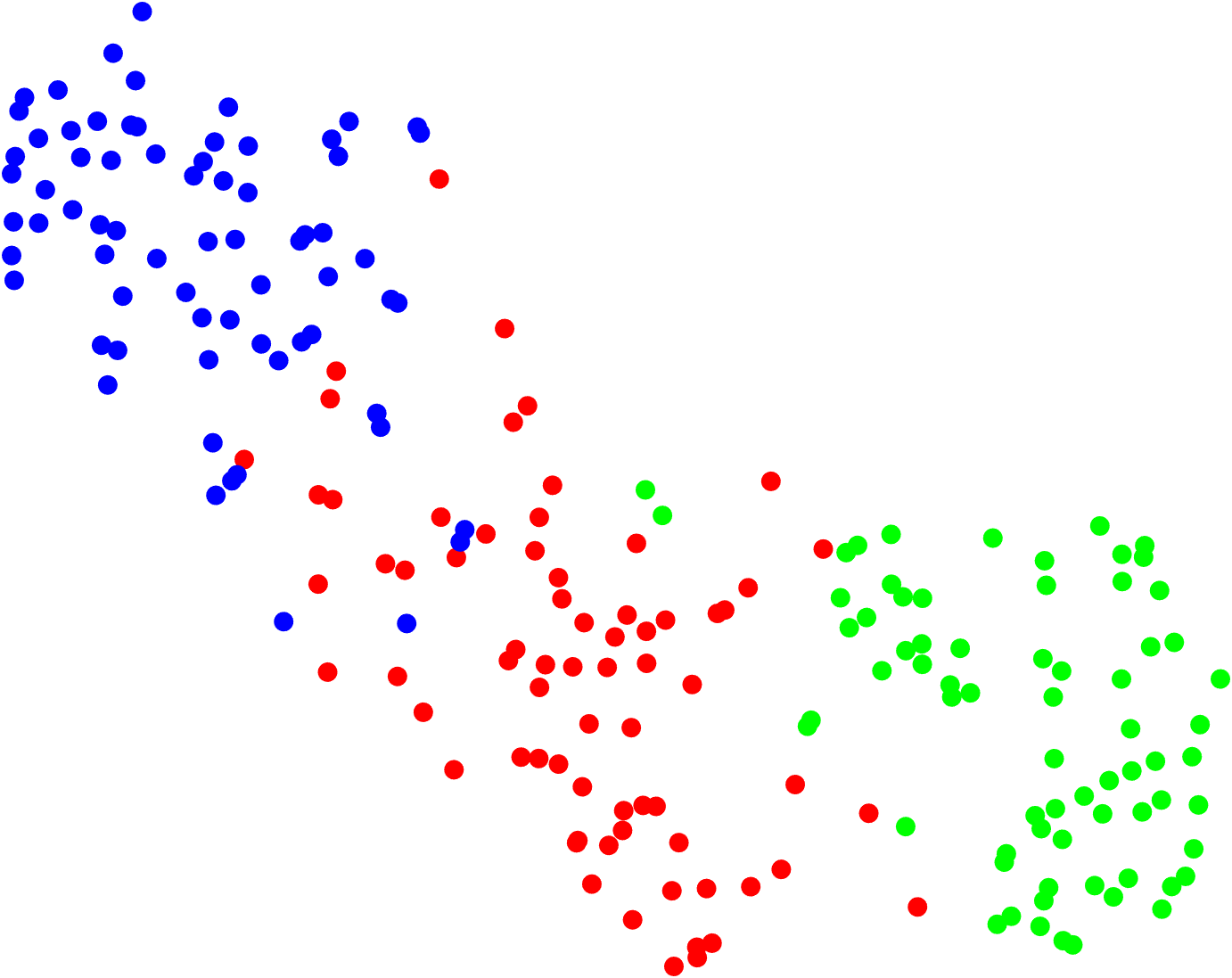}
\caption{}
\end{subfigure}
\hfill
\begin{subfigure}[H!]{0.48\textwidth}
\centering
\includegraphics[width=\textwidth]{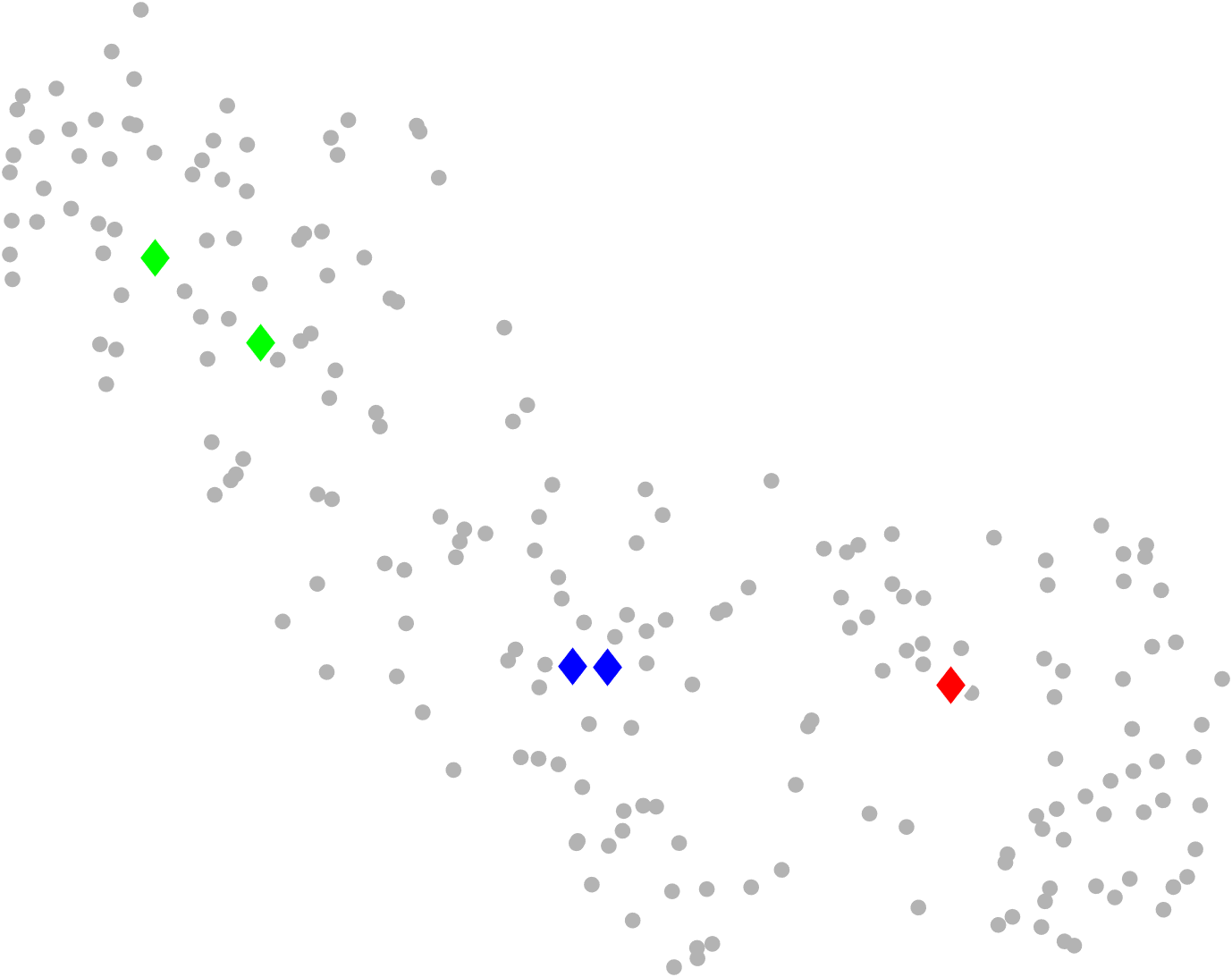}
\caption{}
\end{subfigure}

\begin{subfigure}[H!]{0.48\textwidth}
\centering
\includegraphics[width=\textwidth]{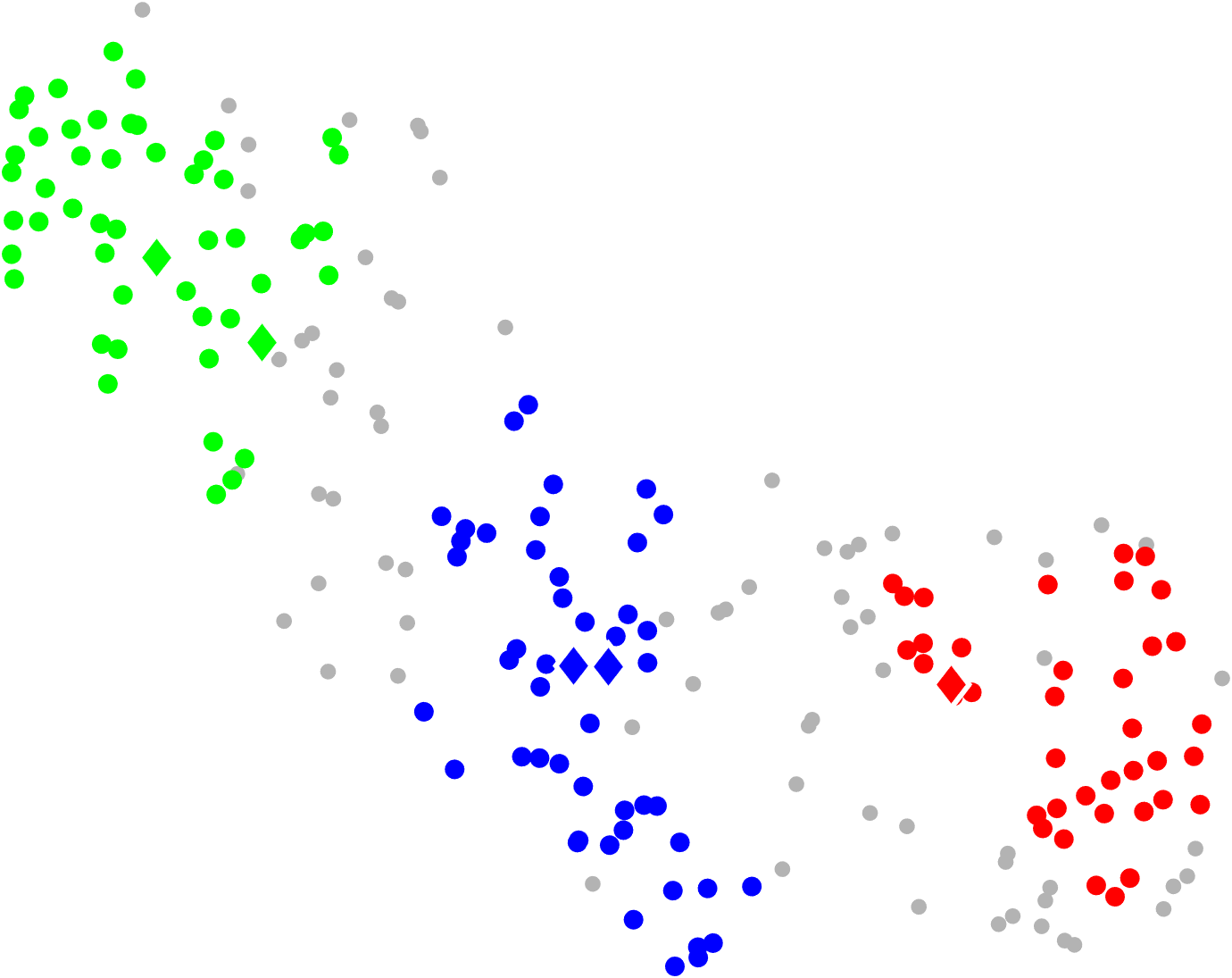}
\caption{}
\end{subfigure}
\hfill
\begin{subfigure}[H!]{0.48\textwidth}
\centering
\includegraphics[width=\textwidth]{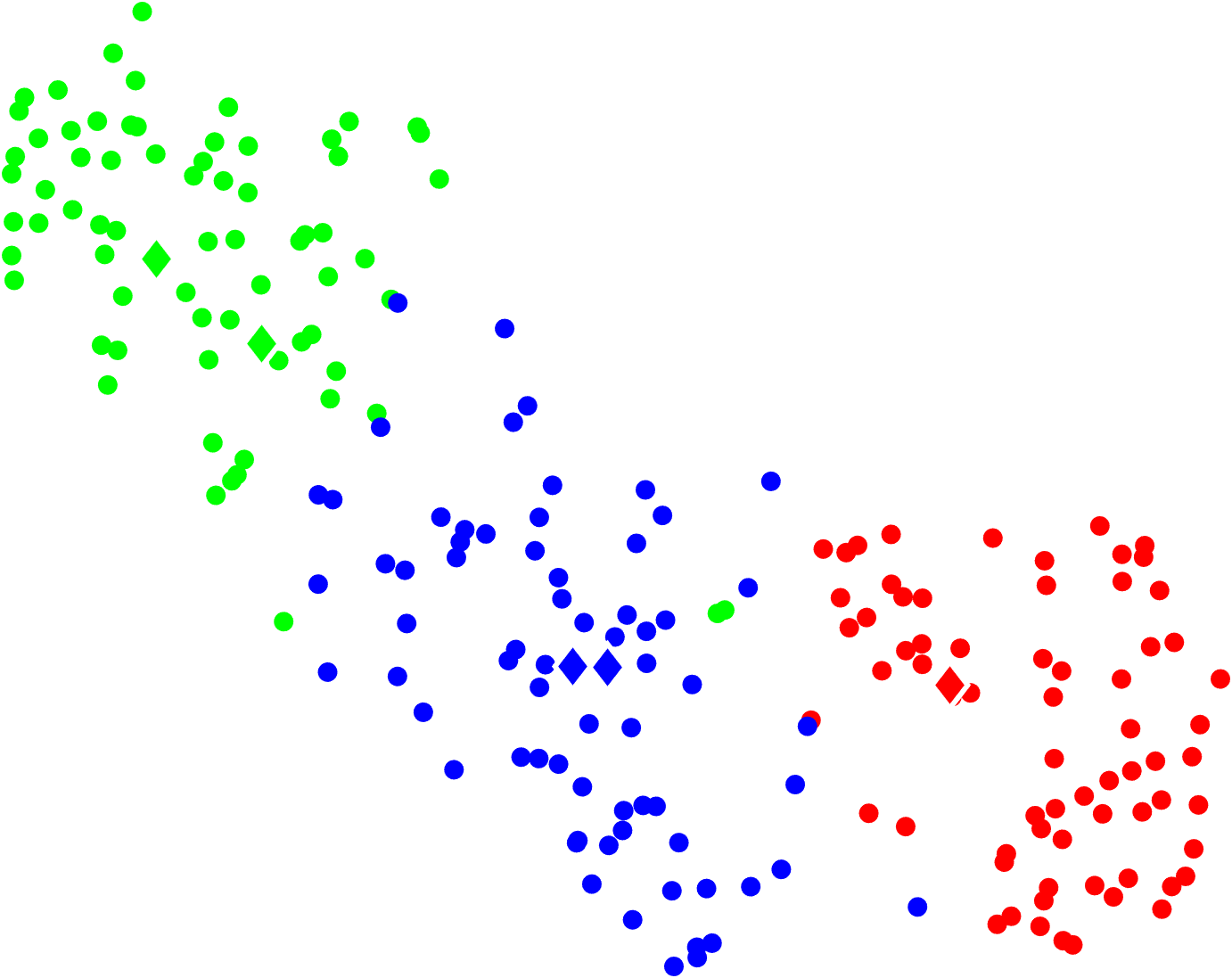}
\caption{}
\end{subfigure}
\caption{DLCC visualization on the \texttt{Seed} dataset. (a) Ground truth labels; (b) Grouped local centers; (c) Temporary clusters; (d) Final DLCC clustering result.}
\label{fig:dlcc-seed}
\end{figure*}

% -------- seg --------
\begin{figure*}[ht]
\centering
\begin{subfigure}[H!]{0.48\textwidth}
\centering
\includegraphics[width=\textwidth]{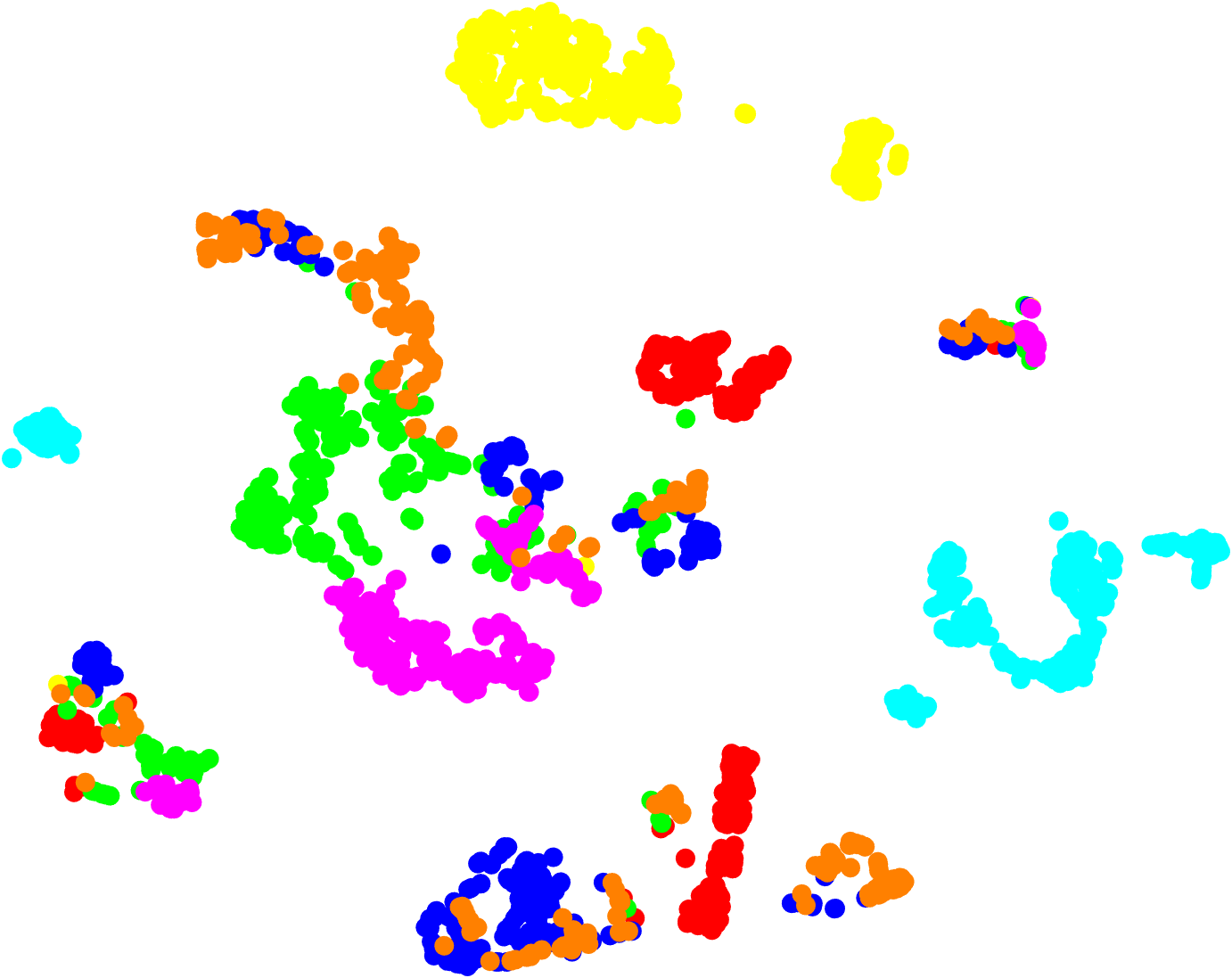}
\caption{}
\end{subfigure}
\hfill
\begin{subfigure}[H!]{0.48\textwidth}
\centering
\includegraphics[width=\textwidth]{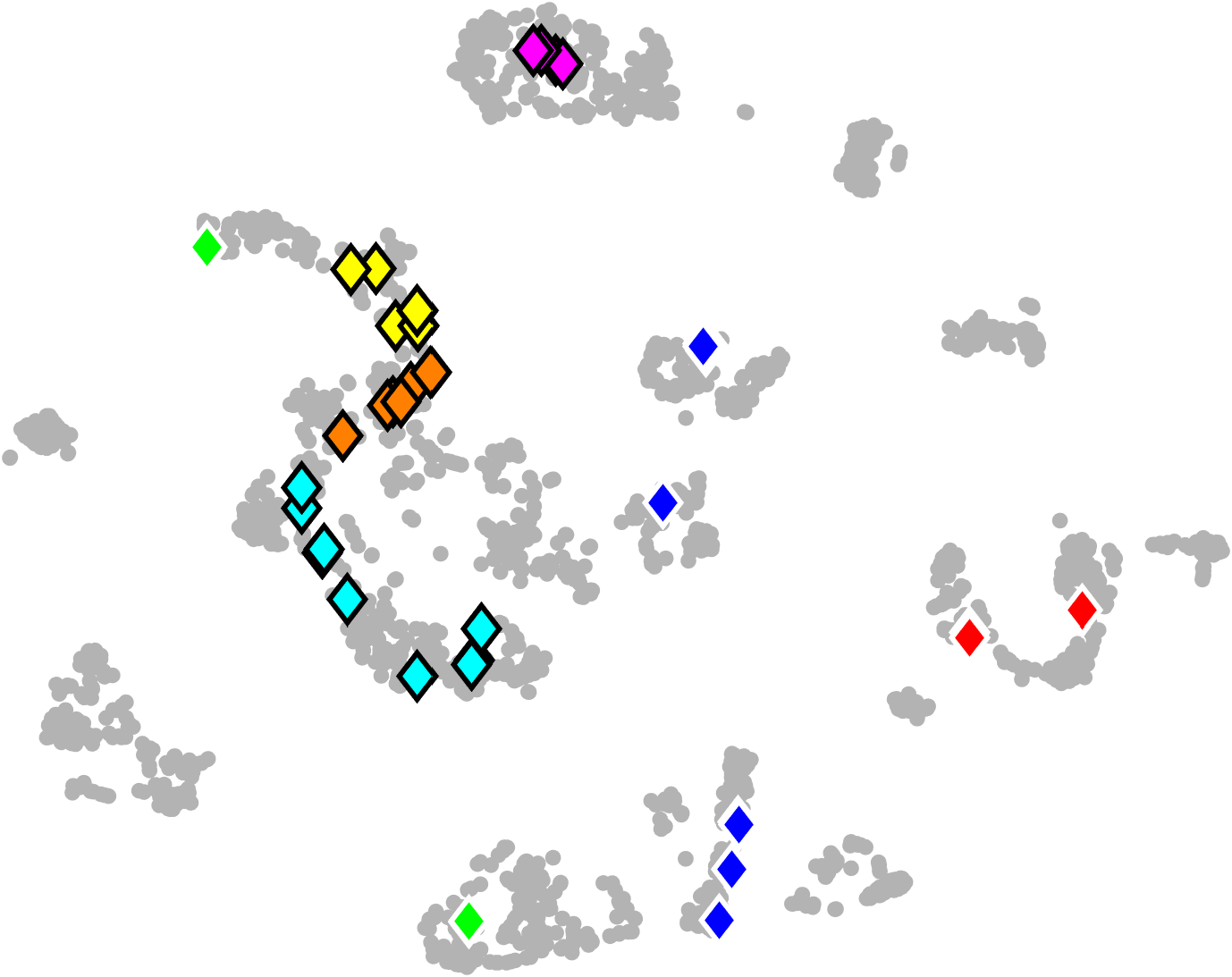}
\caption{}
\end{subfigure}

\begin{subfigure}[H!]{0.48\textwidth}
\centering
\includegraphics[width=\textwidth]{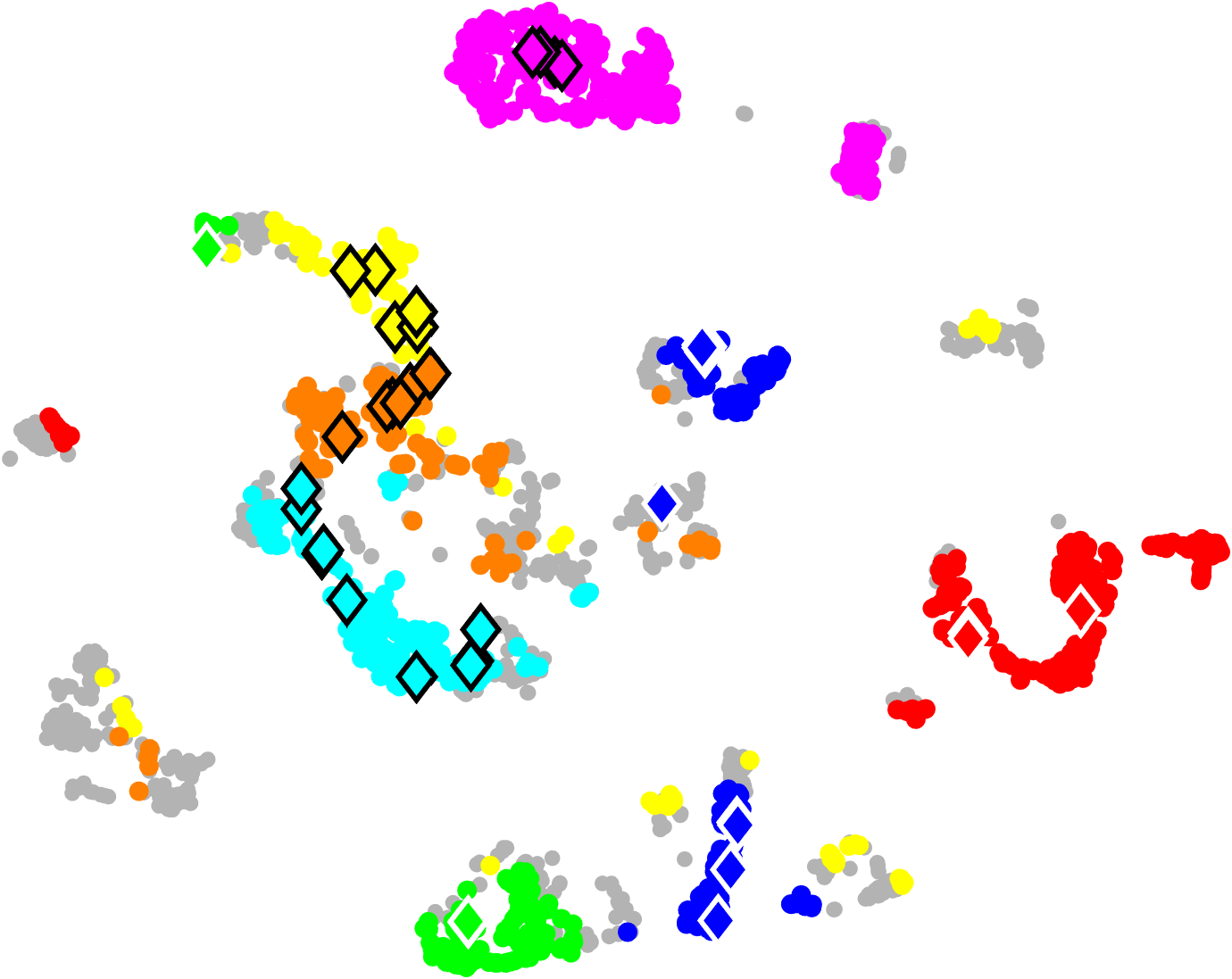}
\caption{}
\end{subfigure}
\hfill
\begin{subfigure}[H!]{0.48\textwidth}
\centering
\includegraphics[width=\textwidth]{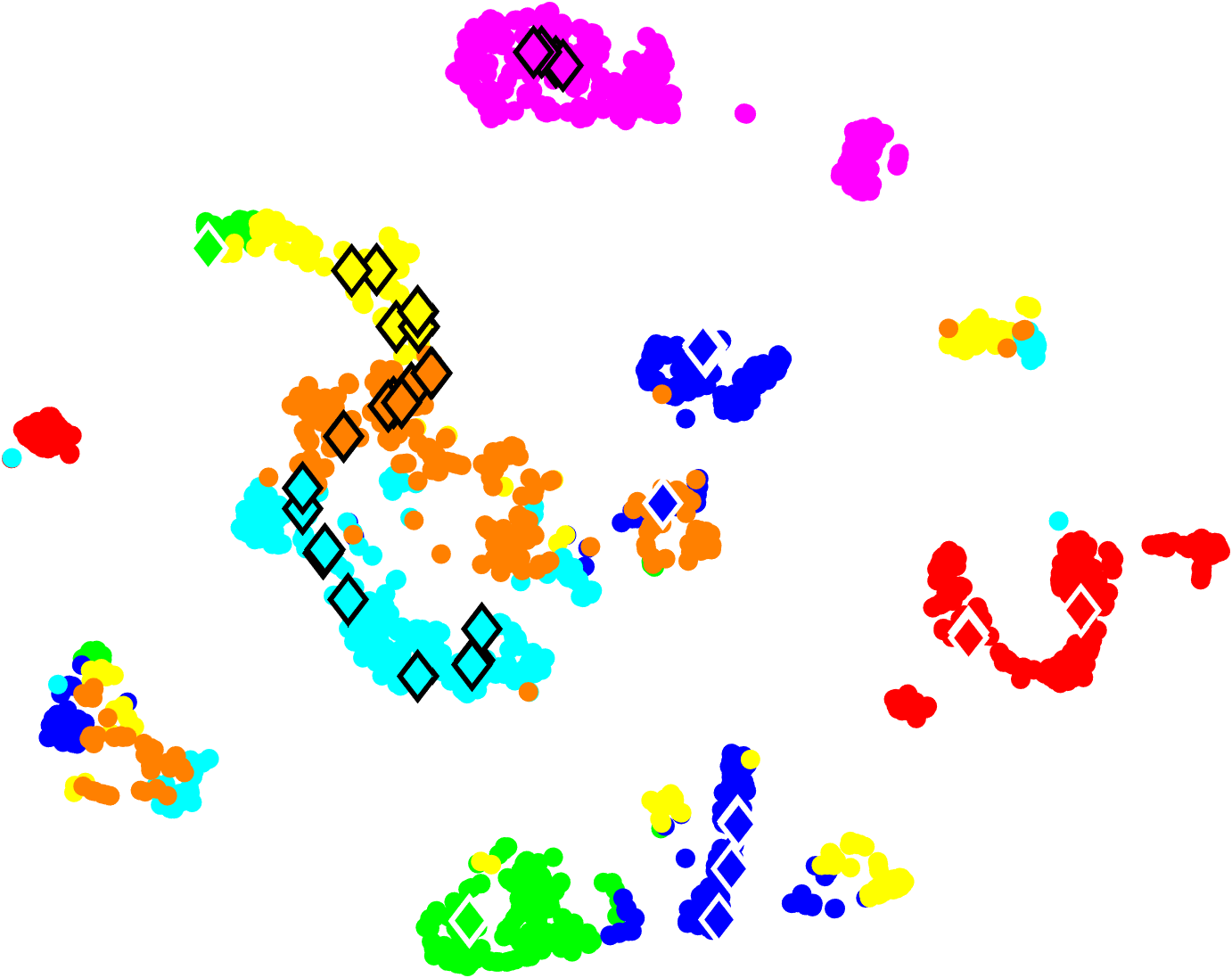}
\caption{}
\end{subfigure}
\caption{DLCC visualization on the \texttt{Segmentation} dataset. (a) Ground truth labels; (b) Grouped local centers; (c) Temporary clusters; (d) Final DLCC clustering result.}
\label{fig:dlcc-seg}
\end{figure*}

% -------- wine --------
\begin{figure*}[ht]
\centering
\begin{subfigure}[H!]{0.48\textwidth}
\centering
\includegraphics[width=\textwidth]{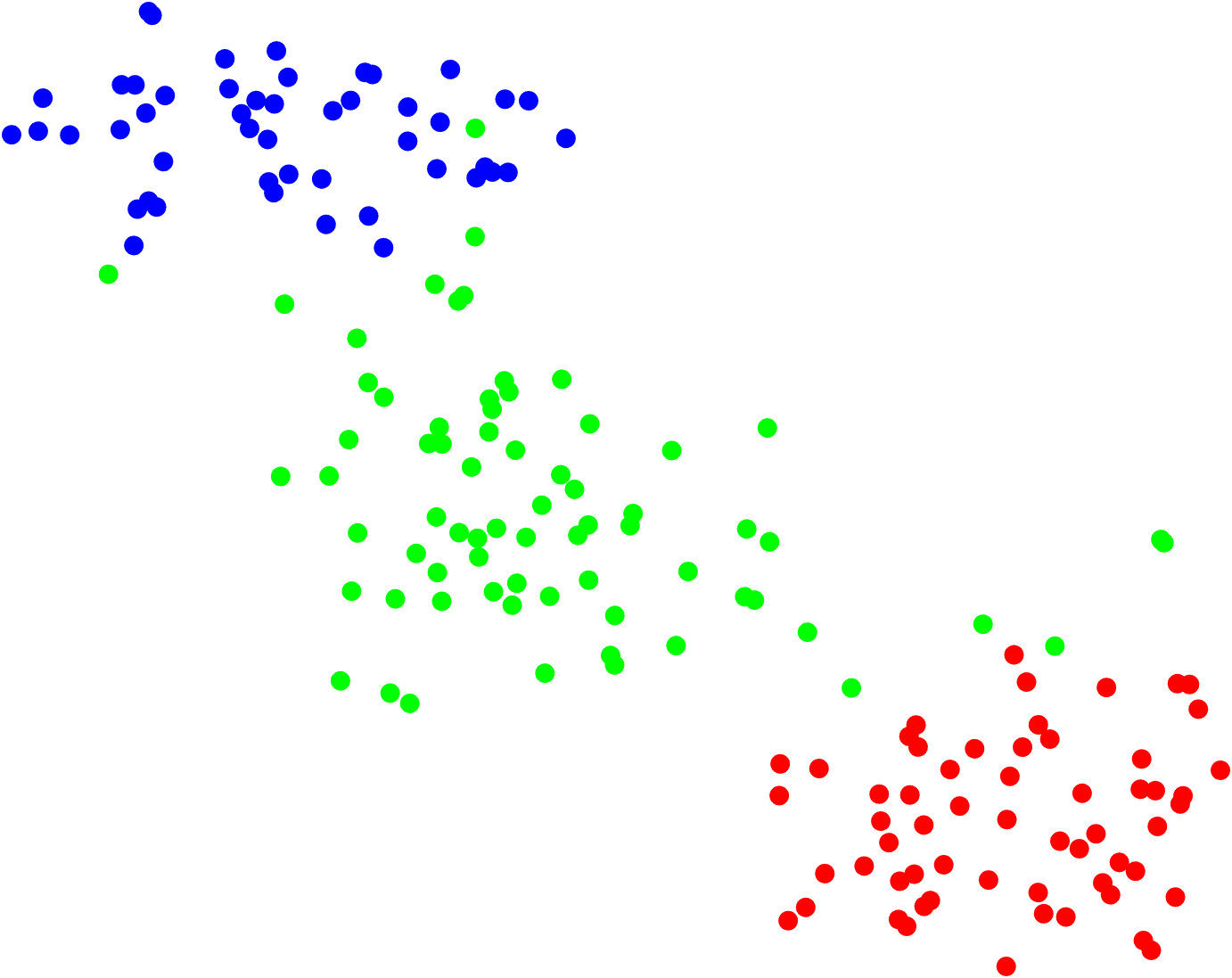}
\caption{}
\end{subfigure}
\hfill
\begin{subfigure}[H!]{0.48\textwidth}
\centering
\includegraphics[width=\textwidth]{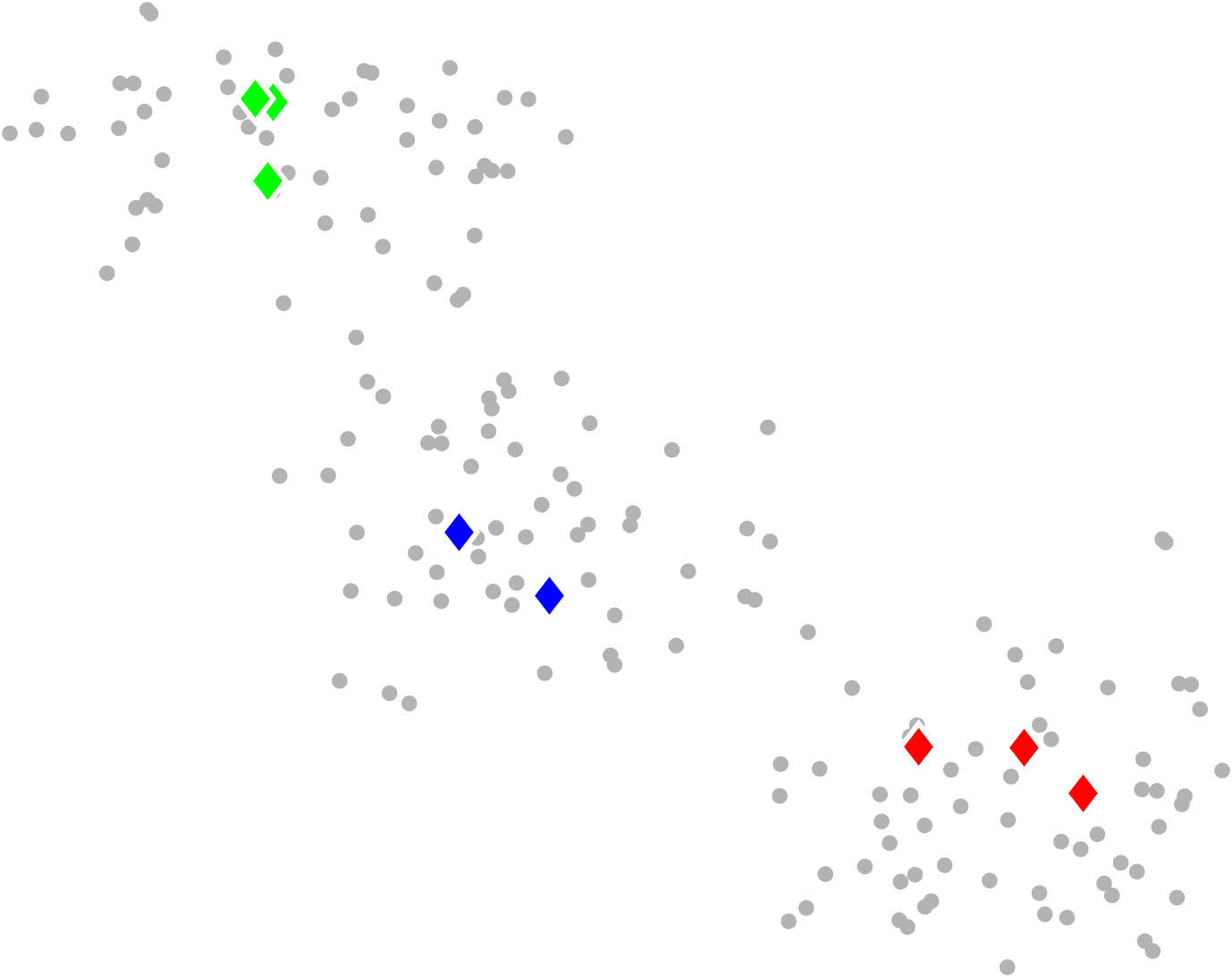}
\caption{}
\end{subfigure}

\begin{subfigure}[H!]{0.48\textwidth}
\centering
\includegraphics[width=\textwidth]{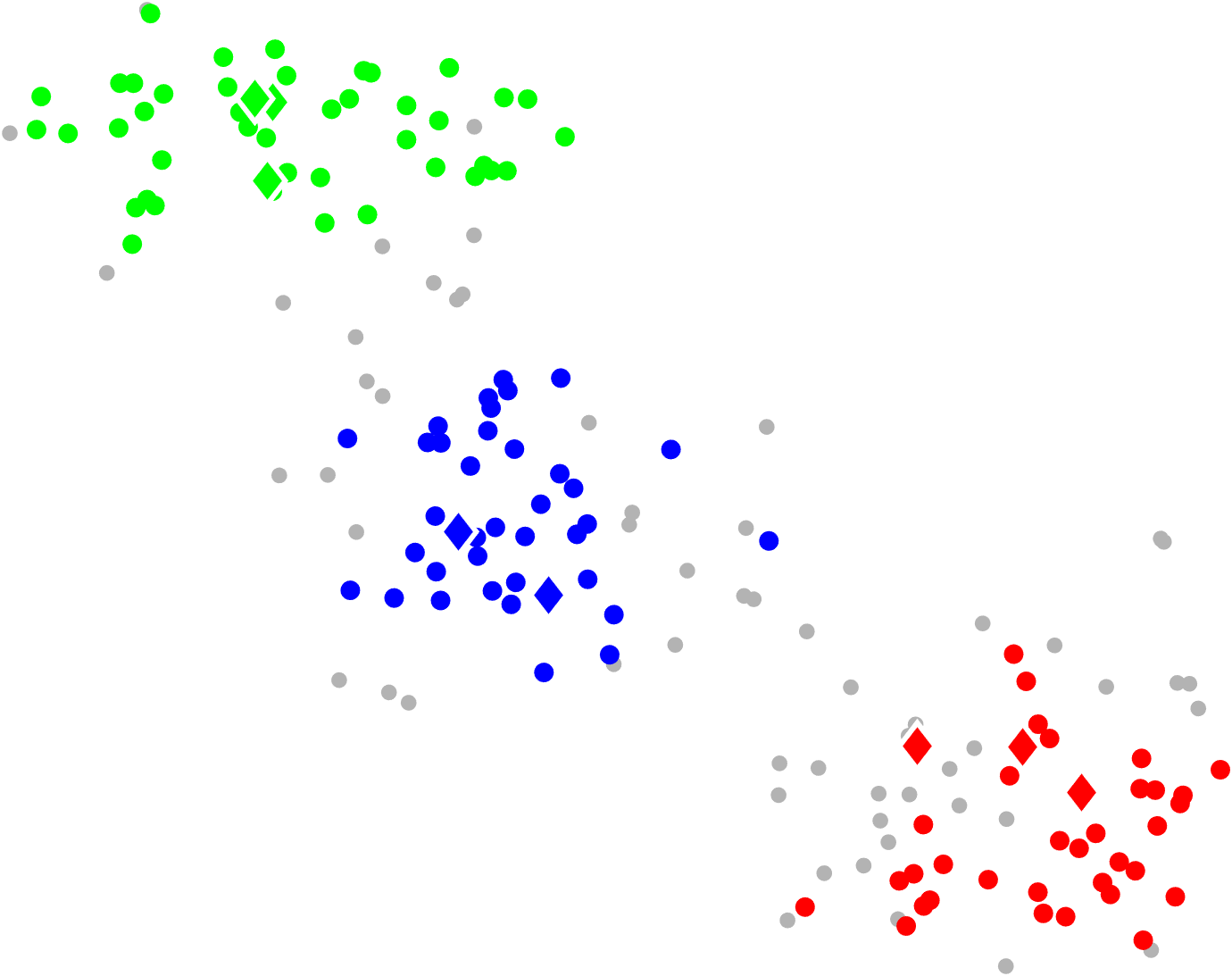}
\caption{}
\end{subfigure}
\hfill
\begin{subfigure}[H!]{0.48\textwidth}
\centering
\includegraphics[width=\textwidth]{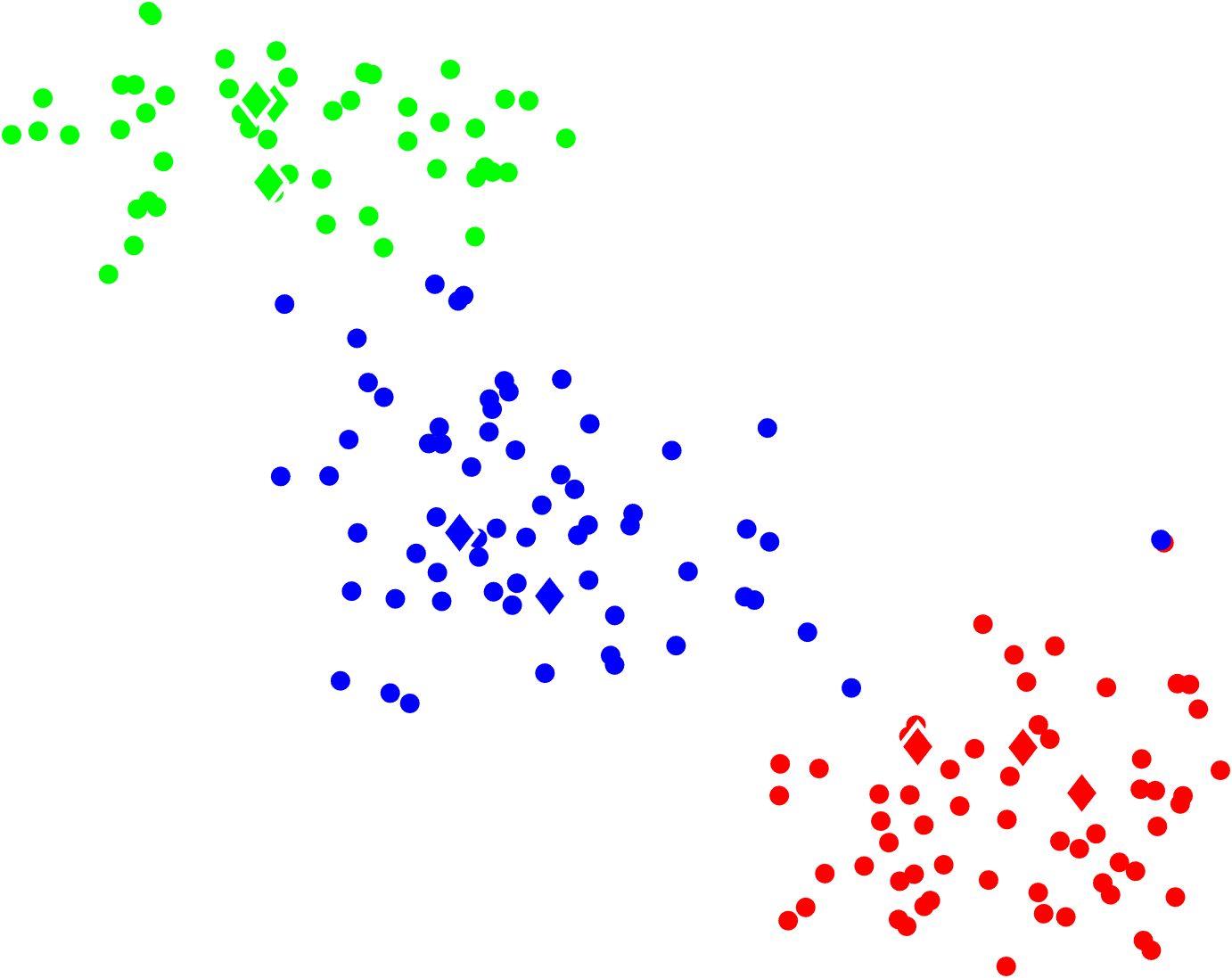}
\caption{}
\end{subfigure}
\caption{DLCC visualization on the \texttt{Wine} dataset. (a) Ground truth labels; (b) Grouped local centers; (c) Temporary clusters; (d) Final DLCC clustering result.}
\label{fig:dlcc-wine}
\end{figure*}

% -------- yale --------
\begin{figure*}[ht]
\centering
\begin{subfigure}[H!]{0.48\textwidth}
\centering
\includegraphics[width=\textwidth]{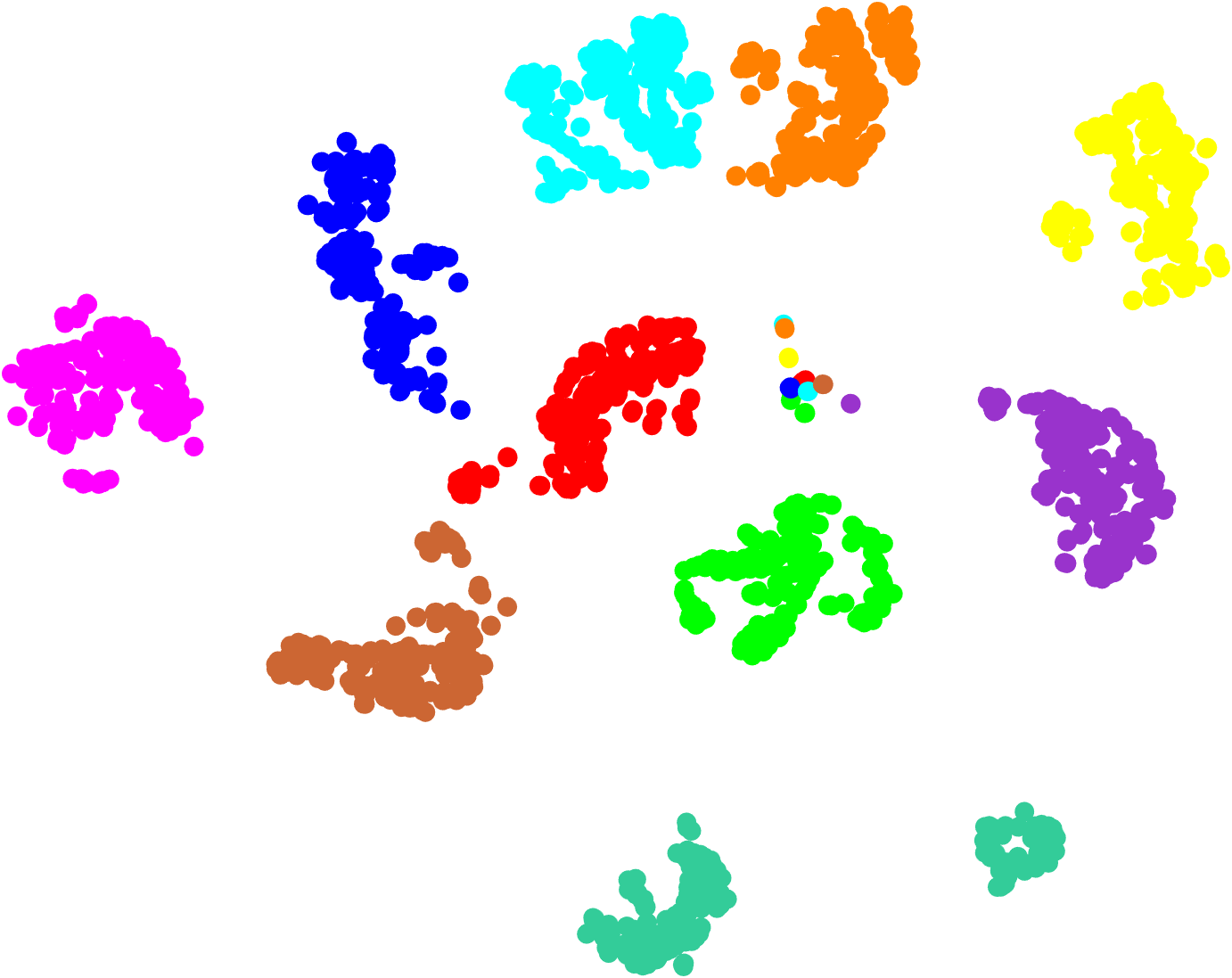}
\caption{}
\end{subfigure}
\hfill
\begin{subfigure}[H!]{0.48\textwidth}
\centering
\includegraphics[width=\textwidth]{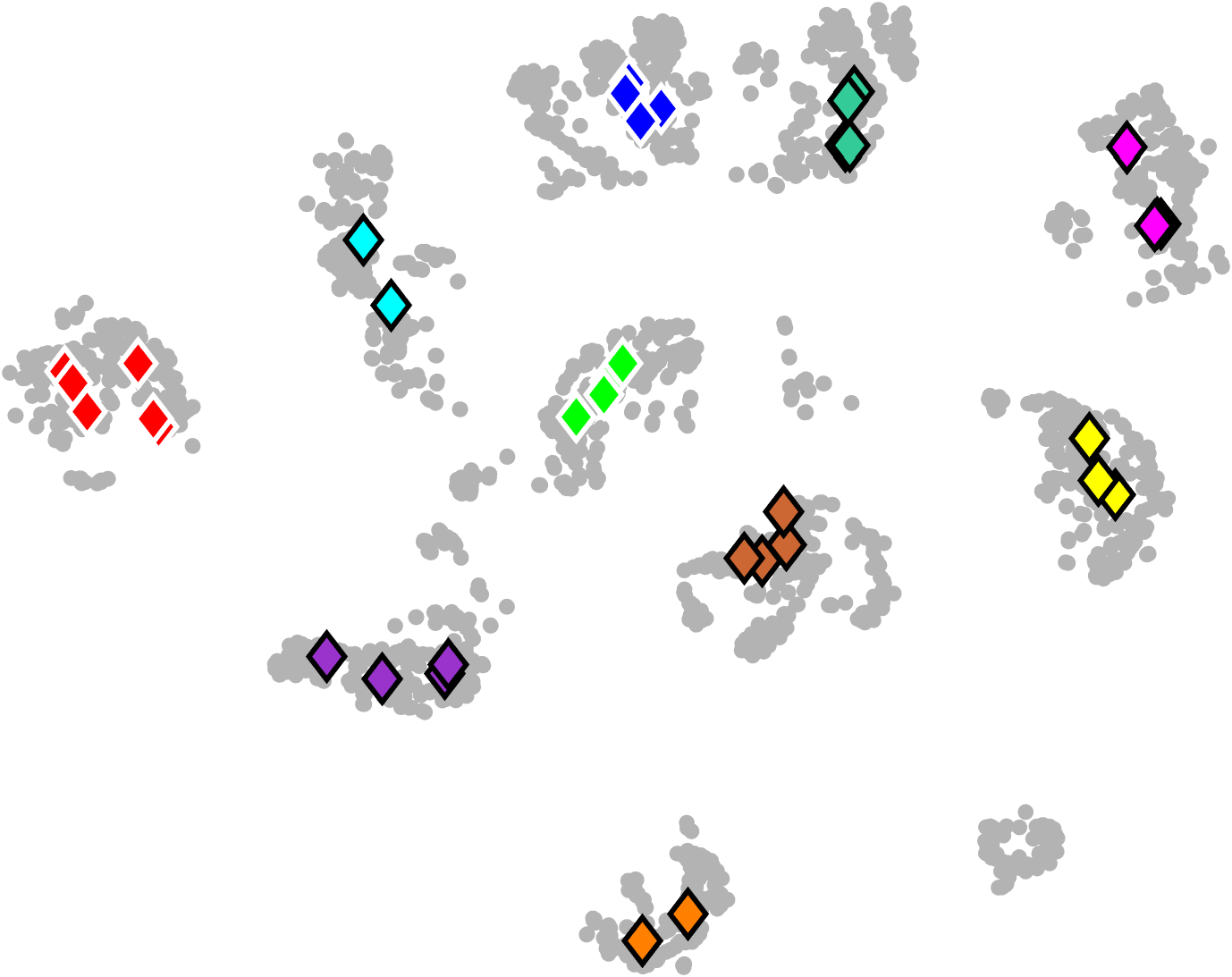}
\caption{}
\end{subfigure}

\begin{subfigure}[H!]{0.48\textwidth}
\centering
\includegraphics[width=\textwidth]{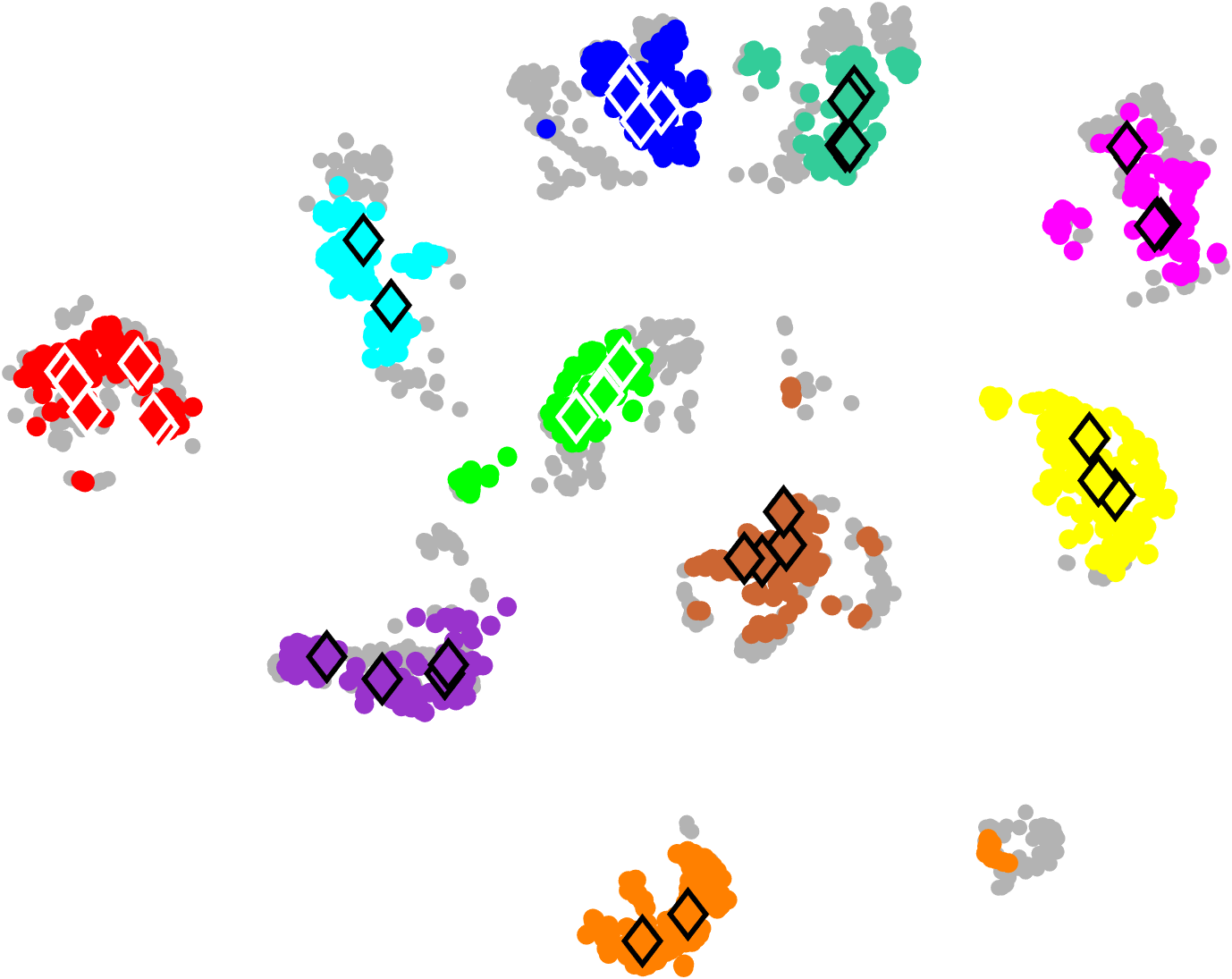}
\caption{}
\end{subfigure}
\hfill
\begin{subfigure}[H!]{0.48\textwidth}
\centering
\includegraphics[width=\textwidth]{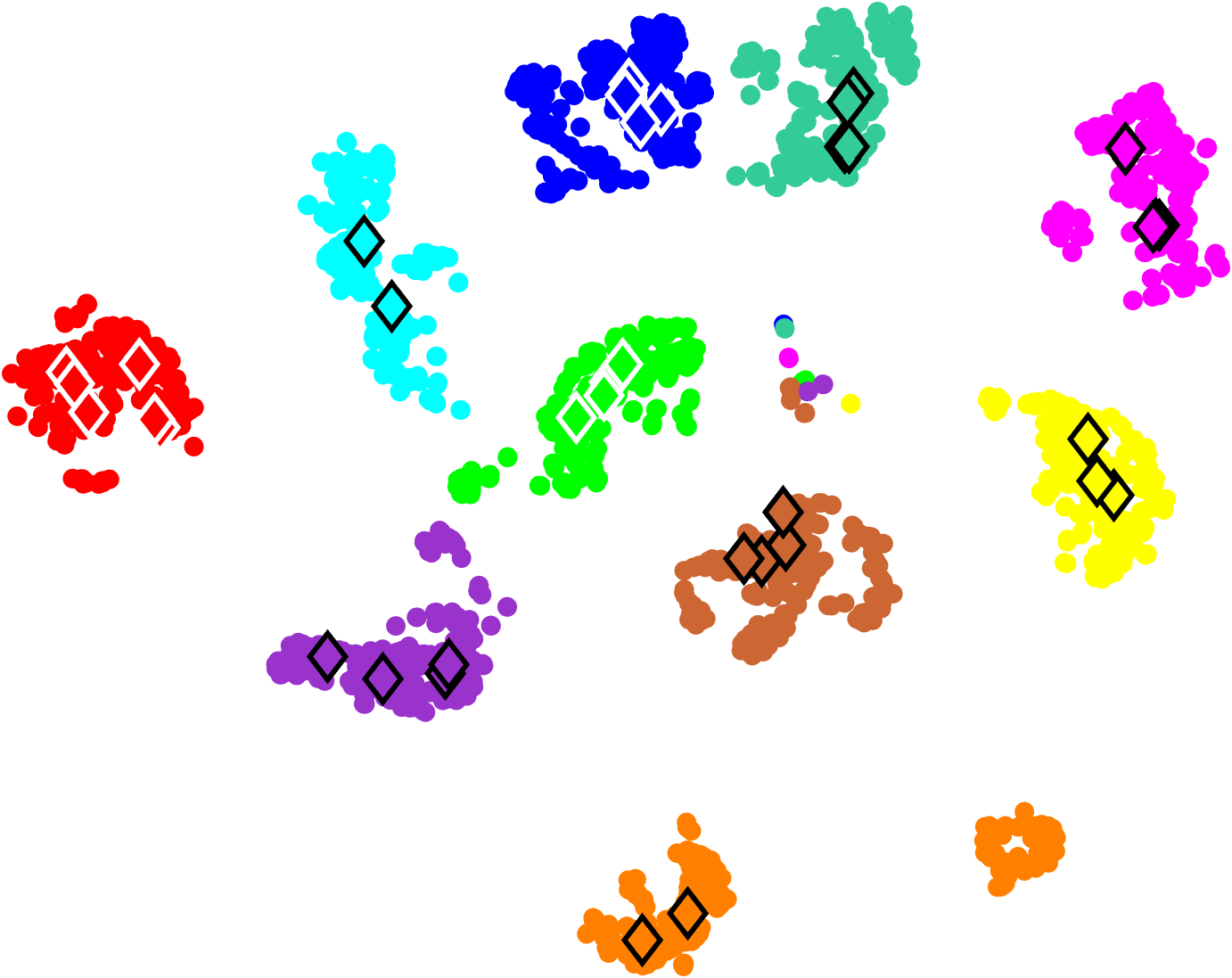}
\caption{}
\end{subfigure}
\caption{DLCC visualization on the \texttt{Yale} dataset. (a) Ground truth labels; (b) Grouped local centers; (c) Temporary clusters; (d) Final DLCC clustering result.}
\label{fig:dlcc-yale}
\end{figure*}

% -------- aggregation --------
\begin{figure*}[ht]
\centering
\begin{subfigure}[H!]{0.48\textwidth}
\centering
\includegraphics[width=\textwidth]{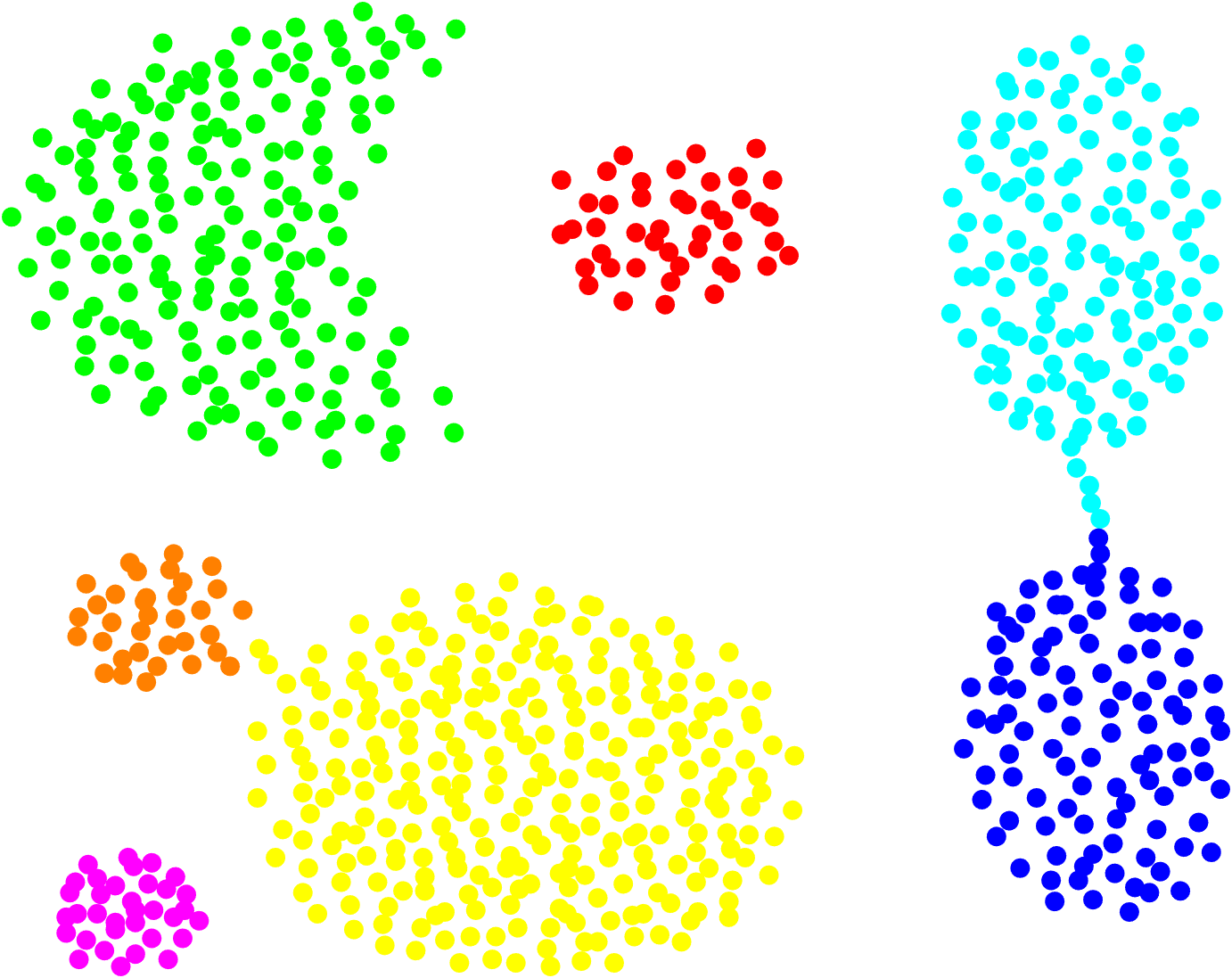}
\caption{}
\end{subfigure}
\hfill
\begin{subfigure}[H!]{0.48\textwidth}
\centering
\includegraphics[width=\textwidth]{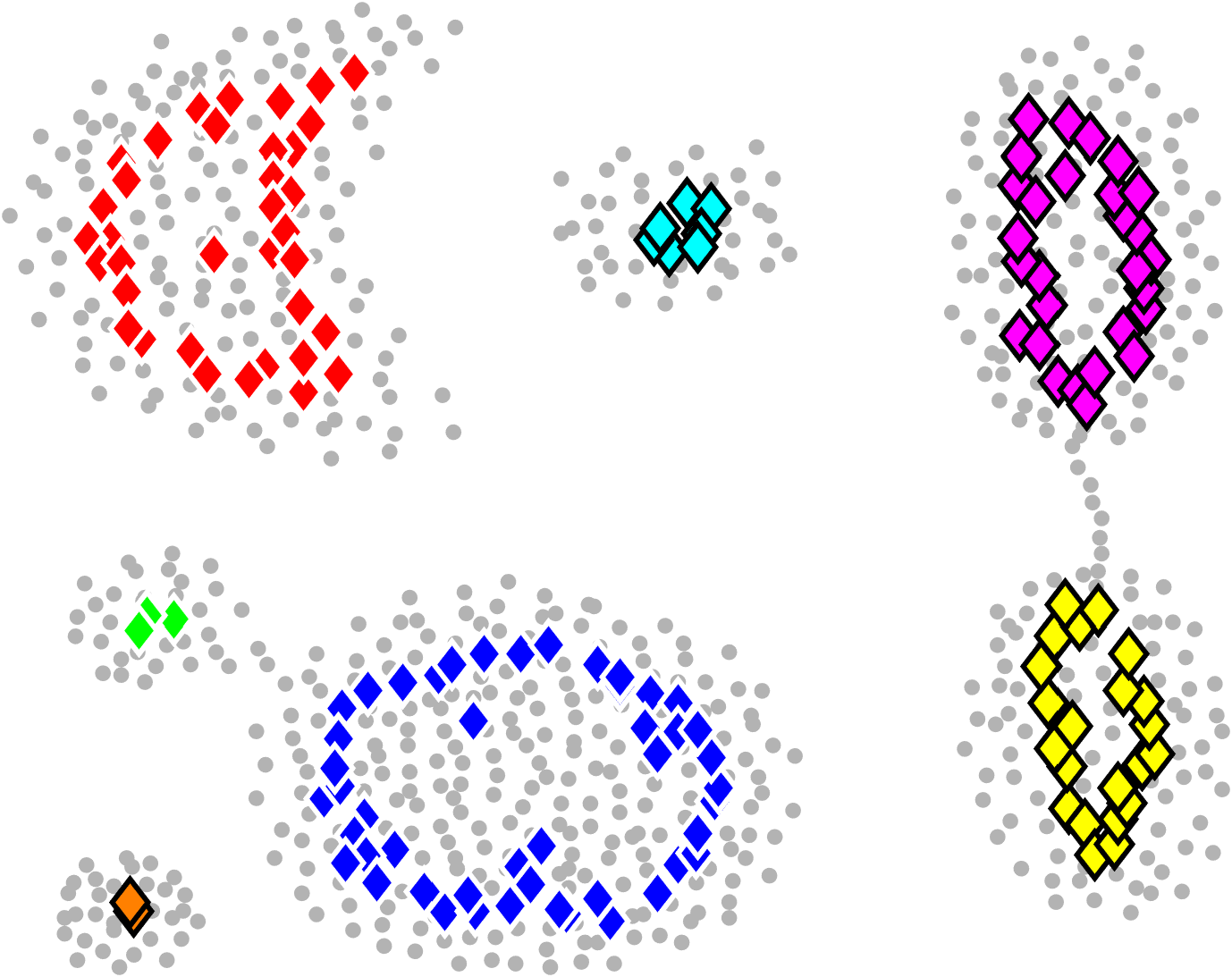}
\caption{}
\end{subfigure}

\begin{subfigure}[H!]{0.48\textwidth}
\centering
\includegraphics[width=\textwidth]{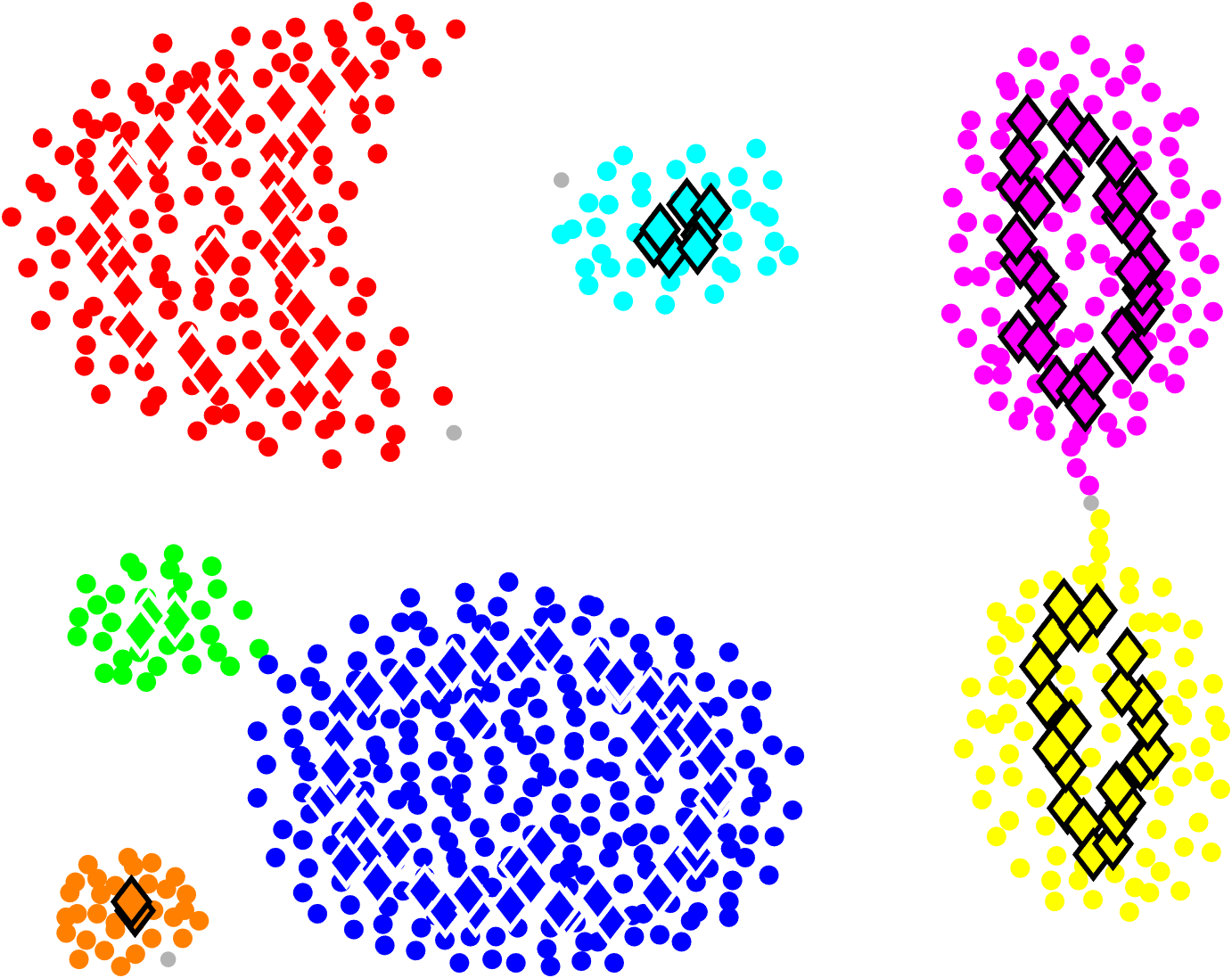}
\caption{}
\end{subfigure}
\hfill
\begin{subfigure}[H!]{0.48\textwidth}
\centering
\includegraphics[width=\textwidth]{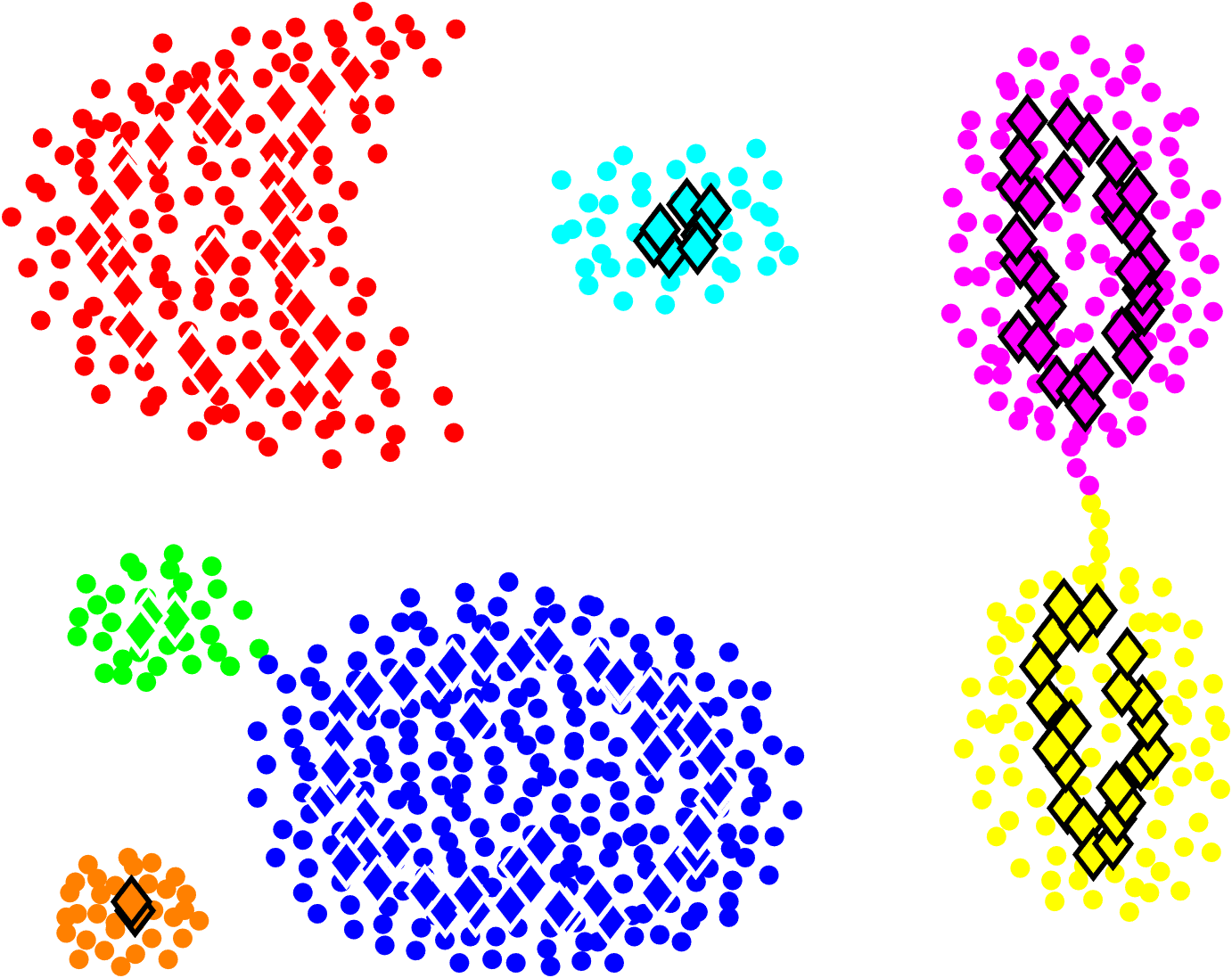}
\caption{}
\end{subfigure}
\caption{DLCC visualization on the \texttt{Aggregation} dataset. (a) Ground truth labels; (b) Grouped local centers; (c) Temporary clusters; (d) Final DLCC clustering result.}
\label{fig:dlcc-aggregation}
\end{figure*}

% -------- atom --------
\begin{figure*}[ht]
\centering
\begin{subfigure}[H!]{0.48\textwidth}
\centering
\includegraphics[width=\textwidth]{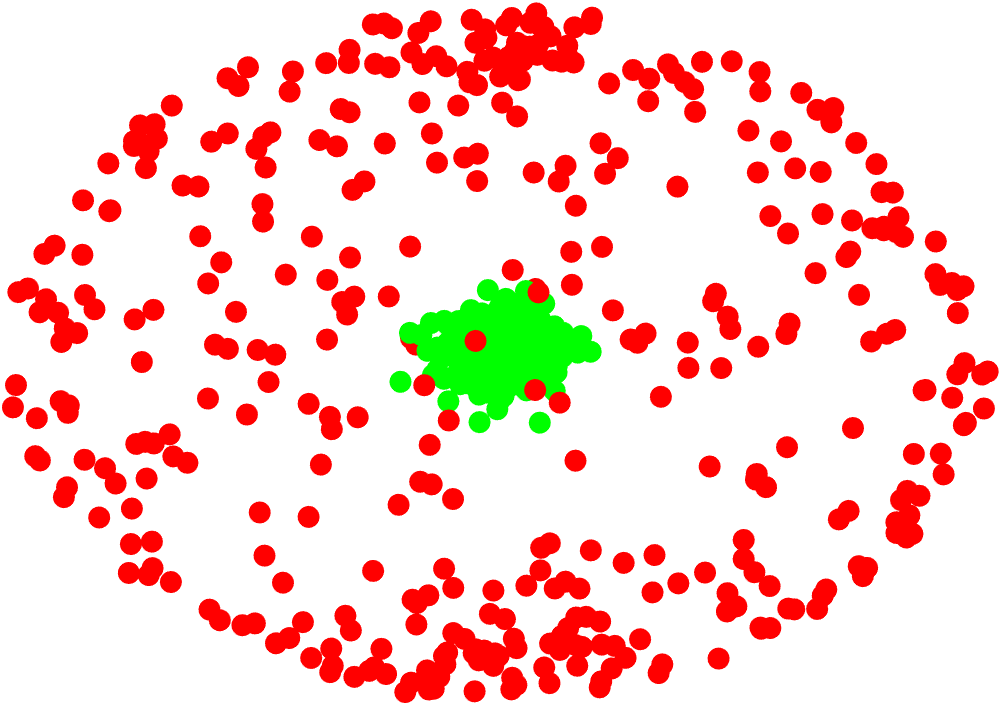}
\caption{}
\end{subfigure}
\hfill
\begin{subfigure}[H!]{0.48\textwidth}
\centering
\includegraphics[width=\textwidth]{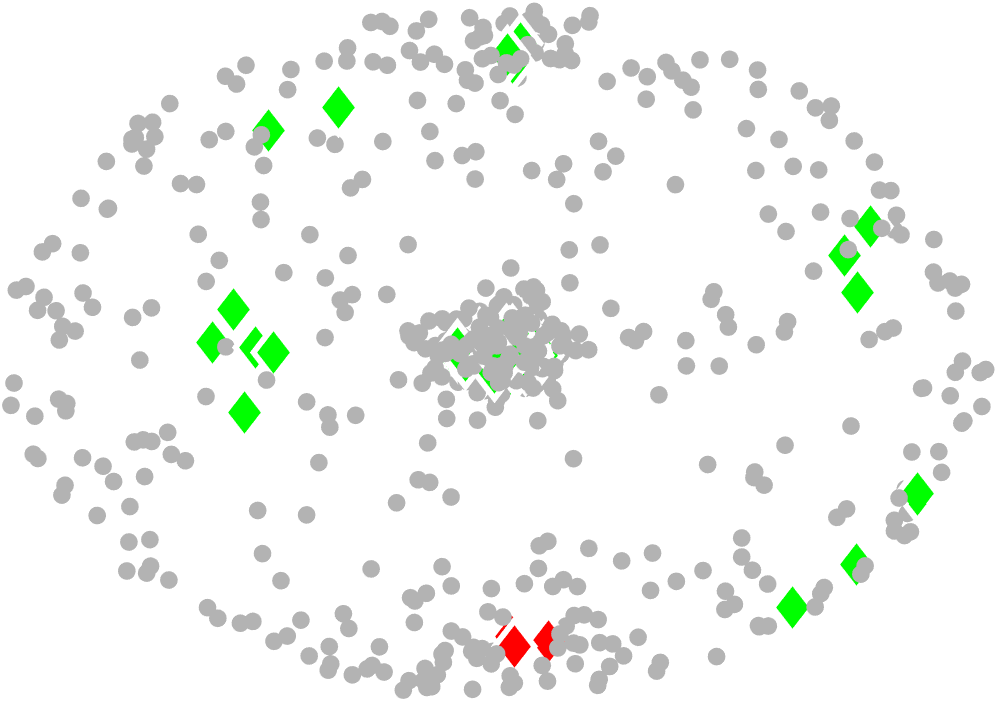}
\caption{}
\end{subfigure}

\begin{subfigure}[H!]{0.48\textwidth}
\centering
\includegraphics[width=\textwidth]{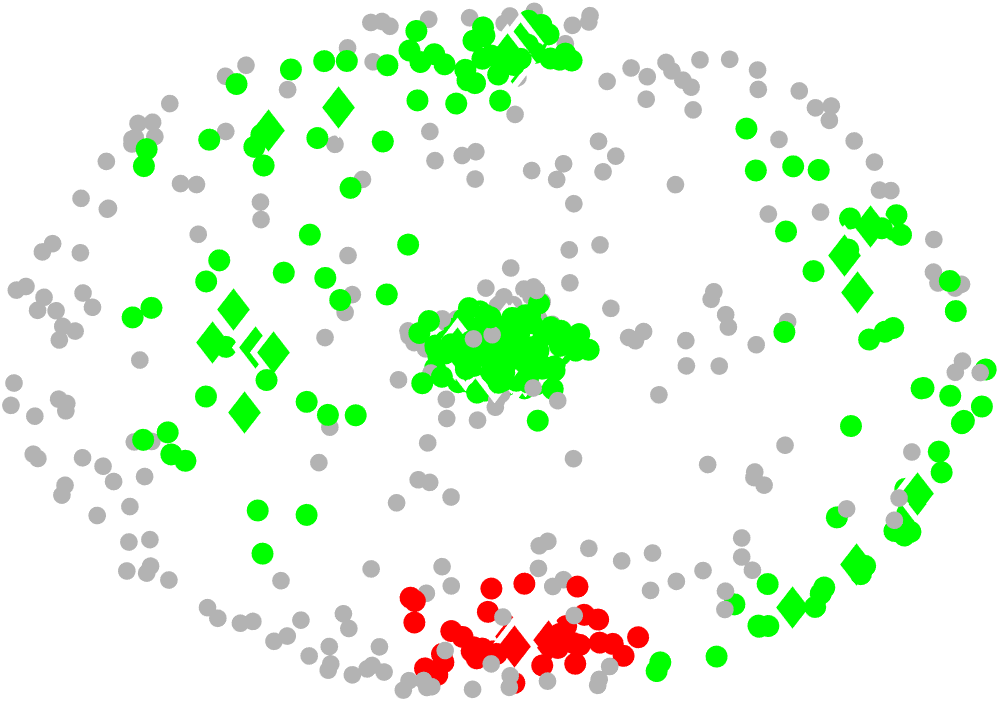}
\caption{}
\end{subfigure}
\hfill
\begin{subfigure}[H!]{0.48\textwidth}
\centering
\includegraphics[width=\textwidth]{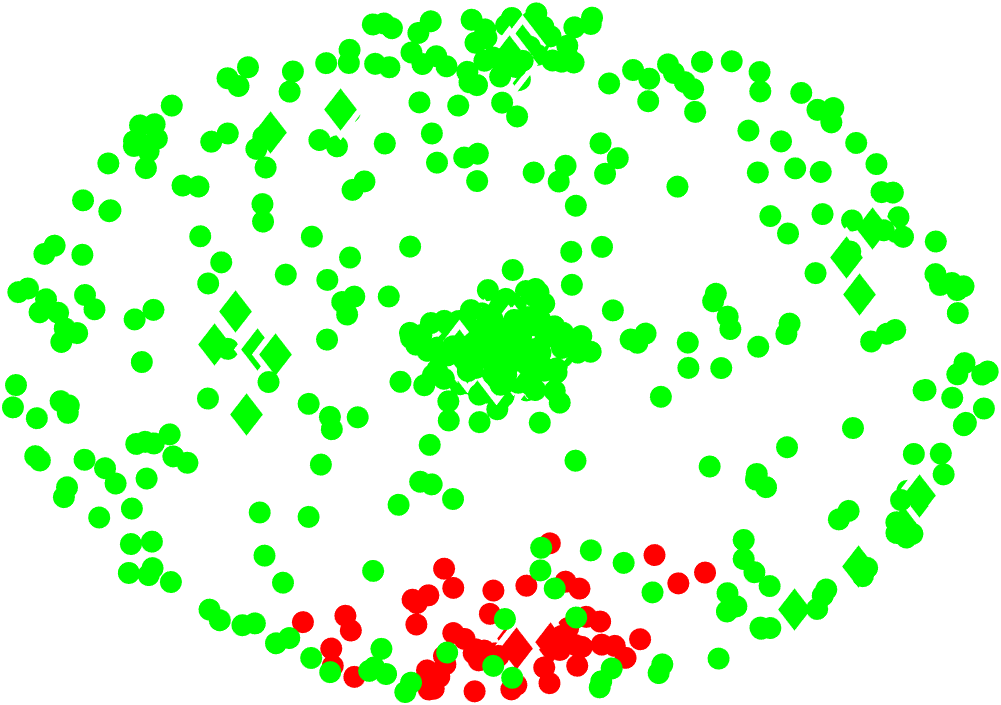}
\caption{}
\end{subfigure}
\caption{DLCC visualization on the \texttt{Atom} dataset. (a) Ground truth labels; (b) Grouped local centers; (c) Temporary clusters; (d) Final DLCC clustering result.}
\label{fig:dlcc-atom}
\end{figure*}

% -------- ba --------
\begin{figure*}[ht]
\centering
\begin{subfigure}[H!]{0.48\textwidth}
\centering
\includegraphics[width=\textwidth]{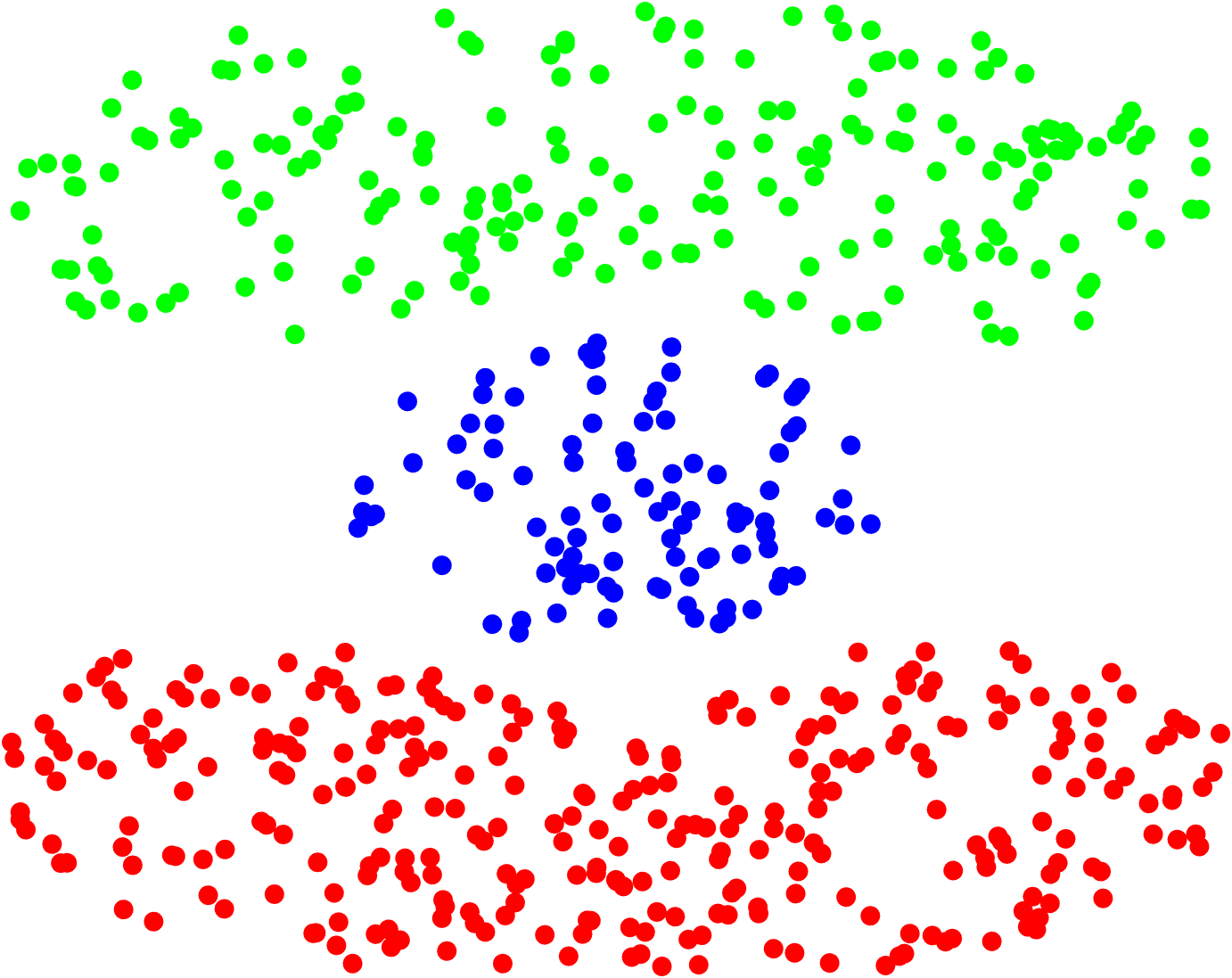}
\caption{}
\end{subfigure}
\hfill
\begin{subfigure}[H!]{0.48\textwidth}
\centering
\includegraphics[width=\textwidth]{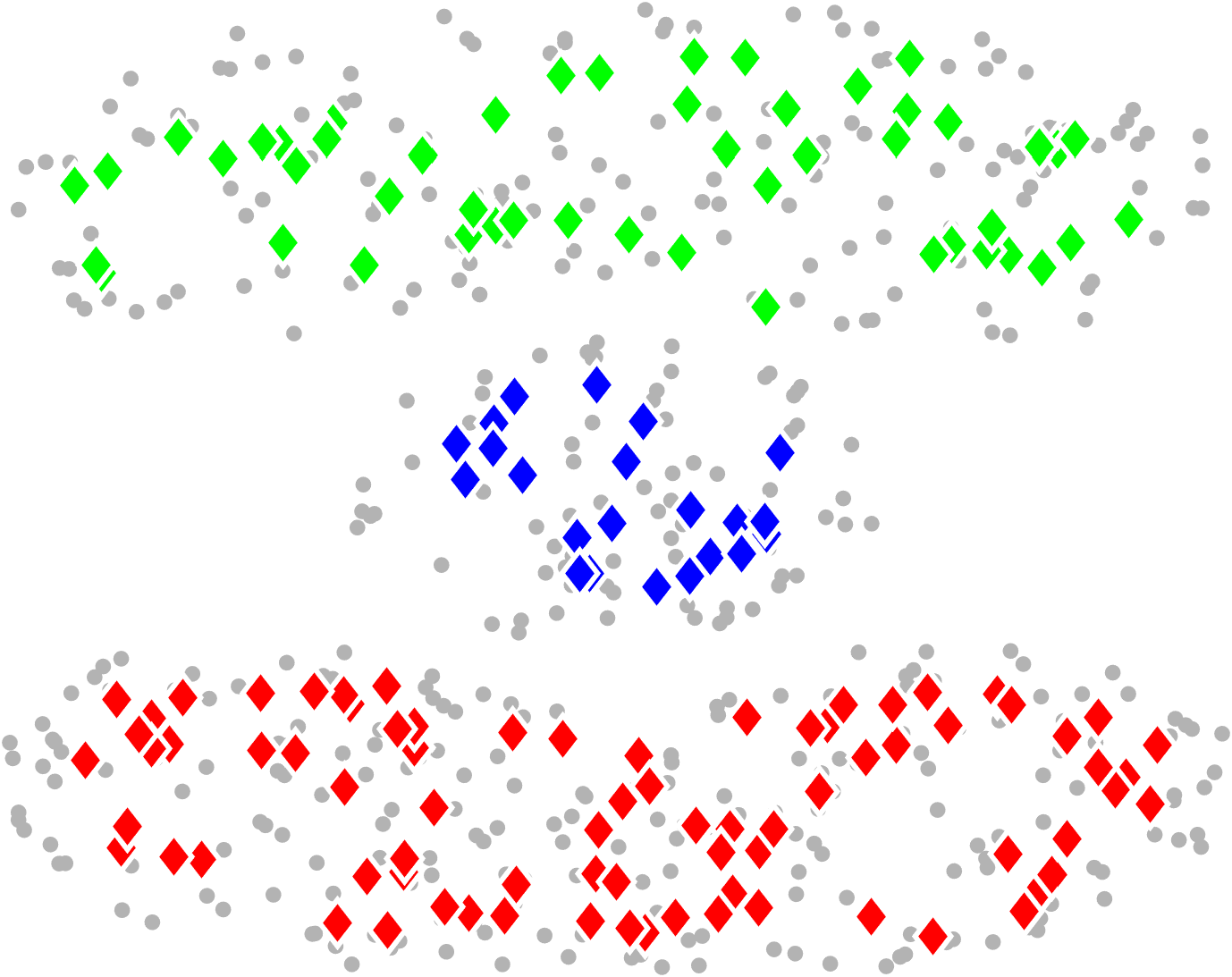}
\caption{}
\end{subfigure}

\begin{subfigure}[H!]{0.48\textwidth}
\centering
\includegraphics[width=\textwidth]{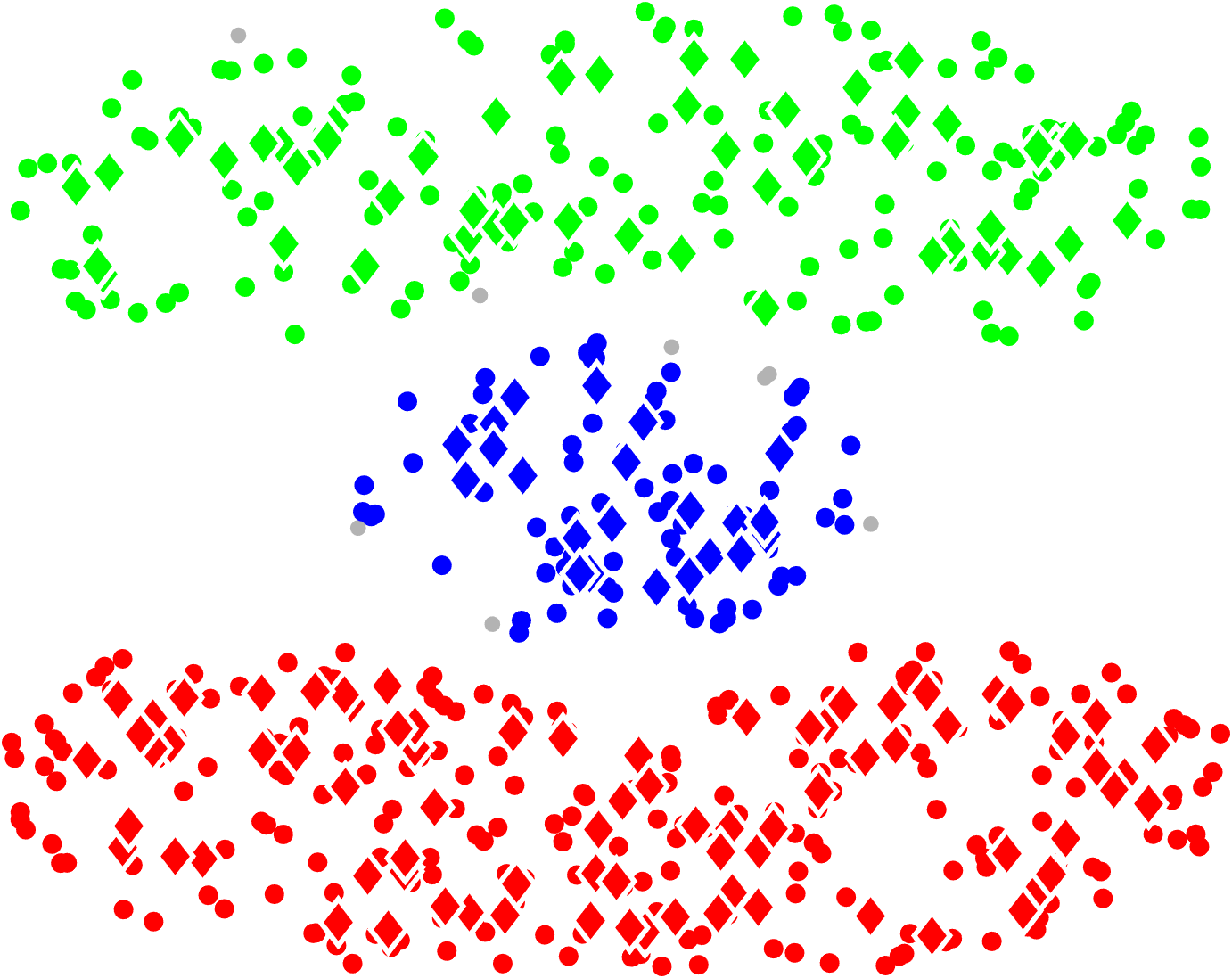}
\caption{}
\end{subfigure}
\hfill
\begin{subfigure}[H!]{0.48\textwidth}
\centering
\includegraphics[width=\textwidth]{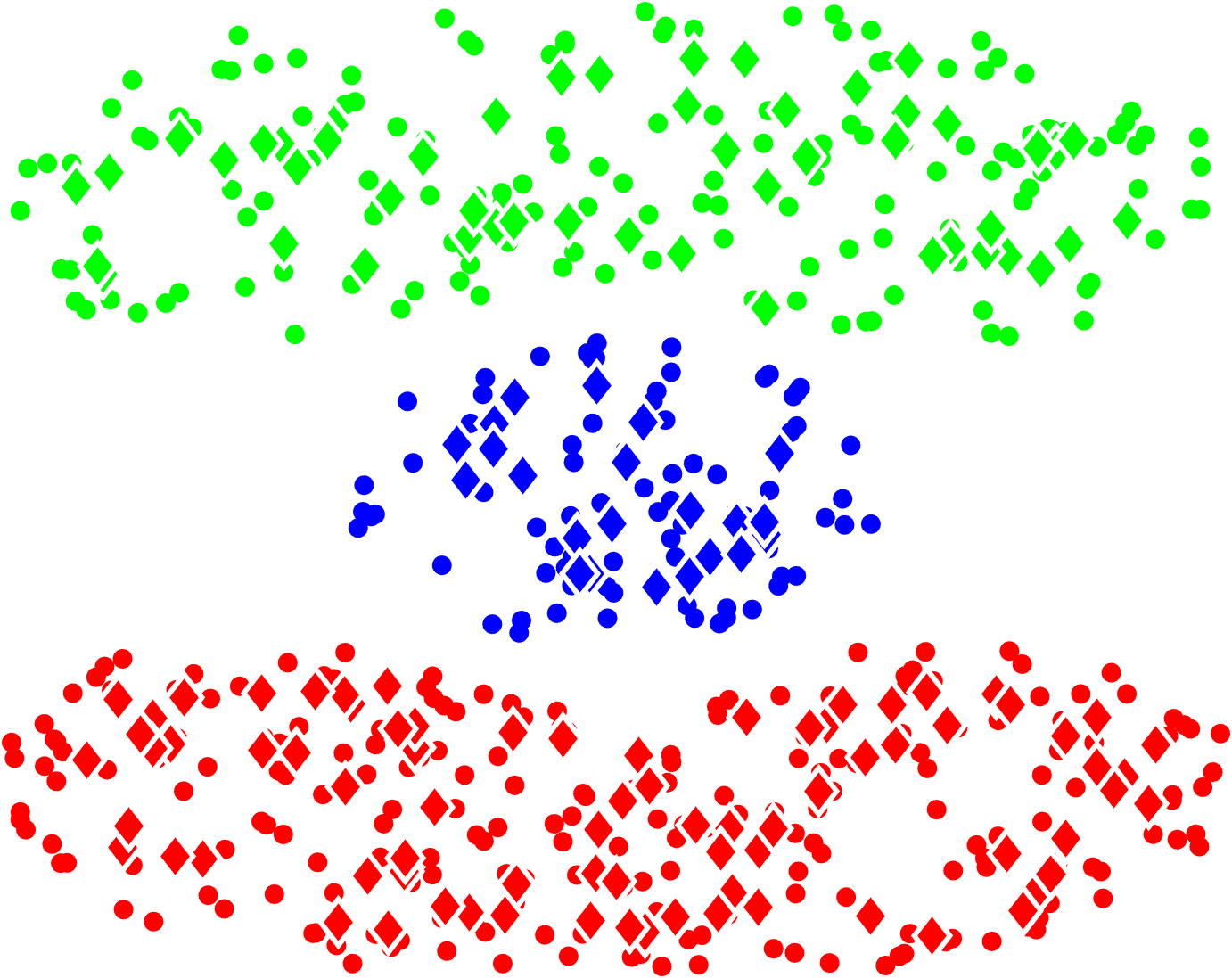}
\caption{}
\end{subfigure}
\caption{DLCC visualization on the \texttt{Bainba} dataset. (a) Ground truth labels; (b) Grouped local centers; (c) Temporary clusters; (d) Final DLCC clustering result.}
\label{fig:dlcc-ba}
\end{figure*}

% -------- blend --------
\begin{figure*}[ht]
\centering
\begin{subfigure}[H!]{0.48\textwidth}
\centering
\includegraphics[width=\textwidth]{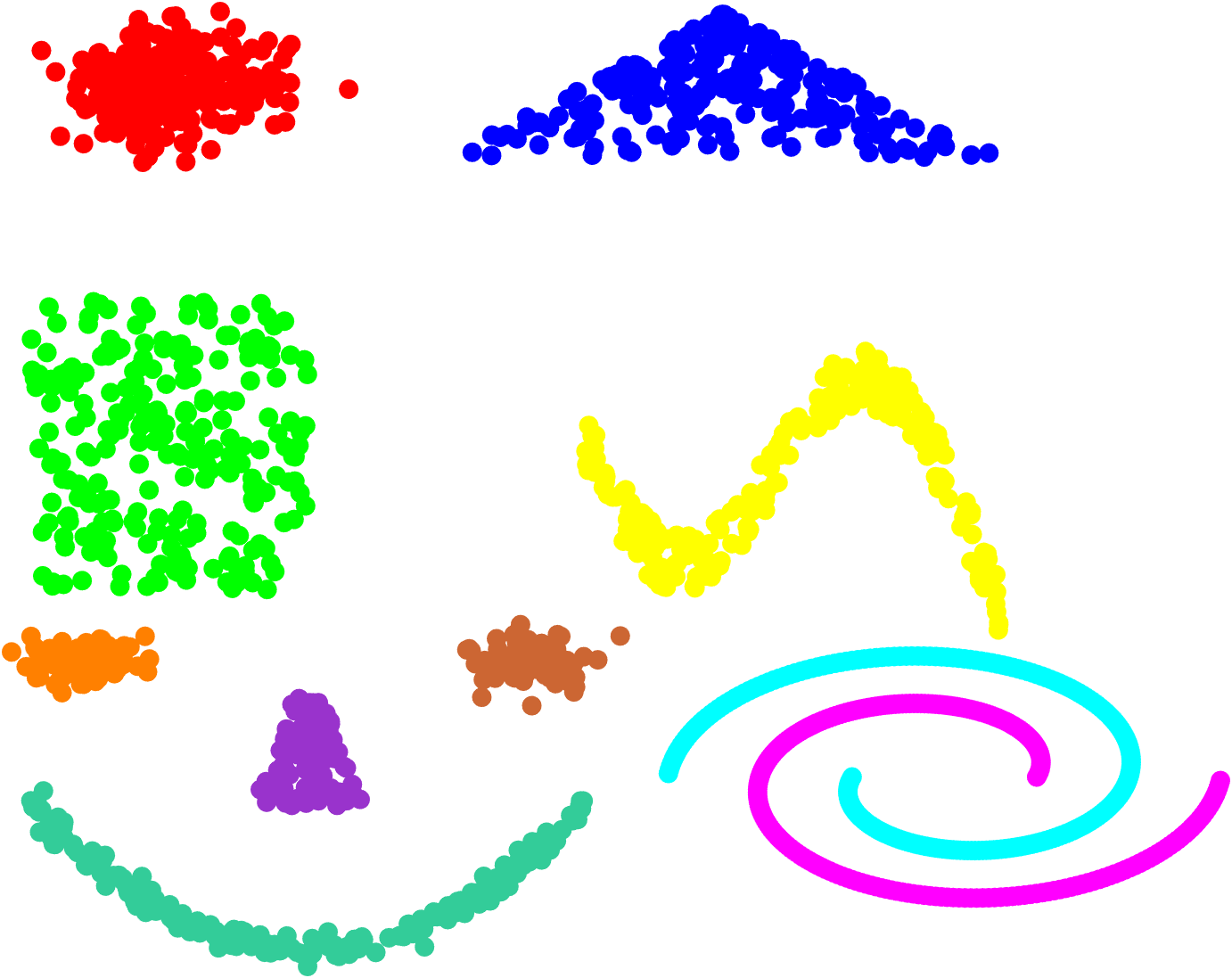}
\caption{}
\end{subfigure}
\hfill
\begin{subfigure}[H!]{0.48\textwidth}
\centering
\includegraphics[width=\textwidth]{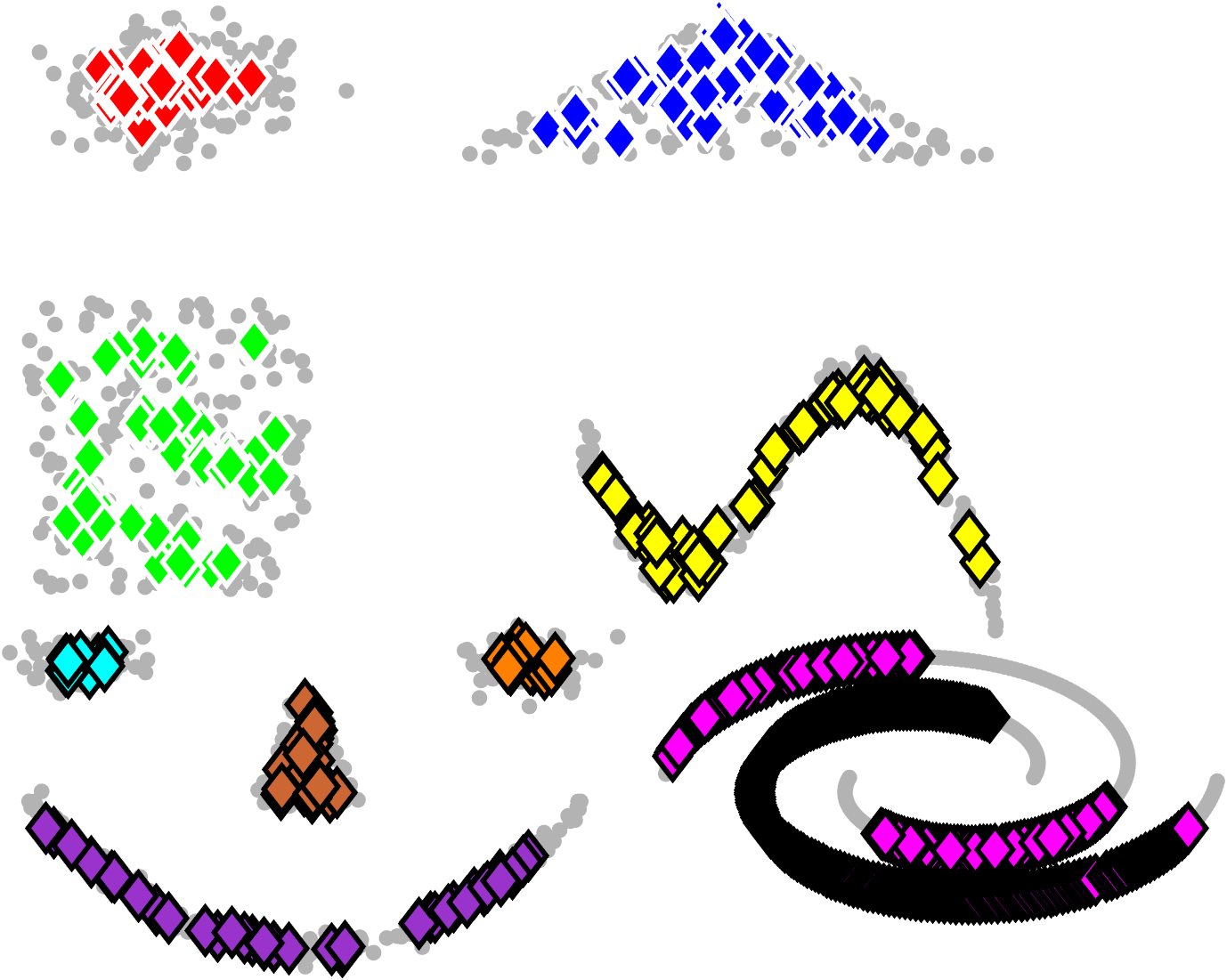}
\caption{}
\end{subfigure}

\begin{subfigure}[H!]{0.48\textwidth}
\centering
\includegraphics[width=\textwidth]{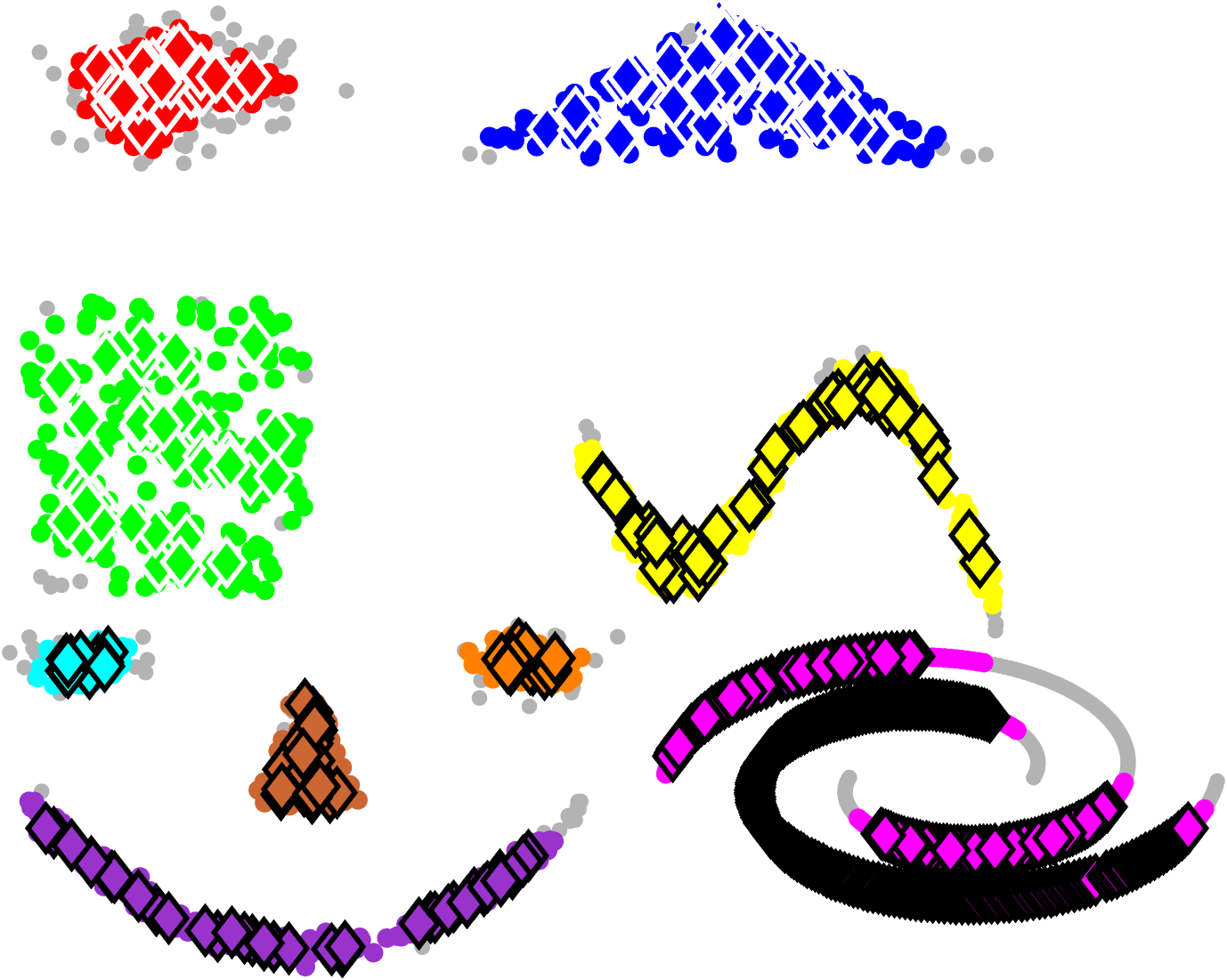}
\caption{}
\end{subfigure}
\hfill
\begin{subfigure}[H!]{0.48\textwidth}
\centering
\includegraphics[width=\textwidth]{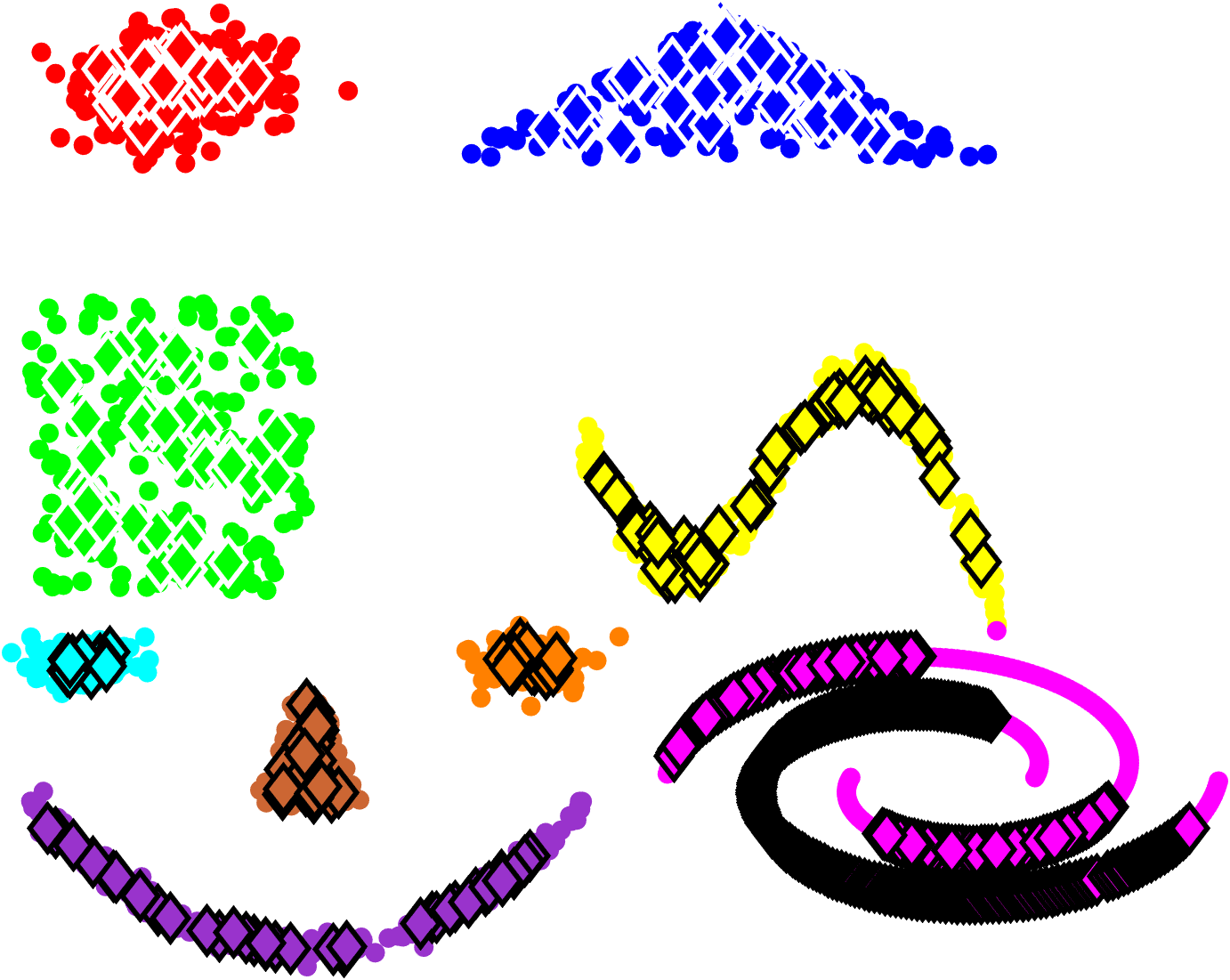}
\caption{}
\end{subfigure}
\caption{DLCC visualization on the \texttt{Blend} dataset. (a) Ground truth labels; (b) Grouped local centers; (c) Temporary clusters; (d) Final DLCC clustering result.}
\label{fig:dlcc-blend}
\end{figure*}

% -------- chainlink --------
\begin{figure*}[ht]
\centering
\begin{subfigure}[H!]{0.48\textwidth}
\centering
\includegraphics[width=\textwidth]{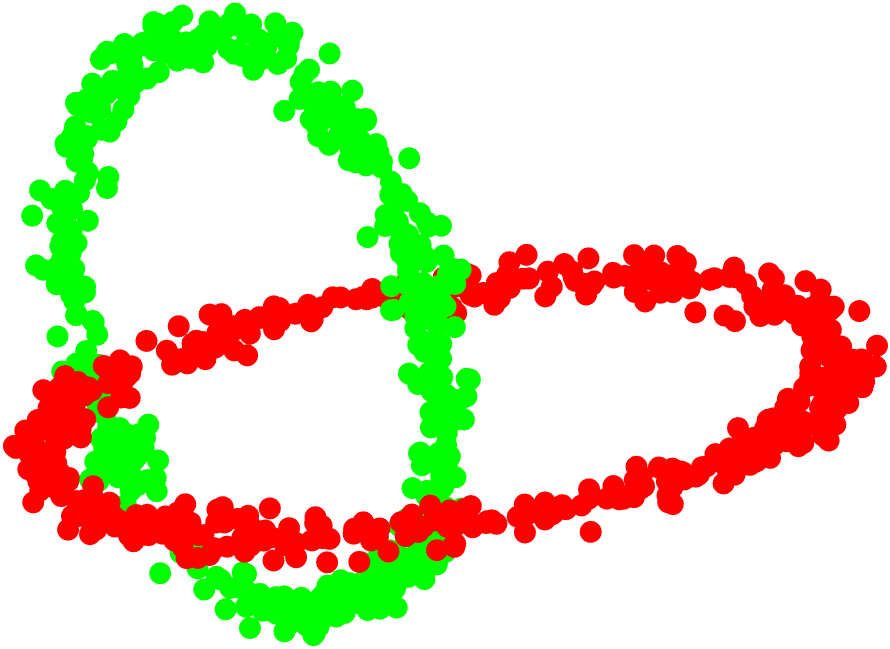}
\caption{}
\end{subfigure}
\hfill
\begin{subfigure}[H!]{0.48\textwidth}
\centering
\includegraphics[width=\textwidth]{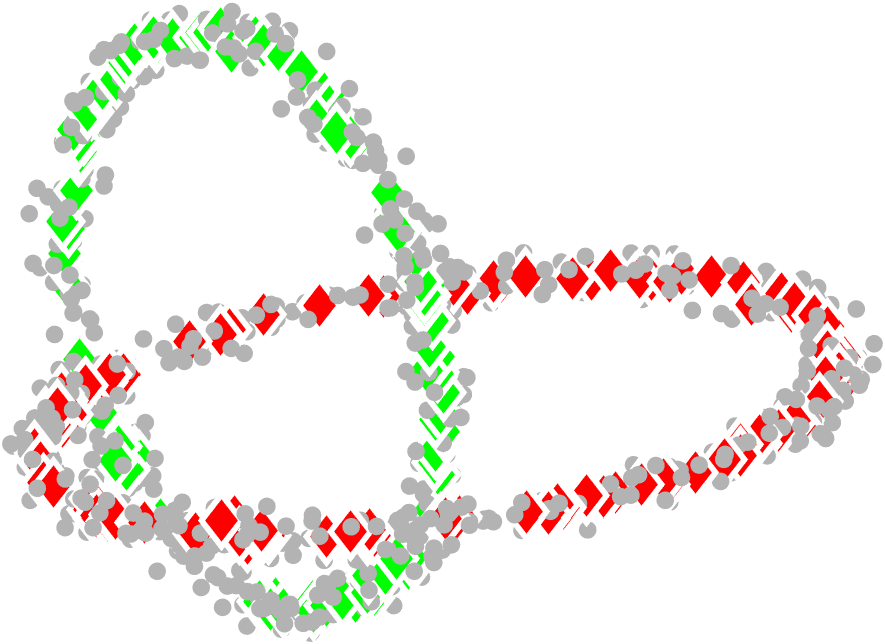}
\caption{}
\end{subfigure}

\begin{subfigure}[H!]{0.48\textwidth}
\centering
\includegraphics[width=\textwidth]{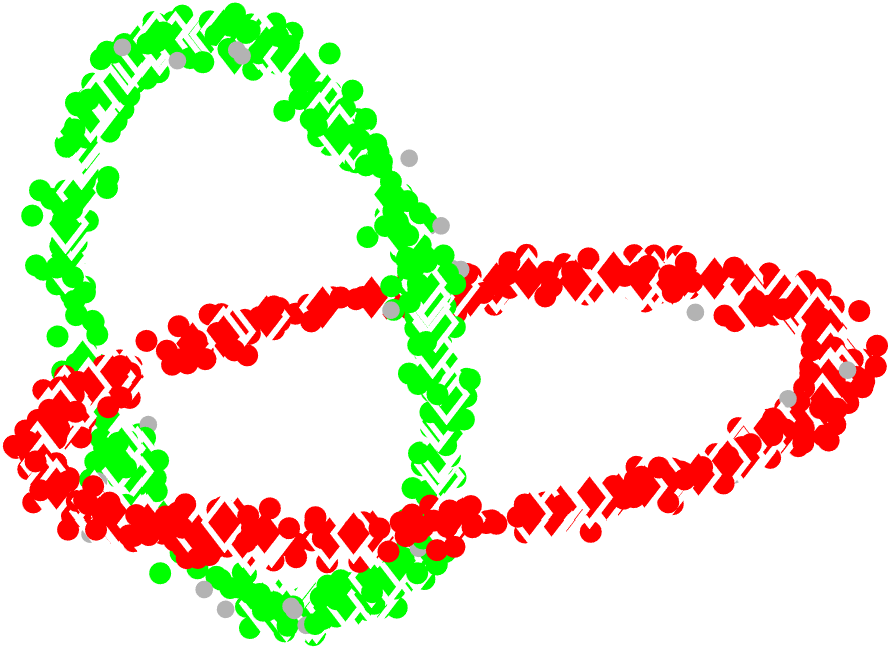}
\caption{}
\end{subfigure}
\hfill
\begin{subfigure}[H!]{0.48\textwidth}
\centering
\includegraphics[width=\textwidth]{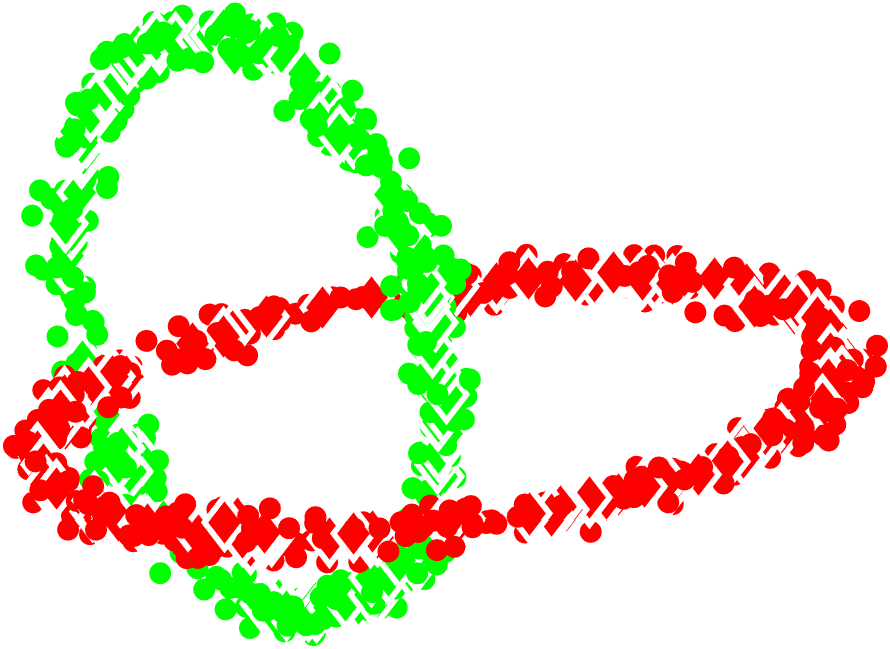}
\caption{}
\end{subfigure}
\caption{DLCC visualization on the \texttt{Chainlink} dataset. (a) Ground truth labels; (b) Grouped local centers; (c) Temporary clusters; (d) Final DLCC clustering result.}
\label{fig:dlcc-chainlink}
\end{figure*}

% -------- cuboids --------
\begin{figure*}[ht]
\centering
\begin{subfigure}[H!]{0.48\textwidth}
\centering
\includegraphics[width=\textwidth]{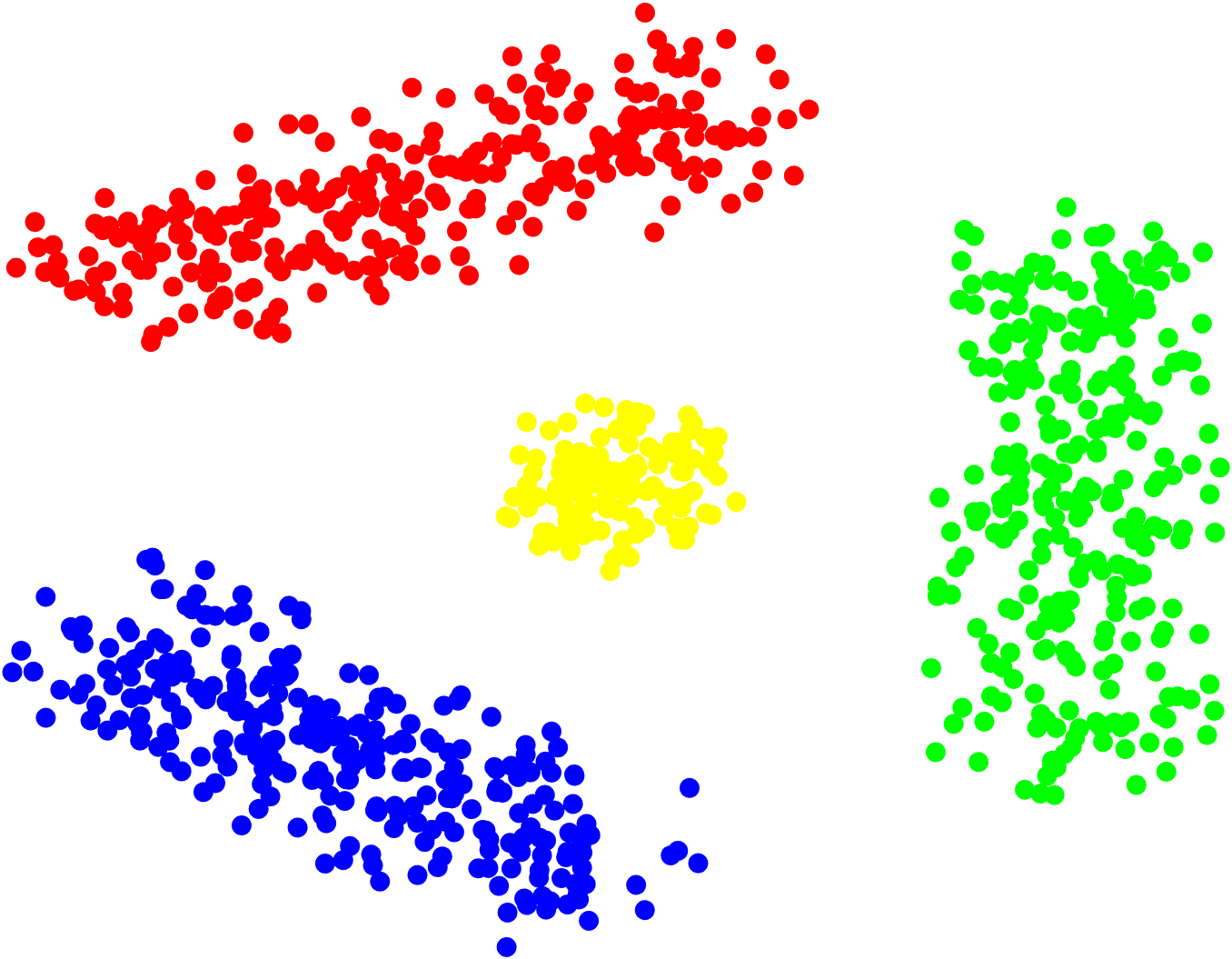}
\caption{}
\end{subfigure}
\hfill
\begin{subfigure}[H!]{0.48\textwidth}
\centering
\includegraphics[width=\textwidth]{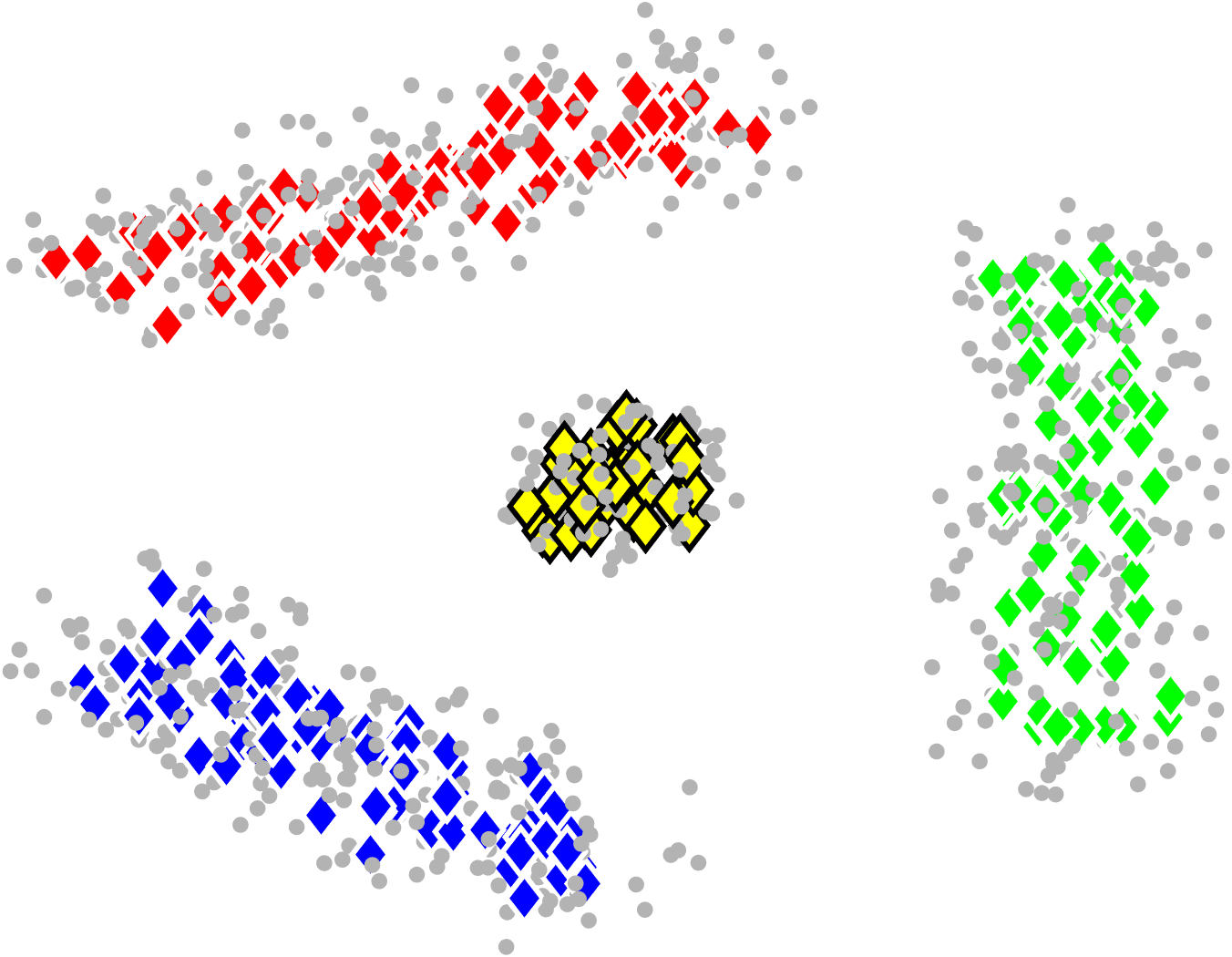}
\caption{}
\end{subfigure}

\begin{subfigure}[H!]{0.48\textwidth}
\centering
\includegraphics[width=\textwidth]{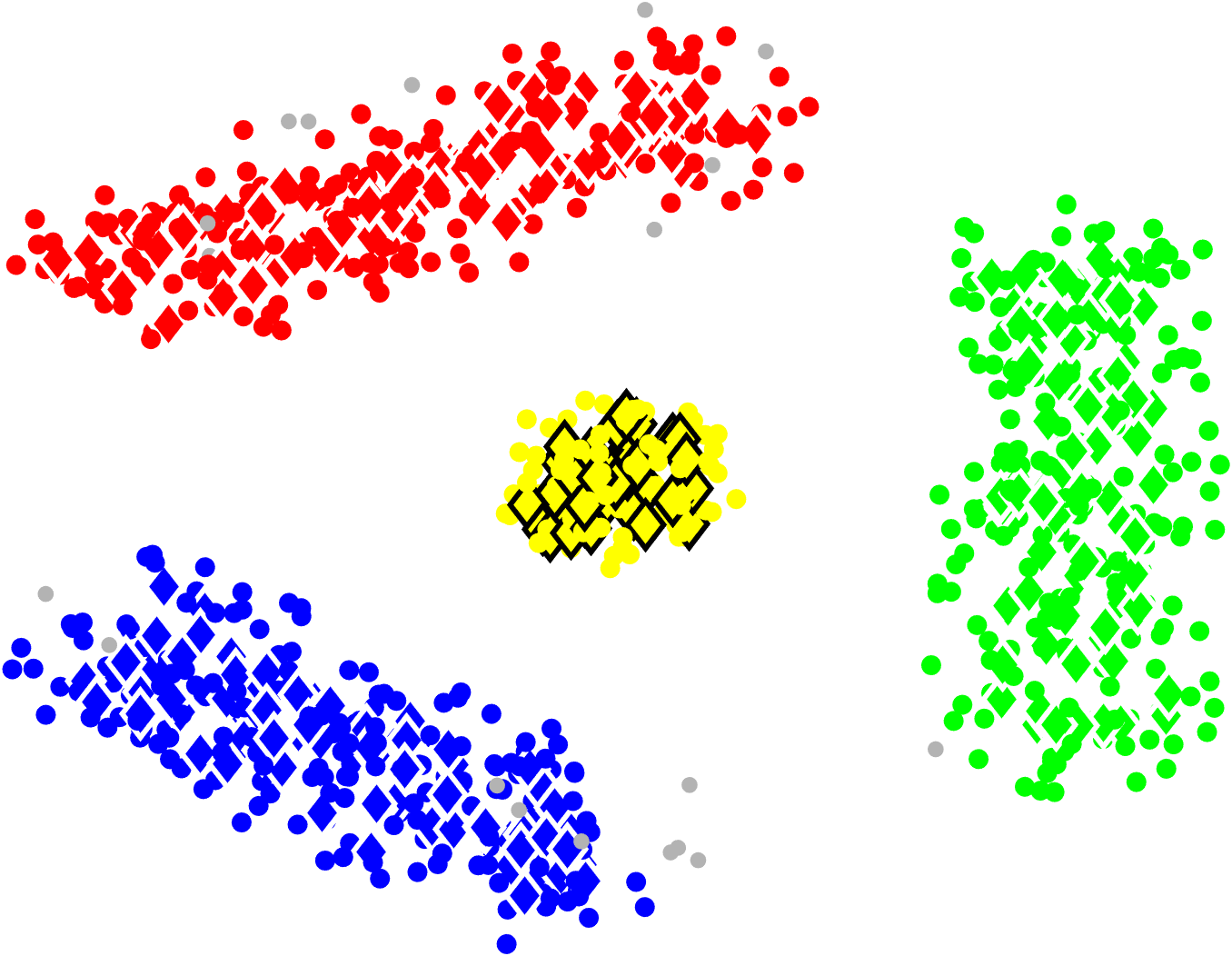}
\caption{}
\end{subfigure}
\hfill
\begin{subfigure}[H!]{0.48\textwidth}
\centering
\includegraphics[width=\textwidth]{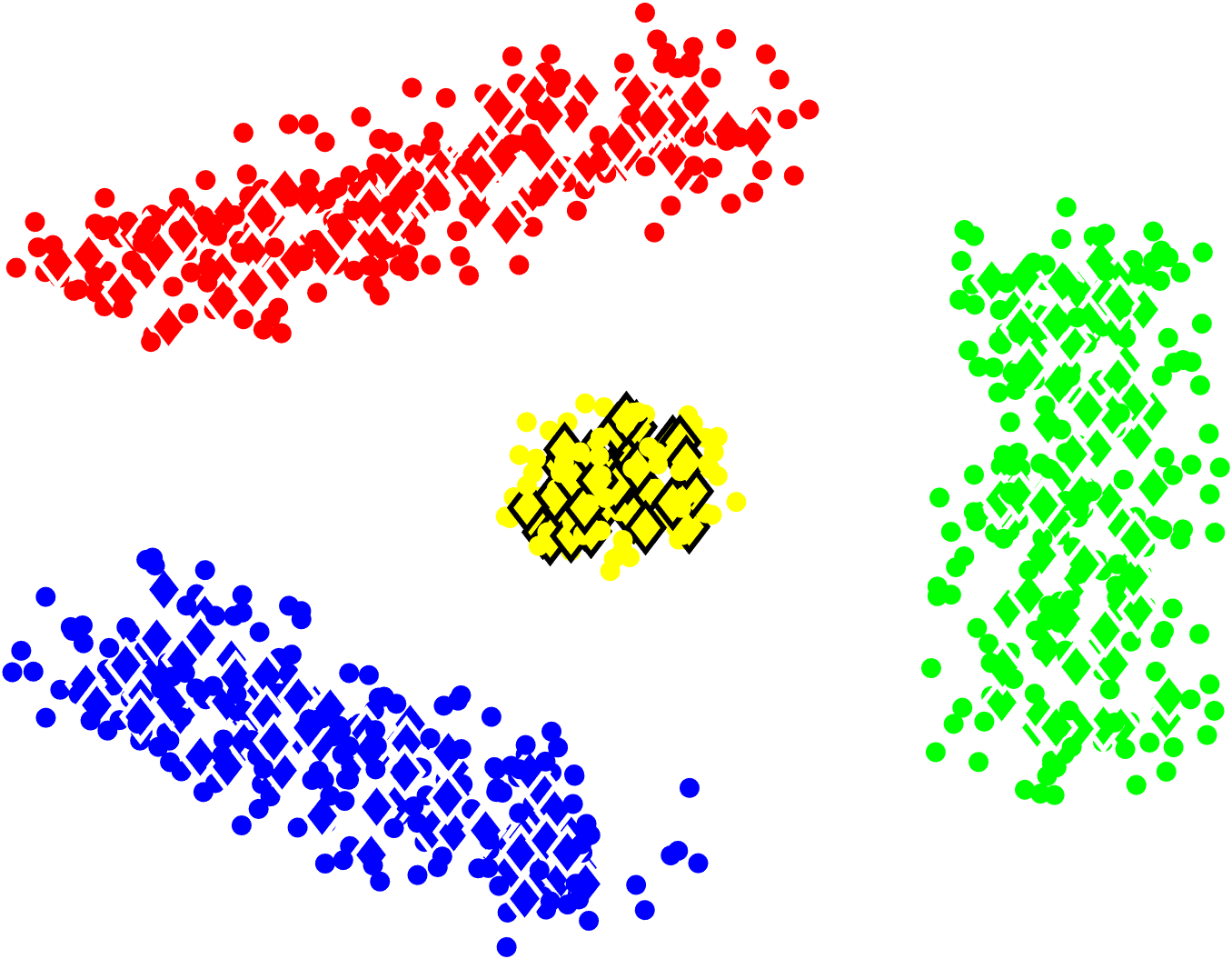}
\caption{}
\end{subfigure}
\caption{DLCC visualization on the \texttt{Cuboids} dataset. (a) Ground truth labels; (b) Grouped local centers; (c) Temporary clusters; (d) Final DLCC clustering result.}
\label{fig:dlcc-cuboids}
\end{figure*}

% -------- flame --------
\begin{figure*}[ht]
\centering
\begin{subfigure}[H!]{0.48\textwidth}
\centering
\includegraphics[width=\textwidth]{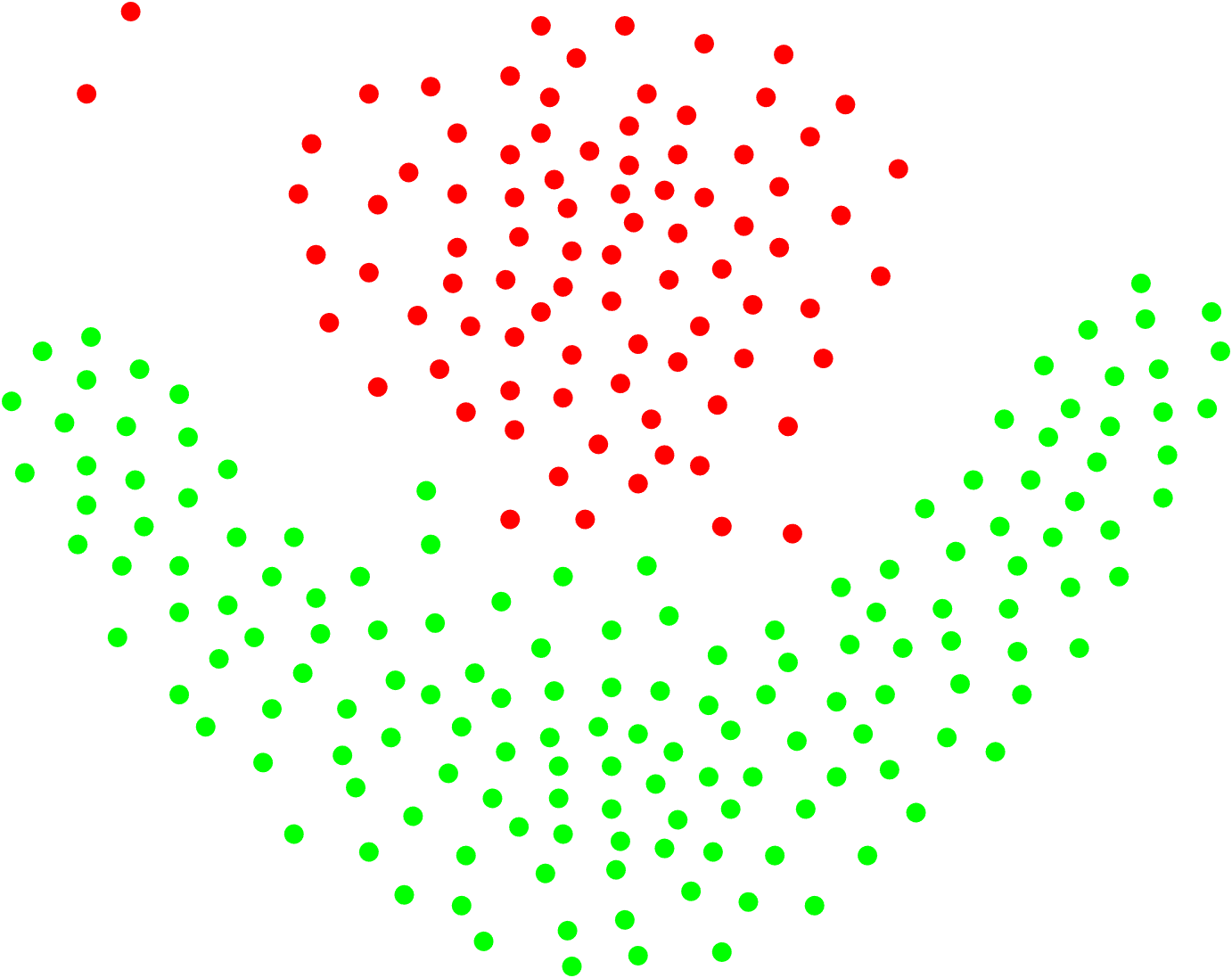}
\caption{}
\end{subfigure}
\hfill
\begin{subfigure}[H!]{0.48\textwidth}
\centering
\includegraphics[width=\textwidth]{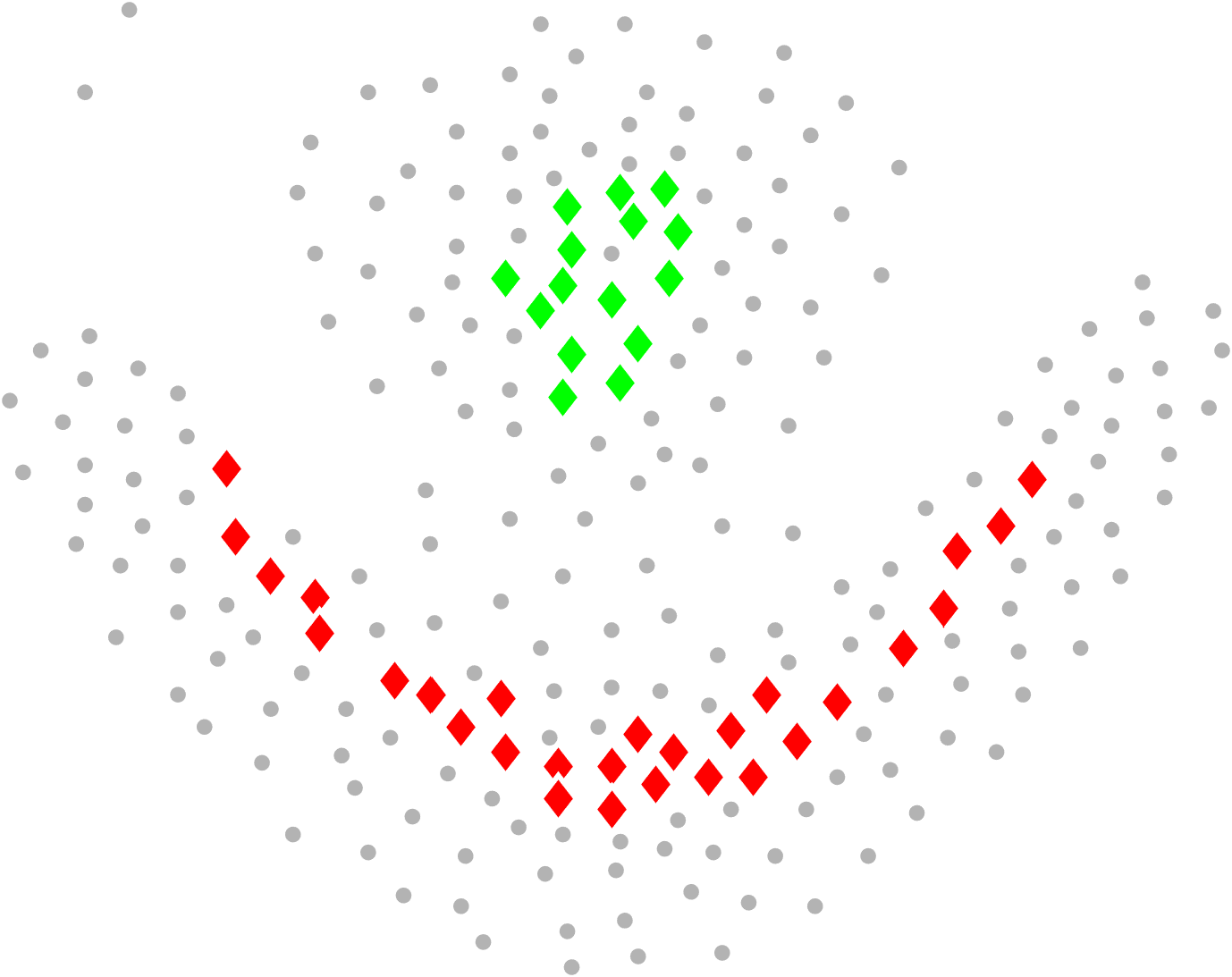}
\caption{}
\end{subfigure}

\begin{subfigure}[H!]{0.48\textwidth}
\centering
\includegraphics[width=\textwidth]{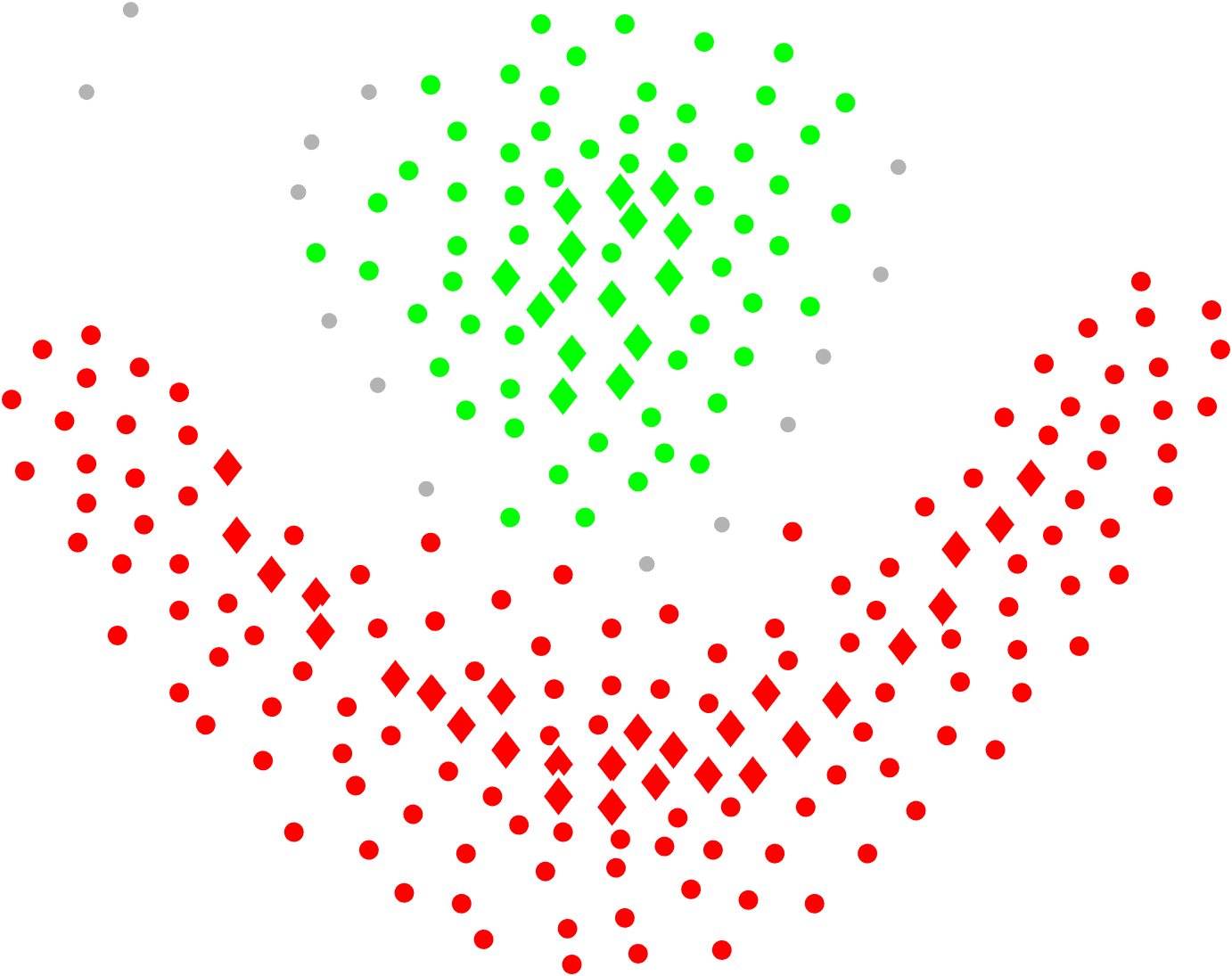}
\caption{}
\end{subfigure}
\hfill
\begin{subfigure}[H!]{0.48\textwidth}
\centering
\includegraphics[width=\textwidth]{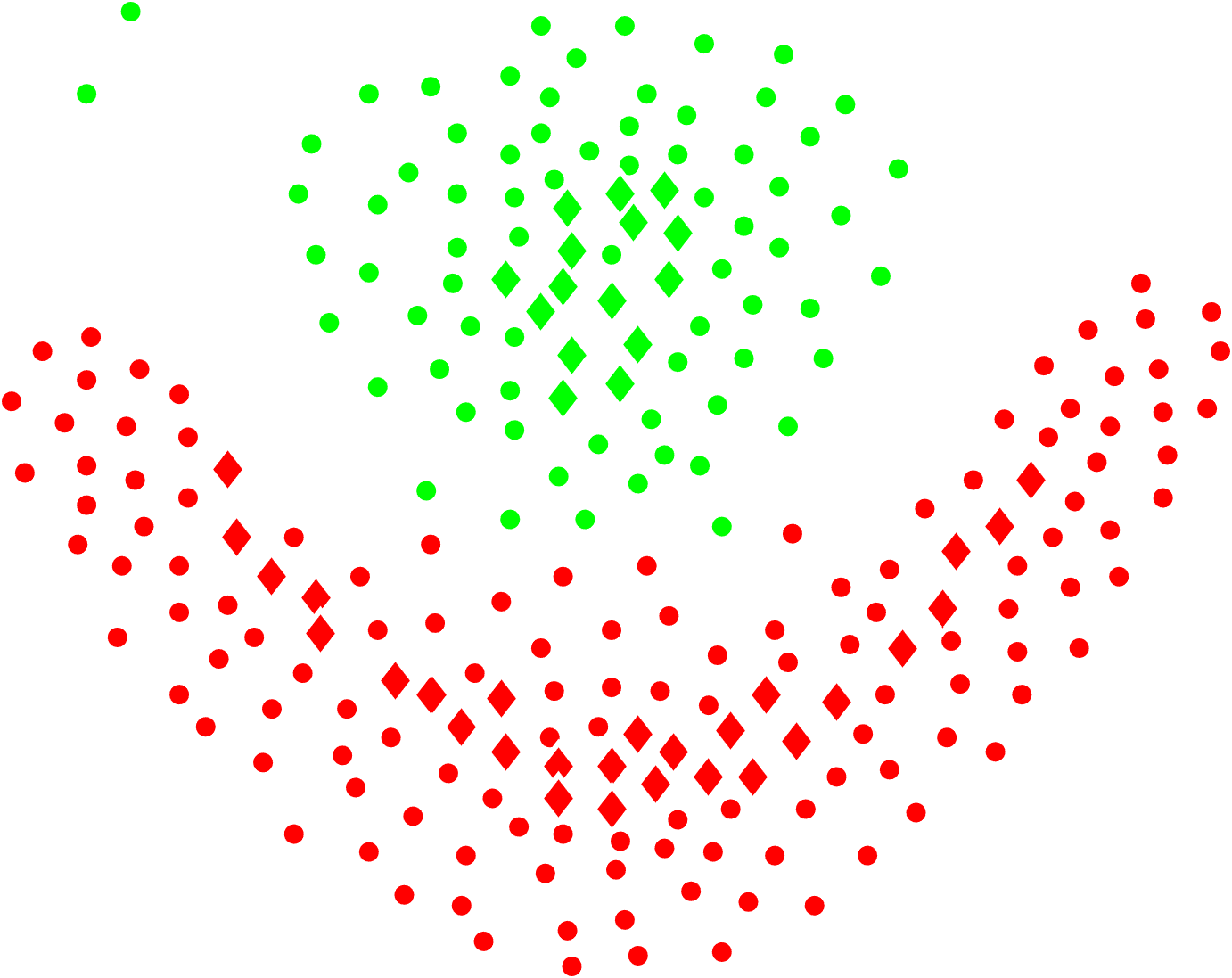}
\caption{}
\end{subfigure}
\caption{DLCC visualization on the \texttt{Flame} dataset. (a) Ground truth labels; (b) Grouped local centers; (c) Temporary clusters; (d) Final DLCC clustering result.}
\label{fig:dlcc-flame}
\end{figure*}

% -------- frog --------
\begin{figure*}[ht]
\centering
\begin{subfigure}[H!]{0.48\textwidth}
\centering
\includegraphics[width=\textwidth]{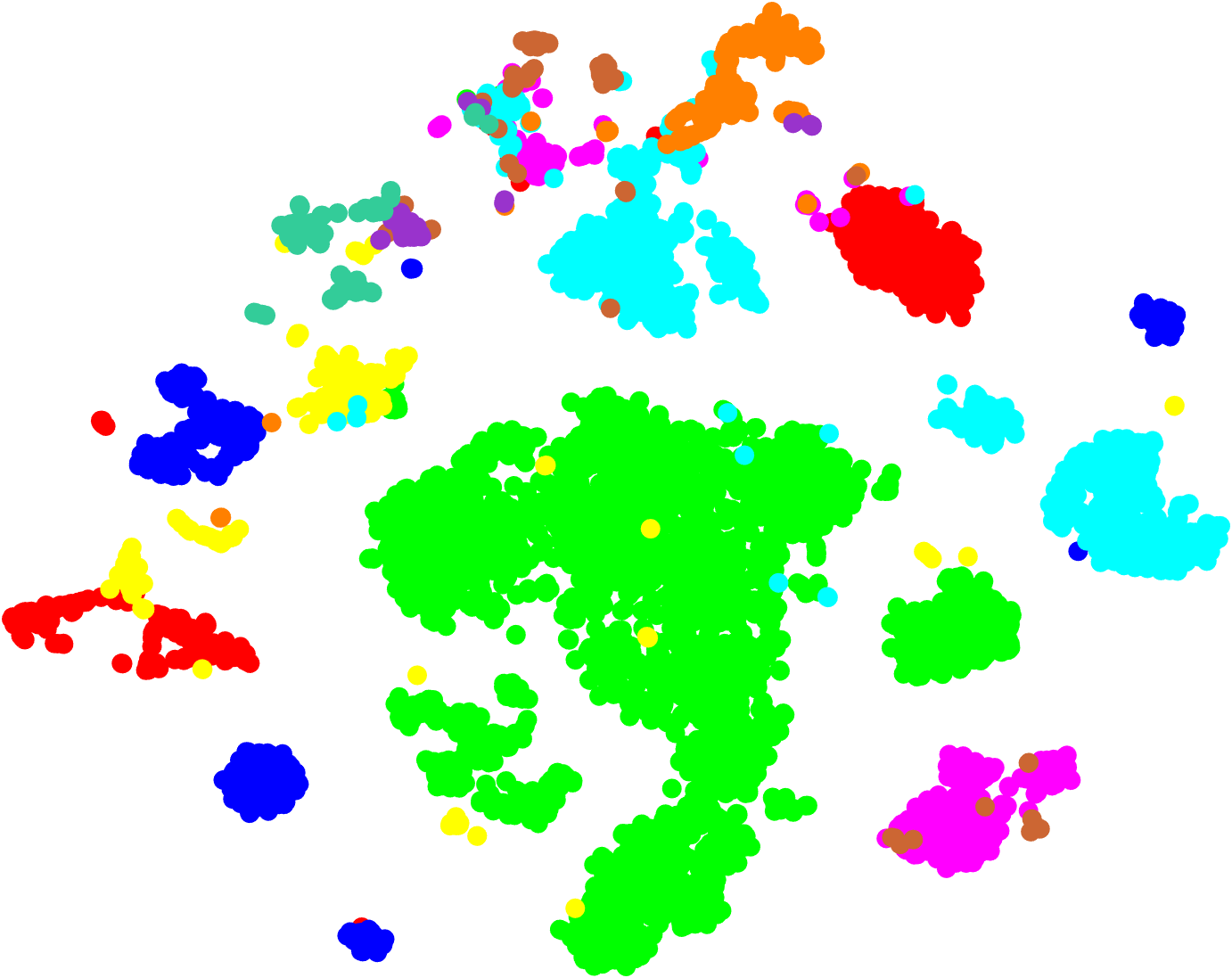}
\caption{}
\end{subfigure}
\hfill
\begin{subfigure}[H!]{0.48\textwidth}
\centering
\includegraphics[width=\textwidth]{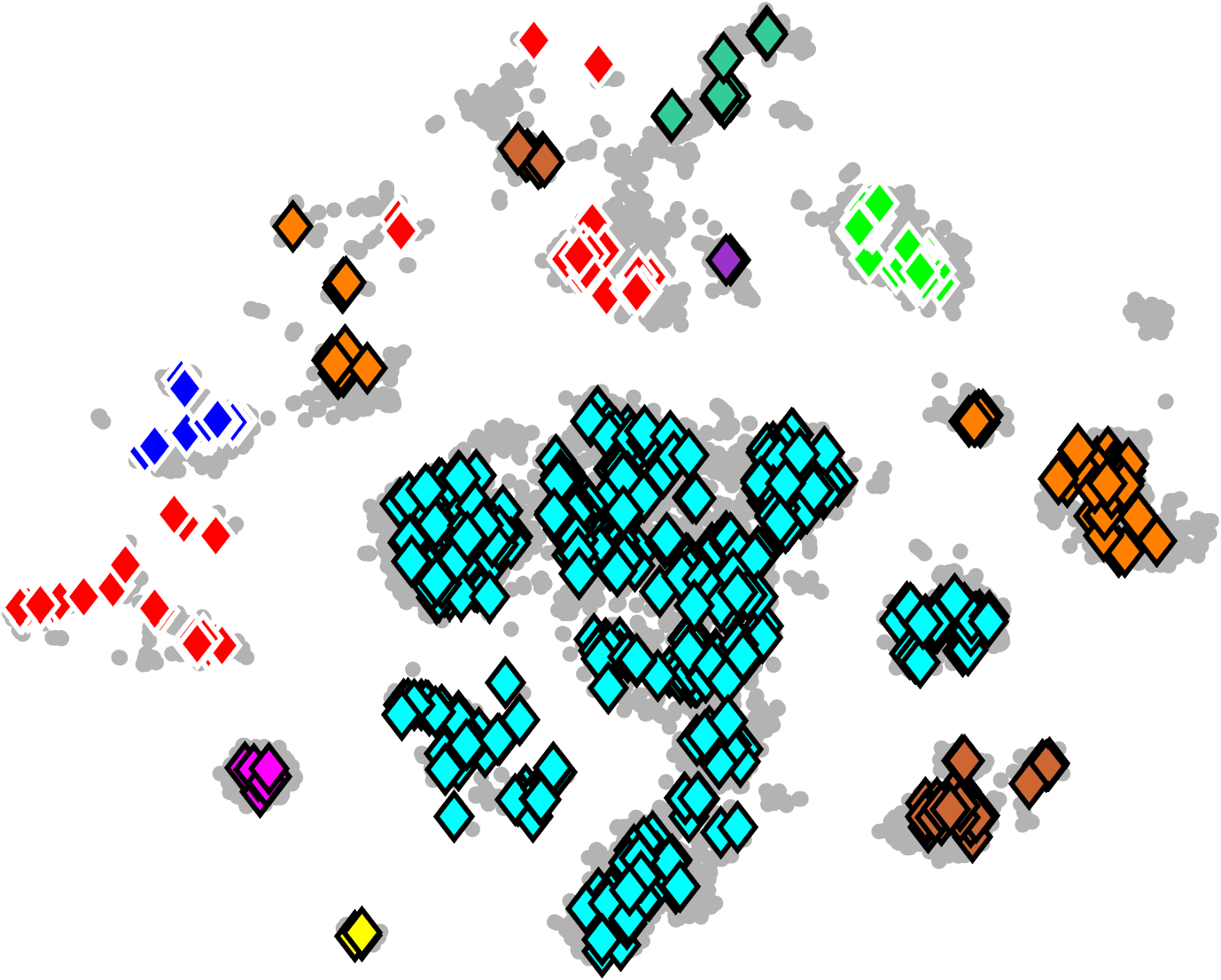}
\caption{}
\end{subfigure}

\begin{subfigure}[H!]{0.48\textwidth}
\centering
\includegraphics[width=\textwidth]{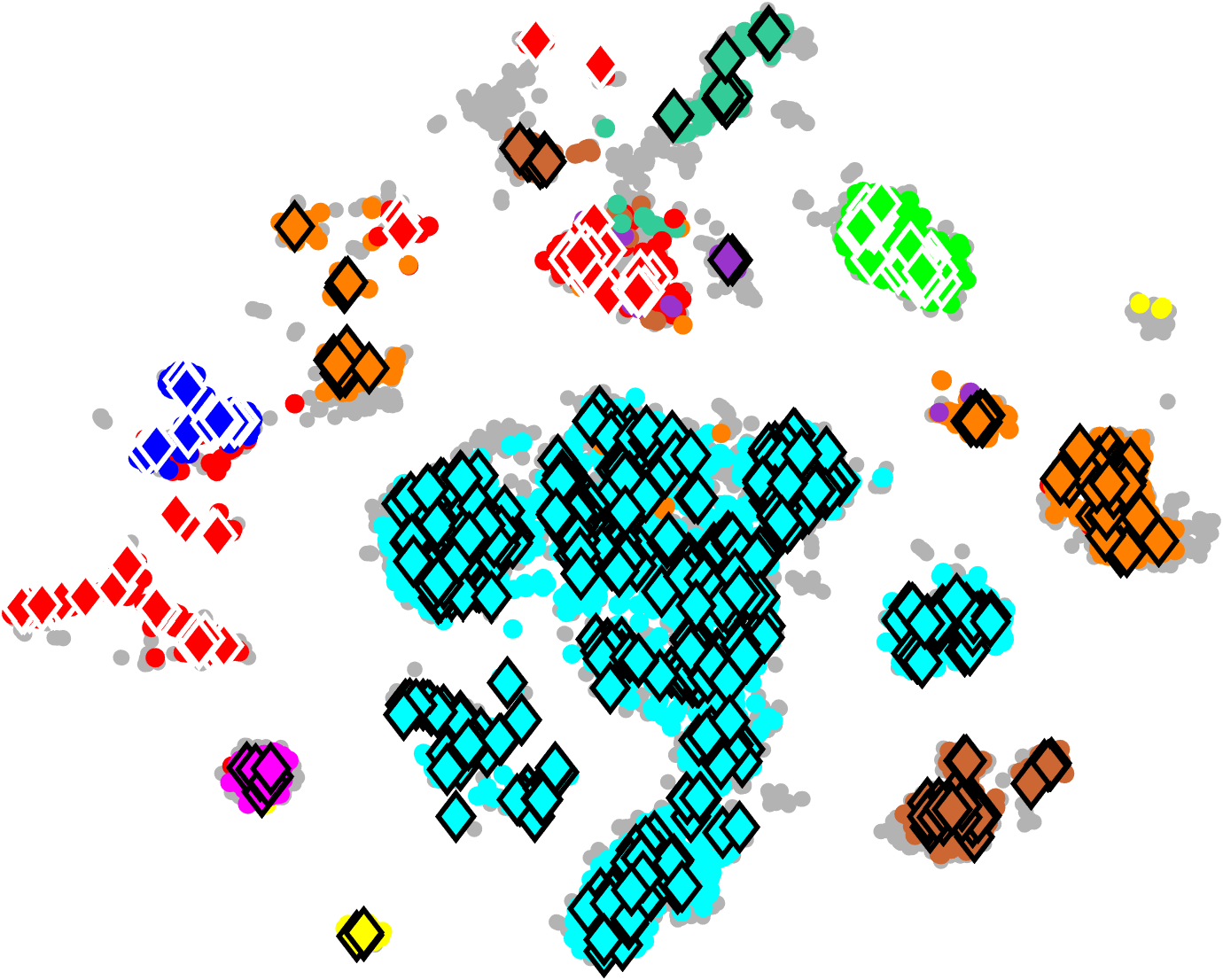}
\caption{}
\end{subfigure}
\hfill
\begin{subfigure}[H!]{0.48\textwidth}
\centering
\includegraphics[width=\textwidth]{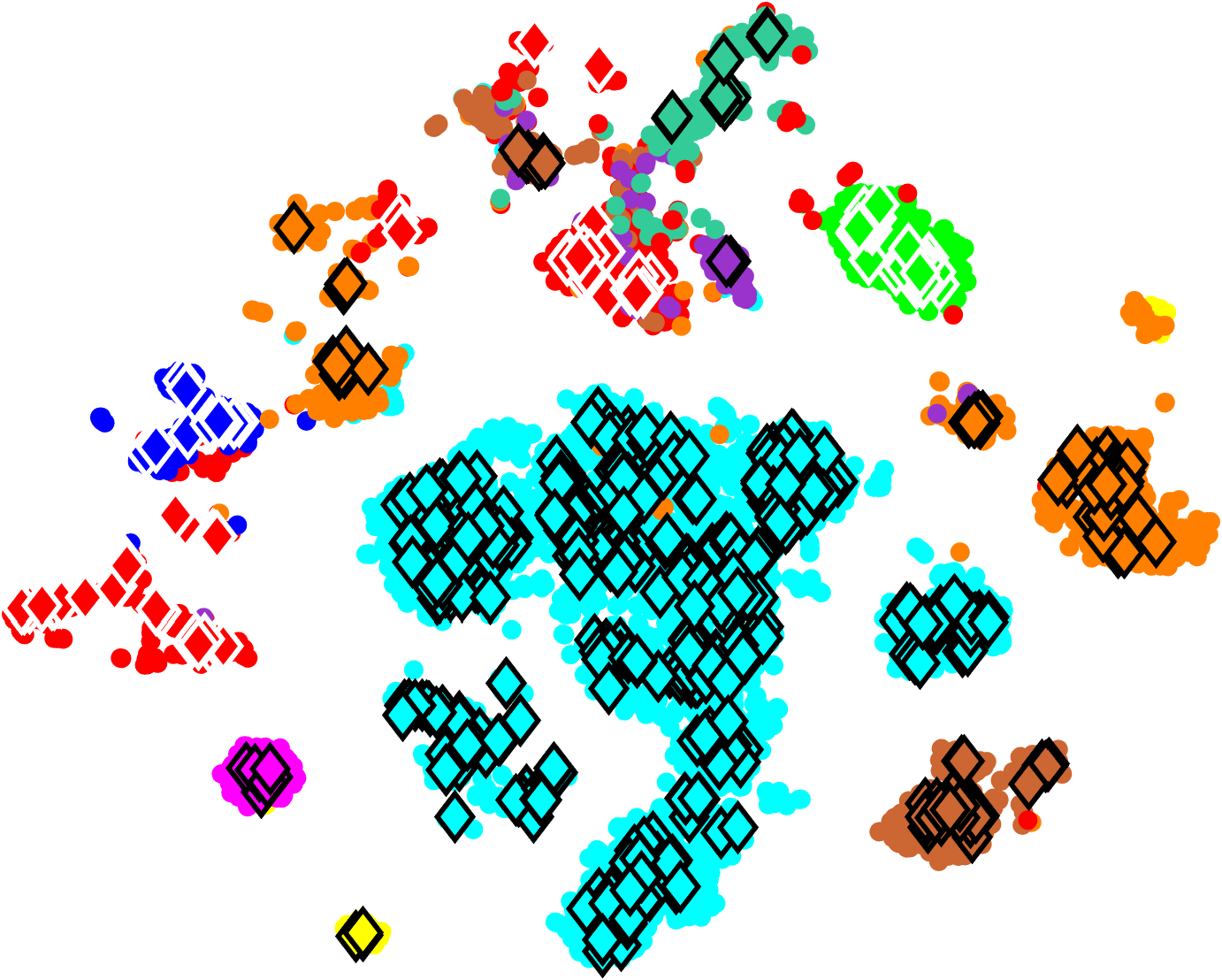}
\caption{}
\end{subfigure}
\caption{DLCC visualization on the \texttt{Anuran calls} dataset. (a) Ground truth labels; (b) Grouped local centers; (c) Temporary clusters; (d) Final DLCC clustering result.}
\label{fig:dlcc-frog}
\end{figure*}

% -------- jain --------
\begin{figure*}[ht]
\centering
\begin{subfigure}[H!]{0.48\textwidth}
\centering
\includegraphics[width=\textwidth]{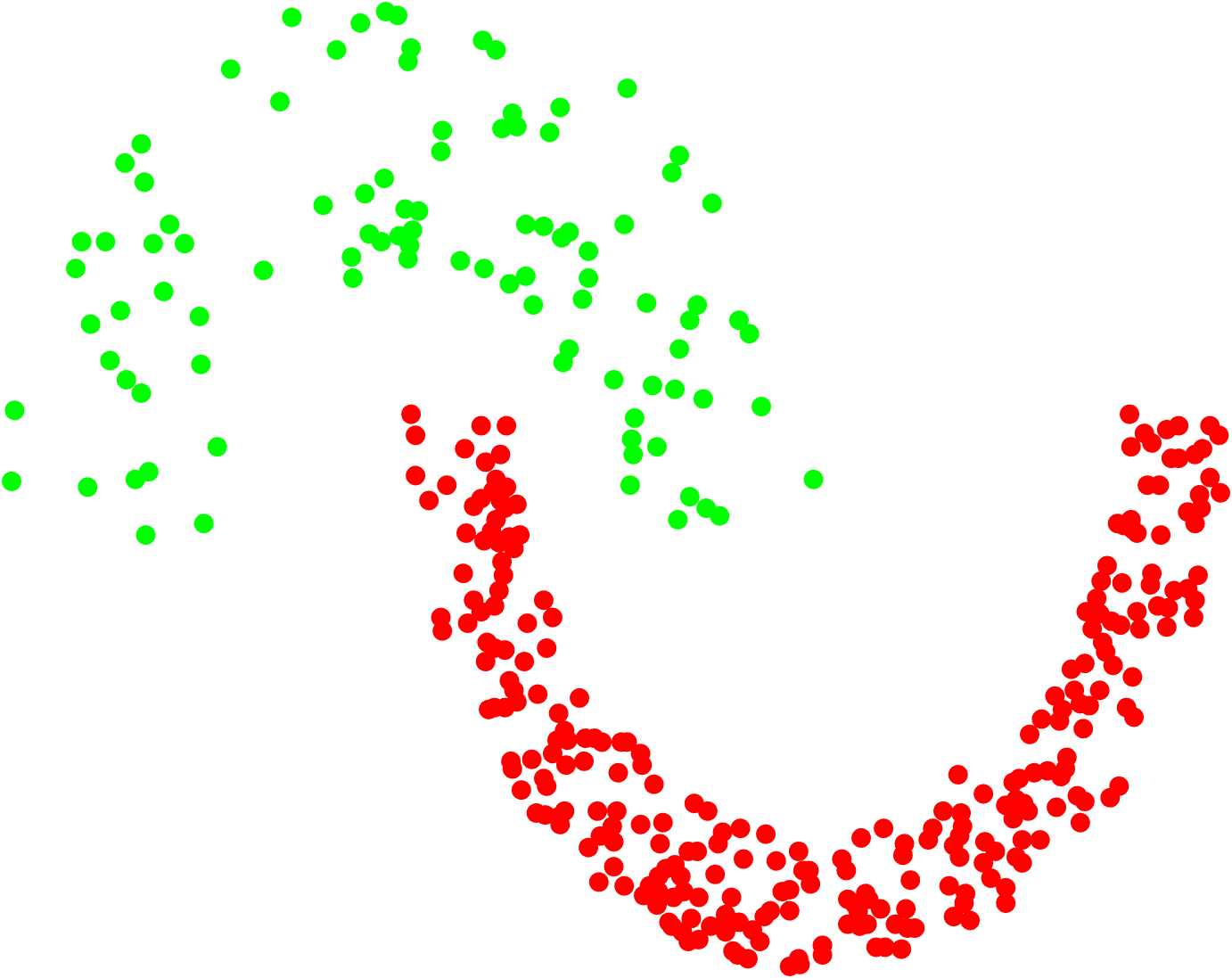}
\caption{}
\end{subfigure}
\hfill
\begin{subfigure}[H!]{0.48\textwidth}
\centering
\includegraphics[width=\textwidth]{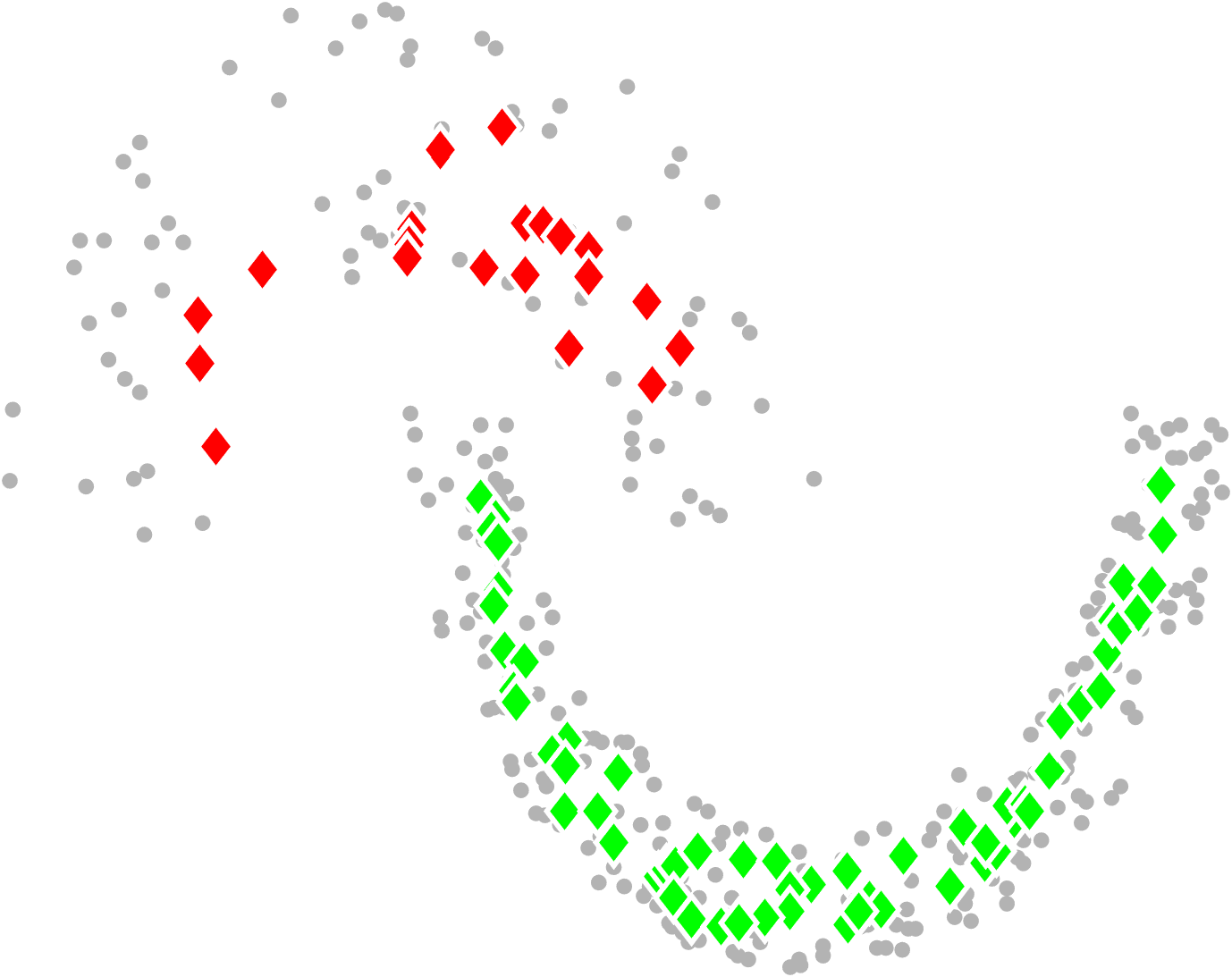}
\caption{}
\end{subfigure}

\begin{subfigure}[H!]{0.48\textwidth}
\centering
\includegraphics[width=\textwidth]{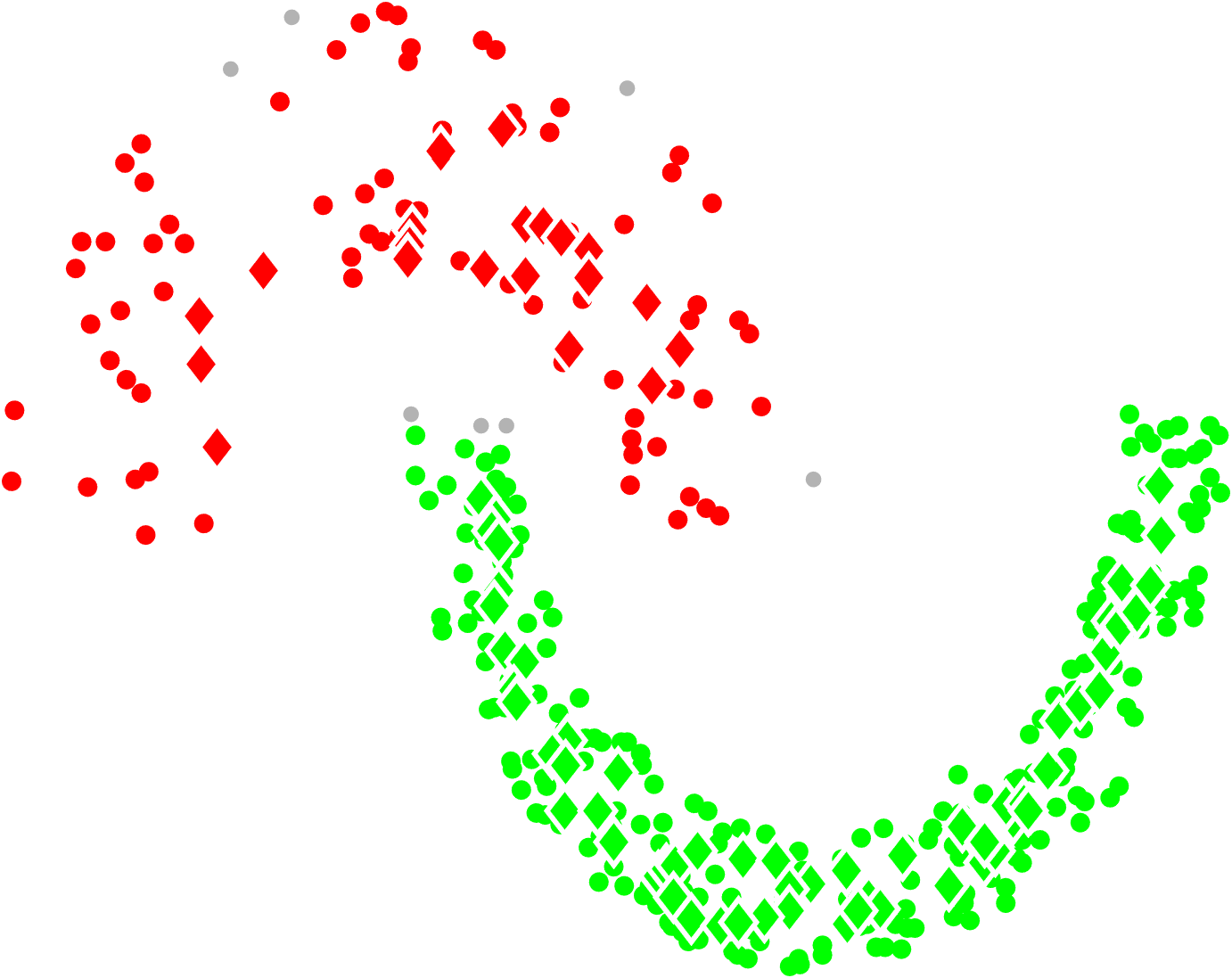}
\caption{}
\end{subfigure}
\hfill
\begin{subfigure}[H!]{0.48\textwidth}
\centering
\includegraphics[width=\textwidth]{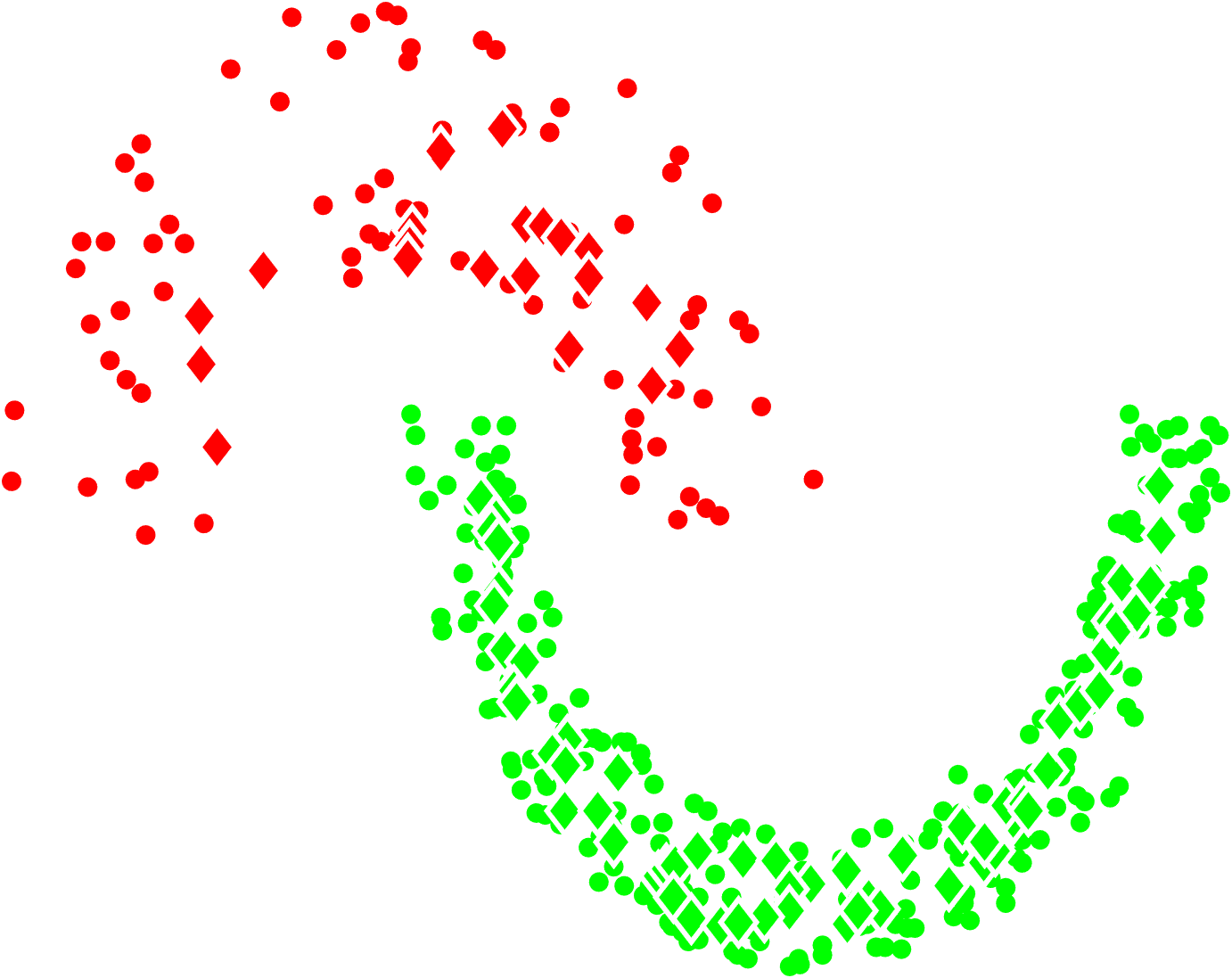}
\caption{}
\end{subfigure}
\caption{DLCC visualization on the \texttt{Jain} dataset. (a) Ground truth labels; (b) Grouped local centers; (c) Temporary clusters; (d) Final DLCC clustering result.}
\label{fig:dlcc-jain}
\end{figure*}

% ------- leukemia ------
\begin{figure*}[ht]
\centering
\begin{subfigure}[H!]{0.48\textwidth}
\centering
\includegraphics[width=\textwidth]{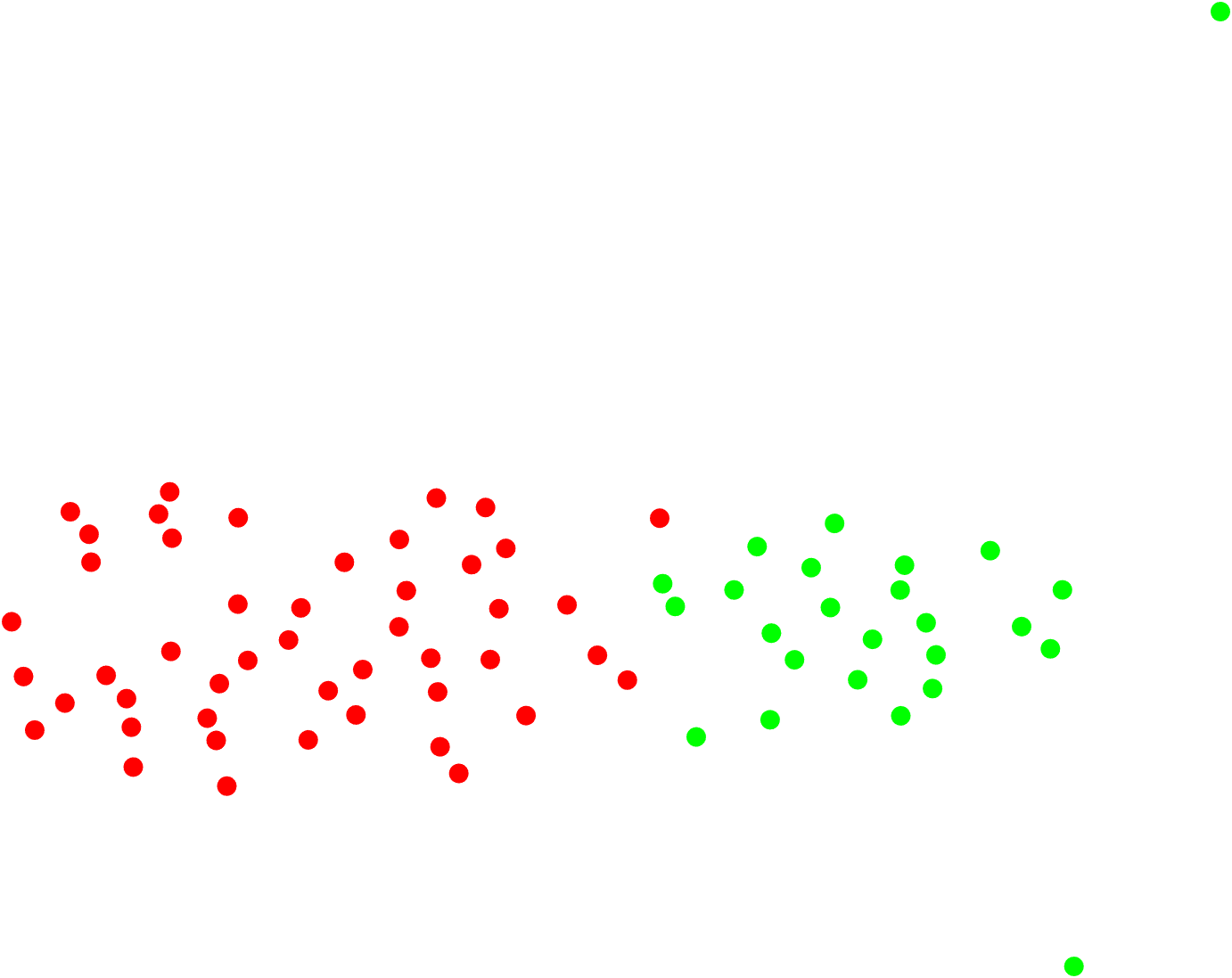}
\caption{}
\end{subfigure}
\hfill
\begin{subfigure}[H!]{0.48\textwidth}
\centering
\includegraphics[width=\textwidth]{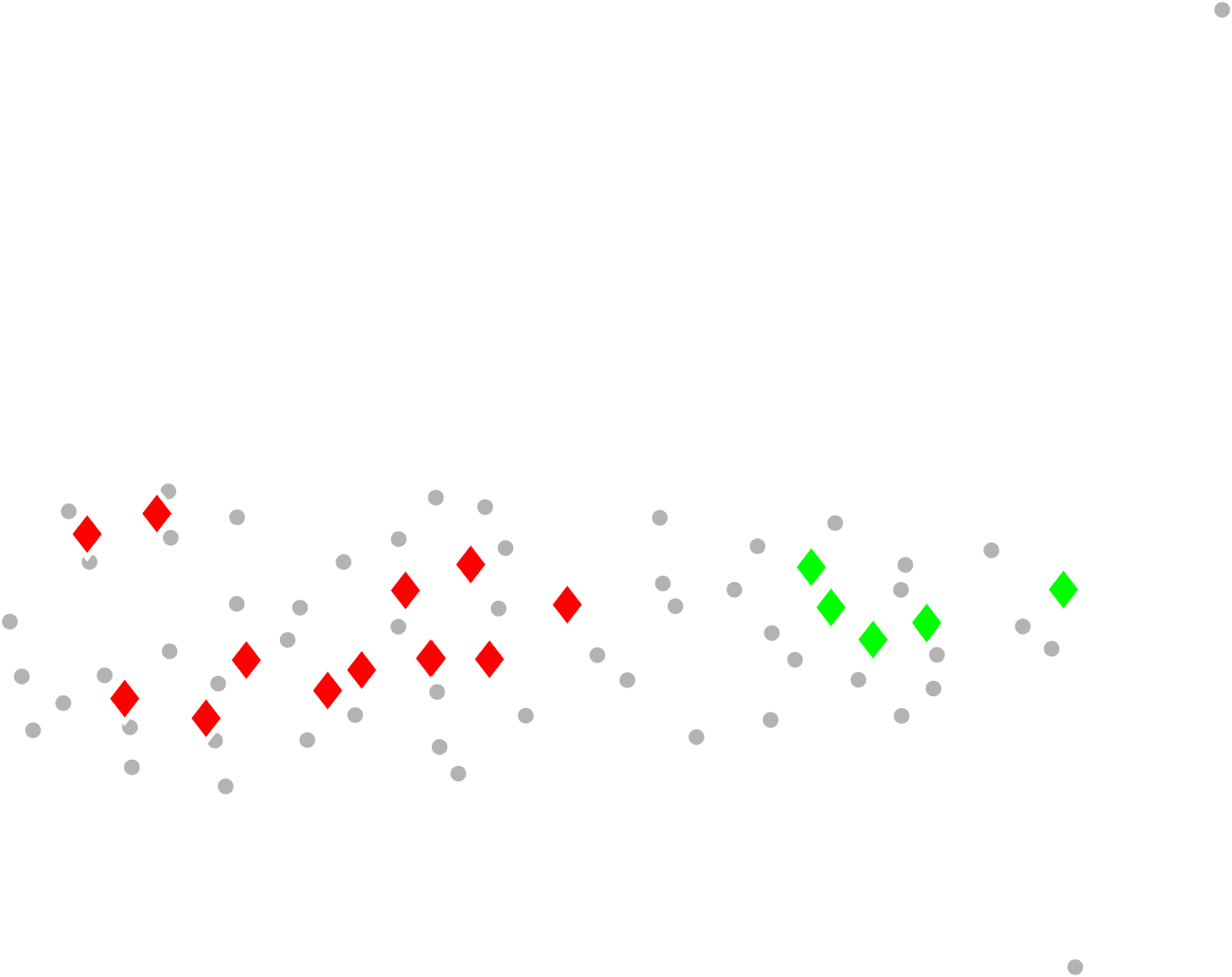}
\caption{}
\end{subfigure}

\begin{subfigure}[H!]{0.48\textwidth}
\centering
\includegraphics[width=\textwidth]{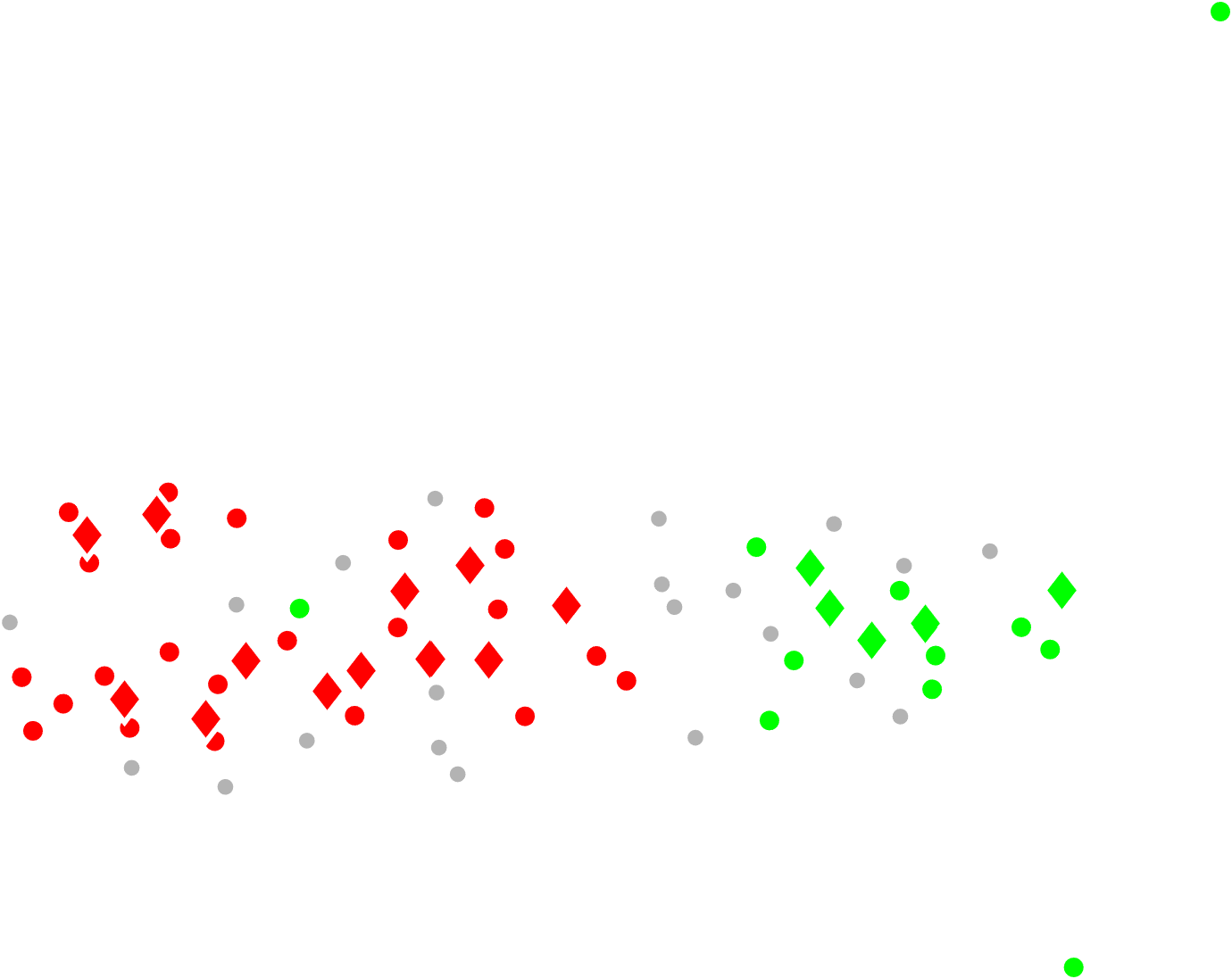}
\caption{}
\end{subfigure}
\hfill
\begin{subfigure}[H!]{0.48\textwidth}
\centering
\includegraphics[width=\textwidth]{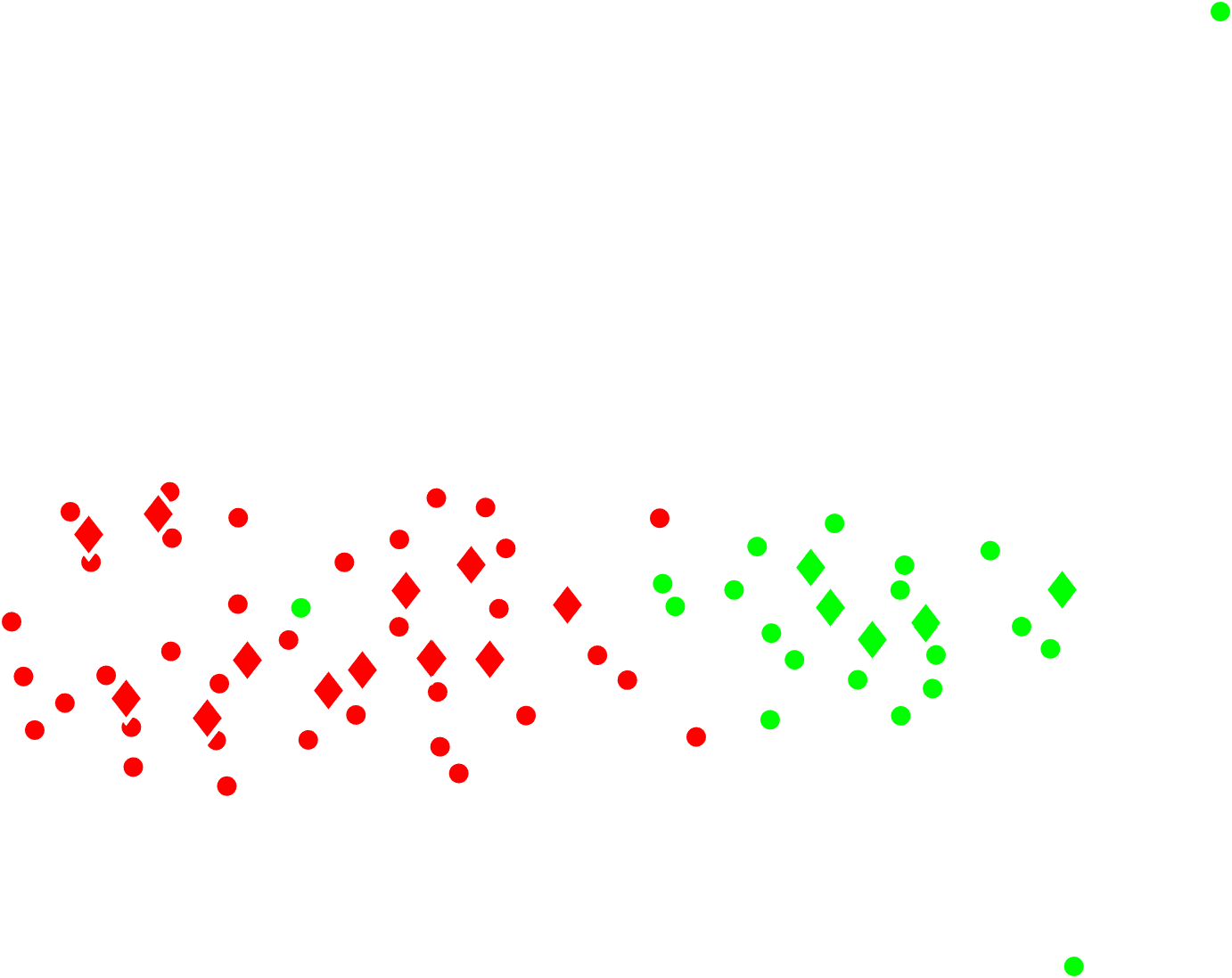}
\caption{}
\end{subfigure}
\caption{DLCC visualization on the \texttt{Leukemia} dataset. (a) Ground truth labels; (b) Grouped local centers; (c) Temporary clusters; (d) Final DLCC clustering result.}
\label{fig:dlcc-leukemia}
\end{figure*}

% -------- pa --------
\begin{figure*}[ht]
\centering
\begin{subfigure}[H!]{0.48\textwidth}
\centering
\includegraphics[width=\textwidth]{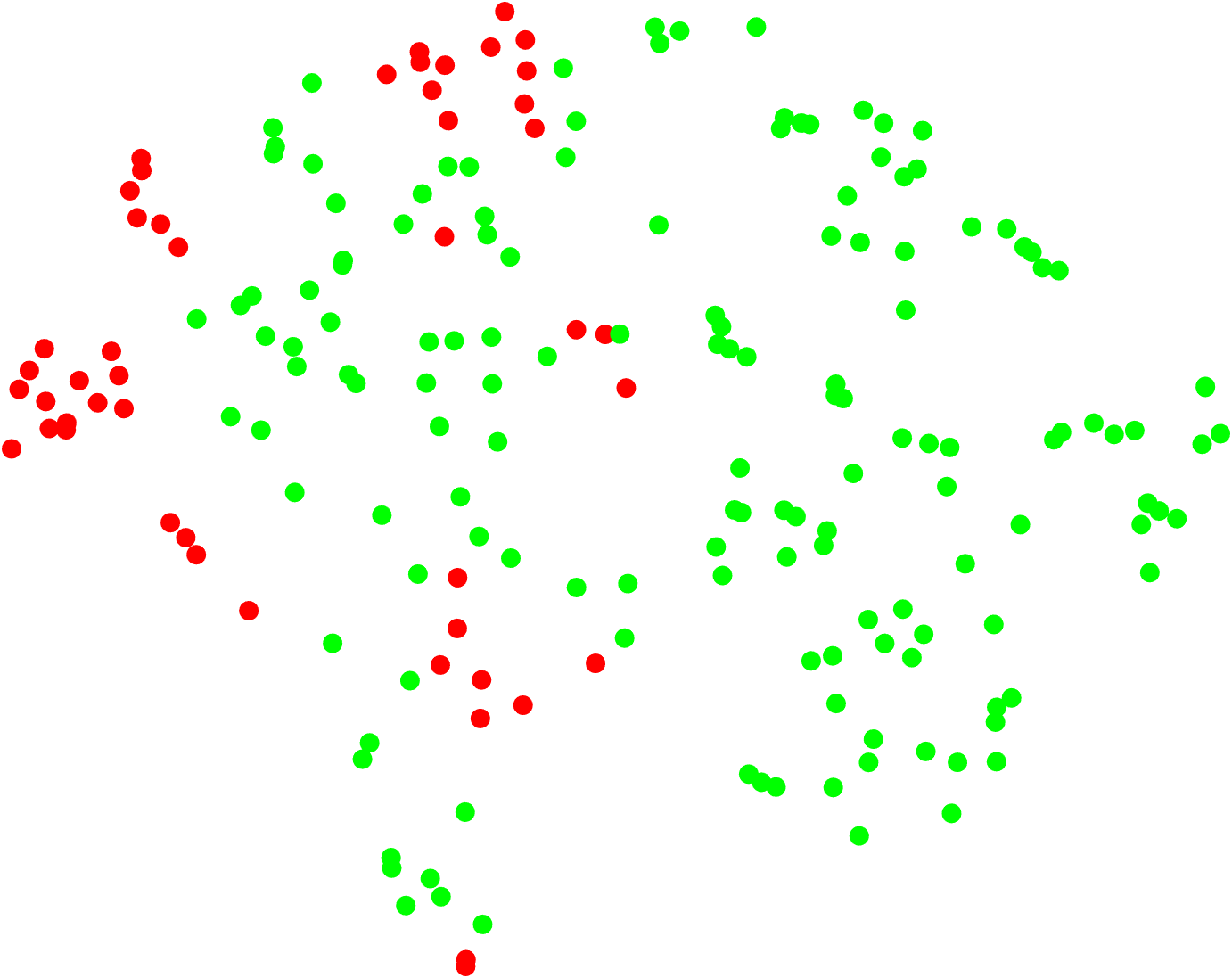}
\caption{}
\end{subfigure}
\hfill
\begin{subfigure}[H!]{0.48\textwidth}
\centering
\includegraphics[width=\textwidth]{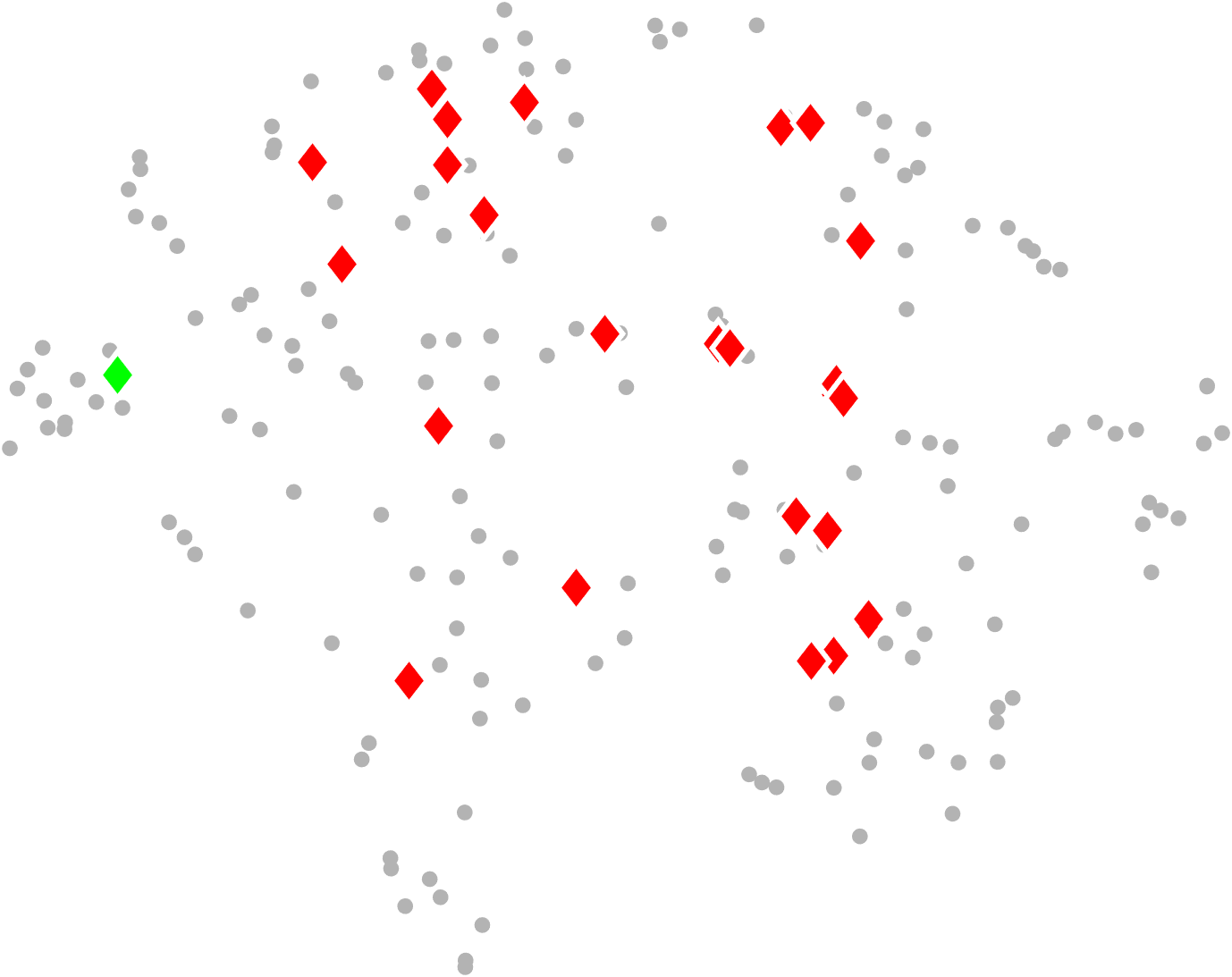}
\caption{}
\end{subfigure}

\begin{subfigure}[H!]{0.48\textwidth}
\centering
\includegraphics[width=\textwidth]{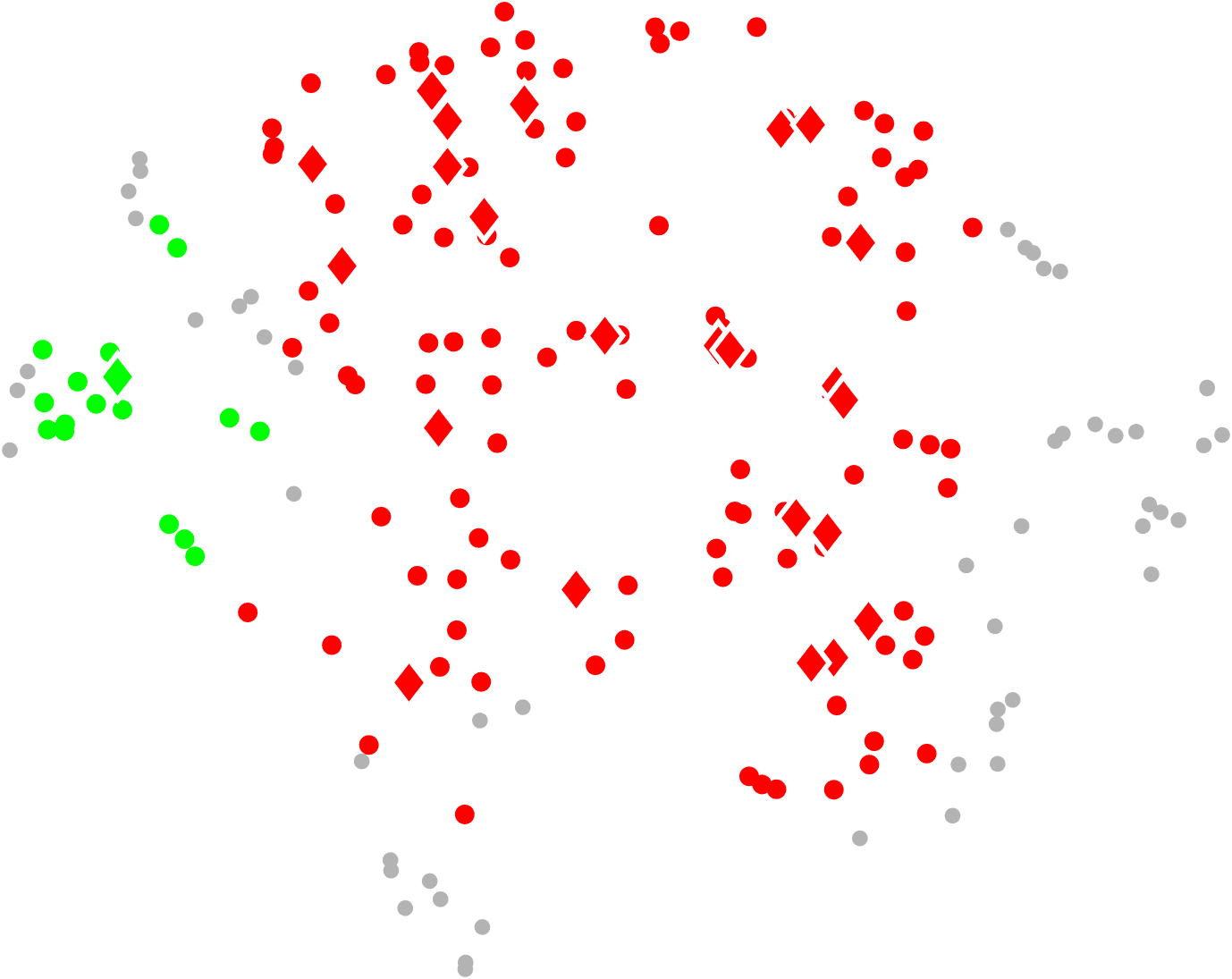}
\caption{}
\end{subfigure}
\hfill
\begin{subfigure}[H!]{0.48\textwidth}
\centering
\includegraphics[width=\textwidth]{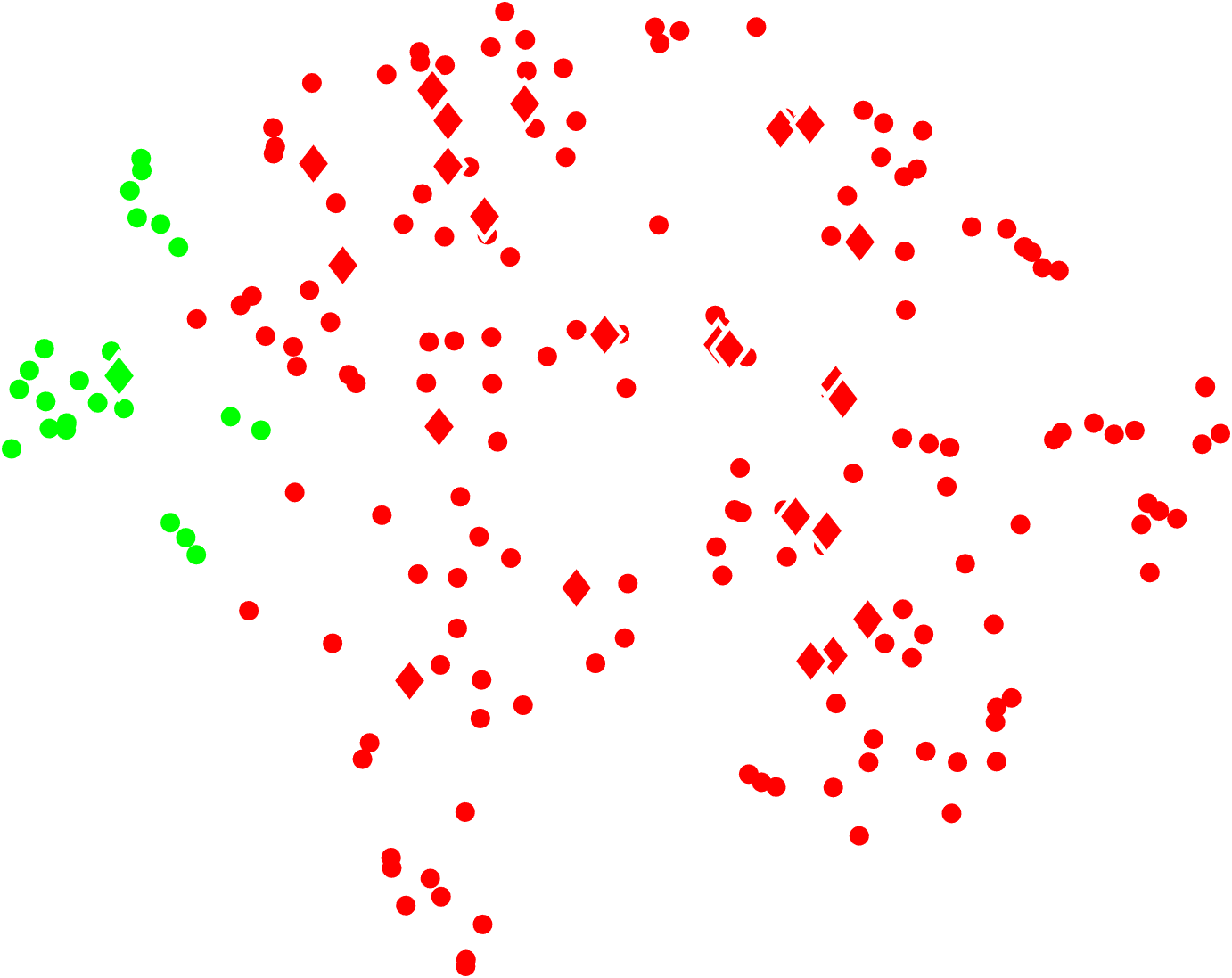}
\caption{}
\end{subfigure}
\caption{DLCC visualization on the \texttt{Parkinsons} dataset. (a) Ground truth labels; (b) Grouped local centers; (c) Temporary clusters; (d) Final DLCC clustering result.}
\label{fig:dlcc-pa}
\end{figure*}

% -------- starbeam --------
\begin{figure*}[ht]
\centering
\begin{subfigure}[H!]{0.48\textwidth}
\centering
\includegraphics[width=\textwidth]{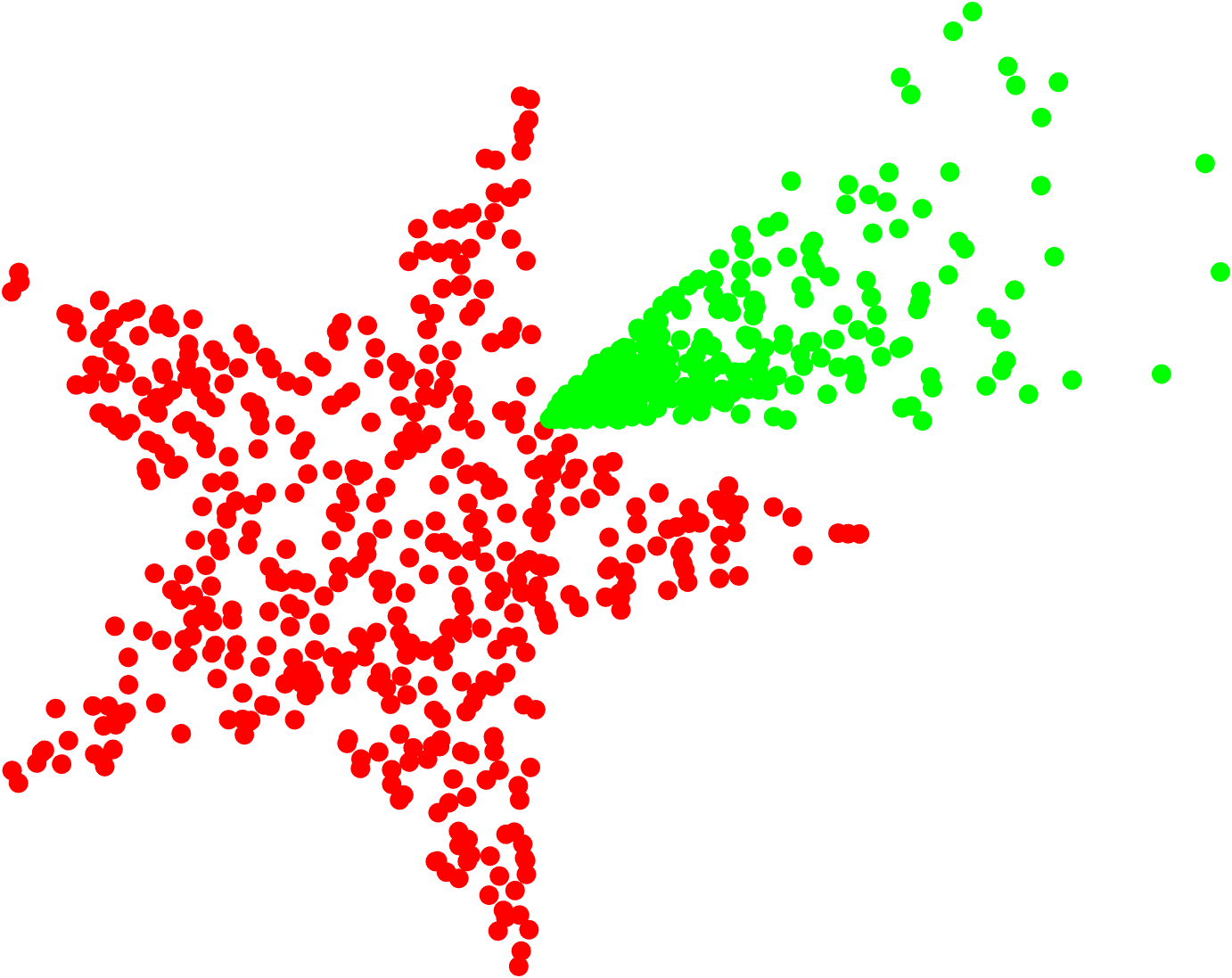}
\caption{}
\end{subfigure}
\hfill
\begin{subfigure}[H!]{0.48\textwidth}
\centering
\includegraphics[width=\textwidth]{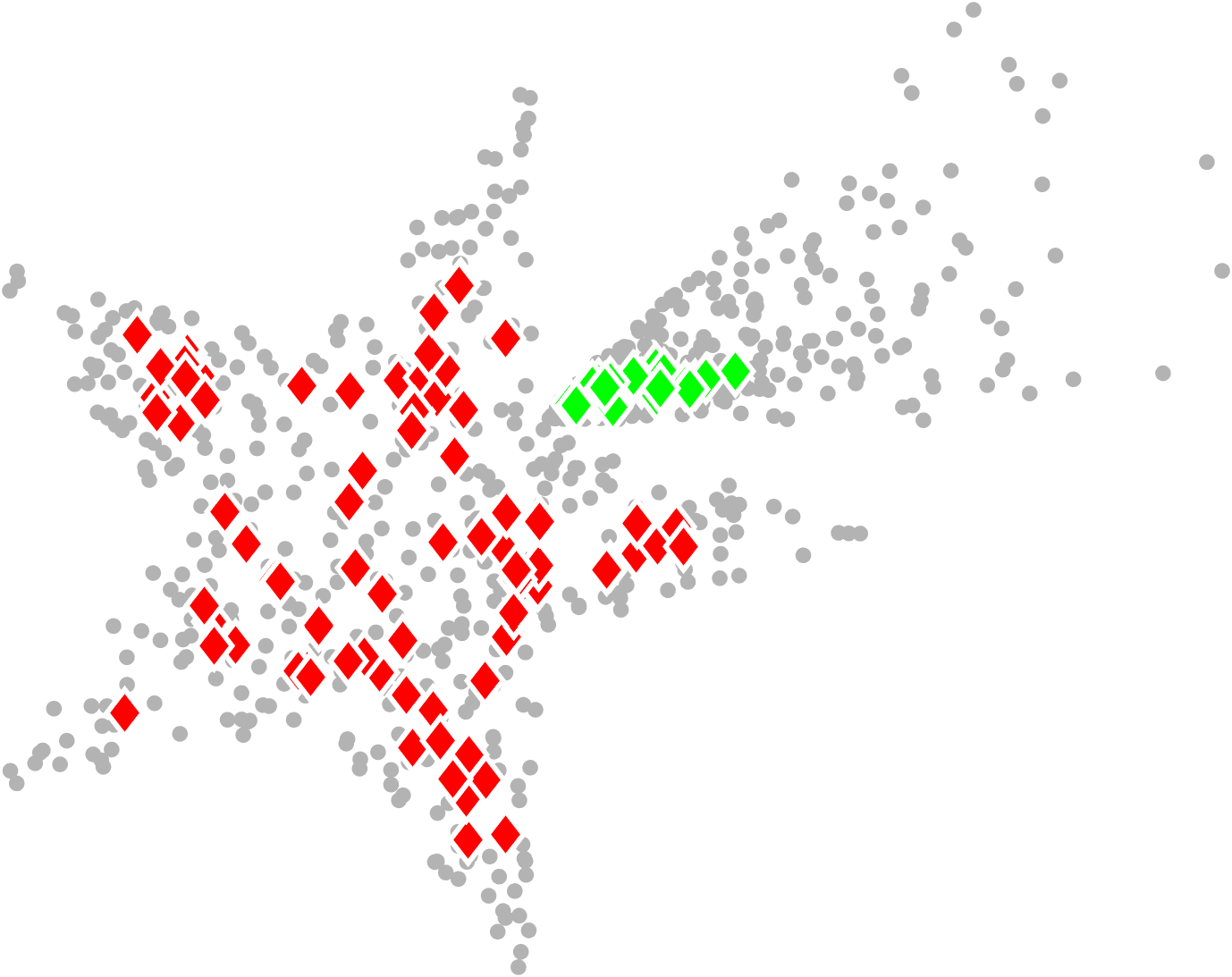}
\caption{}
\end{subfigure}

\begin{subfigure}[H!]{0.48\textwidth}
\centering
\includegraphics[width=\textwidth]{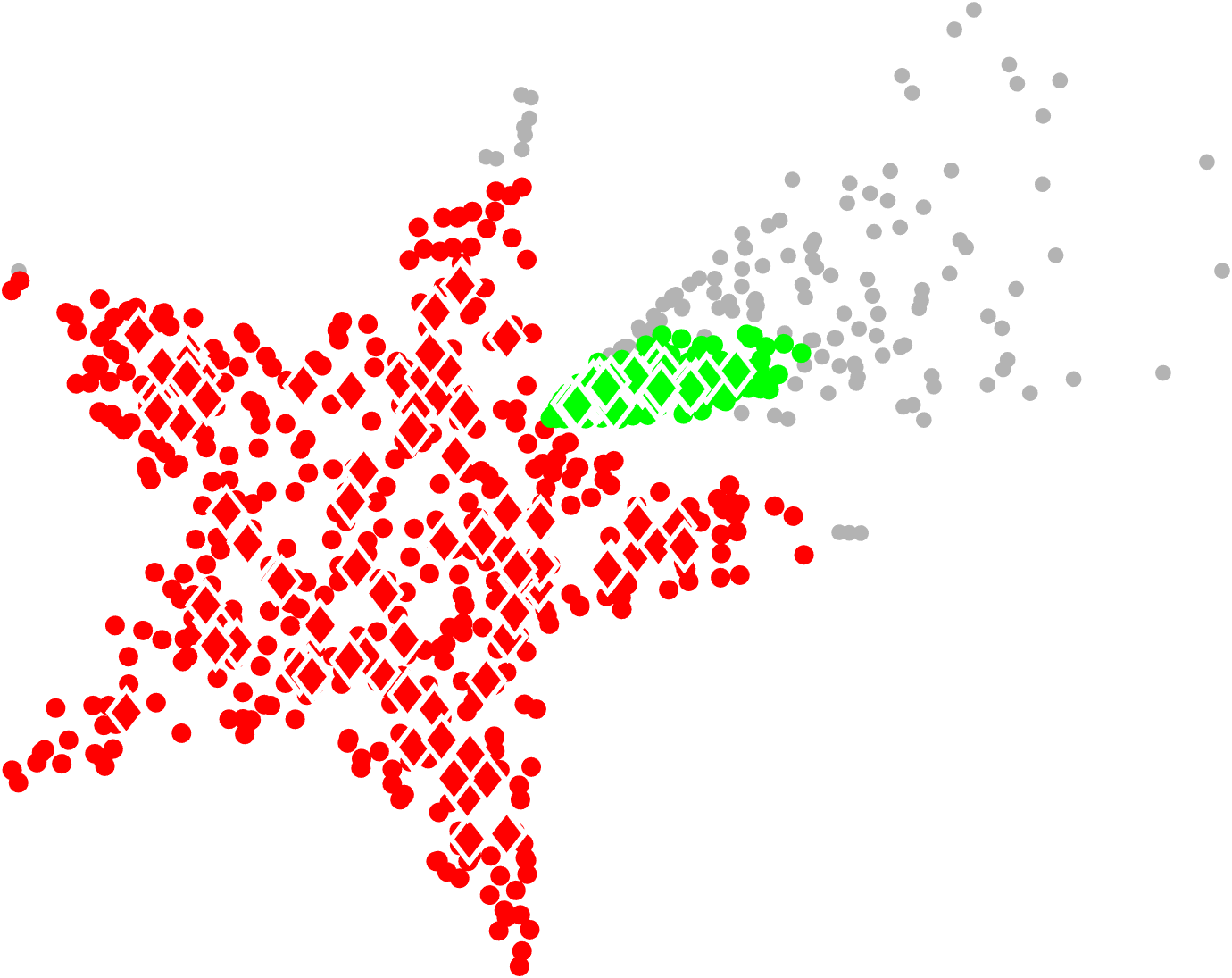}
\caption{}
\end{subfigure}
\hfill
\begin{subfigure}[H!]{0.48\textwidth}
\centering
\includegraphics[width=\textwidth]{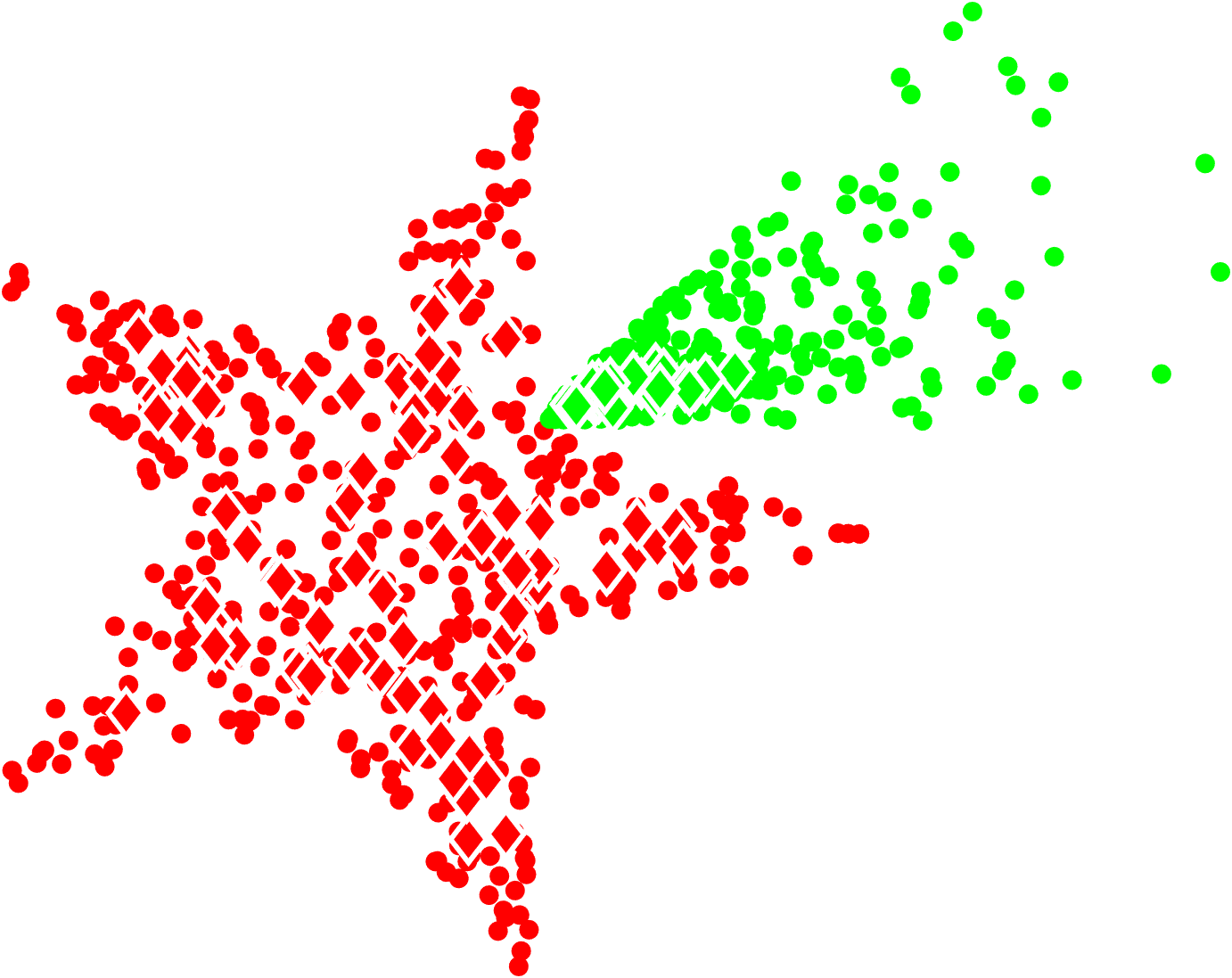}
\caption{}
\end{subfigure}
\caption{DLCC visualization on the \texttt{Starbeam} dataset. (a) Ground truth labels; (b) Grouped local centers; (c) Temporary clusters; (d) Final DLCC clustering result.}
\label{fig:dlcc-starbeam}
\end{figure*}

% -------- SuCancer --------
\begin{figure*}[ht]
\centering
\begin{subfigure}[H!]{0.48\textwidth}
\centering
\includegraphics[width=\textwidth]{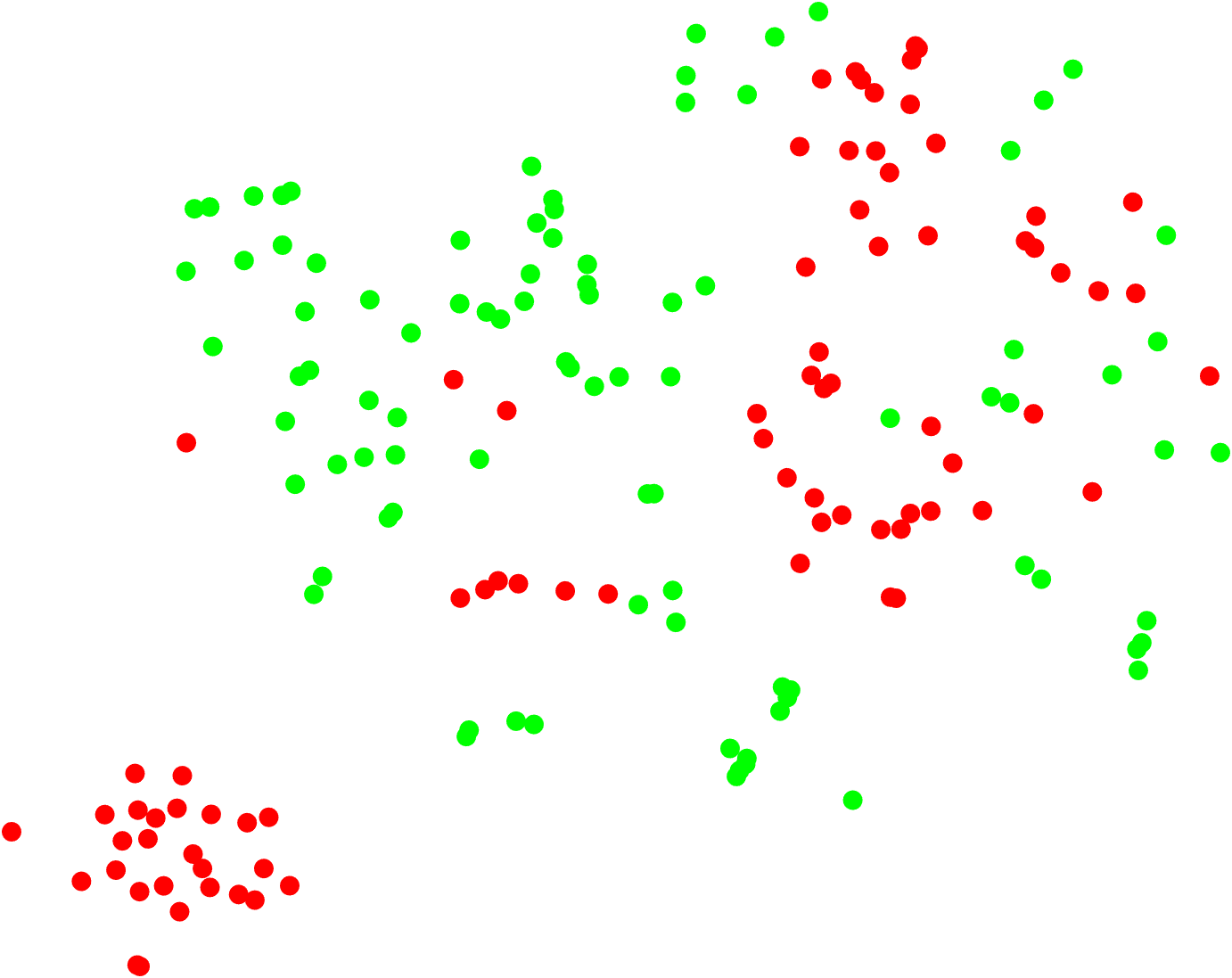}
\caption{}
\end{subfigure}
\hfill
\begin{subfigure}[H!]{0.48\textwidth}
\centering
\includegraphics[width=\textwidth]{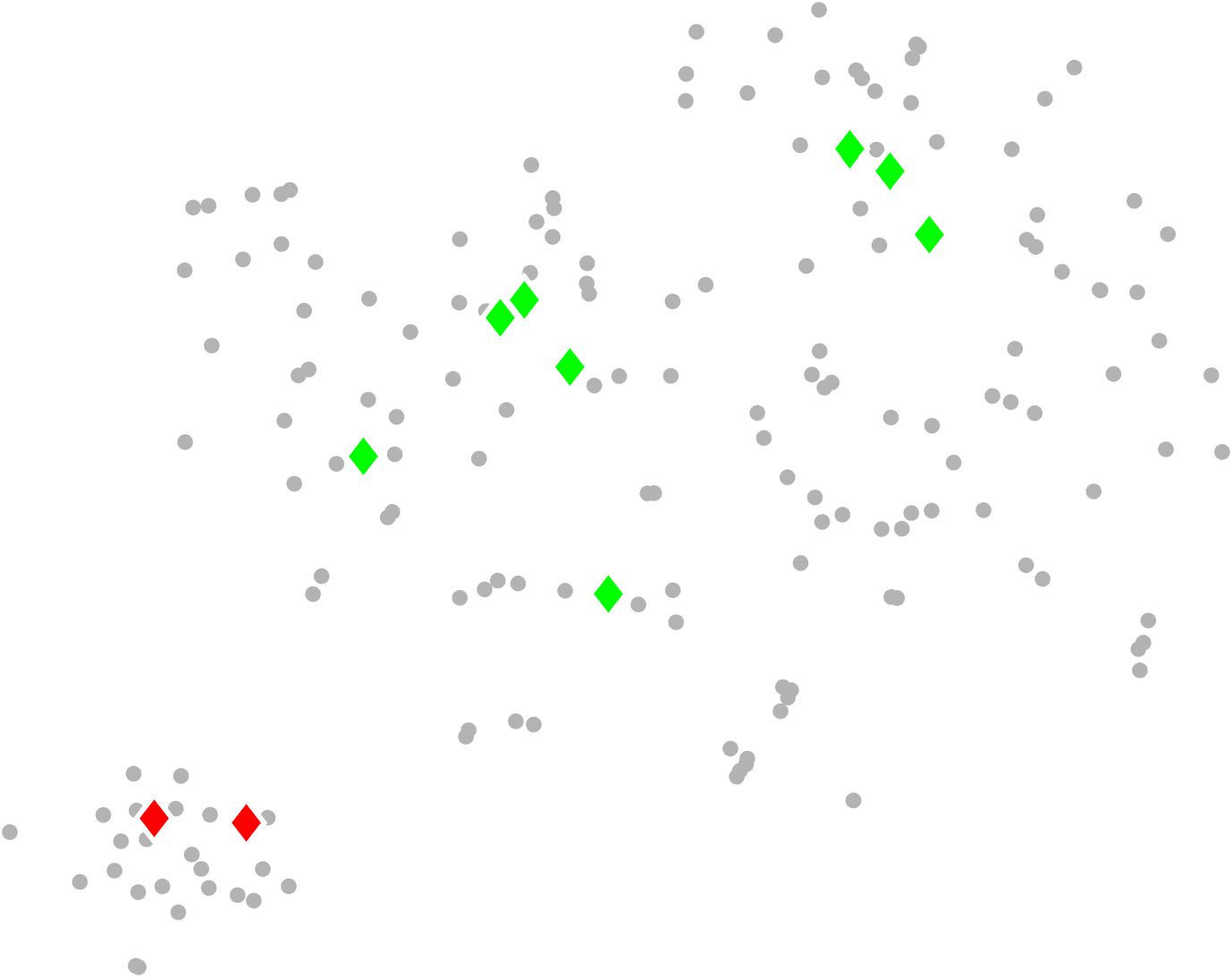}
\caption{}
\end{subfigure}

\begin{subfigure}[H!]{0.48\textwidth}
\centering
\includegraphics[width=\textwidth]{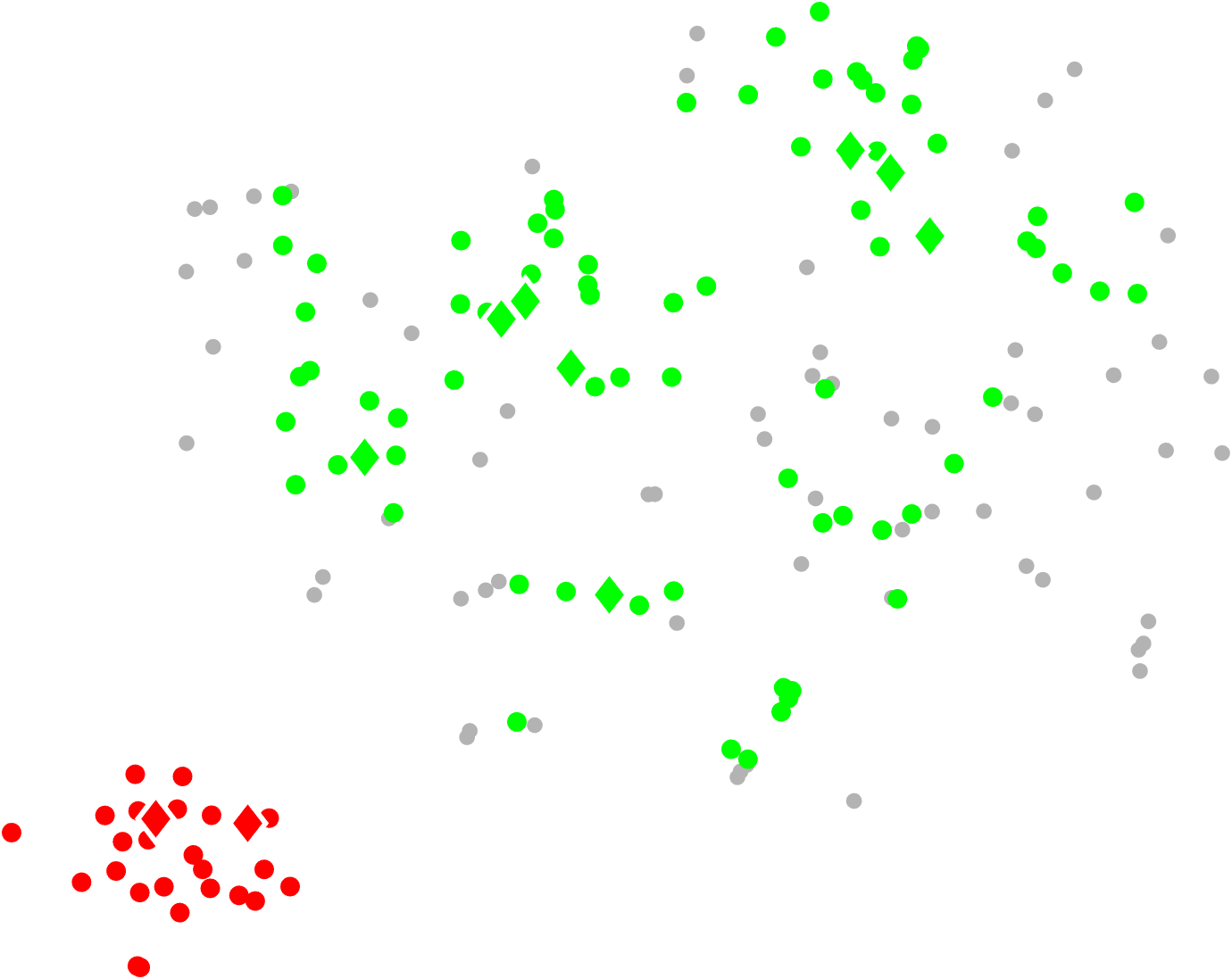}
\caption{}
\end{subfigure}
\hfill
\begin{subfigure}[H!]{0.48\textwidth}
\centering
\includegraphics[width=\textwidth]{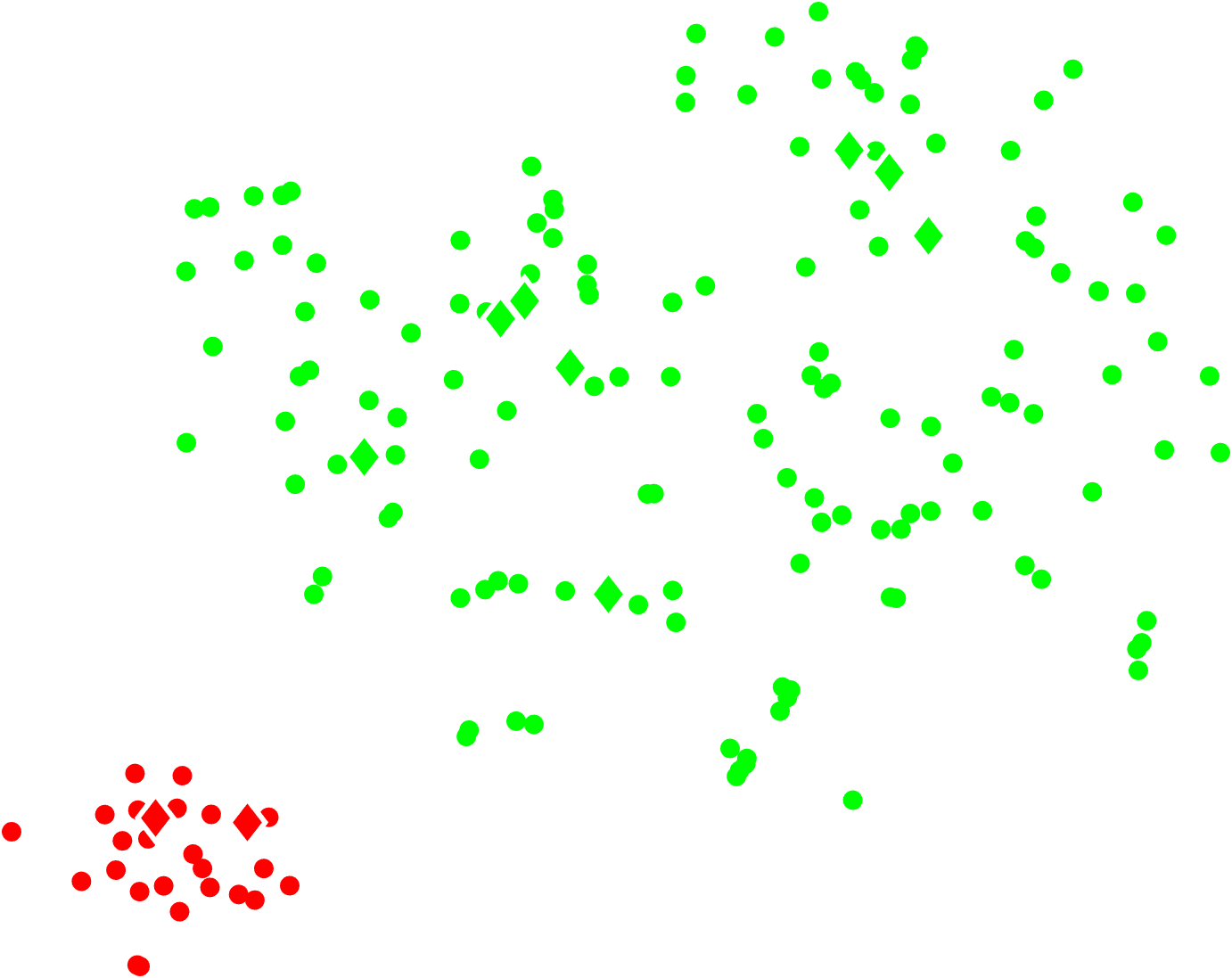}
\caption{}
\end{subfigure}
\caption{DLCC visualization on the \texttt{SuCancer} dataset. (a) Ground truth labels; (b) Grouped local centers; (c) Temporary clusters; (d) Final DLCC clustering result.}
\label{fig:dlcc-SuCancer}
\end{figure*}

\vfill

\end{document}